\documentclass[11pt,letterpaper]{article}

\newcommand{\fullversion}{1}

\usepackage[T1]{fontenc}
\usepackage[utf8]{inputenc}

\ifnum\fullversion=1
  \usepackage[lining]{libertine}
  \usepackage{microtype}
  \usepackage[letterpaper,margin=1in]{geometry}
  \usepackage{amsmath,amsthm,mathtools}
  \usepackage[libertine]{newtxmath}
\else
  \usepackage{lmodern,microtype}
  \usepackage[letterpaper,margin=1in]{geometry}
  \usepackage{amsmath,amssymb,amsthm,mathtools}
\fi

\usepackage{enumitem}
\usepackage{booktabs,array}
\usepackage{xcolor}
\usepackage{tikz}
\usepackage[labelsep=period]{caption}
\usepackage{flafter}
\usepackage[nospace]{varioref}
\usepackage[colorlinks=true,linkcolor=blue!45!black,citecolor=blue!45!black,urlcolor=blue!45!black,backref=page]{hyperref}
\renewcommand*{\backrefalt}[4]{%
  \ifcase #1 %
  \or (Cited on page~#2.)%
  \else (Cited on pages~#2.)%
  \fi}
\usepackage[capitalise, noabbrev]{cleveref}
\usepackage{etoc}   %

\hypersetup{pdftitle={Delegating Computation with Approximate Arithmetic},pdfauthor={Aritra Das, Debayan Gupta, Arup Mondal, Mihir More}}
\AddToHook{env/titlepage/before}{\hypersetup{pageanchor=false}}
\AddToHook{env/titlepage/after}{\hypersetup{pageanchor=true}}
\newtheorem{theorem}{Theorem}[section]
\newtheorem{lemma}[theorem]{Lemma}

\theoremstyle{definition}
\newtheorem{definition}[theorem]{Definition}
\theoremstyle{plain}

\newcommand{\C}{\mathbb C}
\newcommand{\Z}{\mathbb Z}
\newcommand{\bits}{\{0,1\}}
\newcommand{\ii}{\mathrm i}
\newcommand{\eps}{\varepsilon}
\newcommand{\poly}{\operatorname{poly}}
\newcommand{\polylog}{\operatorname{polylog}}

\newcommand{\negl}{\operatorname{negl}}
\newcommand{\Com}{\mathsf{Com}}
\newcommand{\PIRQuery}{\mathsf{Query}}
\newcommand{\PIRAnswer}{\mathsf{Answer}}
\newcommand{\PIRRecover}{\mathsf{Recover}}
\newcommand{\Root}{\mathsf{Root}}
\newcommand{\Round}{\mathsf{round}}   %
\newcommand{\Add}{\mathsf{add}}
\newcommand{\Mul}{\mathsf{mult}}
\newcommand{\eqpoly}{\operatorname{eq}}

\newcommand{\NC}{\mathsf{NC}}
\newcommand{\PSPACE}{\mathsf{PSPACE}}
\newcommand{\BV}{B_V}
\newcommand{\AddRecords}{\mathcal A}
\newcommand{\MulRecords}{\mathcal M}
\newcommand{\Fold}{\mathsf{Eval}}
\newcommand{\Rfold}{R_{\mathrm f}}
\newcommand{\WeightedValue}{\mathcal V}
\newcommand{\Nroot}{N_*}
\newcommand{\GuardBits}{g_{\mathrm p}}
\newcommand{\TableError}{\eta_{\mathrm{val}}}
\newcommand{\errfold}{\mathcal E}
\newcommand{\errvalue}{\mathcal F}
\newcommand{\errgate}{\mathcal G}

\DeclarePairedDelimiter{\normone}{\lVert}{\rVert_1}
\DeclarePairedDelimiter{\ceil}{\lceil}{\rceil}
\allowdisplaybreaks[2]
\numberwithin{equation}{section}

\newcommand{\BW}{B_W}
\newcommand{\GateExtension}{\widetilde V}
\newcommand{\GateValues}{V}
\newcommand{\LayerPoly}{F}
\newcommand{\ParentWeight}{\Psi}
\newcommand{\AddWiring}{\mathsf{add}}
\newcommand{\MulWiring}{\mathsf{mult}}
\newcommand{\AddPoly}{\widetilde{\AddWiring}}
\newcommand{\MulPoly}{\widetilde{\MulWiring}}
\newcommand{\EvaluationString}{\mathsf{Enc}}
\newcommand{\CoefficientBound}{C_*}
\newcommand{\ProverTime}{T_P}
\newcommand{\VerifierTime}{T_V}
\newcommand{\Communication}{\operatorname{Comm}}
\newcommand{\StringEq}{\operatorname{EQ}}

\newcommand{\Expect}{\mathbb E}
\newcommand{\Indicator}{\mathbf 1}

\renewenvironment{abstract}
  {\begin{center}\normalfont\normalsize\bfseries\abstractname\end{center}\quotation\normalsize}
  {\endquotation}

\newcommand{\PCPField}{\mathbb K}
\newcommand{\PCPCode}{\mathcal C}
\newcommand{\PCPDistance}{\operatorname{dist}}
\newcommand{\OracleBits}{b_{\mathrm{or}}}
\newcommand{\PCPCoeffBits}{b_{\mathrm{coef}}}
\newcommand{\DecodedBound}{R_{\mathrm{dec}}}
\newcommand{\ConjugatePoly}{f^{\mathrm c}}

\newcommand{\UnitCircle}{\mathbb T}
\newcommand{\WitnessPositions}{\mathcal I}
\newcommand{\BoolBound}{B_{\mathrm{bool}}}
\newcommand{\TolSym}{\Delta}
\newcommand{\Tol}{\TolSym_0}
\newcommand{\TolC}{\TolSym_{0,\mathrm{out}}}   %
\newcommand{\RunClaim}{Y}   %
\newcommand{\WitLen}{\nu}       %
\newcommand{\DecMap}{\varphi}   %
\newcommand{\TenDim}{t}         %
\newcommand{\LocErr}{\theta}    %
\newcommand{\EntryBound}{U_\pi}  %
\newcommand{\Lag}{\mathsf L}            %
\newcommand{\WitPrec}[1]{p^{\mathrm w}_{#1}}  %
\newcommand{\PCPReps}{r_{\mathrm{pcp}}} %
\newcommand{\Cond}{\operatorname{cond}} %
\newcommand{\Pclass}{\mathsf P}
\newcommand{\WitEntry}[1]{w^{\mathrm{in}}_{#1}}  %
\newcommand{\RoundDown}{\mathsf{floor}}  %

\definecolor{PCPBlue}{RGB}{39,83,109}
\definecolor{PCPAccent}{RGB}{150,66,47}
\tikzset{
  pcp vertex/.style={circle,draw=PCPBlue,fill=white,minimum size=8mm,inner sep=1pt}
}

\title{\bf Delegating Computation with Approximate Arithmetic: Interactive Proofs for Wizards }
\author{%
  Aritra Das$^{1}$\thanks{Email: \texttt{aritra@truthauditlabs.ai}} \quad 
  Debayan Gupta$^{1,2}$\thanks{Email: \texttt{debayan@truthauditlabs.ai}} \quad 
  Arup Mondal$^{1,2,3}$\thanks{Email: \texttt{arup24mondal@gmail.com}} \quad 
  Mihir More$^{1}$\thanks{Email: \texttt{mihir@truthauditlabs.ai}} \\[0.6em]
  \normalsize $^1$Truth Audit Labs \qquad $^2$Ashoka University \qquad $^3$University of Warsaw
}
\date{}

\begin{document}

\ifnum\fullversion=1 \pagenumbering{roman}\fi
\maketitle
\thispagestyle{empty}
\begin{abstract}
We study interactive proofs for arithmetic circuits over the complex numbers $\mathbb{C}$ in which both the prover and the verifier compute approximately (unlike muggles, wizards and witches can work with magic, i.e., floating point numbers). The goal is to prove that a claimed output $y$ is close to the exact output $C(x)$. For thresholds $0 < \Delta_0 < \Delta$, the honest prover should be accepted whenever $|y - C(x)| \le \Delta_0$, and every prover should be rejected with high probability whenever $|y - C(x)| > \Delta$.

Existing proof systems for real-valued data verify a rounded computation over a finite field, but say nothing directly about the exact output. Recently, Bitan et al.~[EUROCRYPT~2026] took a first step towards approximate verification with a sum-check protocol. 
We extend this to arithmetic circuits by generalizing the protocol of Goldwasser, Kalai, and Rothblum (GKR) [STOC~2008]. We give a public-coin interactive proof for any logspace-uniform layered arithmetic circuit over $\mathbb{C}$ with size $S$, depth $L$, and $n$ inputs, given public bounds on the gate values and on the coefficients of the wiring extensions. At constant soundness error, the verifier runs in time $(n+1)\cdot\mathrm{poly}(L, \log S, b)$, the communication is $\mathrm{poly}(L, \log S, b)$, and the prover runs in time $\mathrm{poly}(S, b)$, where $b$ bounds the bit lengths of all numbers involved. To best of our knowledge, these are the first doubly efficient interactive proofs in which both parties compute approximately. The main difficulty is that GKR reduces two claims about a layer to one claim on a random line, whose points generally leave the unit circle, where our error bounds hold. We instead label gates by complex roots of unity and combine the two claims using a random root of unity as the coefficient, so that every evaluation point stays on the unit circle. We also choose the error thresholds backward from the input layer, so that a large output error survives the verifier's rounding at every step. The honest prover computes its messages with $O(S \log S)$ complex operations and only logarithmically many extra bits of precision.
We also obtain: zero-knowledge proofs, based on one-way functions (OWF), with communication nearly linear in the witness length, one-round arguments with polylogarithmic communication, based on private information retrieval (PIR), and interactive probabilistically checkable proofs (PCPs) whose proof length is polynomial in the witness length rather than in the circuit size.

\end{abstract}

\ifnum\fullversion=1
  \newpage
  \tableofcontents
  \newpage
  \pagenumbering{arabic}
\fi

\setcounter{page}{1}

\etocdepthtag.toc{mainmatter}
\section{Introduction}\label{sec:introduction}

Interactive proofs~\cite{GolMicRac89,STOC:Babai85} allow a computationally weak verifier to check a computation performed by an untrusted prover. The protocol of Goldwasser, Kalai, and Rothblum (GKR)~\cite{GKR} makes this efficient for both parties: for a logspace-uniform arithmetic circuit over a finite field of size $S$ and depth $L$ on $n$ inputs, the prover runs in time $\poly(S)$ and the verifier in time $n\cdot\poly(L,\log S)$. Many computations of interest, however, are arithmetic circuits over $\C$ that every party evaluates only approximately. For such a circuit $C$, the right statement to prove is that a claimed output $y$ is close to the exact output $C(x)$, and the right proof system is one in which neither the prover nor the verifier computes exactly.

Unfortunately, such a proof system does not follow by running an exact protocol with finite precision, since rounding error enters on both sides: the honest prover computes only approximately, and so does the verifier. Each check must therefore allow a small error, and a malicious prover can exploit this allowance at every step. Soundness requires that these deviations never accumulate into acceptance of a substantially inaccurate output.

Formally, for thresholds $0<{\Tol}<\Delta$, we consider the promise problem whose YES instances are the triples $(C,x,y)$ with $|y-C(x)|\le{\Tol}$ and whose NO instances are those with $|y-C(x)|>\Delta$. The honest prover must be accepted on every YES instance, and every prover, however powerful, must be rejected with high probability on every NO instance. The guarantee refers to the exact output $C(x)$, and not to the output of a rounding procedure fixed in advance. This is the main difference from existing proof systems for computations on real-valued data, motivated in particular by verifiable machine learning~\cite{ITCS:GRSY21,CCS:LiuXieZha21,CCS:GGJMMP23,EuroSys:CWSK24,CCS:APKP24,USENIX:HCLWZY24,zkLLM24,zkGPT25,cryptoeprint:2026/1112}: they encode scaled and rounded values into finite-field arithmetic, or support floating-point operations and relative-error specifications~\cite{Mystique21,FloatZK22,ZIP25}, and in all of them the proof system itself computes exactly. 

Sum-check~\cite{LFKN} and GKR reduce a claim about the output, layer by layer, to a claim about the input. Over a finite field, soundness rests on counting zeros: a nonzero polynomial of low degree rarely vanishes at a random point. Approximate checks need more, since a nonzero polynomial can be small at a random point, and an adaptive prover can shrink its error a little in every round without ever satisfying an exact identity. Bitan et al.~\cite{ApproxSumcheck26} control this for sum-check using analytic bounds, and leave circuit verification open~\cite[Section~5]{ApproxSumcheck26}. A circuit adds three difficulties: (i) the verifier multiplies claimed values in every gate check, (ii) it must reduce two claims about the next layer to one, and (iii) the errors must not compound across the $L$ layers. The second difficulty is the most delicate, because GKR reduces the two claims to a single claim at a random point on the line through the two evaluation points. The coordinates of this point are in general not roots of unity, whereas our analysis requires them to be so. 

These points motivate us to ask the following question:
\begin{quote}
    \centering
    \emph{Can we obtain doubly efficient interactive proofs for arithmetic circuits \\ when both parties use approximate arithmetic?}
\end{quote}

We answer this question in the affirmative. For every $\Delta>0$ and $0<\eps<1$, we give a public-coin interactive proof for layered arithmetic circuits over $\C$, with perfect completeness for claims within some ${\Tol}<\Delta$ of $C(x)$ and soundness error $\eps$ against computationally unbounded provers for claims more than $\Delta$ away, in which both parties compute with finite precision $b$. For logspace-uniform circuits at constant soundness error, the communication is $\poly(L,\log S,b)$ bits, the verifier uses $(n+1)\poly(L,\log S,b)$ bit operations, and the honest prover uses $\poly(S,b)$. The parameters ${\Tol}$ and $b$ are determined by $\Delta$, $\eps$, the circuit parameters, and bounds on the gate values and on the coefficients of the polynomial extensions of the wiring predicates. To our knowledge, these are the first doubly efficient interactive proofs  in which both parties compute approximately.

Two ideas make this possible. The first is a choice of encoding. We label gates by points whose coordinates are complex roots of unity (as opposed to points of the Boolean hypercube) and we combine the two claims about the next layer by a random linear combination with a root of unity as coefficient, which leaves both evaluation points unchanged. As such, every label, challenge, and evaluation point lies on the unit circle.  To the best of our knowledge, gate labels on complex roots of unity have not been used in GKR-style protocols before. The second idea is to choose the error thresholds backward, starting from the final input check, so that an output error larger than $\Delta$ forces, except with small probability, a claim error above the threshold at every later step, including the verifier's own rounding.

We also give a near-linear algorithm for the prover's messages that needs only logarithmically many additional bits of precision, show that the coefficient norms of the polynomial extensions of the wiring predicates are necessary to control how gate-value errors affect these messages, and extend the construction to zero-knowledge proofs, one-round arguments, and interactive probabilistically checkable proofs (PCPs). We summarize our results in \cref{sec:results} and give a technical overview in \cref{sec:technical-overview}; full proofs appear in the appendices.

\subsection{Our Results}\label{sec:results}

We study interactive proofs for arithmetic circuits over $\C$ in which both the prover's messages and the verifier's computations are approximate. Our main results are as follows:
\begin{itemize}[leftmargin=*]
 \item {\bf Approximate GKR (\cref{sec:overview-agkr} and \cref{sec:agkr}).} We design an approximate version of the GKR protocol, a public-coin interactive proof for layered circuits $C$ over $\C$ of size $S$, depth $L$, and $n$ input gates. For every $\Delta>0$ and $0<\eps<1$, we choose $0<{\Tol}<\Delta$ such that the honest prover is accepted with probability one when $|y-C(x)|\le{\Tol}$, whereas every prover is accepted with probability at most $\eps$ when $|y-C(x)|>\Delta$. As in the bare-bones protocol of GKR, the verifier is given the values of the polynomial extensions of the wiring predicates at one point per layer. With these values, it runs in time $O((n+L)\log S)\cdot\poly(b+\ell)$, and the communication is $O(L\log S\cdot(b+\ell))$ bits, where $b$ bounds the input, message, and working bit lengths and $\ell$ is the bit length of a challenge. For the unique low-degree extensions of the wiring predicates, the coefficient norm is at most $S$. If, in addition, the interpolation set is fixed and $\Delta\le1$, then $\log(\Delta/{\Tol})=O(L\log S\cdot\log(L\log S/\eps)+L\log\BV)$, where $\BV$ bounds the moduli of the gate values, so our sufficient precision bound is linear in the depth, up to a logarithmic factor (\cref{thm:informal-agkr,thm:agkr} and \cref{eq:depth-dependence}).
 \item {\bf Near-linear prover (\cref{sec:overview-agkr} and \cref{sec:agkr}).} For a fixed interpolation-set size $h$ and the unique wiring
extensions of degree below $h$ in each variable, we give an honest
prover that computes all its messages from sufficiently accurate
supplied gate values using $O(S\log S)$ complex arithmetic operations
and approximations of roots of unity. Its working precision exceeds
its message precision by a number of bits logarithmic in the circuit
size and the gate-value bound. This operation count excludes the
cost of computing the supplied gate values. For arbitrary extensions, gate-value errors of size at most $\TableError\le1$ change every coefficient of the prover's round polynomials by at most $\TableError$ times a polynomial in the circuit size, the gate-value bound, and the coefficient norm of the polynomial extensions of the wiring predicates. The dependence on the coefficient norm is necessary, even for circuits with three gates (\cref{thm:informal-stable-thm,thm:informal-conditioning-thm,stable:thm,conditioning:thm}).
 \item {\bf Logspace-uniform circuits (\cref{sec:overview-implementation} and \cref{sec:implementation}).} Using our interactive proofs for space-bounded computation, we remove the assumption that the verifier is given the values of the wiring extensions when the circuit is logspace-uniform: the prover sends approximate values of these extensions and proves that their bits are correct. At constant soundness error, the communication is $\poly(L,\log S,b)$ bits and the verifier runs in time $(n+1)\poly(L,\log S,b)$. In particular, for $S=\poly(n)$ and $L,b=\polylog(n+2)$, the communication is polylogarithmic and the verifier runs in time $(n+1)\polylog(n+2)$. For circuits fixed before the input is known, the verifier can instead compute these values in a private preprocessing phase and store only $O((k+L)\ell+Lb)$ bits per execution, where $k$ is the number of sum-check rounds (\cref{thm:informal-uniform,thm:informal-space-proof,thm:uniform,thm:preprocessing}).
 \item {\bf Zero-knowledge proofs (\cref{sec:overview-zk} and \cref{sec:zk}).} Assuming one-way functions, we make approximate GKR computationally zero-knowledge for inputs that include a private witness $w$, with the promise that either some $w$ satisfies $|y-C(x,w)|\le{\Tol}$ or every $w$ satisfies $|y-C(x,w)|>\Delta$. The resulting public-coin proofs remain sound against computationally unbounded provers. Since the final zero-knowledge proof checks only the acceptance predicate of the finite-precision verifier, and not the circuit, the communication is $(\WitLen+1)\poly(\kappa,L,h,\log S,b)$ for witness length $\WitLen$, security parameter $\kappa$, and interpolation-set size $h$, provided that $w$ is checked and encoded by a Boolean circuit of size $(\WitLen+1)\poly(b)$ and determines $O(\WitLen+1)$ input entries (\cref{thm:informal-zk,thm:zk-direct,thm:zk-uniform}).
 \item {\bf One-round arguments (\cref{sec:overview-one-round} and \cref{sec:one-round}).} Under a private information retrieval (PIR) assumption, we compile approximate GKR into two-message private-coin arguments with the same thresholds, without using Fiat--Shamir. The arguments are sound against nonuniform provers of size at most $2^\kappa$, and the communication is $\poly(\kappa,L,\log S,b)$ (\cref{thm:informal-one-round,thm:one-round-agkr}).
 \item {\bf Interactive PCPs (\cref{sec:overview-ipcp} and \cref{sec:ipcp}).} Finally, for circuits whose inputs include a public affine image of a bit witness $w\in\bits^{\WitLen}$, we give public-coin interactive PCPs with perfect completeness in which the proof string encodes the witness rather than the computation. The proof string has length polynomial in $\WitLen$, $L$, and $\log S$, and at soundness error $\eps$ the verifier reads $\poly(L,\log S,\log(\WitLen+1))\cdot\log(2/\eps)$ of its bits (\cref{thm:informal-ipcp,thm:pcp-native}).
\end{itemize}

\subsection{Related Work}

\medskip\noindent{\bf Delegating circuit computation.}
The sum-check protocol of Lund, Fortnow, Karloff, and Nisan~\cite{LFKN} and the GKR protocol~\cite{GKR} underpin our layer-by-layer verification. Besides doubly efficient interactive proofs for low-depth circuits, GKR develops implementations for logspace-uniform circuits, private preprocessing, zero-knowledge proofs, and one-round arguments. We follow this structure, with a numerical analysis at every reduction and implementation step. Cormode, Mitzenmacher, and Thaler~\cite{CMT} and Thaler~\cite{Thaler} improve the GKR prover for exact arithmetic, in practice and asymptotically. Our near-linear algorithm for the prover's messages adds a bound on the precision that the prover needs, given gate values of a specified accuracy. Soria-Vazquez~\cite{SoriaVazquez22} gives doubly efficient interactive proofs over infinite and noncommutative rings; these handle characteristic-zero domains but require exact computation.

\medskip\noindent{\bf Approximate sum-check.}
Bitan et al.~\cite{ApproxSumcheck26} generalize sum-check to metric integral domains and instantiate it over the reals and the complex numbers using polynomial sublevel-set bounds. They analyze adaptive error contraction, prove round-by-round soundness guarantees relevant to Fiat--Shamir, and construct an approximate fully linear interactive oracle proof for inner products. They identify the extension to GKR-style verification as requiring additional analysis and leave this as an future work~\cite[Section~5]{ApproxSumcheck26}.

\section{Technical Overview}\label{sec:technical-overview}

This section gives a detailed technical overview of our protocols and the main ideas in their analyses. After fixing notation in \cref{sec:overview-notation}, we describe approximate GKR with oracle access to the polynomial extensions of the wiring predicates (\cref{sec:overview-agkr}, with details in \cref{sec:agkr}) and then remove this oracle (\cref{sec:overview-implementation}, with details in \cref{sec:implementation}). The remaining subsections use these protocols to obtain zero-knowledge proofs (\cref{sec:overview-zk}, with details in \cref{sec:zk}), one-round arguments (\cref{sec:overview-one-round}, with details in \cref{sec:one-round}), and interactive PCPs (\cref{sec:overview-ipcp}, with details in \cref{sec:ipcp}). 

\subsection{Notations\texorpdfstring{ (\cref{sec:preliminaries})}{}}\label{sec:overview-notation}
We write $[t]=\{1,\dots,t\}$. For a finite set $\mathcal S$, $|\mathcal S|$ is its size, and $z\leftarrow\mathcal S$ means that $z$ is sampled uniformly at random from $\mathcal S$; for a string $s$, $|s|$ is its length. For a vector $z$, $z_j$ is its $j$th coordinate; in $g_i,u_i,v_i$ and $\rho_{i,b}$ the subscript $i$ instead indexes a layer. Logarithms use base two, and $\log^+t=\max\{0,\log t\}$. We write $\poly(\cdot)$ for a polynomial in the indicated quantities.

We write $\ii$ for the imaginary unit, and $\overline z$ and $|z|$ for the complex conjugate and the modulus of $z\in\C$. Let $\UnitCircle=\{z\in\C:|z|=1\}$, let $\Omega_N=\{\exp(2\pi\ii a/N):0\le a<N\}$ be the set of $N$th roots of unity, and define $\Omega_h$ in the same way. For a univariate polynomial $f$, $[X^a]f$ is the coefficient of $X^a$ in $f$. For any polynomial $f$ over $\C$, $\|f\|_1$ is the sum of the moduli of its coefficients, and $f$ has individual degree at most $d$ if it has degree at most $d$ in each variable. Numbers exchanged or stored by the parties have the form $(A+\ii B)2^{-p}$ with $A,B\in\Z$. The maps $\Round_p$ and $\RoundDown_p$ round the real and imaginary parts of a complex number to the nearest integer multiple of $2^{-p}$ and down to such a multiple, respectively. Precisions are denoted $p$ and $q$; round polynomials always carry a subscript, as in $p_j$ and $q_j$, and generic polynomials are denoted $f$. Running times count bit operations unless stated otherwise.

The computational security parameter is $\kappa$. A function of $\kappa$ is negligible, written $\negl(\kappa)$, if it is $\kappa^{-\omega(1)}$. An algorithm is nonuniform polynomial time if, for each $\kappa$, it is a circuit of size $\poly(\kappa)$; equivalently, it runs in polynomial time with advice of length $\poly(\kappa)$ that depends only on $\kappa$. Additional notation and preliminaries are given in \cref{sec:preliminaries}.

\subsection{Approximate GKR\texorpdfstring{ (\cref{sec:agkr})}{}}\label{sec:overview-agkr}

We develop the approximate GKR protocol for arithmetic circuits over $\C$: a public-coin interactive proof that a claimed value $y$ is close to the exact output $C(x)$ of a circuit $C$ on input $x$. Since both the prover's messages and the verifier's computations are approximate, each check of the GKR verifier~\cite{GKR} is relaxed to accept when the two sides of a polynomial identity agree up to a prescribed error threshold. By choosing these thresholds backward from the input layer, for every $\Delta>0$ and $0<\eps<1$ we obtain $0<{\Tol}<\Delta$ such that the honest prover is accepted with probability one when $|y-C(x)|\le{\Tol}$, whereas every prover is accepted with probability at most $\eps$ when $|y-C(x)|>\Delta$. The honest prover needs only approximate gate values: for the unique low-degree extensions of the wiring predicates, it computes its messages with $O(S\log S)$ complex operations, at a working precision that exceeds the precision of its messages by a number of bits logarithmic in the circuit size and the gate-value bound. For arbitrary extensions, errors in the gate values still have a bounded effect: if each gate value has error at most $\TableError\le1$, then each coefficient of the prover's messages changes by at most $\TableError$ times a polynomial in the circuit size, the gate-value bound, and the coefficient norm of the polynomial extensions of the wiring predicates.

The circuit $C$ is a layered arithmetic circuit with depth $L$ and $S$ gates after padding. Layer $0$ contains the output, layer $L$ contains the $n$ input gates (including constant gates), and every other gate adds or multiplies two gates of the next layer, called its children. As in GKR ~\cite{GKR}, the gate values of each layer $i$ are encoded by a low-degree polynomial $\GateExtension_i$ that agrees with them on the gate labels. For a gate label $g$ of layer $i$ and gate labels $u,v$ of layer $i+1$, let $\Add_i(g,u,v)=1$ if $g$ is an addition gate with ordered children $u,v$, and $\Add_i(g,u,v)=0$ otherwise; $\Mul_i$ is defined in the same way for multiplication gates. The circuit's connections are encoded by low-degree extensions $\AddPoly_i$ and $\MulPoly_i$ of these predicates. As in the bare-bones protocol of GKR, the verifier is given the values of $\AddPoly_i$ and $\MulPoly_i$ at the points it queries; \cref{sec:overview-implementation} shows how to remove this assumption. We assume that every gate value has modulus at most $\BV$ and that the absolute values of the coefficients of $\AddPoly_i$ and $\MulPoly_i$ sum to at most $\BW$. For the unique extensions of individual degree below $h$, the bound $\BW=S$ suffices: each Lagrange basis polynomial on $\Omega_h$ has coefficient norm one, and $\AddPoly_i$ is a sum of at most $S$ products of such polynomials, one for each gate (\cref{sec:complexity}). In the statement below, $\GateExtension_i$, $\AddPoly_i$, and $\MulPoly_i$ also have constant degree in each variable. \Cref{thm:informal-agkr} informally states the approximate GKR protocol; we refer to \cref{sec:agkr} for the formal protocol and its analysis.

\begin{theorem}[Approximate GKR, Informal; \cref{thm:agkr}]\label{thm:informal-agkr}
Let $C$ be a layered arithmetic circuit over $\C$ with depth $L$, $S$ gates after padding, and $n$ input gates. Let $x$ be an input and $y$ a claimed value for the exact output $C(x)$. For every $\Delta>0$ and $0<\eps<1$, there is $0<{\Tol}<\Delta$, depending on $\Delta$, $\eps$, $L$, $S$, $\BV$, and $\BW$, such that the statement ``$y\approx C(x)$'' has a public-coin interactive proof in which the honest prover is accepted with probability $1$ if $|y-C(x)|\le {\Tol}$, and every prover is accepted with probability at most $\eps$ if $|y-C(x)|>\Delta$. The honest prover runs in time $\poly(S,b+\ell)$, the verifier runs in time $O((n+L)\log S)\cdot\poly(b+\ell)$, and the communication is $O(L\log S\cdot(b+\ell))$ bits, where $b$ bounds the bit length of the numbers used and $\ell=O(\log(L\log S/\eps))$ is the bit length of a challenge. These bounds do not include the cost of evaluating $\AddPoly_i$ and $\MulPoly_i$, which the verifier does at one point per layer.
\end{theorem}

\medskip\noindent{\bf The protocol.}
The protocol reduces a claim about the values of layer $i$ to a claim about layer $i+1$.  We fix a power of two $h \ge 2$, and let $H$ be the set of $h$th roots of unity.  For a power of two $N$ chosen in the soundness analysis, the verifier draws all its challenges uniformly from $\Omega_N$. The gates of layer $i$ are labeled by elements of $H^{m_i}$ (where $m_i$ is the number of coordinates of a gate label in layer $i$), and their values are encoded by the unique polynomial $\GateExtension_i$ of degree below $h$ in each variable that agrees with them on the labels. Verification proceeds from the output layer to the input layer. At layer $i$, the verifier holds a claimed value $\widehat{T}_i$ for \begin{equation*}
    T_i = c_{i,0}\,\GateExtension_i(\rho_{i,0}) + c_{i,1}\,\GateExtension_i(\rho_{i,1}),
\end{equation*}
where the points $\rho_{i,b}$ and coefficients $c_{i,b}$ are known to the verifier; the first claimed value is $\widehat T_0=y$, for $T_0=C(x)$. By substituting the addition or multiplication defining each gate, $T_i$ is a sum, over a parent label $g$ and child labels $u,v$, of a low-degree polynomial $\LayerPoly_i(g,u,v)$ built from $\AddPoly_i$, $\MulPoly_i$, and $\GateExtension_{i+1}$. The sum-check protocol reduces this sum to the value of that polynomial at a random point $(g_i,u_i,v_i)$, whose coordinates are the sum-check challenges, one per round. The verifier is given $\AddPoly_i$ and $\MulPoly_i$ at this point, so only the values $\GateExtension_{i+1}(u_i)$ and $\GateExtension_{i+1}(v_i)$ remain unknown. \cref{fig:agkr-layer-reduction} shows one layer of the protocol. 

Let $Z_i$ be the claimed value of this polynomial at $(g_i,u_i,v_i)$ at the end of sum-check. At this point the polynomial equals $w_i^+\bigl(\GateExtension_{i+1}(u_i)+\GateExtension_{i+1}(v_i)\bigr)+w_i^\times\GateExtension_{i+1}(u_i)\GateExtension_{i+1}(v_i)$, where the coefficients $w_i^+$ and $w_i^\times$ depend only on $\AddPoly_i$, $\MulPoly_i$, and the claim, and are known to the verifier. The prover sends claimed values $a_i, b_i$ for $\GateExtension_{i+1}(u_i)$ and
$\GateExtension_{i+1}(v_i)$, and the verifier checks that the residual $Z_i-w_i^+(a_i+b_i)-w_i^\times a_ib_i$ has modulus at most a multiple of ${\Tol}$ that depends only on $\BV$, $\BW$, and $S$.
The verifier then samples $\lambda_i\in\Omega_N$ uniformly at
random and computes a new claimed value $\widehat T_{i+1}$ satisfying
\[
 \left|\widehat T_{i+1}-(a_i+\lambda_i b_i)\right|\le{\Tol},
\]
for the exact quantity
$T_{i+1}=\GateExtension_{i+1}(u_i)
+\lambda_i\GateExtension_{i+1}(v_i)$.
This is a claim of the same form as before, now about layer $i+1$, with points
$\rho_{i+1,0} = u_i$ and $\rho_{i+1,1} = v_i$ and coefficients $c_{i+1,0}=1$ and $c_{i+1,1}=\lambda_i$.
All coordinates of these points and coefficients are roots of unity, so they lie on the unit circle, where our
numerical bounds apply. The prover sends $a_i$ and $b_i$ before $\lambda_i$ is sampled, so it cannot
choose their errors to cancel for that value of $\lambda_i$. If either claimed value has a large error, the combined claim therefore also has a large error, except with small probability. After $L$ such reductions, the verifier holds a claim  $\widehat T_L$ about the input layer. It checks this claim by
evaluating $\GateExtension_L$ at two points itself, which takes time linear in $n$ (up to logarithmic factors). 

\medskip\noindent{\bf Completeness.}
We show that an honest prover is always accepted, even though its messages
and the verifier's checks are computed only approximately. Each challenge is $r = \exp(2\pi \ii a/N)\in\Omega_N$, where the index $a$ is drawn uniformly from
$\{0,\dots,N-1\}$. The value of $r$ is computed only approximately, as some $\tilde r$ with
$|\tilde r - r| \le 2^{-q}$, where $q$ is the number of fractional bits used in the arithmetic. All
error bounds are stated relative to the exact root $r$.

In each round of sum-check, the honest prover sends the univariate polynomial obtained by fixing the earlier variables of the summand to the challenges and summing over all later variables. Let $q_j(X)=\sum_{a\in H^{k_i-j}}\LayerPoly_i(r_1,\dots,r_{j-1},X,a)$ be the true round polynomial, where $k_i=m_i+2m_{i+1}$ is the number of rounds at layer $i$, $r_1,\dots,r_{j-1}\in\Omega_N$ are the earlier challenges, and $X$ is the free $j$th coordinate. It has degree at most $D=\delta+h-1$, where $\delta$ bounds the individual degrees of $\AddPoly_i$ and $\MulPoly_i$. Let $\omega_J=\exp(2\pi \ii/J)$ for a power of two $J$ with $D<J\le 2(D+1)$. Its
coefficients are given in terms of its values at the $J$th roots of unity by $[X^a]\,q_j=J^{-1}\sum_{t=0}^{J-1}q_j(\omega_J^{t})\,\omega_J^{-at}$ for $0\le a\le D$.
The honest prover sends, in round $j$ of layer $i$, a polynomial $p_j$ that approximates the true round
polynomial $q_j$. The prover computes approximations
$\tilde q_j(\omega_J^{t})$ of the evaluations $q_j(\omega_J^{t})$ for $0\le t<J$, and sets $[X^a]\,p_j=\Round_p\bigl(J^{-1}\sum_{t=0}^{J-1}\tilde q_j(\omega_J^{t})\,\omega_J^{-at}\bigr)$ for $0\le a\le D$,
where $\Round_p$ rounds the real and imaginary parts to the nearest integer multiple of $2^{-p}$. With exact
evaluations and no rounding, the inner sum equals $[X^a]\,q_j$, because $\deg q_j<J$. The coefficients
$\omega_J^{-at}/J$ have modulus $1/J$, so their absolute sum is $1$. Hence if
$|\tilde q_j(\omega_J^{t})-q_j(\omega_J^{t})|\le e_{\mathrm{ev}}$ for every $t$, the approximate evaluations contribute error at most
$e_{\mathrm{ev}}$ to each coefficient, regardless of $J$.

Let $\eta_{\mathrm c}={\Tol}/[h(D+1)]$. For every coefficient index $a$, the error of $p_j$ is at most the sum of three
contributions, namely the approximate evaluations, the approximated coefficients $\omega_J^{-at}/J$, and the final rounding, so $|[X^a]p_j-[X^a]q_j|\le e_{\mathrm{ev}}+e_{\mathrm{wt}}+e_{\mathrm{rnd}}\le\eta_{\mathrm c}$.
Here $e_{\mathrm{ev}}$ bounds the error of the approximate evaluations $\tilde q_j(\omega_J^t)$ (including root approximation and
arithmetic), $e_{\mathrm{wt}}$ bounds the error from approximating the coefficients $\omega_J^{-at}/J$, and
$e_{\mathrm{rnd}}$ bounds the error from rounding to $p$ fractional bits. Since $p_j-q_j$ has $D+1$
coefficients, we obtain $|p_j(z)-q_j(z)|\le(D+1)\eta_{\mathrm c}={\Tol}/h$ for all $|z|\le1$. Summing over the $h$
points of $H$, we get $\bigl|\sum_{a\in H}p_j(a)-\sum_{a\in H}q_j(a)\bigr|\le h({\Tol}/h)={\Tol}$.
The verifier accepts the consistency check in round $j$ if the residual $\RunClaim_{j-1}-\sum_{a\in H}p_j(a)$,
that is, the difference between the two sides of the identity, has modulus at most $6{\Tol}$.  Here $\RunClaim_{j-1}$ is the current claim: $\RunClaim_0=\widehat T_i$, and $\RunClaim_j$ is the verifier's computed value of $p_j(r_j)$. Let $e_{j-1}$
be the error of the claim $\RunClaim_{j-1}$ relative to the true sum $\sum_{a\in H}q_j(a)$. By the bound above,
the message contributes at most ${\Tol}$, and the verifier's own arithmetic contributes at most ${\Tol}$, so
 $|\RunClaim_{j-1}-\sum_{a\in H}p_j(a)|\le e_{j-1}+{\Tol}+{\Tol}$.
The first claim of a layer has $e_{j-1}\le 3{\Tol}$ and later claims have $e_{j-1}\le 2{\Tol}$: a later claim is $p_{j-1}(r_{j-1})$ up to arithmetic error ${\Tol}$, and $|p_{j-1}(r_{j-1})-q_{j-1}(r_{j-1})|\le {\Tol}/h$; a first claim is either $y$, with error at most ${\Tol}$, or a combined claim $a_{i-1}+\lambda_{i-1}b_{i-1}$, whose claimed values each have error at most ${\Tol}$ and whose evaluation adds at most ${\Tol}$, so the residual
is at most $5{\Tol}$ in the first round and $4{\Tol}$ in later rounds. Both are below $6{\Tol}$ (a threshold of $5{\Tol}$ would also suffice; $6{\Tol}$ leaves a margin of ${\Tol}$), so an honest prover
passes every consistency check, for every sequence of challenges. The same accounting bounds the honest residuals of the gate check and of the final input check by their thresholds.

\medskip\noindent{\bf Soundness.}
To prove soundness, we must show that the error in an incorrect claim exceeds the required threshold at each reduction from the output
layer to the input layer, except with small probability. In exact sum-check over a finite field $\mathbb{F}$, a nonzero polynomial $f$ of
degree at most $D$ has at most $D$ roots, so
$\Pr_{r \leftarrow \mathbb{F}}[f(r) = 0] \le D/|\mathbb{F}|$.
Over $\C$, this bound does not suffice. The verifier accepts any residual
of modulus at most $6{\Tol}$, so an error can go undetected whenever $|f(r)|$ is of
order ${\Tol}$, not only when $f(r) = 0$. We must therefore bound
the probability that $|f(r)|$ is at most a multiple of ${\Tol}$.

We use an anti-concentration bound over the unit circle. For $f\in\C[X]$, write
$M(f)=\max_{|z|=1}|f(z)|$. Lemma~\ref{lem:circle} shows that if $f$ is nonzero
of degree at most $D\ge1$, then
\[
 \Pr_{r\leftarrow\Omega_N}
 \bigl[|f(r)|\le\alpha M(f)\bigr]
 \le D\bigl(\alpha^{1/D}+N^{-1}\bigr),
 \qquad 0<\alpha<1.
\]
The threshold is relative to $M(f)$ because no absolute threshold can work.
For instance, the polynomial $f(X)={\Tol}X$ satisfies $|f(r)|\le{\Tol}$ at every challenge. To prove the bound, let $\zeta_1,\dots,\zeta_d$ be the roots
of $f$, so that $f=\mathrm{lc}(f)\prod_{j=1}^{d}(X-\zeta_j)$, where $\mathrm{lc}(f)\ne0$ is the leading coefficient and $d\le D$. If
$|f(r)|\le\alpha M(f)$, then $r$ lies in one of at most $d$ arcs of length at
most $2\pi\alpha^{1/D}$, one for each root, and each arc contains at most
$N\alpha^{1/D}+1$ challenges. Indeed, if every factor satisfied $|r-\zeta_j|>\alpha^{1/d}(1+|\zeta_j|)$, then $M(f)\le|\mathrm{lc}(f)|\prod_j(1+|\zeta_j|)$ would give $|f(r)|>\alpha M(f)$. Since $H$ lies on the unit circle,
$M(f)\ge|\sum_{a\in H}f(a)|/h$, so a polynomial with a large sum over $H$ also
has a large maximum on the circle. Thus a
polynomial with a large sum over $H$ is unlikely to be small at a random
challenge.

We apply this bound to the difference between the
prover's round polynomial $p_j$ and the true polynomial
$q_j$. Let $e_{j-1}$ denote the error in the claim
entering round $j$. If the verifier accepts the
consistency check, its tolerance  $6{\Tol}$ and numerical error ${\Tol}$
give
$
    \left|\sum_{a\in H}(p_j-q_j)(a)\right|
    \ge e_{j-1}-7{\Tol}
$.
Since $|H|=h$, we have
$
    M(p_j-q_j)\ge\frac{e_{j-1}-7{\Tol}}{h}
$.
Thus, whenever $e_{j-1}>7{\Tol}$, the anti-concentration
bound implies that, except with probability at most
$D(\alpha^{1/D}+N^{-1})$, the error in the next claim
satisfies $e_j>(\alpha/h)(e_{j-1}-7{\Tol})-{\Tol}$.

At the end of each layer, the verifier checks the residual $Z_i-w_i^+(a_i+b_i)-w_i^\times a_ib_i$ and combines the two claimed evaluations using
a random scalar $\lambda_i$. If this check accepts
despite a sufficiently large error, at least one of the two claimed evaluations must be inaccurate. Writing the two errors in the claimed evaluations of $\GateExtension_{i+1}$
as $e_a$ and $e_b$, that is, $e_a=a_i-\GateExtension_{i+1}(u_i)$ and $e_b=b_i-\GateExtension_{i+1}(v_i)$, the error in their combination is
$e_a+\lambda_i e_b$. For $0<\beta<1$, provided the
errors are not both zero, the degree-one case of the
anti-concentration bound gives $\Pr_{\lambda_i\leftarrow\Omega_N}[|e_a+\lambda_i e_b|\le\beta\max\{|e_a|,|e_b|\}]\le\beta+N^{-1}$.
Hence the combined claim retains at least a $\beta$ fraction of the larger
error in a claimed evaluation, except with probability at most $\beta+N^{-1}$. Both bounds hold
against an adaptive prover. In each case, the polynomial being tested,
$p_j-q_j$ or $e_a+Xe_b$, is fixed by the transcript before the challenge $r_j$
or $\lambda_i$ is sampled. We may therefore condition on the transcript and
apply the bound to a fixed polynomial.

Let $k=\sum_i k_i$ be the total number of sum-check rounds. By a union bound,
the probability that the error falls below the required threshold at some
reduction is at most $kD(\alpha^{1/D}+N^{-1})+L(\beta+N^{-1})$.
We choose $\alpha$, $\beta$, and $N$ so that this probability is at most
$\eps$. Working backward from the final input check, which accepts only if the final claim has error at most $5{\Tol}$, we compute the
error threshold required at each reduction, and we choose ${\Tol}$ small enough
that the threshold at the output layer is at most $\Delta$. A sum-check round multiplies the threshold by at most $9h/\alpha$, and the step across a layer by a factor polynomial in $S$, $\BV$, $\BW$, and $(kD+L)/\eps$. With $\alpha$ and $\beta$ as chosen in \cref{sec:agkr}, this gives
\[
 \log(1/{\Tol})=\log^+(1/\Delta)+O\Bigl((kD+L)\log\frac{kD+L}{\eps}+k\log h+L\log(S\BV\BW)\Bigr).
\]
Since $k\le3L\log S$, the sufficient precision bound is linear in the depth, up to a logarithmic factor. 

With these parameter choices, if
the output error exceeds $\Delta$, the verifier rejects except with
probability at most $\eps$.

\medskip\noindent{\bf A near-linear prover.}
The completeness argument specifies how accurately the prover must
compute its messages. \cref{thm:informal-stable-thm} shows (informally) that, given sufficiently accurate
gate values, the prover can generate these messages using a near-linear number
of operations and only logarithmically many additional precision bits.

\begin{theorem}[Near-Linear Prover, Informal; \cref{stable:thm}]
\label{thm:informal-stable-thm}

Fix a constant power of two $h\ge2$. Suppose that $\AddPoly_i$
and $\MulPoly_i$ are the unique extensions of degree below $h$
in each variable, and let $p\ge0$ be an integer. Given an
approximate value for every gate of $C$, each differing from its
exact value by at most $2^{-p}/(576S^2\BV)$, there is a
deterministic algorithm that outputs transmitted polynomials $p_j$
approximating the exact round polynomials $q_j$, with
\[
 \max_{0\le a\le D}|[X^a](p_j-q_j)|<2^{-p},
\]
and claimed evaluations $a_i,b_i$ satisfying
\[
 |a_i-\GateExtension_{i+1}(u_i)|<2^{-p},\qquad
 |b_i-\GateExtension_{i+1}(v_i)|<2^{-p},
\]
for every sequence of verifier challenges. It uses
$q=p+O(\log S+\log(\BV+1))$ fractional bits, computes each
root of unity with absolute error at most $2^{-q}$, and performs
$O(S\log S)$ complex arithmetic operations and approximations
of roots of unity over all layers. These bounds concern message
generation from the supplied gate values and exclude the cost
of computing those values.
\end{theorem}

Near-linear GKR provers are known in exact arithmetic~\cite{CMT,Thaler}.
Theorem~\ref{thm:informal-stable-thm} shows that a near-linear operation count
remains achievable with approximate gate values and finite-precision
arithmetic, using only $O(\log S+\log(\BV+1))$ bits of precision beyond $p$.
Taking $p=\ceil{\log(h(D+1)/{\Tol})}$ gives every round coefficient the
accuracy ${\Tol}/[h(D+1)]$ required for completeness. The precision required to
compute the gate values themselves may still depend on the circuit depth.
Finally, the near-linear bound counts arithmetic operations, and the bit
complexity is this count multiplied by the cost of one operation at the
chosen precision.

The extra bits are logarithmic because errors grow only polynomially during
computation of the prover's messages. The prover evaluates $\GateExtension_{i+1}$ one coordinate
at a time, starting from the approximate gate values of layer $i+1$, viewed as
a function $\mathsf A\colon H^{m_{i+1}}\to\C$. Fixing the first coordinate to
$z$ gives the function $\mathsf A_z(b)=\sum_{a\in H}\Lag_a(z)\,\mathsf A(a,b)$ for $b\in H^{m_{i+1}-1}$,
where $\Lag_a(z)=\frac1h\sum_{t=0}^{h-1}(z/a)^t$ is the Lagrange basis polynomial for
$a\in H$, evaluated at $z$. Once every coordinate is fixed, the result is the value of
$\GateExtension_{i+1}$ at the chosen point. For $|z|\le1$, orthogonality gives
$\sum_{a\in H}|\Lag_a(z)|^2\le1$, and Cauchy--Schwarz gives
$\sum_{a\in H}|\Lag_a(z)|\le\sqrt h$. Each step therefore amplifies existing
errors by at most $\sqrt h$. The prover stores only the partial evaluations
at the challenges already sampled by the verifier. To evaluate $q_j$ at a $J$th root of
unity, it computes the partial evaluation at that root in a separate array  and discards the
array afterwards. Every stored value thus passes through at most
$m=\max_i m_i$ steps, regardless of how many such evaluations are made, and
the total amplification is at most $h^{m/2}\le\sqrt S$, since $h^m\le S$.
Each value of $q_j$ is then a sum of at most $S$ terms, one per gate, with
bounded coefficients. The prover recovers the coefficients of $q_j$ from these
values by the inverse discrete Fourier transform, and this step does not
amplify errors because the coefficients $\omega_J^{-at}/J$ have absolute sum one.
Altogether the error grows by a factor polynomial in $S$ and $\BV$, so
logarithmically many extra bits suffice.

This argument uses the unique extensions of $\Add_i$ and
$\Mul_i$ of degree below $h$ in each variable. In general, two
extensions of $\Add_i$ agree on $H^{m_i}\times H^{m_{i+1}}\times H^{m_{i+1}}$ but may differ greatly
at the challenge points, and the same holds for $\Mul_i$. For
$f(z)=\sum_{\mathbf e} f_{\mathbf e} z^{\mathbf e}$ and every $z$ whose coordinates have
modulus at most one, $|f(z)|\le\|f\|_1=\sum_{\mathbf e}|f_{\mathbf e}|$. The bound $\BW$
on the coefficient norms of $\AddPoly_i$ and
$\MulPoly_i$ therefore limits how much they amplify errors in
the gate values. The next theorem makes this precise and shows that some such
bound is necessary.

\begin{theorem}[Sensitivity to Gate-Value Errors, Informal; \cref{conditioning:thm}]
\label{thm:informal-conditioning-thm}
Suppose the approximate gate values differ from their exact
values by at most $\TableError\le 1$. For any fixed sequence of
verifier challenges, let $q_j$ and $\widehat q_j$ denote
the round polynomials obtained from the exact and approximate
gate values, respectively. Then $\max_{0\le a\le D}|[X^a](\widehat q_j-q_j)|\le 10\BW S^5\BV\TableError$.
Moreover, no bound depending only on $S$, $D$, and $\BV$
holds for all polynomial extensions, even for a circuit
with only three gates.
\end{theorem}

\subsection{Approximate GKR Protocol Instantiation\texorpdfstring{ (\cref{sec:implementation})}{}}\label{sec:overview-implementation}

We now remove the oracle access to $\AddPoly_i$ and $\MulPoly_i$ assumed in \cref{sec:overview-agkr}. The verifier queries these polynomials once per layer at the point produced by sum-check, and evaluating them directly requires a sum over all gates in the layer, which would dominate the verifier's running time. We instantiate this oracle in three settings. For computations in space $s$, the circuit is regular enough that whether a gate has given children can be decided by a Boolean formula of size $\poly(s)$ on the bits of their labels. Arithmetizing these formulas gives extensions of $\Add_i$ and $\Mul_i$ that the verifier evaluates in $\poly(s)$ time, without enumerating the gates. For logspace-uniform circuits, the prover sends approximate evaluations and proves that their bits equal the output bits of the fixed evaluation algorithm using the space-bounded protocol of \cref{thm:informal-space-proof}. For circuits fixed before the input is known, the verifier computes these evaluations in a private preprocessing phase. In each setting, the evaluations are accurate enough to preserve the completeness and soundness guarantees of the main protocol.

Now, we describe the circuits for space-bounded computation and the resulting interactive proof, the proofs of wiring evaluations for logspace-uniform circuits, and the preprocessing variant for fixed circuits. We refer to \cref{sec:implementation} for the formal constructions and their analysis.

\begin{theorem}[{Circuits for Space-Bounded Computation, Informal; \cref{thm:configuration-circuit}}] Let ${\mathcal M}$ be a nondeterministic Turing machine running in space $s \ge \ceil{\log(n+2)}$ on inputs of length $n$. There is a layered arithmetic circuit $C$ over $\C$ of size $2^{O(s)}$ and depth $O(s^2)$, with constants $-1,0,1$, such that for every $x \in \bits^n$, $C(x)=1$ if ${\mathcal M}$ accepts $x$ and $C(x)=0$ otherwise, and every gate value lies in $\{-1,0,1\}$. Gate labels have $O(s)$ bits, and the addition and multiplication predicates of $C$ are computed by constant-depth, unbounded fan-in Boolean formulas of size $\poly(s)$ that are independent of $x$. The input layer contains $x$ and the constants $-1,0,1$ at fixed positions, and zero elsewhere. \end{theorem}

To construct $C$, we encode the configurations of ${\mathcal M}$ with $n_{\mathrm{cfg}}=O(s)$ bits, add a self-loop at every configuration and an edge from each accepting configuration to a fixed sink, and let $B_0$ be the resulting adjacency matrix; each entry of $B_0$ is $0$, $1$, $x_j$, or $1-x_j$. The circuit computes $B_{t+1}[u,v]=1-\prod_{w\in\bits^{n_{\mathrm{cfg}}}}(1-B_t[u,w]\,B_t[w,v])$, so $B_t[u,v]=1$ if and only if $v$ is reachable from $u$ in at most $2^t$ steps. The output $B_{n_{\mathrm{cfg}}}[a,b]$, for the initial configuration $a$ and the sink $b$, is the acceptance bit. Each reachability iteration has depth $n_{\mathrm{cfg}}+O(1)$, giving depth $O(s^2)$. Although $C$ has exponential size, the children of each gate can be computed directly from its label. A product gate $(u,v,w)$ has children $(u,w)$ and $(w,v)$, and each gate in the product tree over $w$ has the two children obtained by appending one bit to its label. These predicates therefore reduce to equality tests $\StringEq(\mathbf s,\mathbf s')=\prod_j\bigl(\mathbf s_j\mathbf s'_j+(1-\mathbf s_j)(1-\mathbf s'_j)\bigr)$ on label bits. Computing $B_0$ requires checking that $v$ follows from $u$ in one step of ${\mathcal M}$, which reduces to equality and increment tests on their bits. All predicates thus have constant-depth formulas of size $\poly(s)$. This circuit is also used below to verify wiring evaluations for logspace-uniform circuits (see~\cref{eq:wiring-answer-bit}). Take $H=\{1,-1\}$ and arithmetize the formulas, replacing NOT by $1-z$, AND by multiplication, OR by $1-\prod_j(1-z_j)$, and each label bit by $(1-z)/2$. This yields extensions $\AddPoly_i,\MulPoly_i$ of degree $\poly(s)$ and coefficient norm $\BW=2^{\poly(s)}$, computable with $\poly(s)$ operations. They are not unique extensions of degree below $h$ in each variable, but the protocol only requires agreement with the predicates on the gate labels (see~\cref{eq:layer-sum}). Here $L=O(s^2)$ and $m=\max m_i=O(s)$, so
\[
 k=\sum_{i=0}^{L-1}(m_i+2m_{i+1})
 \le3Lm=O(s^3).
\]
With $D=\poly(s)$, $\BV=1$, and $\Delta=1/4$, at constant
soundness error the bound on $\log(1/{\Tol})$
in~\cref{sec:overview-agkr} (see~\cref{eq:precision-choice})
gives $\log(1/{\Tol})=\poly(s)$, and all intermediate values
have modulus $2^{\poly(s)}$. Hence $\poly(s)$-bit arithmetic suffices, which gives the following bounds.

\begin{theorem}[Space-Bounded Computation, Informal; \cref{thm:space-proof}]\label{thm:informal-space-proof}
Let $\mathcal{L}$ be a language decided by a nondeterministic Turing
machine in space $s \ge \ceil{\log(n+2)}$, and let
$r_\eps=\max\{1,\ceil{\log_{10}(1/\eps)}\}$ for $\eps\in(0,1)$. Then
$\mathcal{L}$ has a public-coin interactive proof over $\C$ with perfect
completeness and soundness error $\eps$, in which the prover runs in
time $2^{O(s)}r_\eps$, the verifier runs in time $n\poly(s)r_\eps$ and space
$\poly(s)+O(\log r_\eps)$, and the communication is $\poly(s)r_\eps$ bits. All
bounds count bit operations, including numerical computation.
\end{theorem}
For a general logspace-uniform circuit, evaluating $\AddPoly_i$ and $\MulPoly_i$ directly requires a sum over all gates in the layer. Instead, the prover sends these values and proves them correct. Writing $\widetilde{W}_i$ for either polynomial, the unique extension of degree below $h$ in each variable is $\widetilde{W}_i(z,u,v)=\sum_{g}\chi_g(z)\,\chi_{\mathrm{left}(g)}(u)\,\chi_{\mathrm{right}(g)}(v)$, where $g$ ranges over the gates of the corresponding type. Each summand is a product of $k_i=m_i+2m_{i+1}$ univariate Lagrange basis polynomials: $m_i$ from the parent label and $m_{i+1}$ from each child label. To make the answer a unique string, we approximate each root of unity to $p$ bits by a fixed algorithm, evaluate the sum exactly on these binary fractions, and apply $\RoundDown_q$. For the query $\mathsf q$ specifying the layer, gate type, and evaluation-point indices, write $\EvaluationString_{p,q}(\mathsf q)$ for the resulting binary string. If $A,B\in\Z$ are its encoded integer numerators, its represented value $w_{p,q}=2^{-q}(A+\ii B)$ satisfies $|w_{p,q}-\widetilde{W}_i(z,u,v)| \le 6Sk_ih\cdot 2^{1-p} + 2^{1-q}$, so $p,q=O(\log S+\log(1/\xi))$ give error $\xi$. By Lemma~\ref{lem:evaluation-bits}, each bit of $\EvaluationString_{p,q}(\mathsf q)$ is computable in space $\sigma=O(\log S+\log(p+q+\ell+2))$, where $\ell=\log N$. Logspace uniformity enumerates the gates and their children, and the arithmetic of~\cite{HAB} computes any bit of an iterated sum or product without storing intermediate values. Hence the claim that a given bit has a given value can be decided in space $\sigma$, and the space-bounded proof above, applied with $s=\sigma$, verifies it with communication $\poly(\sigma)$. The verifier uses the answer in the check of $Z_i-w_i^+(a_i+b_i)-w_i^\times a_ib_i$ only after all these interactive proofs reach an accepting state.

\begin{theorem}[Uniform Circuits, Informal; \cref{thm:uniform}]\label{thm:informal-uniform}
Let $C$ be a circuit from a logspace-uniform family, and instantiate
\cref{thm:informal-agkr} with the unique extensions of degree below $h$ in each variable
$\AddPoly_i$ and $\MulPoly_i$. Under
the numerical assumptions of \cref{sec:overview-agkr}, there is a
public-coin interactive proof in which the verifier needs no oracle
access to these polynomials. For the same thresholds ${\Tol}$ and $\Delta$,
it has perfect completeness and soundness error
$\eps_{\mathrm c}+\eps_{\mathrm w}$ for any
$\eps_{\mathrm w}\in(0,1)$, where $\eps_{\mathrm c}$ is the soundness
error of \cref{thm:informal-agkr}. Let $p$, $q$, and $\ell$ be the
maximum root precision, answer precision, and challenge-index length
over all queries. The additional prover time is $\poly(S,p,q,\ell,\log(1/\eps_{\mathrm w}))$, the additional verifier time is $L\cdot\poly(\log S,p,q,\ell,\log(1/\eps_{\mathrm w}))$, and the additional communication is $L\cdot\poly(\log S,p,q,\ell,\log(1/\eps_{\mathrm w}))$ bits.
\end{theorem}

For each query $\mathsf q$, the prover sends the full answer string $\mathsf e$ before the proofs of~\cref{eq:wiring-answer-bit}, and the verifier uses its represented value in the check of $Z_i-w_i^+(a_i+b_i)-w_i^\times a_ib_i$ only if every proof accepts. Each proof runs the space-bounded protocol on a circuit deciding~\cref{eq:wiring-answer-bit}, whose polynomial extensions $\AddPoly_i^{\mathrm{aux}},\MulPoly_i^{\mathrm{aux}}$ of the wiring predicates the verifier evaluates directly. There are $Q=2L$ queries with $b_{\mathrm w}$ bits each, so setting the soundness error of each proof to $\eps_{\mathrm w}/(Qb_{\mathrm w})$ and taking a union bound, an incorrect answer is accepted with probability at most $\eps_{\mathrm w}$. Let $\Pi$ denote the resulting protocol, and let $\Pi'$ be the protocol that also rejects whenever an incorrect answer is accepted. Every check of $Z_i-w_i^+(a_i+b_i)-w_i^\times a_ib_i$ in $\Pi'$ uses the values represented by $\EvaluationString_{p,q}(\mathsf q)$, so $\Pi'$ has soundness error $\eps_{\mathrm c}$. Since $\Pi$ and $\Pi'$ differ only on an event of probability at most $\eps_{\mathrm w}$, the soundness error of $\Pi$ is at most $\eps_{\mathrm c}+\eps_{\mathrm w}$. For logspace-uniform Boolean circuits, arithmetization keeps every gate value in $\{-1,0,1\}$, so $\log(1/{\Tol})=\poly(L,\log S)$ and $p,q,b_{\mathrm w}=\poly(L,\log S)$. At constant soundness error, the prover runs in time $\poly(S)$, the verifier in time $n\poly(L,\log S)$, and the communication and verifier space are $\poly(L,\log S)$. For uniform $\NC$, this gives a verifier running in time $n\polylog(n)$ with $\polylog(n)$ communication and space. These bounds count bit operations, and the near-linear prover bound of \cref{thm:informal-stable-thm} excludes the proofs of~\cref{eq:wiring-answer-bit}. For circuits fixed before the input is known, these interactive proofs can be avoided entirely. The query point $(g_i,u_i,v_i)$ depends only on the verifier's random challenges, not on the input or the prover's messages. In a preprocessing phase, the verifier therefore samples all its challenges and computes the $2L$ evaluations of $\AddPoly_i$ and $\MulPoly_i$ together with the contribution of the constant inputs to the final input check. It keeps these values secret from the prover, since they determine the challenges. When the input arrives, it computes only the contribution of $x$ and runs the remaining checks.

\begin{theorem}[Preprocessing, Informal; \cref{thm:preprocessing}] Let $C$ be a fixed circuit for which the numerical assumptions of \cref{sec:overview-agkr} hold on every input, and let $b$ be the numerical precision and $\ell=\log N$ the challenge-index length. Then the protocol of \cref{thm:informal-agkr} admits an input-independent preprocessing phase running in time $\poly(S,|C|,b,\ell)$, which outputs $O((k+L)\ell+Lb)$ stored bits containing the verifier's challenges and the $2L$ evaluations of $\AddPoly_i$ and $\MulPoly_i$. Using these stored values, the verifier never evaluates these polynomials, and the protocol has the same thresholds ${\Tol}$ and $\Delta$, perfect completeness, and soundness error $\eps_{\mathrm c}$. \end{theorem}

Let $\mathbf r=(r_1,\dots,r_k,\lambda_0,\dots,\lambda_{L-1})$ denote the verifier's challenges, sampled during preprocessing, and let the stored state $\mathsf{st}$ consist of $\mathbf r$ and the stored evaluations. The verifier reveals each $r_j$ only after receiving $p_j$, and each $\lambda_i$ only after receiving $a_i,b_i$. No check performed before this point depends on the unrevealed entries of $\mathbf r$, so conditioned on any transcript prefix, the next challenge is uniform on $\Omega_N$. Hence, for every prover $P^*$, the transcript of $P^*$ with the preprocessing verifier has the same distribution as with the verifier of \cref{thm:informal-agkr}, and completeness and soundness carry over. This requires $\mathsf{st}$ to be used once: in a second execution with the same $\mathbf r$, the prover knows every $r_j$ in advance and can choose $p_j$ to pass each consistency check regardless of the claim.

\subsection{Zero-Knowledge Approximate GKR\texorpdfstring{ (\cref{sec:zk})}{}}\label{sec:overview-zk}
We now make approximate GKR zero-knowledge for inputs that include a private witness. The protocol of \cref{sec:overview-agkr,sec:overview-implementation} verifies a claimed circuit output without evaluating the circuit, but its messages need not preserve the privacy of the input. We consider a public instance $(C,x,y)$ in which the public input $x$ occupies specified positions and a binary witness $w\in\mathcal W\subseteq\bits^{\WitLen}$ determines the other input entries as $\WitEntry{j}=(A_j(w)+\ii B_j(w))2^{-\WitPrec{j}}$, where a fixed Boolean circuit computes the integer encodings from $w$ and the $\WitPrec{j}$ are fixed public parameters. Write $C(x,w)$ for the exact circuit output. As in \cref{sec:overview-agkr}, the numerical parameters depend only on the public circuit and bounds that hold for every $w\in\mathcal W$. The completeness and soundness promises are, respectively, that some $w\in\mathcal W$ satisfies $|y-C(x,w)|\le {\Tol}$, and that every such $w$ satisfies $|y-C(x,w)|>\Delta$. For a Boolean NP relation, setting $y=1$ and $\Delta=1/4$ gives ordinary language membership. We show that, assuming one-way functions, approximate GKR can be made computationally zero-knowledge while retaining statistical soundness against unbounded provers.

Below, we describe the commitments, the Boolean check of the committed transcript, the zero-knowledge proof of acceptance, the security argument, and the extension to logspace-uniform circuits. We refer to \cref{sec:zk} for the formal protocol and its analysis.

\medskip\noindent{\bf Committing to the transcript.} The construction follows the commit-and-prove approach of~\cite[Section~5]{GKR} and~\cite{BGG}, with an important numerical distinction: the final zero-knowledge proof establishes acceptance by the finite-precision implementation of the approximate GKR verifier, without evaluating the delegated circuit $C(x,w)$. Before the first GKR challenge, the prover commits to the entire witness; before each subsequent challenge, it commits to the coefficients of the corresponding sum-check polynomial or the pair of claimed next-layer evaluations. Thus the values tested by any challenge are fixed before the challenge is sampled, even though they are hidden from the verifier. We use the statistically binding, computationally hiding commitment of~\cite{Naor}. Given a pseudorandom generator ${\mathsf G}:\bits^\kappa\to\bits^{3\kappa}$ and a message $R\in\bits^{3\kappa}$ sampled uniformly by the commitment receiver, a bit $\beta$ is committed as ${\mathsf G}(s)\oplus \beta R$ for a fresh seed $s\in\bits^\kappa$. If $R$ admitted openings to both bit values, it would equal ${\mathsf G}(s_0)\oplus {\mathsf G}(s_1)$ for some seeds. There are at most $2^{2\kappa}$ such strings, so a single $R$ makes \emph{all} these commitments binding except with probability $2^{-\kappa}$. Hiding follows from pseudorandomness, including for adaptively chosen messages and malicious verifier choices of $R$.

\medskip\noindent{\bf Checking the committed transcript.}
At the end of the committed GKR interaction, the public transcript consists of the instance, bit commitments $\mathbf d=({\mathsf G}(s_j)\oplus\mathsf m_jR)_j$, where $\mathsf m$ concatenates $w$ and the binary encodings of the prover's GKR messages, and recorded verifier challenges $\mathbf c$. The parties compute $v$ from the public instance and challenges using fixed evaluation and rounding algorithms. Its entries are the approximations $t_a,\widehat\chi_{a,j}$ defined below, the rounded wiring evaluations and challenge values, and the rounded values of $\ParentWeight_i(g_i)$, the factor of $w_i^+$ and $w_i^\times$ that depends on the claim's points and coefficients. We construct a Boolean circuit ${\mathsf B}({\mathsf x},{\mathsf w})$ whose public input is ${\mathsf x}=(C,x,y,R,\mathbf d,\mathbf c,v)$ and whose witness ${\mathsf w}$ consists of the committed bit strings and their opening seeds. The circuit checks $d_j={\mathsf G}(s_j)\oplus\mathsf m_jR$ and $w\in\mathcal W$, then runs the approximate-GKR verifier's numerical checks on the committed messages. The circuit rejects if any encoding is invalid, any magnitude check fails, or any arithmetic operation overflows. These are Boolean checks on finite strings; the original arithmetic identities remain over $\C$.

The final input-layer check depends on $\WitEntry{j}=(A_j(w)+\ii B_j(w))2^{-\WitPrec{j}}$. Using these values, we decompose the exact input-layer claim into public and witness-dependent terms:
\[
 T_L=\sum_{a=0}^1c_a\left(t_a^*+\sum_{j=1}^{n_\mathrm{w}}\chi_{g^{\mathrm w}_j}(\rho_a)\WitEntry{j}\right),
 \qquad
 t_a^*=\sum_{j=1}^{n_\mathrm{p}}\chi_{g^{\mathrm p}_j}(\rho_a)x_j.
\]
Here $\rho_a=\rho_{L,a}$ and $c_a=c_{L,a}$ are the final points and coefficients, and $g^{\mathrm p}_1,\dots,g^{\mathrm p}_{n_\mathrm{p}}$ and $g^{\mathrm w}_1,\dots,g^{\mathrm w}_{n_\mathrm{w}}$ are the labels of the public input entries and of the entries determined by $w$. 
The quantities $t_a^*$ and $\chi_{g^{\mathrm w}_j}(\rho_a)$ depend only on public input entries, their positions, and the challenges. We compute binary approximations to them outside ${\mathsf B}$ using the fixed numerical algorithms, with a total contribution of at most ${\Tol}/2$ to the input-claim error, and allocate the remaining ${\Tol}/2$ to the fixed-point evaluation of the residual. The input check therefore retains the same error budget ${\Tol}$ as in~\cref{sec:overview-agkr}, without introducing a second, independently claimed input value. Likewise, the other entries of $v$ are supplied as public inputs to ${\mathsf B}$ rather than recomputed inside it. 

Let $\mu=O(kDb)$ denote the total length of the approximate GKR prover messages, and let $r_\mathrm{c}=(k+L)\ell$ be the number of challenge bits (see~\cref{thm:agkr}). If $T_{\mathsf G}(\kappa)=\poly(\kappa)$ bounds the Boolean size of ${\mathsf G}$, and $T_{\mathrm{op}}(b+\ell)=\poly(b+\ell)$ bounds one numerical operation used by the verifier on the specified binary encodings, the final verification circuit has size $|{\mathsf B}|=O((\WitLen+\mu)T_{\mathsf G}(\kappa)+(kD+n_\mathrm{w})T_{\mathrm{op}}(b+\ell)+F_{\mathcal W}+r_\mathrm{c})$,
where $F_{\mathcal W}$ accounts for checking and decoding the witness. In particular, $|{\mathsf B}|$ includes the commitment checks and finite-precision verifier, but not evaluation of the delegated circuit or the public polynomial extensions of the wiring predicates. This separation is what makes the final zero-knowledge proof succinct.

\medskip\noindent{\bf Proving acceptance in zero-knowledge.}
To prove that ${\mathsf B}({\mathsf x},{\mathsf w})=1$ for some ${\mathsf w}$ without revealing the openings, we instantiate the MPC-in-the-head transformation of~\cite{IKOS} with a perfectly correct five-party MPC protocol private against any two passive parties, for example~\cite{STOC:BGW88}. The prover XOR-shares ${\mathsf w}$ among five virtual parties, simulates the MPC evaluation of ${\mathsf B}$, and commits to their views. Four uniform challenge bits select one of the ten pairs of views, which the prover opens for local and pairwise consistency checks. On a false statement, some pair must fail: otherwise all five views would define a consistent accepting MPC execution, contradicting perfect correctness. Every pair is selected with probability at least $1/16$, so one repetition accepts a false statement with probability at most $15/16$, conditional on commitment binding. With fresh sharings and MPC randomness in each of $35$ sequential repetitions and one independent receiver message $R'$ for these commitments, the proof for ${\mathsf B}$ has soundness error at most $(15/16)^{35}+2^{-\kappa}<1/8$ and communication $O(|{\mathsf B}|\poly(\kappa))$.

\begin{theorem}[Zero-Knowledge Approximate GKR, Informal; \cref{thm:zk-direct}]\label{thm:informal-zk}
Assume one-way functions secure against nonuniform polynomial-size circuits, and fix $\kappa\ge6$ and the soundness error $0<\eps_\mathrm{c}\le1/8$ of approximate GKR. Under the assumptions on $w\mapsto(A_j(w),B_j(w))_j$ and the uniform numerical bounds above, the promise problem defined above has a public-coin interactive proof with perfect completeness, computational zero-knowledge against classical polynomial-time verifiers with arbitrary auxiliary input, and soundness error at most $\eps_\mathrm{c}+(15/16)^{35}+2^{1-\kappa}<1/2$
against arbitrary provers. Its communication is $O\bigl(\kappa(\WitLen+\mu)+r_\mathrm{c}+|{\mathsf B}|\poly(\kappa)\bigr)$,
and its verifier time is $T_\mathrm{pub}+\poly(|{\mathsf B}|,\kappa)$, where $T_\mathrm{pub}$ is the cost of computing $v$. If $F_{\mathcal W}\le(\WitLen+1)\poly(b)$ and $n_\mathrm{w}=O(\WitLen+1)$, the communication specializes to $(\WitLen+1)\poly(\kappa,L,h,\log S,b)$.
\end{theorem}

For Boolean NP relations computed by polynomial-size, polylogarithmic-depth circuits, setting $y=1$ and $0<{\Tol}<\Delta=1/4$ makes the YES condition equivalent to $\exists w\in\mathcal W:C(x,w)=1$, and the NO condition equivalent to $\forall w\in\mathcal W:C(x,w)=0$. When the stated bounds on $F_{\mathcal W}$ and $n_\mathrm{w}$ hold and the numerical parameters $h,b$ are polylogarithmic in the circuit size, the bound gives communication $(\WitLen+1)\poly(\kappa,\log S)$; without these additional restrictions, the explicit parameter-dependent bound above applies.

\medskip\noindent{\bf Security.}
For soundness, first condition on $\neg\mathsf{Bad}_R$, where $\mathsf{Bad}_R$ is the event that $R={\mathsf G}(s_0)\oplus {\mathsf G}(s_1)$ for some seeds $s_0,s_1$ (see~\cref{sec:zk-security}). Given a cheating prover $P^*$, consider the unbounded prover $P'$ that runs $P^*$, searches for $(\beta,s)$ satisfying $d_j={\mathsf G}(s)\oplus \beta R$ for each commitment $d_j$, and sends each recovered message before its corresponding challenge. In particular, $w$ is fixed before the first GKR challenge; if $w\notin\mathcal W$, ${\mathsf B}$ rejects. If the statement $\exists{\mathsf w}\,{\mathsf B}({\mathsf x},{\mathsf w})=1$ is true on a NO instance, the recovered messages form an accepting approximate GKR transcript for that witness. By~\cref{thm:informal-agkr}, this occurs with probability at most $\eps_\mathrm{c}$. If the NP statement is false, the final proof accepts with probability at most $(15/16)^{35}+2^{-\kappa}$. Adding $\Pr[\mathsf{Bad}_R]$ gives the displayed bound. This reasoning never conditions the GKR challenges on the eventual acceptance of the statement $\exists{\mathsf w}\,{\mathsf B}({\mathsf x},{\mathsf w})=1$.

For zero-knowledge, begin with a genuine honest execution, and replace only the proof for ${\mathsf B}$ by its simulator. For every $w\in\mathcal W$ satisfying $|y-C(x,w)|\le {\Tol}$ and every sequence of challenge indices in the specified sets, the honest messages pass the numerical checks. Thus the honest openings satisfy ${\mathsf B}({\mathsf x},{\mathsf w})=1$. Next replace the entries of $\mathbf d$, one at a time, by commitments to zero. By the computational hiding of the commitments (\cref{def:commitment}), consecutive hybrids are computationally indistinguishable, even when later verifier challenges depend on earlier commitments. The intermediate statements for ${\mathsf B}$ may be false. This hybrid argument uses only the simulator's polynomial running time on every statement. The simulator's zero-knowledge guarantee is used only for true statements, and its rewinding is confined to the proof for ${\mathsf B}$ and does not cross any commitment in $\mathbf d$. The last hybrid is independent of the witness and gives the required simulation.

\medskip\noindent{\bf Logspace-uniform circuits.}
In \cref{thm:informal-zk}, $T_\mathrm{pub}$ includes the cost of the public wiring evaluations, which may require enumerating the gates. We avoid this cost for logspace-uniform circuits using the interactive proofs of wiring evaluations from \cref{sec:overview-implementation} (see~\cref{eq:wiring-answer-bit}). After the GKR challenges have fixed the $Q=2L$ wiring queries $\mathsf q_j$, the prover sends answer strings $\mathsf e_j$. For $j=1,\ldots,Q$ and $\iota=1,\ldots,b_{\mathrm w,j}$, where $b_{\mathrm w,j}=|\EvaluationString_{p,q}(\mathsf q_j)|$, it proves $\mathsf e_j[\iota]=\EvaluationString_{p,q}(\mathsf q_j)[\iota]$ by the space-bounded interactive proof of~\cref{thm:informal-space-proof}. Padding to the common length $b_{\mathrm w}$ is checked directly. For answers of length at most $b_{\mathrm w}$ and bit-algorithm space bound $\sigma$, taking $r_\mathrm{w}=\ceil{\log_{10}(Qb_{\mathrm w}/\eps_\mathrm{w})}$
repetitions per bit bounds the probability that any incorrect answer passes by $\eps_\mathrm{w}$. The verifier stores the strings and, after all these proofs accept, uses \emph{those same strings} as the wiring entries of $v$ in the proof for ${\mathsf B}$. The precision is chosen so that the two wiring-approximation errors contribute at most ${\Tol}/2$ to each computed residual for $Z_i-w_i^+(a_i+b_i)-w_i^\times a_ib_i$, leaving the remaining ${\Tol}/2$ for the other numerical operations. In particular, the parameters ${\Tol}$ and $\Delta$ do not change.

\begin{theorem}[Zero-Knowledge Approximate GKR for Uniform Circuits, Informal; \cref{thm:zk-uniform}]
Under the assumptions above, for a circuit from a fixed logspace-uniform family and $0<\eps_\mathrm{w}\le1/8$ with $\log(1/\eps_\mathrm{w})$ polynomially bounded in $\kappa$, there is a public-coin interactive proof with perfect completeness, computational zero-knowledge, and soundness error at most $\eps_\mathrm{c}+\eps_\mathrm{w}+(15/16)^{35}+2^{1-\kappa}<1/2$. 
Writing $Q=2L$ and using the parameters $b_{\mathrm w},r_\mathrm{w},\sigma$ above and the length $t_{\mathrm w}$ of the public statement for one answer bit, the communication is $O\bigl(\kappa(\WitLen+\mu)+r_\mathrm{c}+|{\mathsf B}|\poly(\kappa)+Qb_{\mathrm w}r_\mathrm{w}\poly(\sigma)\bigr)$,
and the verifier time is $O\bigl((n_\mathrm{p}+n_\mathrm{w})mh+k+Lmh\bigr)\poly(b+\ell)+Qb_{\mathrm w}r_\mathrm{w}t_{\mathrm w}\poly(\sigma)+\poly(|{\mathsf B}|,\kappa)$.
Under the witness-encoding bounds of \cref{thm:informal-zk}, the communication is $(\WitLen+1)\poly(\kappa,L,h,\log S,b)$ and the verifier time is $(n_\mathrm{p}+1)\poly(\WitLen,\kappa,L,h,\log S,b)$.
\end{theorem}

The simulator computes $\mathsf e_j^*=\EvaluationString_{p,q}(\mathsf q_j)$, then runs the honest prover to prove $\mathsf e_j^*[\iota]=\EvaluationString_{p,q}(\mathsf q_j)[\iota]$ for $1\le\iota\le|\mathsf e_j^*|$, as in~\cref{eq:zk-wiring-answer-bit}. These computations take polynomial time and do not use $w$. For soundness, compare with an execution that rejects whenever the proofs for an incorrect answer accept. Except with probability $\eps_\mathrm{w}$, the predicate ${\mathsf B}$ then uses exactly the values represented by $\mathsf e_j^*$, so the preceding soundness reduction applies. We do not condition the earlier GKR challenges on acceptance of the later interactive proofs. Zero-knowledge follows by incorporating the honest prover for~\cref{eq:zk-wiring-answer-bit} into the preceding hybrid argument. Sequential repetition of the complete protocol, with fresh commitment setups and challenges, reduces soundness to any $\eps>0$ without changing the thresholds ${\Tol}$ and $\Delta$.

\subsection{One-Round Arguments for Approximate Computation\texorpdfstring{ (\cref{sec:one-round})}{}}\label{sec:overview-one-round}

We now compile the protocol of \cref{sec:overview-agkr,sec:overview-implementation} into one-round arguments for approximate evaluation of logspace-uniform circuits, with a single verifier message followed by a single prover response. The difficulty is that approximate GKR relies on fixing each prover message before revealing the challenge that tests it. We apply the private-information-retrieval (PIR) transformation of Kalai and Raz~\cite{KR} to retrieve the required responses while hiding the challenges that select them. The main efficiency issue is that the transformation tabulates the honest responses for every assignment to the challenge bits on which they depend. We choose interpolation parameters that keep these databases polynomial in the circuit size and numerical precision, including for the wiring-evaluation proofs. The resulting private-coin argument preserves the numerical thresholds ${\Tol}$ and $\Delta$ while providing computational rather than unconditional soundness under a quantitative PIR assumption.

Below, we state the setting and the result, then describe the PIR compilation, bound the number of verifier bits that determine each response, both for approximate GKR and for the wiring-evaluation proofs, and prove its completeness and soundness. We refer to \cref{sec:one-round} for the formal protocol and its analysis.

\medskip\noindent{\bf The setting.} Let $S_0\ge2$ denote the circuit size before padding, $L$ its depth, and $n$ the number of input gates, including constants. The circuit family and its logspace generator are fixed. Public magnitude bounds hold uniformly over the specified input domain, and $b$ bounds the input, output-claim, message, and working bit lengths, including the wiring-evaluation precisions. For $\Delta>0$, choose ${\Tol}$ as in \cref{sec:overview-agkr} with approximate-GKR soundness error $\eps_\mathrm{c}=1/8$, and set the soundness error of the wiring-evaluation proofs (see~\cref{eq:wiring-answer-bit}) to $\eps_\mathrm{w}=1/8$.

\begin{theorem}[One-Round Approximate GKR, Informal; \cref{thm:one-round-agkr}]\label{thm:informal-one-round}
    Under the numerical conditions above and the quantitative PIR assumption of~\cref{sec:one-round-protocol}, for every integer $\kappa\ge2$, there is a one-round (two-message) private-coin argument for the promise $|y-C(x)|\le {\Tol}$ versus $|y-C(x)|>\Delta$. For every instance $(C,x,y)$ with the specified public protocol parameters fixed independently of the verifier's randomness, the argument has completeness at least $1-2^{-\tau^2}$ and soundness at most $1/4+2^{-\tau^2}<1/2$ against nonuniform provers of size at most $2^\kappa$, where $\kappa\le\tau\le\poly(\kappa,L,\log S_0,b)$ is a suitably chosen PIR security parameter. The protocol communicates $\poly(\kappa,L,\log S_0,b)$ bits, and the verifier and honest prover use $(n+1)\poly(\kappa,L,\log S_0,b)$ and $\poly(S_0,b,\kappa)$ bit operations, respectively. The verifier's first message is generated in $\poly(\kappa,L,\log S_0,b)$ time from only the fixed public protocol parameters and fresh randomness; verification requires its stored random tape, PIR queries, and recovery information.
\end{theorem}

\medskip\noindent{\bf PIR compilation.}
Suppose each honest response bit is determined by the instance $I$, the encoding of $(C,x,y)$, and at most $\Lambda$ earlier verifier bits, at positions fixed by the public protocol parameters. These positions need not be consecutive. Before each response, the verifier resends the verifier bits needed for that response, padded to length $\Lambda$; the original challenge messages are retained, with ignored dummy responses inserted as needed. No challenge is disclosed early, so this transformation preserves completeness and soundness. The honest response in round $j$ of the transformed protocol is now a function $f_j(I,z)$ of the current verifier message $z\in\bits^\Lambda$. The prover can tabulate this function as an $M$-bit database $\mathsf{db}_j[z]=f_j(I,z)$, where $M=2^\Lambda$.
The verifier privately samples its original random tape and computes every normalized message $z_j$. It sends independent PIR queries for the entries $\mathsf{db}_j[z_j]$ in one message. The honest prover constructs the databases, answers all queries, and returns the answers in one response. Using its private recovery information, the verifier recovers the response bits and applies the original acceptance test. Thus the original challenges serve as hidden database indices rather than being revealed to the prover.

For a database of length $M$, the PIR assumption gives query and recovery time and communication $\poly(\tau,\log M)$, and server time $\poly(\tau,M)$. It requires correctness error at most $2^{-\tau^3}$, and that every nonuniform
circuit of size at most $2^{\tau^3}$ have distinguishing advantage at most
$2^{-\tau^3}$ between the queries for any two indices, \ldots, the latter against nonuniform circuits of size at most $2^{\tau^3}$, as for the poly-logarithmic PIR of~\cite[Definition~2.6]{GKR}. Under these quantitative guarantees, \cref{thm:kr} bounds the soundness loss even when a malicious prover generates all replies jointly without constructing any databases. Ordinary polynomial-time PIR privacy is not the assumption used here. The honest prover takes $\poly(\ProverTime,\tau,2^\Lambda)$ time, where $\ProverTime$ is its running time in the original proof. Controlling $\Lambda$, rather than merely the message lengths, is therefore essential.

\medskip\noindent{\bf Bounding the verifier bits needed for each response.}
We choose the interpolation set to encode many gate-label bits in each coordinate. Specifically, let $h=2^{\ceil{\log\max\{2,L,\log(S_0+2)\}}}$.
After padding, the circuit has size $S=O(hS_0)$ and maximum number of coordinates per gate label $m=O(\log S_0/\log h)$. The unique polynomial extensions of the wiring predicates of degree below $h$ in each variable have individual degree $h-1$, giving sum-check degree $D=2(h-1)$ and $k=O(Lm)$ rounds. At constant soundness error, the choice of $N$ in \cref{sec:overview-agkr} (see~\cref{eq:parameters}) gives $N=\poly(h)$, $\ell=\log N=O(\log h)$, and $m\ell=O(\log S_0)$.
At layer $i>0$, the polynomial $q_j$ depends only on the two evaluation points $\rho_{i,0},\rho_{i,1}$, the preceding combining challenge, and the current sum-check prefix. The initial points and coefficients at layer zero are fixed. Thus each GKR response depends on at most $(5m+1)\ell=O(\log S_0)$ challenge bits. The layer, round, and coefficient positions are fixed by the protocol parameters and do not enlarge this bound.

For the transformation, this dependence must hold for the actual binary messages, not merely their exact polynomial counterparts. We therefore use the canonical numerical algorithms of~\cref{sec:completeness}, with fixed evaluation orders, precisions, and rounding rules. They compute each message from the instance and selected challenge indices, without using earlier rounded claims or verifier residuals. Equal selected indices consequently determine equal encoded responses. Increasing the numerical precision changes the cost and length of a GKR response, but does not introduce dependence on additional GKR challenges.

\medskip\noindent{\bf Including the wiring-evaluation proofs.}
The same property must hold for the proofs of~\cref{eq:wiring-answer-bit} that replace direct evaluation of $\AddPoly_i$ and $\MulPoly_i$. Each query is determined by $O(m\ell)$ GKR challenge bits, and $\EvaluationString_{p,q}(\mathsf q)$ is a deterministic function of the query $\mathsf q$ and the public circuit. By~\cref{lem:evaluation-bits}, the algorithm for an answer bit uses space $\sigma=O\bigl(\log S_0+\log(b+2)\bigr)$.
Its space-bounded proof uses a circuit deciding~\cref{eq:wiring-answer-bit} of size $2^{O(\sigma)}$ and depth $\poly(\sigma)$. Here we take $h_\mathrm{aux}$ to be the least power of two at least $\sigma+2$, rather than the choice $h=2$ used in the proof of~\cref{thm:space-proof}. Lemma~\ref{lem:space-circuit} permits this choice. The number of coordinates per gate label becomes $O(\sigma/\log(\sigma+2))$, and challenge indices have length $O(\log(\sigma+2))$. Consequently, each response in this proof depends on only $O(\sigma)$ challenge bits from that proof, in addition to the GKR challenge bits specifying its query.

For each database entry, the honest prover recomputes $\EvaluationString_{p,q}(\mathsf q)$ and its asserted bit from that query; neither an earlier answer nor an earlier rounded claim is an additional index. The verifier still checks the bit actually supplied by the prover. Bit positions and repetition numbers are fixed by the protocol parameters. The complete protocol therefore satisfies $\Lambda=O(m\ell+\sigma)=O(\log S_0+\log(b+2))$, so $2^\Lambda=\poly(S_0,b)$.
Every database entry is computable in $\poly(S_0,b)$ time, so the compiled honest prover remains polynomial time even after including all messages in the proofs of~\cref{eq:wiring-answer-bit}.

\medskip\noindent{\bf Completeness and soundness.}
The underlying public-coin protocol has perfect completeness and soundness at most $\eps_\mathrm{c}+\eps_\mathrm{w}=1/4$. This transformation preserves these guarantees. We choose $\tau$ at least the target security parameter $\kappa$, the communication of the transformed protocol, and the logarithm of the encoded common-input length, as required by~\cref{thm:kr}. The bounds above allow $\tau=\poly(\kappa,L,\log S_0,b)$. The Kalai--Raz transformation adds at most $2^{-\tau^2}$ to soundness against size-$2^\tau$ provers, and hence against size-$2^\kappa$ provers. Completeness loses at most the probability of a PIR recovery error: there are at most $\tau$ recoveries, each failing with probability at most $2^{-\tau^3}$. Perfectly correct PIR eliminates this loss.

The transformation operates on the finite binary transcript and leaves the numerical acceptance test unchanged. In particular, it introduces no additional arithmetic approximation and preserves the parameter ${\Tol}$ and soundness threshold $\Delta$ chosen for the underlying protocol.

When $S_0=\poly(n)$ and $\kappa,L,b=\polylog(n+2)$, the resulting communication is polylogarithmic and verifier time is $(n+1)\polylog(n+2)$. The precision bound remains a separate hypothesis for computation over $\C$; low depth alone does not imply it.

Each execution uses fresh randomness and PIR queries. Although the first message can be prepared in advance, the guarantee is for an instance chosen independently of it and the verifier's private state; challenge reuse and adaptive instance selection are not covered. The argument is not publicly verifiable, and no zero-knowledge guarantee is claimed.

\subsection{Interactive PCPs\texorpdfstring{ (\cref{sec:ipcp})}{}}
\label{sec:overview-ipcp}

We now extend approximate GKR to computations whose inputs
include a witness that the verifier does not know, as in \cref{sec:overview-zk}, but with a fixed proof string in place of commitments. The final input check
can no longer be performed from the public input alone. Following
GKR~\cite[Section~7]{GKR}, the prover supplies a fixed proof string encoding
the witness, from which the verifier reads the required evaluations. The
proof is fixed before any verifier coins are sampled and encodes the
witness (rather than the computation). Over $\C$, testing that this string
is close to a low-degree polynomial is not sufficient: we must also bound
the polynomial's values, check that its witness entries are close to bits,
and account for the change in the output when these entries are rounded.

Below, we describe the encoding of the witness, the low-degree test, the check that the witness entries are close to bits, the completeness and soundness of the protocol, and the choice of parameters. We refer to \cref{sec:ipcp} for the formal protocol and its analysis. \Cref{thm:informal-ipcp} summarizes the resulting protocol.

\begin{theorem}[Interactive PCP, Informal; \cref{thm:pcp-native}]
\label{thm:informal-ipcp}

Let $C$ be a layered arithmetic circuit over $\C$ of size $S$ and depth $L$ that takes a public input $x$ and the inputs ${\DecMap}(w)$ determined by a witness $w\in\bits^{\WitLen}$, where ${\DecMap}$ is a public affine map with finite-binary coefficients; the size and depth include the gates that compute ${\DecMap}$. Assume public bounds on the modulus of every gate value that hold for all $w\in\bits^{\WitLen}$. For every $\Delta>0$ and $\eps\in(0,1)$, there are ${\TolC}\in(0,\Delta)$ and a public-coin interactive PCP $(P,V)$ for the promise problem whose YES instances $(C,x,y)$ satisfy $|y-C(x,{\DecMap}(w))|\le {\TolC}$ for some $w\in\bits^{\WitLen}$, and whose NO instances satisfy $|y-C(x,{\DecMap}(w))|>\Delta$ for every $w\in\bits^{\WitLen}$. The interactive PCP has perfect completeness: on a YES instance witnessed by $w$, the verifier $V$, given oracle access to the honest proof $\pi_w$ and interacting with $P(w)$, accepts with probability $1$. On a NO instance, for every proof string $\pi^*$ and every computationally unbounded prover $P^*$, $V$ accepts with probability at most $\eps$. Let $\PCPReps=\max\{1,\ceil{\log(1/\eps)}\}$. For a suitable $h=\poly(L,\log S,\log({\WitLen}+1))$, the proof has length $\poly({\WitLen},h)$ bits, $V$ makes $\poly(h)\cdot \PCPReps$ bit queries to the proof, and the communication and the verifier's running time are $\poly(h,b_*)\cdot \PCPReps$, where $b_*$ is the bit precision of the two approximate-GKR subprotocols. The running time excludes evaluating the low-degree extensions of the public input and of the wiring predicates of the two circuits. Once the encoding parameters are fixed, $\pi_w$ depends only on $w$ and not on $C$, and the same proof is used across all $\PCPReps$ repetitions.
\end{theorem}

\medskip\noindent{\bf The protocol.} We encode the witness as a low-degree polynomial interpolated on a grid $H^m$, and the proof consists of its evaluations on a larger grid $G^m$. Concretely, let $H = \Omega_h$ and $G = \Omega_N$, where $h, m \ge 2$ and $N$ are powers of two satisfying $h^m\ge {\WitLen}$, $h\mid N$, and $N\ge16mh$. The first condition ensures that $H^m$ has room for the witness, and the second gives $H \subseteq G$. The third ensures that distinct polynomials of individual degree $< h$ disagree on most of $G^m$, which the low-degree test requires. Since approximate GKR draws its challenges from $G$, we also take $N$ large enough for its soundness analysis (see \cref{sec:overview-agkr}). We identify $[{\WitLen}]$ with a fixed set of positions $\WitnessPositions \subseteq H^m$. We then let $f_w$ be the unique polynomial of individual degree $< h$ satisfying $f_w(a)=w_a$ for $a\in\WitnessPositions$ and $f_w(a)=0$ for $a\in H^m\setminus\WitnessPositions$. The honest proof $\pi_w$ lists the values of $f_w$ on the larger grid: $\pi_w[r]=f_w(r)$ for $r\in G^m$. Whenever the verifier needs $f_w$ at a random point $r \in G^m$, it reads the single entry $\pi_w[r]$. Let $\PCPCode$ be the Reed--Solomon code consisting of the evaluations on $G$ of univariate polynomials of degree $< h$. Since $f_w$ has degree $< h$ in each variable, $\pi_w$ is a codeword of the tensor code $\PCPCode^{\otimes m}$. Since $h \mid N$, every interpolation node and evaluation point is an $N$th root of unity, so each entry of $\pi_w$ lies in the cyclotomic field $\mathbb{Q}(\omega_N)$, where $\omega_N=\exp(2\pi\ii/N)$. We represent each entry exactly by its rational coordinates in the basis $1, \omega_N, \dots, \omega_N^{N/2-1}$. By Lemma~\ref{pcp:encoding}, this takes $\OracleBits = \Theta\bigl(N(\log({\WitLen}+1) + m\log h)\bigr)$ bits per entry. Moreover, every string of this length, including one written by a cheating prover, encodes a value of modulus at most ${\EntryBound} = 2N({\WitLen}+1)$. The verifier can therefore check low-degree identities exactly over $\mathbb{Q}(\omega_N)$. It can also approximate each entry to the absolute error required when it is used in approximate GKR. Given a proof $\pi$, which may be dishonest, the verifier checks three properties in turn. It first tests that $\pi$ is close to $\PCPCode^{\otimes m}$, that is, that $\pi$ agrees on most of $G^m$ with some polynomial $f$ of individual degree $< h$. It then runs approximate GKR twice: once to check that the witness values $(f(a))_{a \in \WitnessPositions}$ are close to $\bits^{\WitLen}$, and once to check that $y \approx C(x, {\DecMap}(\cdot))$ on these values. Each invocation ends with claims about $f$ at random points of $G^m$, which the verifier resolves by querying $\pi$ at those points.

\medskip\noindent{\bf Low-degree testing.} To test proximity of $\pi$ to $\PCPCode^{\otimes m}$, the verifier restricts $\pi$ to a random plane obtained by fixing all but two coordinates, using the robustness of tensor codes~\cite{BSS04}. The verifier selects all but two coordinates of $G^m$, according to the schedule described below, and fixes them to uniformly random points of $G$. This determines a plane ${\mathcal P} \cong G^2$ and the restriction $\pi|_{\mathcal P}$. The prover then sends the $h^2$ coefficients of a bivariate polynomial $g$ of individual degree $< h$, which it claims equals $\pi|_{\mathcal P}$. Only after receiving $g$ does the verifier sample a uniform $s \in G^2$ and check that $g(s) = \pi|_{\mathcal P}(s)$, exactly over $\mathbb{Q}(\omega_N)$. Since $g$ is fixed before $s$ is sampled, the test rejects with probability at least $\PCPDistance(\pi|_{\mathcal P}, \PCPCode^{\otimes 2})$, where $\PCPDistance$ denotes relative Hamming distance. It remains to show that a random restriction inherits a large fraction of the distance of $\pi$. The key tool is a robustness property of tensor codes (Lemma~\ref{pcp:tensor}). Let $E \subseteq \mathbb{F}^{n_E}$ be a linear code of relative distance $\delta_E$, and let $v \in \mathbb{F}^{[n_E]^{\TenDim}}$ with ${\TenDim} \ge 3$ and $\delta_E^{\TenDim} \ge 7/8$. Fixing one coordinate of $v$ to a single value gives a subarray, which we call a \emph{coordinate hyperplane} ${\mathcal P}$. The lemma states that for a uniformly random ${\mathcal P}$, $\Expect_{\mathcal P}[\PCPDistance(v|_{\mathcal P},E^{\otimes(\TenDim-1)})]\ge2^{-14}\PCPDistance(v,E^{\otimes \TenDim})$. To prove it, we approximate each hyperplane restriction by a nearby codeword. For each ${\mathcal P}$, let $c_{\mathcal P} \in E^{\otimes(\TenDim-1)}$ be a codeword nearest to $v|_{\mathcal P}$; we call $c_{\mathcal P}$ the \emph{local codeword} of ${\mathcal P}$, since it describes $v$ only on ${\mathcal P}$. Set ${\LocErr} = \Expect_{\mathcal P}\bigl[\PCPDistance(v|_{\mathcal P}, c_{\mathcal P})\bigr]$. It suffices to show that ${\LocErr} \le 2^{-14}$ implies the existence of a \emph{global} codeword $c \in E^{\otimes \TenDim}$ with $\PCPDistance(v, c) \le 8{\LocErr}$. Because each $c_{\mathcal P}$ is chosen independently, two local codewords need not agree where their hyperplanes intersect. The proof shows that a small ${\LocErr}$ forces most of them to agree. Take hyperplanes ${\mathcal P}$ and ${\mathcal Q}$ in different directions. If $c_{\mathcal P}$ and $c_{\mathcal Q}$ differ on ${\mathcal P} \cap {\mathcal Q}$, they are distinct codewords of $E^{\otimes(\TenDim-2)}$ there, and hence disagree on at least a $\delta_E^{\TenDim-2}$ fraction of ${\mathcal P} \cap {\mathcal Q}$. On the other hand, the average disagreement of $c_{\mathcal P}$ and $c_{\mathcal Q}$ on ${\mathcal P} \cap {\mathcal Q}$ is at most $2{\LocErr}$, since at any point where they differ, at least one of them differs from $v$. It follows that only an $O(\sqrt{{\LocErr}})$ fraction of hyperplanes have local codewords inconsistent with many others. Discarding these leaves a consistent family covering all but an $O(\sqrt{{\LocErr}})$ fraction of the array. Because a nonzero codeword of $E$ cannot vanish on more than $(1-\delta_E)n_E$ coordinates, this family extends uniquely to a global codeword $c$. A final incidence argument between points and hyperplanes sharpens the bound to $\PCPDistance(v, c) \le 8{\LocErr}$. Applying the lemma once per coordinate would lose a factor of $2^{-14(m-2)}$. Instead, we fix coordinates in blocks. When $4m'$ coordinates remain, we view $\pi$ as a four-dimensional array over $\PCPCode^{\otimes m'}$, whose relative distance is at least $15/16$ since $N \ge 16mh$, and we fix two of the four blocks. This halves the number of free coordinates and, by two applications of the lemma, preserves a $2^{-29}$ fraction of the expected distance. After $\log m - 1$ halvings, we obtain $\Expect_{\mathcal P}[\PCPDistance(\pi|_{\mathcal P},\PCPCode^{\otimes 2})]\ge m^{-29}\PCPDistance(\pi,\PCPCode^{\otimes m})$. Hence if $\pi$ is $\gamma$-far from $\PCPCode^{\otimes m}$, each repetition rejects with probability at least $\gamma m^{-29}$, even conditioned on earlier repetitions. By Lemma~\ref{pcp:test}, $\pi$ passes $r_{\mathrm{ldt}}$ sequential repetitions with probability at most $\delta_{\mathrm{ldt}}=\exp(-r_{\mathrm{ldt}}\gamma m^{-29})$. Each repetition reads one entry of $\pi$, which costs $\OracleBits$ proof bits under the encoding of Lemma~\ref{pcp:encoding}, and receives $h^2$ coefficients in the same encoding from the prover.

\medskip\noindent{\bf Checking the witness.} Suppose $\pi$ passes the low-degree test, so that it is close to the evaluation table $(f(a))_{a\in G^m}$ for some polynomial $f$ of individual degree $< h$. Write $z = (f(a))_{a \in \WitnessPositions} \in \C^{\WitLen}$ for the values of $f$ at the witness positions. A polynomial of low degree can take arbitrary complex values at the witness positions, so the verifier must also check that $z$ is close to a bit vector. Each expression $z_a(z_a - 1)$ vanishes exactly when $z_a \in \bits$, but over $\C$ their sum may vanish through cancellation. We therefore use the nonnegative quantity $B(f)=\sum_{a\in\WitnessPositions}|z_a(z_a-1)|^2$. Since $0$ and $1$ are at distance one, every $z_a$ satisfies $\max\{|z_a|, |z_a - 1|\} \ge 1/2$. Lemma~\ref{lem:pcp-booleanity} uses this to show $B(f)\ge\frac14\min_{w\in\bits^{\WitLen}}\|z-w\|_2^2$. Thus a small value of $B(f)$ certifies that $z$ is close to a bit vector. An arithmetic circuit cannot compute $|z_a|^2$ directly, because complex conjugation is not a polynomial operation. We avoid this with the coefficientwise conjugate $\ConjugatePoly(X) = \overline{f(\overline{X})}$, which has the same individual degree as $f$. Every $a \in H^m$ lies on the unit torus, so $\overline{a} = a^{-1}$ and $\ConjugatePoly(a^{-1}) = \overline{f(a)}$. Hence \[ B(f) = \sum_{a \in \WitnessPositions} f(a)\bigl(f(a) - 1\bigr)\, \ConjugatePoly(a^{-1})\bigl(\ConjugatePoly(a^{-1}) - 1\bigr). \] This is computed by an arithmetic circuit of size $O({\WitLen})$ and depth $O(\log({\WitLen}+1))$ whose input consists of the values of $f$ and of $\ConjugatePoly$ on $H^m$. At the end of approximate GKR, the verifier must evaluate the low-degree extension of this input at random points, which amounts to evaluating $f$ and $\ConjugatePoly$ at random points of $G^m$. It obtains $f(r)$ by reading $\pi[r]$, and $\ConjugatePoly(r)$ by reading $\pi[r^{-1}]$ and conjugating. Since both values come from the same proof, the second half of the input is forced to be the conjugate of the first, and the prover has no freedom to choose it separately. Fix $0 < \eta < 1/2$. The verifier runs approximate GKR on this circuit with claimed output $0$ and threshold $\Delta_{\mathrm{bool}} < \eta^2/4$. If $z$ is $\eta$-far from $\bits^{\WitLen}$, then $B(f) > \eta^2/4 > \Delta_{\mathrm{bool}}$, and the invocation rejects except with probability $\eps_{\mathrm{bool}}$. Otherwise, there is some $w \in \bits^{\WitLen}$ with $\|z - w\|_2 \le \eta$, but the circuit output on $z$ may still differ from its output on $w$. Lemma~\ref{lem:pcp-perturbation} bounds this difference by propagating the input error through the addition and multiplication gates, which gives a function $\mathcal J$ satisfying $|C(x,{\DecMap}(z))-C(x,{\DecMap}(w))|\le \mathcal J(\eta)$ whenever $\|z-w\|_2\le\eta$. We choose $\eta$ so that $\mathcal J(\eta) \le \Delta/2$ and run the output invocation with threshold $\Delta_{\mathrm{out}} = \Delta/2$. Under the soundness promise, $|y - C(x, {\DecMap}(w))| > \Delta$ for every $w \in \bits^{\WitLen}$, so $|y-C(x,{\DecMap}(z))|>\Delta-\mathcal J(\eta)\ge\Delta_{\mathrm{out}}$, and the output invocation rejects except with probability $\eps_{\mathrm{out}}$. A suitable $\eta$ has bit length polynomial in the circuit depth, the logarithms of the magnitude and decoding bounds, and $\log^+(1/\Delta)$. The parameter ${\TolC}$ in \cref{thm:informal-ipcp} is the completeness threshold ${\Tol}$ of the output invocation.

\medskip\medskip\noindent{\bf Completeness.}
Let $w \in \bits^{\WitLen}$ satisfy $|y - C(x, {\DecMap}(w))| \le {\TolC}$, and suppose the
prover supplies the honest proof $\pi_w$. Since $\pi_w$ is a codeword of
$\PCPCode^{\otimes m}$, the honest prover can always send the true
restriction $\pi_w|_{\mathcal P}$, so every low-degree test accepts. The polynomial
underlying $\pi_w$ is $f_w$, whose witness values are exactly $w$, so
$B(f_w) = 0$. By the completeness of approximate GKR, both the
Booleanity invocation (with claimed output $0$) and the output
invocation (with claimed output $y$) therefore accept for every
challenge sequence. The exact encoding of Lemma~\ref{pcp:encoding}
lets the verifier compute the final input evaluations to the accuracy
these invocations require.

\medskip\medskip\noindent{\bf Soundness.}
Fix an arbitrary proof $\pi$ and prover strategy. Since $\pi$ is fixed
before the verifier samples any coins, we may analyze each case
separately. If $\pi$ is $\gamma$-far from $\PCPCode^{\otimes m}$, the
low-degree test accepts with probability at most $\delta_{\mathrm{ldt}}$. Otherwise,
fix a polynomial $f$ of individual degree $< h$ that agrees with $\pi$
on at least a $1 - \gamma$ fraction of $G^m$.

Approximate GKR requires a bound on the modulus of every input, but $f$
need not inherit the bound ${\EntryBound}$ that holds for the entries of $\pi$:
its coefficients may be arbitrarily large. Lemma~\ref{pcp:decoded}
shows that agreement with ${\EntryBound}$-bounded values on most of $G^m$
nevertheless bounds $f$ on the entire torus $\UnitCircle^m$. The proof starts from a point
where $|f|$ is maximal and replaces its coordinates one at a time by
uniform points of $G$. The inverse-moment estimate of
\cref{lem:circle} (see~\cref{eq:inverse-moment}) shows that $|f|$ does not decrease
much in expectation at each step. If $f$ were large on the torus, it
would therefore be large on most of $G^m$, contradicting its agreement
with $\pi$. For $\gamma \le 1/4$, this gives
$\max_{z\in\UnitCircle^m}|f(z)|\le\DecodedBound={\EntryBound}\,2^{2(h-1)(m+1)}$.
The same bound holds for $\ConjugatePoly$, which supplies the input
magnitude bounds for both invocations. The verifier does not need to compute $f$.

We replace $\pi$ by $f$ using a hybrid argument. Let $V_f$ be the
verifier that behaves exactly like $V$, except that it answers every
final input query using $f$ instead of $\pi$. Let
$\mathcal D = \{r \in G^m : \pi[r] \ne f(r)\}$, so that $|\mathcal D| \le \gamma N^m$.
Each point at which $V$ queries $\pi$ is either a vector of fresh
sum-check challenges or the coordinatewise inverse of one. In both
cases it is uniformly distributed on $G^m$, because inversion is a
bijection of $G^m$. The Booleanity invocation makes at most four
queries and the output invocation at most two, so a union bound gives
$\Pr[\text{some query lies in } \mathcal D]\le6|\mathcal D|/N^m\le6\gamma$.
Unless some query lies in $\mathcal D$, $V$ and $V_f$ receive identical answers
and reach the same decision,
$\Pr[V\text{ accepts}]\le\Pr[V_f\text{ accepts}]+6\gamma$.

It remains to bound the acceptance probability of $V_f$. Its witness
values $z = (f(a))_{a \in \WitnessPositions}$ are determined by $f$,
which is fixed before any verifier coins are sampled. If $z$ is
$\eta$-far from $\bits^{\WitLen}$, the Booleanity claim is false and $V_f$
accepts with probability at most $\eps_{\mathrm{bool}}$. Otherwise, the output claim
is false and $V_f$ accepts with probability at most $\eps_{\mathrm{out}}$. Since
exactly one of these cases applies, the two errors combine by a maximum
rather than a sum, and the overall acceptance probability is at most
$s_0=\max\{\delta_{\mathrm{ldt}},\,6\gamma+\max\{\eps_{\mathrm{bool}},\eps_{\mathrm{out}}\}\}$.

\medskip\medskip\noindent{\bf Efficiency.}
Let $m$ be the least power of two with
$m \ge \max\{2, \ceil{\log_h {\WitLen}}\}$, and let $h \ge 64$ be a power of two
at least the depth and maximum number of coordinates per gate label of both padded circuits.
Setting $N = h^5$ satisfies the challenge-set requirement of
approximate GKR at constant soundness error, as well as
$N \ge 16mh$. Since $m \le 2\max\{2, \log_h {\WitLen} + 1\}$, we have
$h^m \le \max\{h^4, h^2{\WitLen}^2\}$, and the proof length is
$N^m\OracleBits=(h^m)^5\OracleBits=\poly({\WitLen},h)$.
Taking $\gamma = 1/100$, $\eps_{\mathrm{bool}} = \eps_{\mathrm{out}} = 1/20$, and
$r_{\mathrm{ldt}} = \ceil{100\, m^{29} \ln 20}$~\footnote{We use the natural logarithm because $\delta_{\mathrm{ldt}}$ is an exponential: with $\gamma=1/100$, the choice $r_{\mathrm{ldt}}=\ceil{100\,m^{29}\ln 20}$ gives $\delta_{\mathrm{ldt}}\le e^{-\ln 20}=1/20$.}, so that $\delta_{\mathrm{ldt}}\le1/20$, gives $s_0 \le 0.11$. Because this
bound holds conditioned on any preceding transcript, repeating the
interaction $O(\PCPReps)$ times with fresh coins and the same proof reduces the
soundness error to $\eps$. Each repetition reads $r_{\mathrm{ldt}} + 6$ entries of
$\pi$, for a total of $O(\PCPReps m^{29} \OracleBits)$ proof bits.
Communication and the verifier's remaining work are $\poly(h, b_*)\, \PCPReps$.

\bibliographystyle{alpha}
\bibliography{cryptobib,references}

\newcommand{\etalchar}[1]{$^{#1}$}
\begin{thebibliography}{WYX{\etalchar{+}}21}

\bibitem[APKP24]{CCS:APKP24}
Kasra Abbaszadeh, Christodoulos Pappas, Jonathan Katz, and Dimitrios
  Papadopoulos.
\newblock Zero-knowledge proofs of training for deep neural networks.
\newblock In Bo~Luo, Xiaojing Liao, Jun Xu, Engin Kirda, and David Lie,
  editors, {\em ACM CCS 2024: 31st Conference on Computer and Communications
  Security}, pages 4316--4330, Salt Lake City, UT, USA, October~14--18, 2024.
  {ACM} Press.

\bibitem[Bab85]{STOC:Babai85}
L{\'a}szl{\'o} Babai.
\newblock Trading group theory for randomness.
\newblock In {\em 17th Annual {ACM} Symposium on Theory of Computing}, pages
  421--429, Providence, RI, USA, May~6--8, 1985. {ACM} Press.

\bibitem[BCC88]{JCSS:BraChaCre88}
Gilles Brassard, David Chaum, and Claude Cr{\'e}peau.
\newblock Minimum disclosure proofs of knowledge.
\newblock {\em Journal of Computer and System Sciences}, 37(2):156--189,
  October 1988.

\bibitem[BDG{\etalchar{+}}26]{ApproxSumcheck26}
Dor Bitan, Zachary DeStefano, Shafi Goldwasser, Yuval Ishai, Yael~Tauman Kalai,
  and Justin Thaler.
\newblock Sum-check protocol for approximate computations.
\newblock In Joan Daemen and Emmanuel Thom{\'{e}}, editors, {\em Advances in
  Cryptology - {EUROCRYPT} 2026 - 45th Annual International Conference on the
  Theory and Applications of Cryptographic Techniques, Rome, Italy, May 10-14,
  2026, Proceedings, Part {VII}}, volume 16547 of {\em Lecture Notes in
  Computer Science}, pages 209--238. Springer, 2026.

\bibitem[BGG{\etalchar{+}}88]{BGG}
Michael Ben{-}Or, Oded Goldreich, Shafi Goldwasser, Johan H{\aa}stad, Joe
  Kilian, Silvio Micali, and Phillip Rogaway.
\newblock Everything provable is provable in zero-knowledge.
\newblock In Shafi Goldwasser, editor, {\em Advances in Cryptology - {CRYPTO}
  '88, 8th Annual International Cryptology Conference, Santa Barbara,
  California, USA, August 21-25, 1988, Proceedings}, volume 403 of {\em Lecture
  Notes in Computer Science}, pages 37--56. Springer, 1988.

\bibitem[BGW88]{STOC:BGW88}
Michael Ben{-}Or, Shafi Goldwasser, and Avi Wigderson.
\newblock Completeness theorems for non-cryptographic fault-tolerant
  distributed computation (extended abstract).
\newblock In Janos Simon, editor, {\em Proceedings of the 20th Annual {ACM}
  Symposium on Theory of Computing, May 2-4, 1988, Chicago, Illinois, {USA}},
  pages 1--10. {ACM}, 1988.

\bibitem[BM84]{BluMic84}
Manuel Blum and Silvio Micali.
\newblock How to generate cryptographically strong sequences of pseudorandom
  bits.
\newblock {\em {SIAM} Journal on Computing}, 13(4):850--864, 1984.

\bibitem[BSS06]{BSS04}
Eli Ben-Sasson and Madhu Sudan.
\newblock Robust locally testable codes and products of codes.
\newblock {\em Random Structures \& Algorithms}, 28(4):387--402, 2006.
\newblock Preliminary version in APPROX-RANDOM 2004.

\bibitem[CKGS98]{JACM:CGKS98}
Benny Chor, Eyal Kushilevitz, Oded Goldreich, and Madhu Sudan.
\newblock Private information retrieval.
\newblock {\em Journal of the ACM}, 45(6):965--981, November 1998.

\bibitem[CMT12]{CMT}
Graham Cormode, Michael Mitzenmacher, and Justin Thaler.
\newblock Practical verified computation with streaming interactive proofs.
\newblock In Shafi Goldwasser, editor, {\em ITCS 2012: 3rd Innovations in
  Theoretical Computer Science}, pages 90--112, Cambridge, MA, USA,
  January~8--10, 2012. Association for Computing Machinery.
\newblock \url{https://doi.org/10.1145/2090236.2090245}. Full version at
  \url{https://arxiv.org/abs/1105.2003}.

\bibitem[CWSK24]{EuroSys:CWSK24}
Bing-Jyue Chen, Suppakit Waiwitlikhit, Ion Stoica, and Daniel Kang.
\newblock {ZKML}: An optimizing system for {ML} inference in zero-knowledge
  proofs.
\newblock In {\em Proceedings of the Nineteenth European Conference on Computer
  Systems ({EuroSys})}, pages 560--574. ACM, April 2024.

\bibitem[DF04]{DummitFoote2004}
David~S Dummit and Richard~M Foote.
\newblock Abstract algebra. john wile \& sons.
\newblock {\em Inc., Hoboken, NJ}, 2004.

\bibitem[GGJ{\etalchar{+}}23]{CCS:GGJMMP23}
Sanjam Garg, Aarushi Goel, Somesh Jha, Saeed Mahloujifar, Mohammad Mahmoody,
  Guru-Vamsi Policharla, and Mingyuan Wang.
\newblock Experimenting with zero-knowledge proofs of training.
\newblock In Weizhi Meng, Christian~Damsgaard Jensen, Cas Cremers, and Engin
  Kirda, editors, {\em ACM CCS 2023: 30th Conference on Computer and
  Communications Security}, pages 1880--1894, Copenhagen, Denmark,
  November~26--30, 2023. {ACM} Press.

\bibitem[GHL{\etalchar{+}}26]{cryptoeprint:2026/1112}
Nicolas Gailly, Ismael Hishon{-}Rezaizadeh, Tianyi Liu, Nicholas Mainardi,
  Dimitrios Papadopoulos, Charalampos Papamanthou, Christodoulos Pappas,
  Shravan Srinivasan, Zack Youell, and Yupeng Zhang.
\newblock Deepprove: Verifiable end-to-end large language model inference.
\newblock {\em {IACR} Cryptol. ePrint Arch.}, 2026:1112, 2026.

\bibitem[GJJZ22]{FloatZK22}
Sanjam Garg, Abhishek Jain, Zhengzhong Jin, and Yinuo Zhang.
\newblock Succinct zero knowledge for floating point computations.
\newblock In {\em Proceedings of the 2022 ACM SIGSAC Conference on Computer and
  Communications Security}, pages 1203--1216. ACM, 2022.

\bibitem[GKR15]{GKR}
Shafi Goldwasser, Yael~Tauman Kalai, and Guy~N. Rothblum.
\newblock Delegating computation: Interactive proofs for muggles.
\newblock {\em Journal of the {ACM}}, 62(4):27:1--27:64, 2015.
\newblock \url{https://doi.org/10.1145/2699436}. Full version in ECCC TR17-108,
  2017. \url{https://eccc.weizmann.ac.il/report/2017/108/}.

\bibitem[GM84]{GolMic84}
Shafi Goldwasser and Silvio Micali.
\newblock Probabilistic encryption.
\newblock {\em Journal of Computer and System Sciences}, 28(2):270--299, 1984.

\bibitem[GMR89]{GolMicRac89}
Shafi Goldwasser, Silvio Micali, and Charles Rackoff.
\newblock The knowledge complexity of interactive proof systems.
\newblock {\em {SIAM} Journal on Computing}, 18(1):186--208, 1989.

\bibitem[GO94]{JC:GolOre94}
Oded Goldreich and Yair Oren.
\newblock Definitions and properties of zero-knowledge proof systems.
\newblock {\em Journal of Cryptology}, 7(1):1--32, December 1994.

\bibitem[GRSY21]{ITCS:GRSY21}
Shafi Goldwasser, Guy~N. Rothblum, Jonathan Shafer, and Amir Yehudayoff.
\newblock Interactive proofs for verifying machine learning.
\newblock In James~R. Lee, editor, {\em 12th Innovations in Theoretical
  Computer Science Conference, {ITCS} 2021, Virtual Conference, January 6-8,
  2021}, volume 185 of {\em LIPIcs}, pages 41:1--41:19. Schloss Dagstuhl -
  Leibniz-Zentrum f{\"{u}}r Informatik, 2021.

\bibitem[GS86]{STOC:GolSip86}
Shafi Goldwasser and Michael Sipser.
\newblock Private coins versus public coins in interactive proof systems.
\newblock In {\em 18th Annual {ACM} Symposium on Theory of Computing}, pages
  59--68, Berkeley, CA, USA, May~28--30, 1986. {ACM} Press.

\bibitem[HAB02]{HAB}
William Hesse, Eric Allender, and David A.~Mix Barrington.
\newblock Uniform constant-depth threshold circuits for division and iterated
  multiplication.
\newblock {\em Journal of Computer and System Sciences}, 65(4):695--716, 2002.
\newblock \url{https://doi.org/10.1016/S0022-0000(02)00025-9}.

\bibitem[HCL{\etalchar{+}}24]{USENIX:HCLWZY24}
Meng Hao, Hanxiao Chen, Hongwei Li, Chenkai Weng, Yuan Zhang, Haomiao Yang, and
  Tianwei Zhang.
\newblock Scalable zero-knowledge proofs for non-linear functions in machine
  learning.
\newblock In Davide Balzarotti and Wenyuan Xu, editors, {\em USENIX Security
  2024: 33rd {USENIX} Security Symposium}, Philadelphia, PA, USA,
  August~14--16, 2024. {USENIX} Association.

\bibitem[HILL99]{HILL}
Johan H{\aa}stad, Russell Impagliazzo, Leonid~A. Levin, and Michael Luby.
\newblock A pseudorandom generator from any one-way function.
\newblock {\em {SIAM} Journal on Computing}, 28(4):1364--1396, 1999.
\newblock \url{https://www.cs.bu.edu/fac/lnd/pdf/hill.pdf}.

\bibitem[IKOS09]{IKOS}
Yuval Ishai, Eyal Kushilevitz, Rafail Ostrovsky, and Amit Sahai.
\newblock Zero-knowledge proofs from secure multiparty computation.
\newblock {\em {SIAM} Journal on Computing}, 39(3):1121--1152, 2009.
\newblock \url{https://doi.org/10.1137/080725398}. Full version at
  \url{https://web.cs.ucla.edu/~rafail/PUBLIC/77.pdf}.

\bibitem[KO97]{FOCS:KusOst97}
Eyal Kushilevitz and Rafail Ostrovsky.
\newblock Replication is {NOT} needed: {SINGLE} database,
  computationally-private information retrieval.
\newblock In {\em 38th Annual Symposium on Foundations of Computer Science},
  pages 364--373, Miami Beach, Florida, October~19--22, 1997. {IEEE} Computer
  Society Press.

\bibitem[KR08]{KRIPCP08}
Yael~Tauman Kalai and Ran Raz.
\newblock Interactive {PCP}.
\newblock In {\em Automata, Languages and Programming---ICALP 2008, Part II},
  volume 5126 of {\em Lecture Notes in Computer Science}, pages 536--547.
  Springer, 2008.

\bibitem[KR09]{KR}
Yael~Tauman Kalai and Ran Raz.
\newblock Probabilistically checkable arguments.
\newblock In Shai Halevi, editor, {\em Advances in Cryptology --
  {CRYPTO}~2009}, volume 5677 of {\em Lecture Notes in Computer Science}, pages
  143--159, Santa Barbara, CA, USA, August~16--20, 2009. Springer Berlin
  Heidelberg, Germany.
\newblock \url{https://doi.org/10.1007/978-3-642-03356-8_9}. Full version at
  \url{https://www.wisdom.weizmann.ac.il/~ranraz/publications/PPCA.pdf}.

\bibitem[LFKN92]{LFKN}
Carsten Lund, Lance Fortnow, Howard Karloff, and Noam Nisan.
\newblock Algebraic methods for interactive proof systems.
\newblock {\em Journal of the ACM}, 39(4):859--868, 1992.

\bibitem[LXZ21]{CCS:LiuXieZha21}
Tianyi Liu, Xiang Xie, and Yupeng Zhang.
\newblock {zkCNN}: Zero knowledge proofs for convolutional neural network
  predictions and accuracy.
\newblock In Giovanni Vigna and Elaine Shi, editors, {\em ACM CCS 2021: 28th
  Conference on Computer and Communications Security}, pages 2968--2985,
  Virtual Event, Republic of Korea, November~15--19, 2021. {ACM} Press.

\bibitem[Nao91]{Naor}
Moni Naor.
\newblock Bit commitment using pseudorandomness.
\newblock {\em Journal of Cryptology}, 4(2):151--158, January 1991.
\newblock \url{https://doi.org/10.1007/BF00196774}. Full text at
  \url{https://www.wisdom.weizmann.ac.il/~naor/PAPERS/bit.pdf}.

\bibitem[QSL{\etalchar{+}}25]{zkGPT25}
Wenjie Qu, Yijun Sun, Xuanming Liu, Tao Lu, Yanpei Guo, Kai Chen, and Jiaheng
  Zhang.
\newblock {zkGPT}: An efficient non-interactive zero-knowledge proof framework
  for {LLM} inference.
\newblock In {\em 34th USENIX Security Symposium}, pages 2045--2063. USENIX
  Association, 2025.

\bibitem[RWB{\etalchar{+}}25]{ZIP25}
Arman Riasi, Haodi Wang, Rouzbeh Behnia, Viet Vo, and Thang Hoang.
\newblock Zero-knowledge {AI} inference with high precision.
\newblock In {\em Proceedings of the 2025 ACM SIGSAC Conference on Computer and
  Communications Security}, pages 1053--1067. ACM, 2025.

\bibitem[SLZ24]{zkLLM24}
Haochen Sun, Jason Li, and Hongyang Zhang.
\newblock {zkLLM}: Zero knowledge proofs for large language models.
\newblock In {\em Proceedings of the 2024 ACM SIGSAC Conference on Computer and
  Communications Security}, pages 4405--4419. ACM, 2024.

\bibitem[SV22]{SoriaVazquez22}
Eduardo Soria-Vazquez.
\newblock Doubly efficient interactive proofs over infinite and non-commutative
  rings.
\newblock In {\em Theory of Cryptography---TCC 2022}, volume 13747 of {\em
  Lecture Notes in Computer Science}, pages 497--525. Springer, 2022.

\bibitem[Tha13]{Thaler}
Justin Thaler.
\newblock Time-optimal interactive proofs for circuit evaluation.
\newblock In Ran Canetti and Juan~A. Garay, editors, {\em Advances in
  Cryptology - {CRYPTO} 2013 - 33rd Annual Cryptology Conference, Santa
  Barbara, CA, USA, August 18-22, 2013. Proceedings, Part {II}}, volume 8043 of
  {\em Lecture Notes in Computer Science}, pages 71--89. Springer, 2013.

\bibitem[WYX{\etalchar{+}}21]{Mystique21}
Chenkai Weng, Kang Yang, Xiang Xie, Jonathan Katz, and Xiao Wang.
\newblock Mystique: Efficient conversions for zero-knowledge proofs with
  applications to machine learning.
\newblock In {\em 30th USENIX Security Symposium}, pages 501--518. USENIX
  Association, 2021.

\bibitem[Yao82]{FOCS:Yao82a}
Andrew Chi-Chih Yao.
\newblock Theory and applications of trapdoor functions (extended abstract).
\newblock In {\em 23rd Annual Symposium on Foundations of Computer Science},
  pages 80--91, Chicago, Illinois, November~3--5, 1982. {IEEE} Computer Society
  Press.

\end{thebibliography}

\newpage

\begin{center}
    \huge{\textbf{Appendix}}
\end{center}

\appendix

\renewcommand{\theHsection}{appendix.\Alph{section}}
\renewcommand{\theHsubsection}{appendix.\Alph{section}.\arabic{subsection}}
\crefalias{section}{appendix}
\Crefname{appendix}{Appendix}{Appendices}
\crefalias{subsection}{subappendix}
\crefalias{subsubsection}{subsubappendix}

\etocdepthtag.toc{appendix}

\ifnum\fullversion=0
\begingroup
\etocsettagdepth{mainmatter}{none}
\etocsettagdepth{appendix}{subsection}
\etocsettocstyle{\section*{Contents}}{}
\tableofcontents
\endgroup

\newpage
\fi

\section{Acknowledgment on AI Usage}
In this work, we used GPT-6 Astra and GPT-5.6 Pro for literature review, for discussing and checking the correctness of theorems. We used Claude Opus 5.5 to check for grammatical issues. We take responsibility for the final content of this work.

\section{Preliminaries}\label{sec:preliminaries}

We use the notation of \cref{sec:overview-notation} throughout. This appendix adds the notation and definitions needed in \cref{sec:agkr,sec:implementation,sec:zk,sec:one-round,sec:ipcp}. \Cref{sec:prelim-notation} fixes notation for distributions, \cref{sec:prelim-ip} defines interactive proofs, arguments, zero-knowledge, and interactive PCPs, and \cref{sec:prelim-crypto} recalls the cryptographic primitives used in \cref{sec:zk,sec:one-round}. Their concrete instantiations and parameters are given where they are used.

\subsection{Additional Notation}\label{sec:prelim-notation}
We write $U_\ell$ for the uniform distribution on $\bits^\ell$. Two families of distributions $\{X_\kappa\}$ and $\{Y_\kappa\}$ are \emph{computationally indistinguishable}~\cite{GolMic84,FOCS:Yao82a}, written $X_\kappa\approx_{\mathrm c}Y_\kappa$, if every nonuniform polynomial-time distinguisher $\mathcal D$ satisfies $|\Pr[\mathcal D(X_\kappa)=1]-\Pr[\mathcal D(Y_\kappa)=1]|=\negl(\kappa)$.

\subsection{Interactive Proofs and Arguments}\label{sec:prelim-ip}
A promise problem is a pair $(\mathcal Y,\mathcal N)$ of disjoint sets of instances; no requirement is placed on instances outside $\mathcal Y\cup\mathcal N$. For thresholds $0<{\Tol}<\Delta$, \emph{approximate circuit evaluation} is the promise problem whose YES instances are the triples $(C,x,y)$ with $|y-C(x)|\le{\Tol}$ and whose NO instances are the triples with $|y-C(x)|>\Delta$. Here $C$ is a layered arithmetic circuit over $\C$, $x$ is its input, and $y\in\C$ is the claimed output.

For interactive algorithms $P$ and $V$ and an instance $I$, we write $\langle P,V\rangle(I)\in\bits$ for the output of $V$ after its interaction with $P$ on common input $I$, where $1$ means that $V$ accepts. Probabilities are over the coins of both parties.

\begin{definition}[Interactive proofs and arguments~\cite{GolMicRac89,JCSS:BraChaCre88}]\label{def:ip}
Let $0\le s<c\le1$. A pair $(P,V)$ of interactive algorithms is an \emph{interactive proof} for a promise problem $(\mathcal Y,\mathcal N)$ with completeness $c$ and soundness error $s$ if the following hold.
\begin{itemize}[leftmargin=*]
 \item \textbf{Completeness.} For every $I\in\mathcal Y$, $\Pr[\langle P,V\rangle(I)=1]\ge c$.
 \item \textbf{Soundness.} For every $I\in\mathcal N$ and every prover strategy $P^*$, $\Pr[\langle P^*,V\rangle(I)=1]\le s$.
\end{itemize}
The completeness is \emph{perfect} if $c=1$. The proof is \emph{public coin}~\cite{STOC:Babai85,STOC:GolSip86} if every verifier message is a fresh uniformly random string and the verifier's decision is a deterministic function of $I$ and the transcript. An \emph{argument} with soundness error $s$ against provers of a given size is defined in the same way, except that soundness is required only for provers $P^*$ implemented by nonuniform circuits of at most that size; each statement specifies the size bound.
\end{definition}

The communication of a protocol is the total number of bits exchanged, and running times count bit operations. When the instances in $\mathcal Y$ have witnesses, as in \cref{sec:zk}, the honest prover also receives a witness for $I$; we then write $P(w)$ for the honest prover with witness $w$.

\begin{definition}[Computational zero-knowledge~\cite{GolMicRac89,JC:GolOre94}]\label{def:zk}
Let $(P,V)$ be an interactive proof for $(\mathcal Y,\mathcal N)$ in which the honest prover receives a witness $w$ for $I\in\mathcal Y$, and let $\kappa$ be the security parameter. For a verifier $V^*$ with auxiliary input $\mathsf{aux}$, let $\mathrm{View}_{V^*}\langle P(w),V^*(\mathsf{aux})\rangle(I)$ denote its view, consisting of $\mathsf{aux}$, its random tape, and the messages it receives. The proof is \emph{computational zero-knowledge} if for every probabilistic polynomial-time verifier $V^*$ there is a probabilistic polynomial-time simulator $\mathsf{Sim}$ such that, for all sequences of instances $I_\kappa\in\mathcal Y$ with witnesses $w_\kappa$ and auxiliary inputs $\mathsf{aux}_\kappa$, all of length polynomial in $\kappa$,
\[
 \mathrm{View}_{V^*}\langle P(w_\kappa),V^*(\mathsf{aux}_\kappa)\rangle(I_\kappa)\approx_{\mathrm c}\mathsf{Sim}(I_\kappa,\mathsf{aux}_\kappa).
\]
\end{definition}

\begin{definition}[Interactive PCPs~\cite{KRIPCP08}]\label{def:ipcp}
An \emph{interactive PCP} for $(\mathcal Y,\mathcal N)$ with completeness $c$ and soundness error $s$ is a pair $(P,V)$ that runs in two phases on an instance $I$. First, $P$ outputs a proof string $\pi$. Then $P$ and $V^{\pi}$ interact, where $V$ has oracle access to $\pi$, and $V$ accepts or rejects. We require the following.
\begin{itemize}[leftmargin=*]
 \item \textbf{Completeness.} For every $I\in\mathcal Y$, the honest proof $\pi$ and the honest prover make $V$ accept with probability at least $c$.
 \item \textbf{Soundness.} For every $I\in\mathcal N$, every proof string $\pi^*$, and every prover strategy $P^*$ for the second phase, $\Pr[\langle P^*,V^{\pi^*}\rangle(I)=1]\le s$.
\end{itemize}
The \emph{proof length} is $|\pi|$, the \emph{query complexity} is the maximum number of bits of $\pi$ that $V$ reads, and the communication is that of the second phase. The interactive PCP is \emph{public coin} if every verifier message in the second phase is a fresh uniformly random string and the verifier's decision is a deterministic function of $I$, the transcript, and the bits of $\pi$ that it reads.
\end{definition}

Since $\pi$ is sent first, it is fixed before the verifier samples any coins. The interactive PCPs of \cref{sec:ipcp} are public coin.

\subsection{Cryptographic Primitives}\label{sec:prelim-crypto}

\begin{definition}[Pseudorandom generators~\cite{BluMic84,FOCS:Yao82a}]\label{def:prg}
Let $\ell(\kappa)>\kappa$. A deterministic polynomial-time algorithm ${\mathsf G}$ mapping $\bits^\kappa$ to $\bits^{\ell(\kappa)}$ is a \emph{pseudorandom generator} if ${\mathsf G}(U_\kappa)\approx_{\mathrm c}U_{\ell(\kappa)}$.
\end{definition}

One-way functions (OWF) secure against nonuniform polynomial-size circuits yield pseudorandom generators with the same security and any polynomial stretch~\cite{HILL}. \Cref{sec:zk} uses $\ell(\kappa)=3\kappa$.

\begin{definition}[Commitments]\label{def:commitment}
A bit commitment scheme with a receiver's first message is a deterministic polynomial-time algorithm $\Com$ that, on a first message $R\in\bits^{\poly(\kappa)}$, a bit $\beta\in\bits$, and randomness $s\in\bits^\kappa$, outputs a commitment $\Com_R(\beta;s)$. The receiver sends $R$, the sender commits by sending $\Com_R(\beta;s)$ for a fresh uniform $s$, and an opening reveals $(\beta,s)$, which the receiver checks by recomputing the commitment.
\begin{itemize}[leftmargin=*]
 \item \textbf{Statistical binding.} $\Pr_R\bigl[\exists\,s_0,s_1:\Com_R(0;s_0)=\Com_R(1;s_1)\bigr]=\negl(\kappa)$, where $R$ is uniform.
 \item \textbf{Computational hiding.} For every sequence of first messages $R_\kappa$, $\Com_{R_\kappa}(0;U_\kappa)\approx_{\mathrm c}\Com_{R_\kappa}(1;U_\kappa)$.
\end{itemize}
\end{definition}

Binding holds even against computationally unbounded senders, and hiding holds for every first message, including one chosen by a malicious receiver. \Cref{sec:zk-protocol} uses Naor's scheme~\cite{Naor}, built from a pseudorandom generator with $\ell(\kappa)=3\kappa$; its binding fails with probability at most $2^{-\kappa}$.

\begin{definition}[Five-party MPC]\label{def:mpc}
Let $f$ be a functionality of five inputs. In a five-party protocol for $f$, party $i\in[5]$ has input $\mathsf w_i$ and a private random tape. The \emph{view} $\mathrm{View}_i$ of party $i$ consists of $\mathsf w_i$, its random tape, the messages it sends and receives, and its output.
\begin{itemize}[leftmargin=*]
 \item \textbf{Perfect correctness.} For all inputs and random tapes, every party outputs $f(\mathsf w_1,\ldots,\mathsf w_5)$.
 \item \textbf{Perfect $2$-privacy in the semi-honest model.} For every set $\mathcal T\subseteq[5]$ with $|\mathcal T|\le2$, there is a probabilistic polynomial-time simulator $\mathsf{Sim}_{\mathcal T}$ such that, for all inputs, $\mathsf{Sim}_{\mathcal T}\bigl((\mathsf w_i)_{i\in\mathcal T},f(\mathsf w_1,\ldots,\mathsf w_5)\bigr)$ and $(\mathrm{View}_i)_{i\in\mathcal T}$ are identically distributed.
\end{itemize}
\end{definition}

The protocol of Ben-Or, Goldwasser, and Wigderson~\cite{STOC:BGW88} satisfies \cref{def:mpc}. The MPC-in-the-head transformation of~\cite{IKOS} combines such a protocol with a commitment scheme to give a zero-knowledge proof for NP; \cref{sec:zk-protocol} gives the instantiation.

Private information retrieval (PIR) was introduced in~\cite{JACM:CGKS98}, and the one-server computational setting used here goes back to~\cite{FOCS:KusOst97}.

\begin{definition}[One-server PIR]\label{def:pir}
A one-server PIR scheme consists of probabilistic algorithms $\PIRQuery$ and $\PIRAnswer$ and a deterministic algorithm $\PIRRecover$. For a security parameter $\tau$, a database $\mathsf{db}\in\bits^M$, and an index $z\in[M]$, the client computes $(\mathsf{qry},\mathsf{st})\leftarrow\PIRQuery(\tau,M,z)$, the server computes $\mathsf{ans}\leftarrow\PIRAnswer(\tau,\mathsf{db},\mathsf{qry})$, and the client outputs $\PIRRecover(\tau,M,z,\mathsf{qry},\mathsf{st},\mathsf{ans})$. For functions $\mathrm{err}$, $\mathrm{adv}$, and $\mathrm{size}$ of $\tau$, we require the following.
\begin{itemize}[leftmargin=*]
 \item \textbf{Correctness.} For all $M$, $\mathsf{db}\in\bits^M$, and $z\in[M]$, the output differs from $\mathsf{db}[z]$ with probability at most $\mathrm{err}(\tau)$.
 \item \textbf{Privacy.} For all $M$ and $z,z'\in[M]$, every nonuniform circuit of size at most $\mathrm{size}(\tau)$ distinguishes the query $\mathsf{qry}$ generated for $z$ from the query generated for $z'$ with advantage at most $\mathrm{adv}(\tau)$.
\end{itemize}
\end{definition}

\Cref{sec:one-round-protocol} assumes a scheme with $\mathrm{err}(\tau)=\mathrm{adv}(\tau)=2^{-\tau^3}$ and $\mathrm{size}(\tau)=2^{\tau^3}$, with query generation and recovery in time $\poly(\tau,\log M)$, answering in time $\poly(\tau,M)$, and communication $\poly(\tau,\log M)$, as in~\cite[Definition~3]{KR}.

\section{Approximate GKR}\label{sec:agkr}

The Goldwasser--Kalai--Rothblum (GKR) protocol~\cite{GKR} verifies a circuit's output by reducing a claim about each layer to a claim about the next, until the verifier reaches the known input. We carry out this reduction over $\C$ using finite-precision messages and approximate checks, so the verifier need not repeat the circuit computation. The main numerical task is to separate honest rounding errors from a substantially incorrect output: the tolerances must accommodate the former without allowing the latter to disappear during verification. We first describe the protocol, assuming access to evaluations of the polynomial extensions of the wiring predicates; \cref{sec:implementation} supplies this access. \crefrange{sec:completeness}{sec:complexity} establish completeness, soundness, and polynomial-time implementation. \cref{stable:section} then gives a faster prover and a sharper precision bound for the unique extensions of degree below $h$ in each variable, while \cref{sec:conditioning} explains why more general extensions require a separate bound on their coefficient sizes.

\subsection{Protocol}\label{sec:protocol}

The prover claims that a circuit $C$, evaluated on input $x$, has output $y$. Let $C$ be a layered arithmetic circuit of depth $L\ge1$, with the single output at layer $0$ and the inputs at layer $L$. Each noninput gate adds or multiplies the values of two gates in the next layer, called its \emph{children}. Computation therefore proceeds from layer $L$ toward layer $0$, whereas verification proceeds in the opposite direction. To represent gate labels algebraically, fix a power of two $h\ge2$ and let
\[
 H=\left\{\exp(2\pi\ii a/h):0\le a<h\right\}.
\]
The root with index $a$ represents the $\log h$ bits of that index, so a label in $H^{m_i}$ represents a binary label of $m_i\log h$ bits. We set $m_0=0$ for the output's empty label and require $m_i\ge1$ for $i>0$. Pad binary labels to a multiple of $\log h$ bits, assign to each unused label a gate whose value is zero for every input in the specified domain, and implement copies by adding zero. Layer $i$ then has $h^{m_i}$ labels; let $S$ be the total size after padding. The values of the $n$ input gates, including constant gates, are at known labels; unused labels in the input layer have value zero. Their real and imaginary parts have finite binary expansions. Throughout, $C(x)$ denotes the exact output, before numerical approximation. The elements of $H$ are label coordinates, not gate values, and all arithmetic is over $\C$.

We encode the vector of exact gate values $(\GateValues_i(g))_{g\in H^{m_i}}$ by a polynomial $\GateExtension_i$ that agrees with it at every label. For $a\in H$ and $g\in H^{m_i}$, define
\begin{equation}\label{eq:lagrange}
 \begin{aligned}
 \Lag_a(z)&=\frac1h\sum_{t=0}^{h-1}(z/a)^t,
 &\chi_g(z)&=\prod_j\Lag_{g_j}(z_j),\\
 \GateExtension_i(z)&=\sum_{g\in H^{m_i}}\GateValues_i(g)\chi_g(z).
 \end{aligned}
\end{equation}
The geometric sum makes $\Lag_a$ a Lagrange basis polynomial, wherin it equals one at $a$ and zero at the other points of $H$. The product $\chi_g$ equals one at $g$ and zero at every other label, giving $\GateExtension_i(g)=\GateValues_i(g)$. For example, when $h=2$, the Lagrange basis polynomials are $\Lag_1(z)=(1+z)/2$ and $\Lag_{-1}(z)=(1-z)/2$. The extension has degree at most $h-1$ in each coordinate and is unique with these degree bounds: a univariate polynomial of degree below $h$ cannot vanish at all $h$ labels unless it is zero, and applying this fact one coordinate at a time proves the multivariate statement. Away from $H^{m_i}$, $\GateExtension_i$ gives a polynomial evaluation rather than the value of an individual gate.

For gate labels $g$ of layer $i$ and $u, v$ of layer $i+1$, let $\mathrm{add}_i(g, u, v) = 1$ if $g$ is an addition gate with left child $u$ and right child $v$, and $\mathrm{add}_i(g, u, v) = 0$ otherwise; define $\mathrm{mult}_i$ in the same way for multiplication gates. We call $\mathrm{add}_i$ and $\mathrm{mult}_i$ the \emph{wiring predicates} of layer $i$. On gate labels, $\Add_i(g,u,v)$ equals one exactly when $g$ is an addition gate with ordered children $u,v$, and is zero otherwise; define $\Mul_i$ analogously for multiplication gates. Let $\AddPoly_i,\MulPoly_i$ be extensions of these indicators, each of degree at most a nonnegative integer $\delta$ in every variable. They need to agree with the indicators only on $H^{m_i}\times H^{m_{i+1}}\times H^{m_{i+1}}$. There are unique extensions of degree below $h$ in each coordinate, but we also allow higher-degree extensions for the constructions in \cref{sec:implementation}; their values away from the grid affect the required numerical precision. For now, both parties may query these polynomials at tuples of roots of unity to any requested absolute error. In particular, Lemma~\ref{lem:space-circuit} supplies suitable extensions for space-bounded computations by recovering the bits encoded in the labels. Write
\[
 m=\max_i m_i,\qquad D=\delta+h-1,\qquad
 k_i=m_i+2m_{i+1},\qquad k=\sum_{i=0}^{L-1}k_i.
\]
Here $m$ is the maximum number of coordinates in a gate label, $D$ will bound each sum-check message's degree, $k_i$ is the number of rounds at layer $i$, and $k$ is the total number of rounds.

Each addition or multiplication gate uses the values of two gates in the next layer. To prevent the number of claims from doubling at every layer, the verifier maintains one scalar claim about a weighted sum of two evaluations. At layer $i$, it holds a number $\widehat T_i$ that claims to approximate
\begin{equation}\label{eq:claim}
 T_i=c_{i,0}\GateExtension_i(\rho_{i,0})
       +c_{i,1}\GateExtension_i(\rho_{i,1}).
\end{equation}
The evaluation points and coefficients are known to the verifier; the value of $T_i$ is not yet established. Initially, $\widehat T_0=y$, both points are empty, and $(c_{0,0},c_{0,1})=(1,0)$, so $T_0=C(x)$. At later layers, the coefficients have the form $(1,\lambda)$ with $|\lambda|=1$, and every coordinate of the evaluation points lies on the unit circle. The random combination changes the coefficients without moving the evaluation points.

To reduce this claim to the next layer, first substitute the addition or multiplication defining each gate. Interpolation gives a coefficient for each gate value; the following polynomials extend those coefficients away from $H^{m_i}$:
\begin{align}
 \eqpoly_H(\rho,z)&=\sum_{a\in H}\Lag_a(\rho)\Lag_a(z)
 =\frac1h\left(1+\sum_{t=1}^{h-1}\rho^tz^{h-t}\right),\label{eq:kernel}\\
 \eqpoly(\rho,g)&=\prod_j\eqpoly_H(\rho_j,g_j),\qquad
 \ParentWeight_i(g)=\sum_{b=0}^1c_{i,b}\eqpoly(\rho_{i,b},g).
 \label{eq:parent-weight}
\end{align}
The formula for $\eqpoly_H$ follows by expanding the Lagrange basis polynomials and using $\sum_{a\in H}a^{-t}=h$ when $h\mid t$, and zero otherwise. At a label $g\in H^{m_i}$, we have $\eqpoly(\rho,g)=\chi_g(\rho)$, so $\ParentWeight_i(g)$ is the coefficient of $\GateValues_i(g)$ in $T_i$. Substituting the gate operations yields
\begin{align}
 T_i&=\sum_{g\in H^{m_i}}\sum_{u,v\in H^{m_{i+1}}}\LayerPoly_i(g,u,v),
 \label{eq:layer-sum}\\
 \begin{split}
 \LayerPoly_i(g,u,v)&=\ParentWeight_i(g)\bigl[
 \AddPoly_i(g,u,v)\bigl(\GateExtension_{i+1}(u)+\GateExtension_{i+1}(v)\bigr)\\
 &\hspace{37mm}+\MulPoly_i(g,u,v)\GateExtension_{i+1}(u)\GateExtension_{i+1}(v)\bigr].
 \end{split}\label{eq:layer-poly}
\end{align}
For a fixed parent label, the gate indicators select its ordered pair of children. For a fixed parent label $g$, the wiring predicates are zero unless $u$ and $v$
are the children of $g$, so
\[
  \sum_{u, v \in H^{m_{i+1}}}
  \Bigl(
    \widetilde{\mathrm{add}}_i(g, u, v)\bigl(\widetilde{\GateValues}_{i+1}(u)
      + \widetilde{\GateValues}_{i+1}(v)\bigr)
    + \widetilde{\mathrm{mult}}_i(g, u, v)\,\widetilde{\GateValues}_{i+1}(u)\,
      \widetilde{\GateValues}_{i+1}(v)
  \Bigr)
  = \GateValues_i(g).
\]
Weighting by $\Psi_i(g)$ and summing over $g$ then gives $T_i$.Since $g$, $u$, and $v$ range only over gate labels, this identity holds for
every extension of the wiring predicates that agrees with them on labels. Each coordinate of $\LayerPoly_i$ has degree at most $D=\delta+h-1$: a coordinate of $g$ occurs in $\ParentWeight_i$ and in $\AddPoly_i$ or $\MulPoly_i$, while a coordinate of $u$ or $v$ occurs in $\AddPoly_i$ or $\MulPoly_i$ and in one evaluation of $\GateExtension_{i+1}$.

Sum-check reduces \cref{eq:layer-sum} to one evaluation by removing one summation variable per round. In round $j$, after challenges $r_1,\ldots,r_{j-1}$ have been chosen, define
\[
 q_j(X)=\sum_{a\in H^{k_i-j}}
 \LayerPoly_i(r_1,\ldots,r_{j-1},X,a).
\]
That is, the first $j-1$ variables of $F_i$ are set to the challenges
$r_1, \dots, r_{j-1}$, the $j$th variable is left free as $X$, and the last
$k_i - j$ variables are summed over $H$. Thus $q_j$ has degree at most $D$. Summing $q_1$ over $H$ gives $T_i$; in later rounds, summing $q_j$ gives $q_{j-1}(r_{j-1})$. The prover sends an approximation $p_j$ as a coefficient vector. The verifier checks its sum against the current claim, then evaluates it at a fresh random challenge to obtain the next claim; it does not compute $q_j$ itself. Challenges are drawn from
\[
 \Omega_N=\left\{\exp(2\pi\ii a/N):0\le a<N\right\},
\]
where $N$ is a power of two chosen in \cref{sec:soundness}. Unlike $H$, which supplies labels and summation points, $\Omega_N$ supplies random test points. A uniform integer index specifies an exact root; only its numerical evaluation is approximate, and all error bounds refer to the exact root specified by that index. Each polynomial is sent before its challenge is sampled. After $k_i$ rounds, the verifier holds a claim $Z_i$ about $\LayerPoly_i(g_i,u_i,v_i)$ at the challenge vector. It can evaluate $\ParentWeight_i(g_i)$ and query $\AddPoly_i,\MulPoly_i$ there, leaving only $\GateExtension_{i+1}(u_i)$ and $\GateExtension_{i+1}(v_i)$ unknown. The prover supplies claims $a_i,b_i$ for them. After checking the residual $Z_i-w_i^+(a_i+b_i)-w_i^\times a_ib_i$, the verifier samples a fresh $\lambda_i$ and combines the claimed evaluations as $a_i+\lambda_i b_i$. Both claimed evaluations must be fixed first; \cref{sec:soundness} shows why this timing prevents their errors from canceling with high probability.

We now specify the bounds needed to implement the reduction with finite
precision. Let $\BV, \BW \ge 1$ be integers, known to both parties, such that
every exact gate value has modulus at most $\BV$ and $\|\AddPoly_i\|_1,\|\MulPoly_i\|_1\le\BW$. Here $\|f\|_1$ is the sum of the absolute values of the coefficients; it bounds $|f(z)|$ whenever every coordinate of $z$ has modulus at most one. Since each Lagrange basis polynomial has coefficient norm one, $\|\GateExtension_i\|_1\le h^{m_i}\BV$. For $|\rho|\le1$, \cref{eq:kernel} has coefficient norm at most one in $z$, giving $\|\ParentWeight_i\|_1\le2$. Define
\begin{equation}\label{eq:bounds}
 \begin{gathered}
 R=h^m\BV,\qquad U=R+1,\qquad W=2\BW,\\
 \CoefficientBound=h^{3m}W(2R+R^2)+1,\qquad K=W(2+U+R).
 \end{gathered}
\end{equation}
The bound $R$ controls $\GateExtension_{i+1}(u)$ and $\GateExtension_{i+1}(v)$, and $U$ leaves one unit for rounding. The bound $W$ controls the coefficient norms of $\ParentWeight_i\AddPoly_i$ and $\ParentWeight_i\MulPoly_i$, so $W(2R+R^2)$ bounds $\|\LayerPoly_i\|_1$. The polynomial $q_j$ sums at most $h^{k_i}\le h^{3m}$ restrictions of $\LayerPoly_i$; hence $\CoefficientBound$ bounds its coefficients, again with one unit for rounding. Finally, $K$ bounds how much errors in the two evaluations of $\GateExtension_{i+1}$ can change $w_i^+(a_i+b_i)+w_i^\times a_ib_i$, as shown in \cref{eq:gate-error}. The prover represents a complex number as $(A+\ii B)2^{-p}$, with $A,B\in\Z$. For each position occupied by a complex number in a prover message, fix $p$ and the bit lengths of $A,B$ before the interaction. Round polynomials are represented by their $D+1$ coefficients, including zeros. The verifier rejects incorrect lengths or coefficient counts, coefficients of modulus greater than $\CoefficientBound$, claimed evaluations of modulus greater than $U$, and an initial claim of modulus greater than $\BV+1$. These restrictions bound all messages on which the verifier performs arithmetic, including dishonest ones.

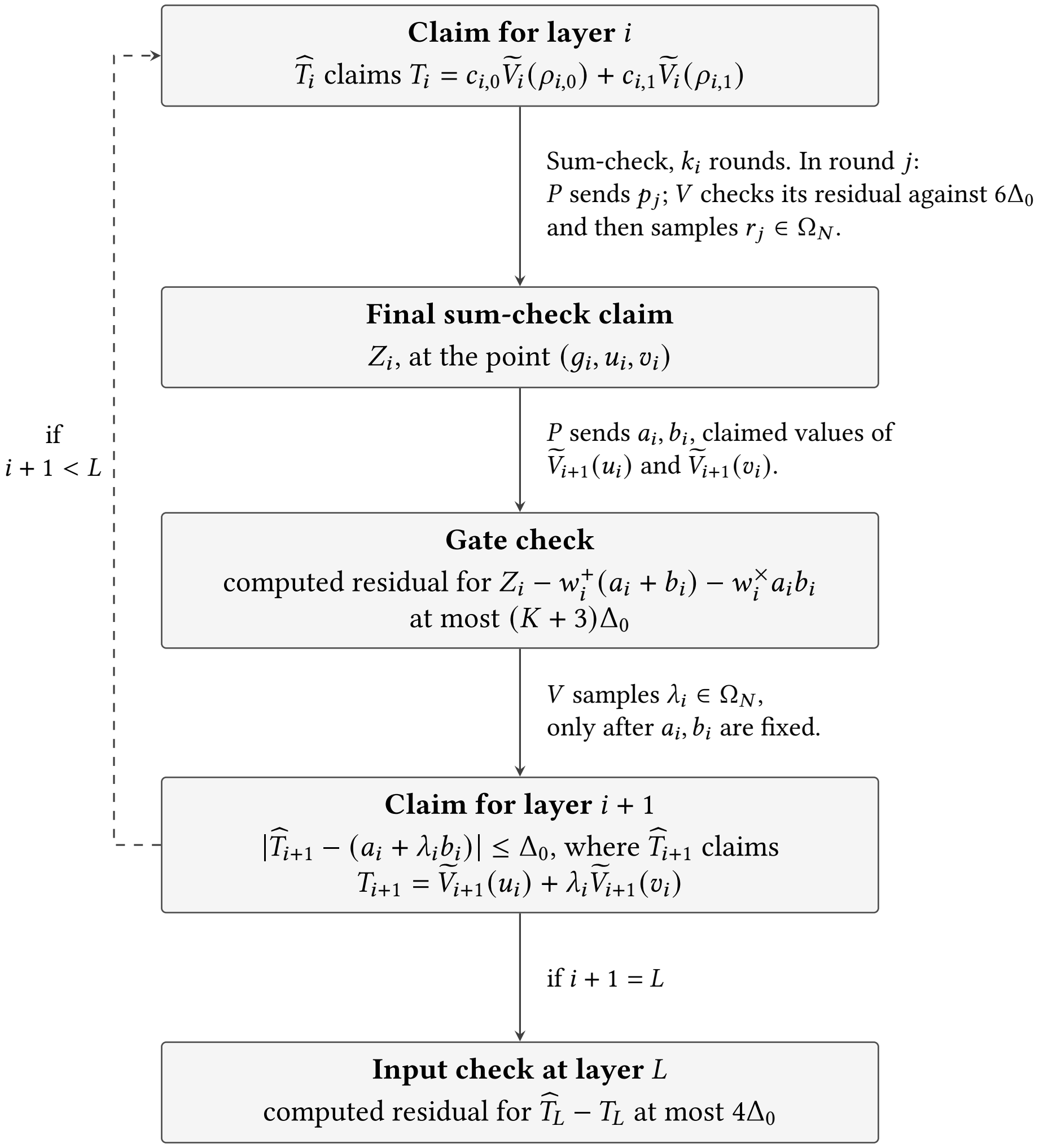
\begin{figure}[ht!]
\centering
\begin{tikzpicture}[
  x=1cm,y=1cm,>=stealth,font=\small,
  claim/.style={draw=black!70,fill=black!4,line width=.5pt,rounded corners=1.5pt,
    minimum width=8.4cm,align=center,inner xsep=6pt,inner ysep=5pt},
  flow/.style={->,draw=black!75,line width=.6pt},
  act/.style={anchor=west,align=left,font=\footnotesize,xshift=5pt}
]
\node[claim] (c0) at (0,0)
  {\textbf{Claim for layer $i$}\\[2pt]
   $\widehat T_i$ claims $T_i=c_{i,0}\GateExtension_i(\rho_{i,0})+c_{i,1}\GateExtension_i(\rho_{i,1})$};
\node[claim] (c1) at (0,-3.3)
  {\textbf{Final sum-check claim}\\[2pt]
   $Z_i$, at the point $(g_i,u_i,v_i)$};
\node[claim] (c2) at (0,-6.15)
  {\textbf{Gate check}\\[2pt]
   computed residual for $Z_i-w_i^+(a_i+b_i)-w_i^\times a_ib_i$\\
   at most $(K+3){\Tol}$};
\node[claim] (c3) at (0,-9.25)
  {\textbf{Claim for layer $i+1$}\\[2pt]
   $|\widehat T_{i+1}-(a_i+\lambda_ib_i)|\le{\Tol}$, where $\widehat T_{i+1}$ claims\\
   $T_{i+1}=\GateExtension_{i+1}(u_i)+\lambda_i\GateExtension_{i+1}(v_i)$};
\node[claim] (c4) at (0,-12.15)
  {\textbf{Input check at layer $L$}\\[2pt]
   computed residual for $\widehat T_L-T_L$ at most $4{\Tol}$};
\draw[flow] (c0.south) -- (c1.north)
  node[midway,act]{Sum-check, $k_i$ rounds. In round $j$:\\
   $P$ sends $p_j$; $V$ checks its residual against $6{\Tol}$\\
   and then samples $r_j\in\Omega_N$.};
\draw[flow] (c1.south) -- (c2.north)
  node[midway,act]{$P$ sends $a_i,b_i$, claimed values of\\
   $\GateExtension_{i+1}(u_i)$ and $\GateExtension_{i+1}(v_i)$.};
\draw[flow] (c2.south) -- (c3.north)
  node[midway,act]{$V$ samples $\lambda_i\in\Omega_N$,\\only after $a_i,b_i$ are fixed.};
\draw[flow] (c3.south) -- (c4.north)
  node[midway,act]{if $i+1=L$};
\draw[flow,dashed] (c3.west) -- ++(-0.55,0) |- (c0.west)
  node[pos=0.25,left,align=center,font=\footnotesize]{if\\$i+1<L$};
\end{tikzpicture}
\caption{One layer of approximate GKR. Boxes show claims and checks; arrows show the prover's ($P$) and verifier's ($V$) steps. After the gate check, the verifier combines the two claimed evaluations with a fresh $\lambda_i\in\Omega_N$, so the claim for the next layer again concerns points with coordinates in $\Omega_N$. The steps repeat for $i=0,\dots,L-1$, and the verifier checks the last claim directly from the inputs.}
\label{fig:agkr-layer-reduction}
\end{figure}

Fix $0<{\Tol}\le1$. The honest prover computes every round coefficient within ${\Tol}/[h(D+1)]$ and every claimed value of $\GateExtension_{i+1}$ within ${\Tol}$. Each verifier evaluation, linear combination, and residual computation has absolute error at most ${\Tol}$, where a \emph{residual} is the difference between the two sides of a checked identity. Every claim $Y_j$ computed during sum-check has modulus at most $(D+1)\CoefficientBound + 1$, and every combined claim $\widehat{T}_{i+1}$ has modulus at most $2U + 1$. \cref{sec:completeness} shows how to attain the required accuracy on every transcript passing the encoding, length, and magnitude checks. Errors in the verifier's approximate evaluations
of the wiring polynomials $\AddPoly_i$ and
$\MulPoly_i$ are included in the ${\Tol}$ error
allowance for the gate-check residual.
If each of these two evaluations has absolute
error at most $\xi$, the resulting change in
$Z_i-w_i^+(a_i+b_i)-w_i^\times a_ib_i$
is at most $2(2U+U^2)\xi$.
Reserve at most half of this allowance for
errors in these evaluations and the remainder
for finite-precision arithmetic.
Once the computed residual is represented as
a binary fraction, compare its squared modulus
with the squared threshold using exact integer
arithmetic. With these conventions, initialize $\widehat T_0=y$ and the claim points and coefficients as above, and perform the following steps for each $i=0,\ldots,L-1$, rejecting any message that fails an encoding, length, or magnitude check.
\begin{enumerate}
 \item \emph{Reduce the sum to one evaluation.} Set $\RunClaim_0=\widehat T_i$. At round $j=1,\ldots,k_i$, receive a degree-at-most-$D$ polynomial $p_j$ and reject if the computed residual for
 \[
 \RunClaim_{j-1}-\sum_{a\in H}p_j(a)
 \]
 has modulus greater than $6{\Tol}$. Compute the sum directly from the coefficients using
 \[
 \sum_{a\in H}p_j(a)
   =h\sum_{\substack{0\le t\le D\\h\mid t}}[X^t]p_j(X),
 \]
 where $[X^t]p_j(X)$ denotes the coefficient of $X^t$. The root-sum identity used in \cref{eq:kernel} gives this formula, avoiding evaluations at all $h$ labels. If the check passes, sample a fresh uniform $r_j\in\Omega_N$ and compute $\RunClaim_j$ with $|\RunClaim_j-p_j(r_j)|\le {\Tol}$. After the last round, write the challenge vector as $(g_i,u_i,v_i)$ and set $Z_i=\RunClaim_{k_i}$.
 \item \emph{Check the claimed evaluations.} Receive the claimed evaluations $a_i,b_i$, evaluate $\ParentWeight_i(g_i)$, and query $\AddPoly_i$ and $\MulPoly_i$ once each at $(g_i,u_i,v_i)$. Writing the exact coefficients as
 \[
 \begin{aligned}
 w_i^+&=\ParentWeight_i(g_i)\AddPoly_i(g_i,u_i,v_i),\\
 w_i^\times&=\ParentWeight_i(g_i)\MulPoly_i(g_i,u_i,v_i),
 \end{aligned}
 \]
 reject if the computed residual for
 \[
 Z_i-w_i^+(a_i+b_i)-w_i^\times a_ib_i
 \]
 has modulus greater than $(K+3){\Tol}$.
 \item \emph{Form the next layer's claim.} Only after $a_i,b_i$ are fixed, sample a fresh uniform $\lambda_i\in\Omega_N$. Compute $\widehat T_{i+1}$ within ${\Tol}$ of $a_i+\lambda_i b_i$ and set
 \[
 (\rho_{i+1,0},\rho_{i+1,1})=(u_i,v_i),\qquad
 (c_{i+1,0},c_{i+1,1})=(1,\lambda_i).
 \]
 The resulting claim has the form of \cref{eq:claim} at layer $i+1$.
\end{enumerate}
At layer $L$, evaluate the two input extensions directly from the values of the $n$ input gates, including constant gates; unused labels contribute zero. Accept if no earlier check has rejected and the computed residual for $\widehat T_L-T_L$ has modulus at most $4{\Tol}$. Figure~\ref{fig:agkr-layer-reduction} summarizes one layer's message order. The protocol uses two error thresholds with different roles: ${\Tol}$ bounds the allowed error in an honest output claim and sets the numerical tolerances, whereas $\Delta$ specifies how far an output must be from correct for soundness to apply. \cref{sec:soundness} chooses ${\Tol}$ sufficiently small relative to $\Delta$ and the protocol parameters.

\begin{theorem}[Approximate GKR]\label{thm:agkr}
Fix the circuit and the polynomial extensions $\AddPoly_i,\MulPoly_i$ of the wiring predicates above, with the stated bounds on gate values and coefficients, and let $y=(A+\ii B)2^{-p}$, where $A,B\in\Z$, with $p$ and the integer encoding lengths fixed for the output claim. For rational parameters $\Delta>0$ and $0<\eps<1$, the protocol admits choices of $N$, $0<{\Tol}\le1$, and finite precision such that an honest prover is accepted with probability one whenever $|y-C(x)|\le {\Tol}$, while every prover is accepted with probability at most $\eps$ whenever $|y-C(x)|>\Delta$. The protocol is public coin. The verifier queries $\AddPoly_i$ and $\MulPoly_i$ once each per layer and evaluates the input extension twice at the end. If $b$ bounds the input, message, and working bit lengths and $\ell=\log N$, communication is
\[
 O\bigl(kDb+(k+L)\ell\bigr).
\]
With the costs of evaluating the polynomial extensions $\AddPoly_i,\MulPoly_i$ of the wiring predicates charged separately, the honest prover takes $\poly(S,D,b+\ell)$ time, and the verifier takes $O(nmh+kD+Lmh)\poly(b+\ell)$ time and $O(Db+m\ell)$ space.
\end{theorem}

\subsection{Completeness}\label{sec:completeness}

An honest prover satisfies \cref{eq:layer-sum} exactly. To prove that its finite-precision messages also pass, we first bound the errors introduced when computing those messages and then compare the resulting residuals with the verifier's tolerances. Model each numerical computation as a directed acyclic graph of arithmetic operations, with depth at most $d_\mathrm{num}$ and exact intermediate values of modulus at most $B_\mathrm{num}\ge1$. For the prover, this includes gate values and messages; for the verifier, the bound must hold for every message passing the encoding, length, and magnitude checks. Approximate roots and supplied evaluations of the polynomial extensions of the wiring predicates are treated as approximate inputs. If each input approximation and arithmetic operation introduces absolute error at most $e_\mathrm{op}$, a computation of depth $d$ has output error at most $e_\mathrm{op}(2B_\mathrm{num}+3)^d$, provided this quantity is at most one. To see this, induct on depth. At depth zero the error is at most $e_\mathrm{op}$. If both operand errors are at most $e_\mathrm{in}\le1$, addition contributes at most $2e_\mathrm{in}+e_\mathrm{op}$, while multiplication contributes at most $2B_\mathrm{num}e_\mathrm{in}+e_\mathrm{in}^2+e_\mathrm{op}\le(2B_\mathrm{num}+1)e_\mathrm{in}+e_\mathrm{op}$. With $A_\mathrm{num}=2B_\mathrm{num}+3$ and $e_\mathrm{in}=e_\mathrm{op}A_\mathrm{num}^{j-1}\ge e_\mathrm{op}$, either error is at most $e_\mathrm{op}A_\mathrm{num}^j$. The assumed upper bound ensures $e_\mathrm{in}\le1$ throughout.

The preceding error bound can be attained
with fixed-point arithmetic.
We represent $(A+\ii B)2^{-p}$ by the pair of
fixed-width signed integers $(A,B)$.
Compute integer products exactly before
rescaling, and round each component to the
nearest multiple of $2^{-p}$.
The resulting absolute error in the complex
value is at most $2^{-p}/\sqrt{2}$. To attain an error target $\eta\le1$, it suffices to use
\[
 p\ge\ceil{\log(1/\eta)+d_\mathrm{num}\log(2B_\mathrm{num}+3)}+3
\]
fractional bits and $O(1+\log B_\mathrm{num})$ integer bits. Compute roots to three extra fractional bits using \cref{lem:root-bits}, then round them; the resulting error is less than $(1/4+1/\sqrt2)2^{-p}<2^{-p}$. Long sums may instead be accumulated sequentially: summing $t$ terms, each accurate within $e_\mathrm{in}$, with addition error at most $e_\mathrm{op}$, gives total error at most $te_\mathrm{in}+(t-1)e_\mathrm{op}$. Consequently, if $B_\mathrm{num}\le2^{b_\mathrm{num}}$, the working bit length is
\[
 O\bigl(1+b_\mathrm{num}+\log(1/\eta)+d_\mathrm{num}(b_\mathrm{num}+1)\bigr).
\]
Polynomially bounded precision parameters therefore give polynomial bit complexity for a polynomial-size arithmetic computation.

To generate a round message, the prover evaluates $q_j$ at roots of unity and recovers its coefficients by Fourier inversion. For a polynomial $f$ of degree at most $D$, choose a power of two $J>D$ with $J\le2(D+1)$ and set $\omega_J=\exp(2\pi\ii/J)$. Then
\begin{equation}\label{eq:fourier}
 [X^a]f(X)=\frac1J\sum_{t=0}^{J-1}f(\omega_J^t)\omega_J^{-at},
 \qquad 0\le a\le D.
\end{equation}
Expanding $f$ and using the root sum $\sum_{t=0}^{J-1}\omega_J^{ts}=J$ when $J\mid s$, and zero otherwise, leaves only the desired coefficient because $\deg f<J$. Each evaluation of $q_j$ sums values of $\LayerPoly_i$ over the remaining labels. Starting from the original gate values, the preceding rounding procedure computes these evaluations and the Fourier sums using balanced addition trees. The coefficients $J^{-1}\omega_J^{-at}$ have absolute sum one, so errors of at most $e_\mathrm{ev}$ in the evaluations contribute at most $e_\mathrm{ev}$ to each recovered coefficient, independently of $J$  Compute a coefficient within ${\Tol}/[2h(D+1)]$ and round it with error at most the same amount. The total coefficient error is then at most ${\Tol}/h$, and the transmitted polynomial differs from $q_j$ by at most ${\Tol}/h$ everywhere on $|z|\le1$. Evaluations of $\GateExtension_{i+1}$ are computed within ${\Tol}$ in the same manner. If $\normone{\LayerPoly_i}\le\LayerPoly_i^*$, restricting coordinates to the unit disk and summing the remaining labels bounds every true round coefficient by $h^{k_i}\LayerPoly_i^*$. Adding one covers rounding when ${\Tol}\le1$, so each coefficient needs $O(1+\log(1+h^{k_i}\LayerPoly_i^*))$ integer bits and $\ceil{\log(2h(D+1)/{\Tol})}$ fractional bits. Fixing the evaluation order, precision, and rounding rule makes the honest prover deterministic on finite binary strings.

These message bounds imply acceptance for every challenge sequence. In sum-check, the transmitted polynomial's sum over $H$ differs from the true sum by at most ${\Tol}$, and evaluating it and rounding the result gives a next-claim error of at most ${\Tol}/h+{\Tol}\le2{\Tol}$. The first claim at each layer has error at most $3{\Tol}$: the initial claim already meets this bound, and later combined claims have at most $2{\Tol}$ of message error and ${\Tol}$ of arithmetic error. The first computed consistency residual is therefore at most $3{\Tol}+{\Tol}+{\Tol}=5{\Tol}$, and later ones are at most $4{\Tol}$, both below the $6{\Tol}$ tolerance. For the check of $Z_i-w_i^+(a_i+b_i)-w_i^\times a_ib_i$, let $e_a=a_i-\GateExtension_{i+1}(u_i)$ and $e_b=b_i-\GateExtension_{i+1}(v_i)$ be the errors in the received values, and $e=\max\{|e_a|,|e_b|\}$. The bounds in \cref{eq:bounds} give
\begin{equation}\label{eq:gate-error}
 |w_i^+(e_a+e_b)+w_i^\times(a_ib_i-\GateExtension_{i+1}(u_i)\GateExtension_{i+1}(v_i))|\le Ke.
\end{equation}
Indeed, $a_ib_i-\GateExtension_{i+1}(u_i)\GateExtension_{i+1}(v_i)=a_ie_b+\GateExtension_{i+1}(v_i)e_a$, with $|a_i|\le U$ and $|\GateExtension_{i+1}(v_i)|\le R$. This is the role of $K$: it controls the effect of errors in the claimed evaluations on $w_i^+(a_i+b_i)+w_i^\times a_ib_i$. The honest computed residual for $Z_i-w_i^+(a_i+b_i)-w_i^\times a_ib_i$ is at most $2{\Tol}+K{\Tol}+{\Tol}=(K+3){\Tol}$, and the final input residual is at most $3{\Tol}+{\Tol}=4{\Tol}$. All honest messages also meet the magnitude bounds in \cref{eq:bounds}, so every check accepts, regardless of the challenges.

\subsection{Soundness}\label{sec:soundness}

For soundness, we follow an incorrect claim from the output toward the input. At each reduction, either a check rejects or the error in the next claim exceeds its recursively defined threshold, except with small probability. The numerical setting requires more than a bound on the number of zeros of a polynomial: a nonzero value can still be too small for an approximate check to detect. We therefore begin with a bound on how often a polynomial is small relative to its maximum on the unit circle. This will control both sum-check, through the difference $p_j-q_j$, and the random linear combination that merges the claimed evaluations.

\begin{samepage}
\begin{lemma}[Anti-concentration]\label{lem:circle}
Let $f\in\C[X]$ be nonzero of degree at most $D\ge1$, and let $M(f)=\max_{|z|=1}|f(z)|$. For $0<\alpha<1$,
\begin{equation}\label{eq:small-ball}
 \Pr_{r\leftarrow\Omega_N}[|f(r)|\le\alpha M(f)]
 \le D(\alpha^{1/D}+N^{-1}).
\end{equation}
Moreover, $M(f)\ge|\sum_{a\in H}f(a)|/h$. For fixed $e_a,e_b\in\C$ and $e=\max\{|e_a|,|e_b|\}>0$,
\begin{equation}\label{eq:linear}
 \Pr_{\lambda\leftarrow\Omega_N}[|e_a+\lambda e_b|\le\beta e]
 \le\beta+N^{-1}\qquad(0<\beta<1).
\end{equation}

There is a set $\mathcal Z\subseteq\Omega_N$ of at most $\deg f$ points,
containing every zero of $f$ in $\Omega_N$, such that for every $0<s<1/D$,
\begin{equation}\label{eq:inverse-moment}
 \frac1N\sum_{r\in\Omega_N\setminus \mathcal Z}
 \left(\frac{M(f)}{|f(r)|}\right)^s
 \le\frac1{1-Ds}.
\end{equation}
\end{lemma}
\end{samepage}
\begin{proof}[Proof of~\cref{lem:circle}]
The geometric observation is that a small normalized distance to one root restricts a challenge to a short arc. For $\zeta=|\zeta|e^{\ii\psi_0}$ and $0<t<1$,
\[
 \frac{|e^{\ii\psi}-\zeta|^2}{(1+|\zeta|)^2}
 =\left(\frac{1-|\zeta|}{1+|\zeta|}\right)^2
 +\frac{4|\zeta|}{(1+|\zeta|)^2}\sin^2\frac{\psi-\psi_0}{2}
 \ge\sin^2\frac{\psi-\psi_0}{2}.
\]
Thus $|e^{\ii\psi}-\zeta|\le t(1+|\zeta|)$ confines $\psi$ to an arc of length at most $4\arcsin t\le2\pi t$. Since the challenges are equally spaced, this arc contains at most $Nt+1$ points of $\Omega_N$. Now factor a nonconstant polynomial as $f=\mathrm{lc}(f)\prod_{j=1}^d(X-\zeta_j)$. The bound $M(f)\le|\mathrm{lc}(f)|\prod_j(1+|\zeta_j|)$ shows that the event in \cref{eq:small-ball} requires at least one factor to satisfy $|r-\zeta_j|\le\alpha^{1/d}(1+|\zeta_j|)$: otherwise the product would exceed $\alpha M(f)$. Applying the arc bound to each root and taking a union bound gives $d(\alpha^{1/d}+N^{-1})\le D(\alpha^{1/D}+N^{-1})$. For a nonzero constant polynomial the event is empty. Finally, every point of $H$ lies on the unit circle, so $|\sum_{a\in H}f(a)|\le hM(f)$. For the linear claim, $M(e_a+Xe_b)=|e_a|+|e_b|\ge e$, and \cref{eq:linear} follows from the degree-one case, with $e_b=0$ immediate.

For the inverse-moment bound, first suppose that $f$ has degree
$d\ge1$. For each root $\zeta_j$, choose a nearest grid point
$\zeta^\circ_j\in\Omega_N$ and put $\mathcal Z=\{\zeta^\circ_1,\ldots,\zeta^\circ_d\}$.
This set contains every grid zero of $f$. For $r\ne \zeta^\circ_j$, write
\[
 \Upsilon_j(r)=\frac{1+|\zeta_j|}{|r-\zeta_j|}\ge1.
\]
For $\varpi>1$, the points with $\Upsilon_j(r)>\varpi$ lie in an arc of length
at most $2\pi/\varpi$. If any grid point satisfies this inequality,
so does the nearest point $\zeta^\circ_j$. Deleting it removes the extra
point in the arc-counting bound, giving
\[
 \frac1N\#\{r\in\Omega_N\setminus\{\zeta^\circ_j\}:\Upsilon_j(r)>\varpi\}
 \le\frac1\varpi.
\]
Integrating this tail, for $0<s'<1$ we obtain
\[
 \frac1N\sum_{r\ne \zeta^\circ_j}\Upsilon_j(r)^{s'}
 \le 1+s'\int_1^\infty \varpi^{s'-2}\,d\varpi
 =\frac1{1-s'},
\]
where the sum is over the remaining grid points. Since
$M(f)/|f(r)|\le\prod_{j=1}^d\Upsilon_j(r)$ outside $\mathcal Z$, H\"older's
inequality with exponent $d$ gives
\[
 \frac1N\sum_{r\notin \mathcal Z}\left(\frac{M(f)}{|f(r)|}\right)^s
 \le\prod_{j=1}^d
 \left(\frac1N\sum_{r\notin \mathcal Z}\Upsilon_j(r)^{ds}\right)^{1/d}
 \le\frac1{1-ds}\le\frac1{1-Ds}.
\]
All sums are over $\Omega_N$. For constant $f$, take $\mathcal Z=\varnothing$.
The left-hand side is one, so the same bound holds.
\end{proof}

Figure~\ref{fig:agkr-small-arc} illustrates the arc argument. To apply the lemma to sum-check, as in~\cite{ApproxSumcheck26}, consider the difference $p_j-q_j$. If a substantially incorrect claim passes the consistency check, this difference has a large sum over $H$, hence a large maximum on the unit circle. Condition on the preceding transcript and the prover's message $p_j$.
Then $p_j-q_j$ is fixed, and the next challenge $r_j$ is uniform over $\Omega_N$.

\begin{figure}[tbp]
\centering
\begin{tikzpicture}[x=1cm,y=1cm,>=stealth,font=\small]
\begin{scope}[shift={(3.0,2.4)},scale=1.3]
  \fill[black!9] (0,0) -- (-5:1.9)
    arc[start angle=-5,end angle=75,radius=1.9] -- cycle;
  \draw[->,black!40,line width=.4pt] (-2.25,0) -- (2.60,0)
    node[right,font=\footnotesize] {$\operatorname{Re}$};
  \draw[->,black!40,line width=.4pt] (0,-2.15) -- (0,2.30)
    node[above,font=\footnotesize] {$\operatorname{Im}$};
  \draw[black!55,line width=.6pt] (0,0) circle[radius=1.9];
  \draw[black!55,densely dashed,line width=.4pt]
    (0,0) -- (-5:1.9) (0,0) -- (75:1.9);
  \draw[black!45,densely dashed,line width=.4pt] (0,0) -- (35:2.45);
  \draw[black,line width=2pt] (-5:1.9)
    arc[start angle=-5,end angle=75,radius=1.9];
  \foreach \angle in {0,15,...,345}
    \draw[fill=white,draw=black,line width=.45pt]
      (\angle:1.9) circle[radius=1.3pt];
  \foreach \angle in {0,15,...,75}
    \fill[black] (\angle:1.9) circle[radius=1.3pt];
  \coordinate (root) at (35:2.45);
  \draw[line width=.8pt] (root) ++(-.06,-.06) -- ++(.12,.12);
  \draw[line width=.8pt] (root) ++(-.06,.06) -- ++(.12,-.12);
  \node[anchor=south west,inner sep=2pt] at (root)
    {$\zeta=|\zeta|e^{\ii\psi_0}$};
  \node[anchor=north east,font=\footnotesize,inner sep=3pt] at (0,0) {$0$};
  \node[anchor=east,font=\footnotesize] at (-1.60,-1.25) {$|z|=1$};
  \draw[black!55,line width=.4pt] (-1.57,-1.25) -- (220:1.9);
  \node[font=\footnotesize] at (0,-2.4)
    {Equally spaced dots: challenges in $\Omega_N$};
\end{scope}
\node[anchor=north west,align=left,text width=7.1cm] at (7.6,5.2)
  {\textbf{A small factor restricts the challenge}\\[4pt]
   $|r-\zeta|\le t(1+|\zeta|)$ implies that $r$ lies in the highlighted arc.};
\node[anchor=north west,align=left,text width=7.1cm] at (7.6,3.15)
  {\textbf{Count the possible challenges}\\[4pt]
   Arc length is at most $2\pi t$, so at most $Nt+1$ challenges lie in the arc.};
\node[anchor=north west,align=left,text width=7.1cm] at (7.6,1.15)
  {\textbf{Apply this to every root}\\[4pt]
   For $d$ roots, use $t=\alpha^{1/d}$ and take a union bound.};
\end{tikzpicture}
\caption{The arc argument in the proof of \cref{lem:circle}. On the unit circle, the set where $|f(z)|\le\alpha M(f)$ is covered by short arcs, one per root of $f$. The figure shows one root $\zeta$ (cross), its arc (bold, with the shaded sector), and the challenges in $\Omega_N$ (dots, here $N=24$); the filled dots are the challenges in the arc.}
\label{fig:agkr-small-arc}
\end{figure}

\begin{samepage}
\begin{lemma}[Approximate sum-check]\label{lem:sumcheck}
Let $f$ have individual degree at most $D$ in $n_\mathrm v\ge1$ variables. Suppose the initial claim differs from $\sum_{z\in H^{n_\mathrm v}}f(z)$ by more than $t_0$. For a target terminal error $t_{n_\mathrm v}\ge0$ and $0<\alpha<1$, define
\begin{equation}\label{eq:sc-backward}
 t_{j-1}=7{\Tol}+\frac{h(t_j+{\Tol})}{\alpha},
 \qquad j=n_\mathrm v,\ldots,1.
\end{equation}
Then sum-check, with the tolerance and arithmetic error specified above, rejects or ends with error greater than $t_{n_\mathrm v}$, except with probability at most $n_\mathrm vD(\alpha^{1/D}+N^{-1})$.
\end{lemma}
\end{samepage}
\begin{proof}[Proof of~\cref{lem:sumcheck}]
Fix the preceding transcript, let $p_j$ be the prover's polynomial in round $j$ and $q_j(X)=\sum_{a\in H^{n_\mathrm v-j}}f(r_1,\ldots,r_{j-1},X,a)$ the true round polynomial, and let $e_{j-1}$ be the modulus of the error in the preceding claim. Acceptance with computed tolerance $6{\Tol}$ implies an exact residual of at most $7{\Tol}$, allowing ${\Tol}$ for numerical evaluation. Since $\sum_{a\in H}q_j(a)$ is the preceding true sum,
\[
 \left|\sum_{a\in H}(p_j-q_j)(a)\right|\ge e_{j-1}-7{\Tol}.
\]
When $e_{j-1}>7{\Tol}$, Lemma~\ref{lem:circle} implies that, outside a conditional event of probability $D(\alpha^{1/D}+N^{-1})$, the next error satisfies
\[
 e_j>(\alpha/h)(e_{j-1}-7{\Tol})-{\Tol}.
\]
The factor $\alpha/h$ accounts for the passage from a lower bound on the modulus of the sum to a lower bound on the modulus of the evaluation, and the additive losses account for the consistency test and the next numerical evaluation. The recurrence in \cref{eq:sc-backward} therefore ensures that $e_{j-1}>t_{j-1}$ implies $e_j>t_j$ except when the check accepts but $e_{j-1}>t_{j-1}$ and $e_j\le t_j$. At each reached round, condition on the transcript and bound the probability that this is the first round with these properties. Summing the conditional bounds over the $n_\mathrm v$ rounds proves the claim, including for adaptive provers.
\end{proof}

It remains to show that a large error in the final sum-check claim leads to a
large error in the claim about the next layer. If the terminal claim is still far from correct but the check of $Z_i-w_i^+(a_i+b_i)-w_i^\times a_ib_i$ accepts, at least one claimed evaluation must be inaccurate; a fresh random linear combination then prevents cancellation with high probability. To quantify both steps, work backward from the input check, which can accept only when the exact error is at most $5{\Tol}$, consisting of its $4{\Tol}$ tolerance and at most ${\Tol}$ numerical error. Fix $0<\beta<1$, set $E_L=5{\Tol}$, and define
\begin{align}
 t_{i,k_i}&=(K+4){\Tol}+K\frac{E_{i+1}+{\Tol}}{\beta},\label{eq:layer-threshold}\\
 t_{i,j-1}&=7{\Tol}+\frac{h(t_{i,j}+{\Tol})}{\alpha}\quad(j=k_i,\ldots,1),
 \qquad E_i=t_{i,0}.\label{eq:layer-backward}
\end{align}
Here $E_i$ is the error required at the start of layer $i$, and $t_{i,j}$ is the threshold that the error must exceed after round $j$. Suppose $|\widehat T_i-T_i|>E_i$. By Lemma~\ref{lem:sumcheck}, except on the event that all its checks accept but the terminal error is at most $t_{i,k_i}$, the verifier rejects or the terminal error exceeds $t_{i,k_i}$. If the check of $Z_i-w_i^+(a_i+b_i)-w_i^\times a_ib_i$ accepts, its exact residual is at most $(K+4){\Tol}$. The perturbation bound in \cref{eq:gate-error} and the definition of $t_{i,k_i}$ then force at least one error in a claimed evaluation to exceed $(E_{i+1}+{\Tol})/\beta$. Both errors in the claimed evaluations of $\GateExtension_{i+1}$ are fixed before $\lambda_i$ is sampled, so \cref{eq:linear} shows that their random combination has error greater than $E_{i+1}$, even after ${\Tol}$ of arithmetic error, except with conditional probability $\beta+N^{-1}$. If none of these implications fails, the final error exceeds $E_L=5{\Tol}$, contradicting acceptance. Summing the probabilities that the error first falls at or below its required threshold while the corresponding check accepts, over all rounds and layers, bounds acceptance by
\begin{equation}\label{eq:soundness}
 kD(\alpha^{1/D}+N^{-1})+L(\beta+N^{-1})
\end{equation}
whenever $|y-C(x)|>E_0$. The argument conditions only on messages already sent and places no restriction on the prover's computational power or adaptive strategy.

We now choose the parameters so that the probability bounded in \cref{eq:soundness} is small and every output error above $\Delta$ exceeds the starting threshold $E_0$. First set
\begin{equation}\label{eq:parameters}
 \vartheta=\frac{\eps}{4(kD+L)},\qquad
 \alpha=\vartheta^D,\qquad\beta=\vartheta,
\end{equation}
and take $N$ to be the least power of two at least $1/\vartheta$. The bound in \cref{eq:soundness} is then at most $\eps/2$. Moreover, $h\le D+1\le kD$ and $N>4kD$, so both $H$ and the Fourier grid in \cref{eq:fourier} are subsets of $\Omega_N$. Next collect the worst-case backward error losses into $\Gamma$ and choose
\begin{equation}\label{eq:precision-choice}
 \Gamma=6\left(\frac{5(1+K)}\beta\right)^L
            \left(\frac{9h}\alpha\right)^k,
 \qquad
 {\Tol}=2^{-\max\{0,\ceil{\log(\Gamma/\Delta)}\}}.
\end{equation}
These choices give $E_0\le {\Tol}\Gamma\le\Delta$. For the first inequality, divide each threshold in \cref{eq:layer-threshold,eq:layer-backward} by ${\Tol}$ and add one. A backward gate step increases this quantity by a factor of at most $5(1+K)/\beta$, and a backward sum-check step by at most $9h/\alpha$. It starts at $6$ at the input layer, so $E_0/{\Tol}+1\le\Gamma$. The choice of ${\Tol}$ gives the second inequality. Thus every output error greater than $\Delta$ is covered by the soundness argument. Together with completeness, this proves Theorem~\ref{thm:agkr}.

\subsection{Complexity}\label{sec:complexity}

The verifier's work consists of processing the round polynomials, checking $Z_i-w_i^+(a_i+b_i)-w_i^\times a_ib_i$ once per layer, and reading the input-gate values at the end. Communication comprises $k$ polynomials with $D+1$ coefficients each, $2L$ claimed evaluations, and $k+L$ random indices. Evaluating each round polynomial by Horner's
rule and computing its sum over $H$ directly
from the coefficients each require $O(D)$
arithmetic operations per round. For the check of $Z_i-w_i^+(a_i+b_i)-w_i^\times a_ib_i$, evaluating $\ParentWeight_i$ costs $O(mh)$ operations, in addition to one query each to $\AddPoly_i$ and $\MulPoly_i$. The two final input evaluations cost $O(nmh)$ operations, since only the values of the input gates, including constant gates, contribute. Each Lagrange basis polynomial or factor in \cref{eq:kernel} can be evaluated by Horner's rule or successive powers using $O(h)$ operations and a constant number of registers. Storing the current coefficient vector takes $O(Db)$ bits, and storing the preceding and current challenge vectors takes $O(m\ell)$ bits; the root-approximation algorithm of \cref{lem:root-bits} fits within this bound. As in Theorem~\ref{thm:agkr}, the time and workspace for evaluating the polynomial extensions of the wiring predicates are charged separately.

A direct prover suffices for the theorem's polynomial-time guarantee. It computes sufficiently accurate gate values, evaluates each $q_j$ at $O(D)$ roots of unity, and recovers the coefficients using \cref{eq:fourier}. Each evaluation is a sum of at most $h^{k_i}\le S^3$ terms, each requiring polynomial work and queries to $\AddPoly_i$ and $\MulPoly_i$. To translate these operation counts into bit complexity, \cref{eq:precision-choice} gives
\[
 \begin{aligned}
 \log(1/{\Tol})=O\bigl(&1+\log^+(1/\Delta)
 +(kD+L)\log((kD+L)/\eps)\\
 &+k\log h+L\log(1+K)\bigr),
 \end{aligned}
\]
where $\log^+t=\max\{0,\log t\}$. Together with \cref{eq:bounds} and the rounding analysis of \cref{sec:completeness}, this supplies sufficient working precision $b$. The bit complexity therefore depends on intermediate magnitudes as well as circuit size and depth.

\medskip\noindent{\bf Dependence on the depth.} For the unique extensions of individual degree below $h$, every padded layer has at most $S$ labels, so $h^{m_i}\le S$ and $k_i\le3\log S/\log h$. Each of $\AddPoly_i$ and $\MulPoly_i$ is then a sum of at most $S$ products of Lagrange basis polynomials, one for each gate, and each such product has coefficient norm one; hence $\BW=S$ suffices (see also \cref{sec:uniform}). In this case $R\le S\BV$ and $1+K=O(S^2\BV)$ by \cref{eq:bounds}, and $k\le3L\log S$. If, in addition, $h$ is a constant, then $D=2(h-1)$ is a constant, and for $\Delta\le1$ the choice in \cref{eq:precision-choice} gives
\begin{equation}\label{eq:depth-dependence}
 \log(\Delta/{\Tol})\le1+\log\Gamma=O\bigl(L\log S\cdot\log(L\log S/\eps)+L\log\BV\bigr).
\end{equation}
Thus the sufficient precision bound is linear in the depth, up to a logarithmic factor.

Two features of the transcript will be useful when implementing the queries to the polynomial extensions of the wiring predicates. Each query depends only on public coins; before conditioning on whether it is reached, its coordinates are uniform on the relevant product of $\Omega_N$. Also, an honest message depends only on the public instance and the challenges at the current and preceding layers. Oracle access to the input extension would similarly replace the final scan by two evaluations, with their errors included in the final ${\Tol}$ budget. The direct prover described above expands sums over all label triples. When the polynomial extensions of the wiring predicates have individual degree below $h$, it can instead sum over actual gates. The next subsection analyzes this implementation, separating the accuracy required of its given vector of approximate gate values from rounding during computation of the prover's messages.

\subsection{Prover precision}\label{stable:section}

We now analyze computation of the prover's messages from a given vector of approximate gate values. There are two distinct error sources: inaccuracies already present in that vector and rounding introduced by the prover's arithmetic. For the unique polynomial extensions of the wiring predicates of degree below $h$ in each variable, the prover can partially evaluate the interpolating polynomial one coordinate at a time, sum over the gates to evaluate each $q_j$, and recover its coefficients by Fourier inversion. The prover stores the array obtained by partial evaluation at the challenges already sampled by the verifier. Computing $q_j$ at roots of unity does not update this array. Each array entry consequently undergoes at most $m$ successive partial evaluations, regardless of the number of evaluations of $q_j$. This allows logarithmically many extra fractional bits, called \emph{guard bits}, for arithmetic used to compute the prover's messages. The accuracy required of the given vector is a separate hypothesis.

Use the notation of \cref{eq:lagrange} and the circuit-model bounds $S\ge h$, $m\le\log S$, $Lh\le S$, and $\BV\ge1$; all logarithms in this subsection are to base two. For each gate at layer $i$, write $\mathsf r=(a_{\mathsf r},b_{\mathsf r},c_{\mathsf r})$, where $a_{\mathsf r}$ is its label and $b_{\mathsf r},c_{\mathsf r}$ are its ordered input labels. Let $\AddRecords_i$ and $\MulRecords_i$ be the sets of these triples for addition and multiplication gates, respectively. Throughout this subsection, take
\begin{equation}\label{stable:predicate-extensions}
 \begin{aligned}
 \AddPoly_i(g,u,v)&=\sum_{\mathsf r\in\AddRecords_i}\chi_{a_{\mathsf r}}(g)\chi_{b_{\mathsf r}}(u)\chi_{c_{\mathsf r}}(v),\\
 \MulPoly_i(g,u,v)&=\sum_{\mathsf r\in\MulRecords_i}\chi_{a_{\mathsf r}}(g)\chi_{b_{\mathsf r}}(u)\chi_{c_{\mathsf r}}(v).
 \end{aligned}
\end{equation}
Thus $\delta=h-1$ and $D=2(h-1)$. Retain $\eqpoly_H$, $\ParentWeight_i$, and $\LayerPoly_i$ from \cref{eq:kernel,eq:parent-weight,eq:layer-poly}, and abbreviate the current claim's points and coefficients by $\rho_b,c_b$, with $\rho_b\in\Omega_N^{m_i}$ and $c_b\in\{0\}\cup\Omega_N$. Process coordinates in the block order $g,u,v$: first the coordinates of $g$, then those of $u$, and then those of $v$. As before, $q_j$ fixes earlier coordinates, leaves the current coordinate free, and sums the later coordinates over $H$.

\begin{theorem}\label{stable:thm}
Let $p\ge0$ be an integer, and suppose the stored gate values satisfy
\begin{equation}\label{stable:table-accuracy}
 \max_{i,a}|\widehat\GateValues_i(a)-\GateValues_i(a)|\le\TableError,
 \qquad \TableError\le\frac{2^{-p}}{576S^2\BV}.
\end{equation}
There is a deterministic algorithm that computes every honest round coefficient and evaluation of $\GateExtension_{i+1}$ with absolute error less than $2^{-p}$, for every challenge sequence. It suffices to use
\begin{equation}\label{stable:guard}
 \begin{aligned}
 q&=p+\GuardBits,\\
 \GuardBits&=\ceil{\log\bigl(6000(m+1)hS^2\BV^2\bigr)}
 =O\bigl(\log S+\log(D+1)+\log(\BV+1)\bigr)
 \end{aligned}
\end{equation}
fractional bits for stored complex operands and $O(\log S+\log(D+1)+\log(\BV+1))$ integer bits per component. Each computed root must have absolute error at most $2^{-q}$, and final messages are rounded to nearest with $p+1$ fractional bits per component.

Over all layers, the algorithm uses
\begin{equation}\label{stable:cost}
 O\bigl((D+1)Smh\bigr)
\end{equation}
complex arithmetic operations and calls to $\Root_{q+1}$ (\cref{lem:root-bits}) with the specified precision and integer indices, with workspace for $O(S+D)$ complex numbers. This excludes circuit labels, challenge indices, and the integer workspace used by $\Root_{q+1}$; exact arithmetic on labels and root indices is charged separately. For the common root order $\Nroot$ defined below, the bit running time is bounded by \cref{stable:cost} times $\poly(q+\log\Nroot)$, in addition to reading the circuit and supplied values. The algorithm computing $\Root_{q+1}$ uses at most $\poly(q+\log\Nroot)$ integer workspace bits. Before final message rounding, the coefficient error is at most
\begin{equation}\label{stable:main-error}
 144S^2\BV\TableError+1500(m+1)hS^2\BV^2\,2^{-q}.
\end{equation}
This is an absolute error bound, including when the exact coefficient is zero.
\end{theorem}

The theorem assumes an accurate vector of gate values; computing that vector from the circuit input may require additional work and precision. In particular, depth does not appear in $\GuardBits$, but it may affect the cost of meeting \cref{stable:table-accuracy}. Once the vector is available, computation of the prover's messages uses $O(S\log S)$ operations for fixed $h$, and $S\polylog S$ when $D=\polylog S$.

\begin{proof}[Proof of~\cref{stable:thm}]
We describe the sums over $\AddRecords_i,\MulRecords_i$ and then follow their errors through interpolation, evaluation of $q_j$, and Fourier recovery. Write $e_q=2^{-q}$. Each complex addition or multiplication has absolute error at most $e_q$: form full integer products and then round both components to nearest on $2^{-q}\Z$. Measure \cref{stable:table-accuracy} after storing the given vector with this precision. The first issue is that the values $\Lag_a(z)$ need not have absolute sum one, so partial evaluation can amplify array entries and their errors. Orthogonality gives, for every $|z|\le1$,
\begin{equation}\label{stable:l2}
 \sum_{a\in H}|\Lag_a(z)|^2
 =\frac1h\sum_{t=0}^{h-1}|z|^{2t}\le1.
\end{equation}
To obtain the identity, expand the squared moduli and use $\sum_{a\in H}a^{s-t}=h\,\Indicator_{s=t}$ for $0\le s,t<h$. Hence $|\Lag_a(z)|\le1$ and, by Cauchy--Schwarz, $\sum_a|\Lag_a(z)|\le\sqrt h$. For an array $\mathsf A(a,b)$, where $a\in H$ is the coordinate to be fixed and $b$ indexes the remaining coordinates, define the \emph{partial evaluation}
\[
 (\Fold_z\mathsf A)(b)=\sum_{a\in H}\Lag_a(z)\mathsf A(a,b).
\]
One step can increase the largest entry modulus by at most $\sqrt h$. After $j$ steps, exact entries are bounded by $h^{j/2}\BV$, and thus by
\begin{equation}\label{stable:fold-bound}
 \Rfold=\sqrt S\,\BV.
\end{equation}
The amplification is real: for $h=2$ and $z=\ii$, the sum $\sum_{a\in H}|\Lag_a(z)|$ equals $\sqrt2$.

The expansions in \cref{stable:predicate-extensions} express an evaluation of $q_j$ as sums over $\AddRecords_i$ and $\MulRecords_i$. For a label $a\in H^d$ and a fixed prefix $\mathbf z=(z_1,\ldots,z_t)$, define
\[
 s_a(\mathbf z)=\prod_{j=1}^t\Lag_{a_j}(z_j),\qquad
 \WeightedValue(a,\mathbf z)=s_a(\mathbf z)\GateExtension_{i+1}(\mathbf z,a_{t+1},\ldots,a_d).
\]
Here $s_a(\mathbf z)$ is the product of the displayed Lagrange basis values, and $\WeightedValue(a,\mathbf z)$ multiplies this product by the indicated evaluation of $\GateExtension_{i+1}$. The corresponding parent quantity is
\[
 \mathcal P(a,\mathbf z)=s_a(\mathbf z)\ParentWeight_i(\mathbf z,a_{t+1},\ldots,a_d),
\]
with an empty prefix allowed. To evaluate $q_j$ at a candidate point, append that point to the prefix of its block. Denoting the three resulting prefixes by $\mathbf z_g,\mathbf z_u,\mathbf z_v$, the resulting value is
\begin{align}
 Q(\mathbf z_g,\mathbf z_u,\mathbf z_v)
 ={}&\sum_{\mathsf r\in\AddRecords_i}\mathcal P(a_{\mathsf r},\mathbf z_g)
 \bigl[\WeightedValue(b_{\mathsf r},\mathbf z_u)s_{c_{\mathsf r}}(\mathbf z_v)+s_{b_{\mathsf r}}(\mathbf z_u)\WeightedValue(c_{\mathsf r},\mathbf z_v)\bigr]\notag\\
 &+\sum_{\mathsf r\in\MulRecords_i}\mathcal P(a_{\mathsf r},\mathbf z_g)\WeightedValue(b_{\mathsf r},\mathbf z_u)\WeightedValue(c_{\mathsf r},\mathbf z_v).
 \label{stable:sparse}
\end{align}
Indeed, substitute \cref{stable:predicate-extensions} into $\LayerPoly_i$ and exchange the finite sums. Each coordinate still summed over $H$ is selected by $\Lag_a(b)=\Indicator_{a=b}$, leaving one contribution per gate. This formula uses only the values in layer $i+1$: approximate vectors in adjacent layers need not satisfy each gate's addition or multiplication equation exactly, since each vector is compared with its own exact gate values. Because $\eqpoly_H$ is an average of roots of unity at the points used here,
\[
 |s_a(\mathbf z)|\le1,\qquad |\ParentWeight_i|\le2,\qquad |\mathcal P(a,\mathbf z)|\le2,
 \qquad |\WeightedValue(a,\mathbf z)|\le\Rfold.
\]
There are at most $S$ gate triples, so every exact evaluation of $q_j$ is bounded by
\begin{equation}\label{stable:sample-bound}
 Q_*:=2S(2\Rfold+\Rfold^2)\le6S^2\BV^2.
\end{equation}
While processing the coordinates of $u$ or of $v$, retain the array partially evaluated at the challenges already sampled by the verifier. To compute $q_j$ at a root of unity, evaluate one additional coordinate in a separate array, sum over $\AddRecords_i$ and $\MulRecords_i$ using \cref{stable:sparse}, and discard that array. Update the retained array only after receiving the next challenge, and restart the coordinates of $v$ from the original given values. Every evaluation of $q_j$ then starts from the same retained array, and no entry undergoes more than $m$ successive partial evaluations.

We next bound the errors in $\Lag_a(z)$, $s_a(\mathbf z)$, $\ParentWeight_i$, and $\mathcal P(a,\mathbf z)$. Let $J$ be the least power of two greater than $D$, so $J\le2(D+1)$, and set $\Nroot=\max\{N,h,J\}$. All interpolation points, challenges, and Fourier points lie in $\Omega_{\Nroot}$. For any product or power of these roots, compute its index exactly modulo $\Nroot$ and approximate the resulting root once. Apply this rule to every term of $\Lag_a(z)$ and $\eqpoly_H(\rho,z)$, avoiding repeated multiplication of approximate roots. Both formulas are normalized sums: root errors contribute at most $e_q$, the $h-1$ addition errors contribute less than $e_q$ after division by $h$, and division and rounding contribute at most $e_q$. Each factor is therefore accurate within $3e_q$. For a product of $t\le m$ such factors, each of modulus at most one, two subproducts with errors $e',e''$ produce a rounded-product error satisfying $1+e\le(1+e')(1+e'')+e_q$. Induction, in any multiplication order, gives
\[
 (1+3e_q)^t(1+e_q)^{t-1}-1\le\exp(4te_q)-1\le8me_q.
\]
The last step requires $4me_q\le1/2$, which follows from the stronger bound $e_q\le1/[100(m+1)]$ implied by \cref{stable:guard}. Empty products are exact. Including the two claim coefficients and their sum gives
\begin{equation}\label{stable:weight-errors}
 \begin{aligned}
 |\widehat s-s|&\le8me_q,
 &|\widehat\ParentWeight_i-\ParentWeight_i|&\le20(m+1)e_q,\\
 |\widehat{\mathcal P}-\mathcal P|&\le50(m+1)e_q,
 &|\widehat{\mathcal P}|&\le3.
 \end{aligned}
\end{equation}
For the third bound, use $2(8me_q)+(1+8me_q)20(m+1)e_q+e_q\le50(m+1)e_q$. Every product defining $s_a(\mathbf z)$ or a term of $\ParentWeight_i$ contains at most $m$ factors. Update these products by multiplication, never by dividing by a factor that might be zero.

Let $\eta_j$ be the largest error in an array entry after $j$ partial evaluations, with $\eta_0\le\TableError$. Each evaluation uses $h$ products and at most $h-1$ additions. Combining the errors in the array entries, the values $\Lag_a(z)$, and the arithmetic using \cref{stable:l2,stable:weight-errors} gives
\begin{equation}\label{stable:fold-recurrence}
 \eta_j\le(\sqrt h+3he_q)\eta_{j-1}
             +3he_q\,h^{(j-1)/2}\BV+2he_q.
\end{equation}
The three terms respectively account for previous errors in the array entries, inaccurate values of $\Lag_a(z)$ applied to exact entries, and rounding in the arithmetic. Normalize by the possible growth of exact values, writing $\eta'_j=\eta_j/h^{j/2}$. Then
\[
 \eta'_j\le(1+3\sqrt h\,e_q)\eta'_{j-1}+5h\BV e_q.
\]
Using $3m\sqrt h\,e_q\le1$ and $(1+a)^m\le\exp(ma)$ to iterate this recurrence yields
\begin{equation}\label{stable:fold-error}
 \eta_j\le3\sqrt S\,\TableError+15mh\sqrt S\,\BV e_q=:\errfold.
\end{equation}
Multiplication by $s_a(\mathbf z)$ and rounding then give
\begin{equation}\label{stable:weighted-value-error}
 \begin{aligned}
 |\widehat{\WeightedValue}-\WeightedValue|
 &\le2\errfold+9(m+1)\Rfold e_q\\
 &\le6\sqrt S\,\TableError+40(m+1)h\Rfold e_q=:\errvalue.
 \end{aligned}
\end{equation}
Thus $\errfold$ bounds errors in the partially evaluated array, while $\errvalue$ bounds errors in $\WeightedValue(a,\mathbf z)$ used in the sums in \cref{stable:sparse}. The choices in \cref{stable:table-accuracy,stable:guard} ensure $\errfold\le1$, $\errvalue\le1$, and all the upper bounds on $e_q$ used above. For example, substitution gives
\[
 \begin{aligned}
 \errfold&\le\frac{2^{-p}}{S^{3/2}\BV}
       \left(\frac3{576}+\frac{15}{6000}\right)<1,\\
 \errvalue&\le\frac{2^{-p}}{S^{3/2}\BV}
       \left(\frac6{576}+\frac{40}{6000}\right)<1.
 \end{aligned}
\]

We now propagate these errors through a gate contribution in \cref{stable:sparse}. For an addition gate, each product $\WeightedValue s$ has error at most $(1+8me_q)\errvalue+8m\Rfold e_q+e_q$. Adding the two products and allowing the final rounding error, with $\errvalue\le1\le\Rfold$, bounds the bracket error by $2\errvalue+35(m+1)\Rfold e_q$. For a multiplication gate, the error is at most
\[
 2\Rfold\errvalue+\errvalue^2+e_q\le3\Rfold\errvalue+e_q.
\]
Both gate types are therefore covered by $\errgate:=4\Rfold\errvalue+40(m+1)\Rfold e_q$. The exact bracket has modulus at most $2\Rfold^2$, so multiplication by $\widehat{\mathcal P}$ gives error at most
\[
 3\errgate+100(m+1)\Rfold^2e_q+e_q
 \le12\Rfold\errvalue+221(m+1)\Rfold^2e_q.
\]
Sum over at most $S$ gates, include the addition errors, and substitute \cref{stable:weighted-value-error} to obtain
\begin{equation}\label{stable:sample-error}
 e_\mathrm{ev}\le72S^2\BV\TableError+702(m+1)hS^2\BV^2e_q.
\end{equation}
Since this estimate sums absolute errors, it remains valid under cancellation, including when the exact evaluation of $q_j$ is zero.

We use the Fourier inversion formula of~\cref{eq:fourier} to
bound the errors in the computed coefficients. Let $\widehat Q_t$
be the computed approximation to $q_j(\omega_J^t)$, so that
$|\widehat Q_t-q_j(\omega_J^t)|\le e_\mathrm{ev}$.
For each coefficient index $a$, compute the unnormalized root
$\omega_J^{-at}$ from its exact integer index, obtaining
$\widehat\omega_{a,t}$ with
$|\widehat\omega_{a,t}-\omega_J^{-at}|\le e_q$.
Form the products $\widehat Q_t\widehat\omega_{a,t}$, sum them,
and only then divide by $J$. Round each product, each addition,
and the final division to the working precision.

Before these arithmetic roundings, the coefficient error is at most
\[
 \left|\frac1J\sum_{t=0}^{J-1}
       \widehat Q_t\widehat\omega_{a,t}-[X^a]q_j\right|
 \le (1+e_q)e_\mathrm{ev}+Q_*e_q.
\]
Indeed, $|\widehat\omega_{a,t}|\le1+e_q$ and
$|q_j(\omega_J^t)|\le Q_*$. The $J$ rounded products contribute
at most $Je_q/J=e_q$ after normalization; the $J-1$ rounded
additions contribute at most $(J-1)e_q/J<e_q$; and the final
division and rounding contribute at most $e_q$.
Thus, by~\cref{stable:sample-bound,stable:sample-error}, each
computed coefficient has error at most
\[
 \begin{aligned}
 &(1+e_q)e_\mathrm{ev}+Q_*e_q+3e_q\\
 &\qquad\le144S^2\BV\TableError
       +1500(m+1)hS^2\BV^2e_q.
 \end{aligned}
\]
This proves~\cref{stable:main-error}. Its terms for errors in the given gate values and in the arithmetic are each at most $2^{-p}/4$ under the theorem's hypotheses. Final rounding to $p+1$ fractional bits per component adds at most $2^{-(p+1)}/\sqrt2<2^{-p}/2$, so the total error is strictly less than $2^{-p}$. For an evaluation of $\GateExtension_{i+1}$, fix every coordinate of $\GateExtension_{i+1}$ and apply \cref{stable:fold-error}, followed by the same final rounding. Fixing the arithmetic order, algorithm computing $\Root_{q+1}$, and rounding tie rule makes the algorithm deterministic, and none of these estimates requires a favorable challenge sequence.

For the operation count, first reuse the values $\Lag_a(z)$ and products $s_a(\mathbf z)$. At each point, computing all $h$ values $\Lag_a(z)$ as $h$ normalized sums costs $O(h^2)$ calls to $\Root_{q+1}$ and arithmetic operations. Save $\Lag_a(z)$ for fixed coordinates and each gate's product $s_a(\mathbf z)$ in each block, updating only once per verifier challenge. Evaluating $q_j$ at a root of unity multiplies by the corresponding $\Lag_a(z)$ without changing the saved product, so no division by a possibly zero factor is needed. The computation of $\ParentWeight_i$ admits similar reuse: for each point in the current parent claim, precompute the suffix products
\[
 \mathsf S_{b,t}(a_{t+1},\ldots,a_d)
   =\prod_{j>t}\Lag_{a_j}(\rho_{b,j}),\qquad d=m_i.
\]
There are $\sum_{r=0}^d h^r\le2h^d$ entries. Build them recursively with one multiplication per nonempty suffix, and retain the two common prefix products of $\eqpoly_H(\rho_{b,j},z_j)$. Since $\eqpoly_H(\rho,a)=\Lag_a(\rho)$ on $H$, evaluating $\ParentWeight_i$ then uses only a prefix product, the current factor, and a saved suffix entry. All parent coordinates are fixed before the coordinates of $u$ or $v$ are processed, so each gate contribution takes $O(1)$ operations. These products combine disjoint sets of coordinates and contain at most $m$ factors, preserving the preceding error estimate; multiplications by empty-product identities are omitted.

For one round and one evaluation of $q_j$ at a root of unity, partial evaluation in a separate array costs $O(h^{m_{i+1}})$ operations and summing over $\AddRecords_i,\MulRecords_i$ costs $O(h^{m_i})$. Including computation of $\Lag_a(z)$ and coefficient recovery, the cost at layer $i$ is
\[
 O\bigl(k_i(D+1)(h^{m_i}+h^{m_{i+1}}+h^2)+k_i(D+1)^2\bigr),
 \qquad k_i=m_i+2m_{i+1}.
\]
The final term accounts for evaluating the sums in~\cref{eq:fourier} directly. Summing over layers and using $k_i\le3m$, $\sum_i(h^{m_i}+h^{m_{i+1}})\le2S$, $Lh\le S$, and $D+1\le2h$ proves \cref{stable:cost}. Process the evaluations of $q_j$ one at a time, retaining the array evaluated at the preceding verifier challenges, one array for the candidate point, and $O(D)$ evaluation values or coefficient accumulators. Partial evaluation reduces array lengths geometrically, giving workspace for $O(S+D)$ complex numbers, excluding circuit labels, challenge indices, and the integer workspace used by $\Root_{q+1}$ as stated.

Finally, the integer precision is controlled by the unnormalized sums: the sums defining $\Lag_a(z)$ have magnitude $O(h)$, partial gate sums have magnitude $O(S^2\BV^2)$, and Fourier sums have magnitude $O((D+1)S^2\BV^2)$. The error estimates also bound the computed values, so $O(\log S+\log(D+1)+\log(\BV+1))$ integer bits suffice. The root of unity specified by an integer index can be approximated deterministically in $\poly(q+\log\Nroot)$ bit operations by approximating $\pi$ and evaluating truncated sine and cosine series. Full-width integer arithmetic followed by rounding has polynomial cost in the same parameters, giving the claimed bit-cost bound.
\end{proof}

To use this algorithm in the protocol, a round coefficient must be accurate within ${\Tol}/[h(D+1)]$. Choose
\[
 p=\ceil{\log\frac{h(D+1)}{\Tol}}.
\]
Theorem~\ref{stable:thm} attains this accuracy with the guard bits in \cref{stable:guard}; with ${\Tol}$ chosen by \cref{eq:precision-choice}, the completeness proof applies unchanged. If a fixed algorithm computes the vector of approximate gate values independently of the challenges, each message depends only on the current claim's points and coefficients $c_{i,0},c_{i,1}$ and the current layer's challenge prefix. The arrays retained after partial evaluation therefore introduce no dependence on earlier rounded claims. Near-linear algebraic GKR provers appear in~\cite{CMT,Thaler}; Thaler discusses computation over $\C$ in a unit-cost model~\cite[Section~5.5.1]{Thaler}, and approximate sum-check uses roots of unity and Fourier recovery~\cite[Section~3.1.1]{ApproxSumcheck26}. 

\subsection{Sensitivity to gate-value errors}\label{sec:conditioning}

The preceding precision bound uses the unique polynomial extensions of the wiring predicates of individual degree below $h$. For the more general extensions allowed in \cref{sec:protocol}, where $D=\delta+h-1$, agreement on $H^{m_i}\times H^{m_{i+1}}\times H^{m_{i+1}}$ does not control values elsewhere: one may add an arbitrarily large multiple of a polynomial that vanishes at every label. Such terms leave the circuit connections unchanged but can amplify errors in the evaluations of $\GateExtension_{i+1}$ at challenge points. We now bound this amplification in terms of $\BW$ and show that some dependence on a coefficient bound is necessary. Throughout this subsection, we study the exact map from gate values to message coefficients
Fix a layer $i$, its polynomial extensions of the wiring predicates, and the points and coefficients in the current claim, with every coordinate and coefficient of modulus at most one. Let $n_\mathrm{c}=h^{m_{i+1}}$. For a vector $\GateValues_{i+1}$ of gate values in layer $i+1$, write $\LayerPoly_i[\GateValues_{i+1}]$ for \cref{eq:layer-poly} with both occurrences of $\GateExtension_{i+1}$ interpolated from $\GateValues_{i+1}$. In round $1\le j\le k_i$, fix the preceding coordinates $r\in\C^{j-1}$ with $|r_t|\le1$. Define
\begin{equation}\label{conditioning:marginal}
 q_j[\GateValues_{i+1}](X)=\sum_{a\in H^{k_i-j}}\LayerPoly_i[\GateValues_{i+1}](r_1,\ldots,r_{j-1},X,a).
\end{equation}
Only $\GateValues_{i+1}$ varies; the polynomial extensions of the wiring predicates, claim points, coefficients, and preceding challenges stay fixed. Let $\Phi_j(\GateValues_{i+1})$ be the vector of all $D+1$ coefficients of $q_j[\GateValues_{i+1}]$, including zeros. Its \emph{absolute condition number} measures the largest local ratio of coefficient error to error in $\GateValues_{i+1}$:
\begin{equation}\label{conditioning:definition}
 \Cond_\mathrm{abs}(\Phi_j,\GateValues_{i+1})
 =\limsup_{\varpi\downarrow0}\ 
 \sup_{0<\|\mathbf e\|_\infty\le\varpi}
 \frac{\|\Phi_j(\GateValues_{i+1}+\mathbf e)-\Phi_j(\GateValues_{i+1})\|_\infty}{\|\mathbf e\|_\infty}.
\end{equation}
The vector norm is the maximum modulus of an entry. Using absolute rather than relative error keeps this definition meaningful when an exact coefficient is zero.

\begin{theorem}[Sensitivity to gate-value errors]\label{conditioning:thm}
Suppose $\|\AddPoly_i\|_1,\|\MulPoly_i\|_1\le\BW$, $\|\GateValues_{i+1}\|_\infty\le\BV$, and $\|\widehat\GateValues_{i+1}-\GateValues_{i+1}\|_\infty\le\TableError$. For every round and fixed prefix as above,
\begin{equation}\label{conditioning:perturbation}
 \|q_j[\widehat\GateValues_{i+1}]-q_j[\GateValues_{i+1}]\|_1
 \le2\BW h^{k_i-j}
 \bigl(2n_\mathrm{c}\TableError
       +2n_\mathrm{c}^2\BV\TableError
       +n_\mathrm{c}^2\TableError^2\bigr).
\end{equation}
Consequently,
\begin{equation}\label{conditioning:upper}
 \Cond_\mathrm{abs}(\Phi_j,\GateValues_{i+1})
 \le4\BW h^{k_i-j}n_\mathrm{c}(1+n_\mathrm{c}\BV).
\end{equation}
No finite function of $S,D,\BV$ alone bounds these condition numbers for all polynomial extensions satisfying the assumptions of~\cref{sec:agkr}, even when $S=3$, $D=3$, and $\BV=2$.
\end{theorem}

\begin{proof}[Proof of~\cref{conditioning:thm}]
First bound the effect of a perturbation. Let $\GateExtension_{i+1}$ be the polynomial interpolated from $\GateValues_{i+1}$, and let $\Xi=\widehat\GateExtension_{i+1}-\GateExtension_{i+1}$ be the change in this polynomial, where $\widehat\GateExtension_{i+1}$ is interpolated from $\widehat\GateValues_{i+1}$. Each product of Lagrange basis polynomials in \cref{eq:lagrange} has coefficient norm one, so
\[
 \|\GateExtension_{i+1}\|_1\le n_\mathrm{c}\BV,\qquad
 \|\Xi\|_1\le n_\mathrm{c}\TableError.
\]
The addition term changes by $\Xi(u)+\Xi(v)$, and the multiplication term changes by
\[
 \GateExtension_{i+1}(u)\Xi(v)+\Xi(u)\GateExtension_{i+1}(v)+\Xi(u)\Xi(v).
\]
Using the fact that the coefficient norm of a product is at most the product of the coefficient norms, together with the wiring bound and $\|\ParentWeight_i\|_1\le2$, bounds the coefficient norm of $\LayerPoly_i[\widehat\GateValues_{i+1}]-\LayerPoly_i[\GateValues_{i+1}]$ by
\[
 2\BW\bigl(2n_\mathrm{c}\TableError
       +2n_\mathrm{c}^2\BV\TableError
       +n_\mathrm{c}^2\TableError^2\bigr).
\]
Fixing a coordinate of modulus at most one cannot increase this norm, because each new coefficient is a sum of old coefficients multiplied by powers of that coordinate. Summing the $h^{k_i-j}$ restrictions in \cref{conditioning:marginal} therefore proves \cref{conditioning:perturbation}. Apply this estimate to $\GateValues_{i+1}+\mathbf e$, set $\TableError=\|\mathbf e\|_\infty$, divide by the perturbation size, and let it tend to zero. The quadratic term vanishes, giving \cref{conditioning:upper}.

For the lower bound, take $h=2$ and a circuit with two input gates, labeled $1,-1$, and one output addition gate whose two children both have label $1$. Thus $S=3$. For an integer $G\ge1$, choose
\begin{equation}\label{conditioning:extension}
 \begin{gathered}
 \AddPoly_G(u,v)=\Lag_1(u)\Lag_1(v)+G(u^2-1),\\
 \MulPoly_G(u,v)=0,\qquad \Lag_1(z)=\frac{1+z}{2}.
 \end{gathered}
\end{equation}
The extra term vanishes on $H^2$, so all these extensions describe the same connections. Their values away from the grid can nevertheless grow with $G$, while their individual degree stays two and $D=3$ stays fixed. Assign the same value $\upsilon$ to both inputs. The polynomial $\GateExtension_1$ is then the constant $\upsilon$, the initial coefficient $\ParentWeight_0$ is one, and the polynomial for the first sum-check round is
\begin{equation}\label{conditioning:example}
 q_{G,\upsilon}(u)
 =\sum_{v\in\{1,-1\}}\AddPoly_G(u,v)(\upsilon+\upsilon)
 =2\upsilon\Lag_1(u)+4\upsilon G(u^2-1).
\end{equation}
Perturb both entries from $\upsilon=1/2$ to $\upsilon'=1/2+\varpi$, where $0<\varpi\le1/4$. Every gate value in either exact circuit evaluation has modulus at most $3/2$, hence at most $\BV=2$, and the inputs have finite binary expansions whenever $\varpi$ does. Each input-gate value changes by only $\varpi$, but the quadratic coefficient changes by
\begin{equation}\label{conditioning:lower}
 \bigl|[u^2](q_{G,\upsilon'}-q_{G,\upsilon})\bigr|=4G\varpi.
\end{equation}
Letting $\varpi\downarrow0$ shows that the condition number at the constant vector $1/2$ is at least $4G$. Since $S,D,\BV$ remain fixed while $G$ grows, they alone cannot bound the sensitivity. The coefficient norm does record this growth:
\[
 \|\AddPoly_G\|_1=|1/4-G|+3/4+G=2G+1/2.
\]
The amplification scales linearly with this
coefficient norm, and $\BW=2G+1$ bounds
$\|\AddPoly_G\|_1$.
\end{proof}

The upper bound also gives a single accuracy requirement on the given gate values sufficient for all rounds. For $\TableError\le1$, combine \cref{conditioning:perturbation} with $n_\mathrm{c}\le S$ and $h^{k_i-j}\le S^3$ to obtain
\begin{equation}\label{conditioning:uniform}
 \|q_j[\widehat\GateValues_{i+1}]-q_j[\GateValues_{i+1}]\|_1
 \le10\BW S^5\BV\TableError.
\end{equation}
Indeed, the three terms are bounded by $4\BW S^4\TableError$, $4\BW S^5\BV\TableError$, and $2\BW S^5\TableError^2$; their sum satisfies the displayed bound because $S,\BV\ge1$ and $\TableError\le1$. For an integer $p\ge0$, it therefore suffices to require
\[
 \TableError\le\frac{2^{-p}}{10\BW S^5\BV}
\]
to ensure coefficient error at most $2^{-p}$. This costs $O(\log\BW+\log S+\log(\BV+1))$ additional bits of gate-value accuracy. Conversely, for \cref{conditioning:extension}, a guarantee covering every perturbation of size at most $2^{-(p+g_\mathrm{val})}\le1/4$, where $g_\mathrm{val}$ is the number of additional bits of gate-value accuracy, must satisfy
\[
 g_\mathrm{val}\ge\log(4G)
\]
by \cref{conditioning:lower}. Thus some dependence on a coefficient bound is necessary even for a fixed, tiny circuit. This conclusion concerns the exact map $\GateValues_{i+1}\mapsto q_j[\GateValues_{i+1}]$ applied to approximate gate values. Further, It does not change the coefficient and arithmetic bounds in Lemma~\ref{lem:space-circuit}. The distinction is that agreement on circuit labels alone does not control sensitivity away from those labels.

\section{Implementing the Approximate GKR  Protocol}\label{sec:implementation}

The verifier protocol described in \cref{sec:agkr} evaluates $\AddPoly_i$ and $\MulPoly_i$ at random points. Direct evaluation may require summing over a whole layer. Following~\cite[Section~4]{GKR}, we give three implementations of these evaluations, one for each of the following settings. For space-bounded computations (\cref{subsec:sp-bound-comp}), the wiring predicates have small arithmetic formulas. For uniform circuits (\cref{sec:uniform}), the prover supplies the evaluations and proves their binary encodings correct. For fixed nonuniform circuits (\cref{subsec:non-uniform-circ}), the verifier computes them privately before receiving the input.

\subsection{Space-bounded Computation}\label{subsec:sp-bound-comp}

The input length $n$ and space bound $s$ are public. Acceptance is reachability from the initial configuration to an accepting one. Repeated Boolean matrix squaring expresses this test as an arithmetic circuit whose wiring predicates can be evaluated without enumerating the gates.

\begin{theorem}\label{thm:configuration-circuit}
Fix a nondeterministic Turing machine using space
$s\ge\ceil{\log(n+2)}$ on $n$ input bits. Its accepting bit is
computed by a layered arithmetic circuit over~$\C$ of size $2^{O(s)}$ and depth
$O(s^2)$, with constants $-1,0,1$. On Boolean inputs, every gate value
lies in $\{-1,0,1\}$. Gate labels have $O(s)$ bits. The addition and
multiplication predicates have constant-depth Boolean formulas with
unbounded fan-in and size $\poly(s)$, independent of the input values. The input bits and
three constants occupy fixed positions in the input layer, and all
other positions are zero.
\end{theorem}

\begin{proof}[Proof of~\cref{thm:configuration-circuit}] 
The circuit decides reachability in the configuration graph by repeatedly squaring its Boolean adjacency matrix.
Encode configurations with $n_{\mathrm{cfg}}=O(s)$ bits. Add an accepting sink, an edge from each accepting configuration to that sink, and a self-loop at every label. Invalid labels have no other edges. If $a$ is the initial label and $b$ is the sink, the machine accepts exactly when $b$ is reachable from $a$. A reachable vertex has a simple path of length less than $2^{n_{\mathrm{cfg}}}$.

Let $B_0$ denote the Boolean adjacency matrix. To double the maximum path length at each iteration, form
\begin{equation}\label{eq:space-square}
 B_{t+1}[u,v]
 =1-\prod_{w\in\bits^{n_{\mathrm{cfg}}}}\bigl(1-B_t[u,w]B_t[w,v]\bigr).
\end{equation}
For each midpoint $w$, the product $B_t[u,w]B_t[w,v]$ tests both halves of a path. Taking their OR and using self-loops to extend shorter paths shows inductively that $B_t$ records paths of length at most $2^t$. The output is $B_{n_{\mathrm{cfg}}}[a,b]$.

Each iteration uses $2^{3n_{\mathrm{cfg}}}$ pair products. Negate each product and add one, multiply the results in a balanced depth-$n_{\mathrm{cfg}}$ tree, and finally negate and add one again. This uses $O(2^{3n_{\mathrm{cfg}}})$ gates and depth $n_{\mathrm{cfg}}+5$. Every intermediate value is Boolean or its negative. Copies of $-1,0,1$ are propagated through the layers, and an output gate computes $B_{n_{\mathrm{cfg}}}[a,b]+0$.

A multiplication gate $(u,v,w)$ has children $(u,w),(w,v)$. In each product tree, gate $(u,v,z)$ has children $(u,v,z0),(u,v,z1)$. Negation and addition of one preserve the indices and use a constant input. Bits identifying gate types and layers complete the labels. The wiring predicates therefore reduce to equality tests.

To construct $B_0$, we test whether two configurations are related by a machine step. The tests have constant-depth Boolean formulas with unbounded fan-in and determine the connections of the fan-in-two arithmetic circuit. For equal-length bit strings $\mathbf s,\mathbf s'$, equality is given by
\[
 \StringEq(\mathbf s,\mathbf s')=\prod_j\bigl(\mathbf s_j\mathbf s'_j+(1-\mathbf s_j)(1-\mathbf s'_j)\bigr).
\]
Each factor compares one pair of bits. Equality with a fixed string is a conjunction of bits or their complements. With the least significant bit first, increment without overflow is given by
\[
 \operatorname{INC}(\mathbf s,\mathbf s')=
 \bigvee_j\left[(1-\mathbf s_j)\mathbf s'_j
    \prod_{k<j}\mathbf s_k(1-\mathbf s'_k)
    \prod_{k>j}\StringEq(\mathbf s_k,\mathbf s'_k)\right].
\]
Each term specifies where the carry ends. Lower bits change from one to zero, the chosen bit changes from zero to one, and higher bits stay unchanged. Thus the formula tests $\mathbf s'=\mathbf s+1$. Reversing its arguments tests decrement, and equality tests a stationary head. These formulas have constant depth and polynomial size.

A binary work-tape alphabet changes the space bound by only a constant factor. Write $\operatorname{st}(u),\operatorname{pos}(u),\operatorname{ipos}(u),\operatorname{tape}(u)$ for the state, work-head position, input-head position, and work-tape contents. For a rule $\varrho$ with source state $q_\varrho$ and target state $q'_\varrho$ and a work-head position $\alpha$, define
\begin{align*}
 \operatorname{Tr}_{\varrho,\alpha}(u,v)={}&\operatorname{Val}(u)\operatorname{Val}(v)
 \StringEq(\operatorname{st}(u),q_\varrho)\StringEq(\operatorname{st}(v),q'_\varrho)\\
 &\cdot\StringEq(\operatorname{pos}(u),\alpha)\StringEq(\operatorname{pos}(v),\alpha+d_\varrho)
 \operatorname{Move}_{d^{\mathrm{in}}_\varrho}(\operatorname{ipos}(u),\operatorname{ipos}(v))\\
 &\cdot\StringEq(\operatorname{tape}(u)_\alpha,\varsigma_\varrho)
 \StringEq(\operatorname{tape}(v)_\alpha,\varsigma'_\varrho)
 \prod_{\alpha'\ne\alpha}\StringEq(\operatorname{tape}(u)_{\alpha'},\operatorname{tape}(v)_{\alpha'}).
\end{align*}
The first, second, and third lines verify configuration encodings and states, head movements, and tape updates, respectively. Cells not under the head must remain unchanged. Here $d_\varrho,d^{\mathrm{in}}_\varrho\in\{-1,0,1\}$ are the head movements, $\varsigma_\varrho,\varsigma'_\varrho$ are the work-tape symbols, and $\operatorname{Move}$ tests increment, decrement, or equality. $\operatorname{Val}$ checks the type and state encodings and that each head points to a cell of its tape. Terms for which $\alpha+d_\varrho$ leaves the work-head range are omitted. Each test has constant depth and size $\poly(s)$.

For each tape symbol $\varsigma_{\mathrm{in}}$ read by the head,
form the disjunction of $\operatorname{Tr}_{\varrho,\alpha}$
over all rules $\varrho$ that read $\varsigma_{\mathrm{in}}$
and all admissible positions $\alpha$, and include the
self-loop and accepting-sink edges. For fixed $u,v$, the entry depends only on the scanned bit and is one of $0,1,x_{\operatorname{ipos}(u)},1-x_{\operatorname{ipos}(u)}$. At an endmarker it is constant. Two arithmetic layers with constants $-1,0,1$ compute these entries, with connections determined by the labels and machine rules.

Place $x_1,\ldots,x_n$ and one copy of each constant $-1,0,1$ at fixed input-layer positions. A type bit distinguishes variables from constants, and the index $\operatorname{ipos}(u)$ identifies an input gate. Other input entries are zero, and padding gates select zeros. The verifier can therefore omit the known zero entries when evaluating the input polynomial.

The $n_{\mathrm{cfg}}$ reachability iterations give size $2^{O(s)}$ and depth $O(s^2)$. Each label has $O(s)$ bits. The transition tests and equality tests above give constant-depth formulas of size $\poly(s)$ for the addition and multiplication predicates. Position and rule counters of $O(\log s)$ bits generate these formulas from the public length and layer parameters.

\end{proof}

To use the labels from \cref{sec:agkr}, we replace each label bit by a polynomial in a root of unity. Substitution into the Boolean formulas gives the required polynomial extensions.

\begin{lemma}\label{lem:space-circuit}
Let $h=2^t$ with $1\le t=O(s)$ and $H=\Omega_h$. The circuit of
\cref{thm:configuration-circuit} can be padded so that each layer
is labelled by a power of $H$, while retaining size $2^{O(s)}$, depth
$O(s^2)$, and gate values in $\{-1,0,1\}$ on Boolean inputs.
Its addition and multiplication predicates have polynomial extensions
of individual degree at most $(h-1)F_{\mathrm{leaf}}$, where $F_{\mathrm{leaf}}=\poly(s)$, which gives $\delta=(h-1)F_{\mathrm{leaf}}$ in \cref{sec:agkr}.
They can be evaluated with $\poly(s,h)$ arithmetic operations, and every
intermediate coefficient norm is at most $(2h)^{O(F_{\mathrm{leaf}})}$.
For $h=\poly(s)$, the degree and evaluation cost are $\poly(s)$,
and the coefficient norms are $2^{\poly(s)}$.
\end{lemma}

\begin{proof}[Proof of~\cref{lem:space-circuit}]
Divide the label bits into blocks of $t=\log h$ bits. In layer $i$, prepend zeros to the binary gate label of length $\beta_i$ until their number is $t\ceil{\beta_i/t}$, and set $m_i=\ceil{\beta_i/t}$. Encode each block $a\in\{0,\ldots,h-1\}$ by $\omega^a\in H$, where $\omega=\exp(2\pi\ii/h)$.

To apply a Boolean predicate to these labels, we must recover the bits of each block. Define
\begin{equation}\label{eq:grid-selectors}
 \Lag_{\omega^a}(z)=\frac1h\sum_{d=0}^{h-1}(z\omega^{-a})^d,
 \qquad
 \widetilde{\operatorname{bit}}_j(z)=\sum_{a=0}^{h-1}\operatorname{bit}_j(a)\Lag_{\omega^a}(z)
 \quad(0\le j<t).
\end{equation}
Here $\operatorname{bit}_j(a)$ denotes a binary digit. For $a'\in\{0,\ldots,h-1\}$, the geometric sum is $h$ when $a=a'$ and zero otherwise. Thus $\Lag_{\omega^a}(\omega^{a'})$ tests equality and $\widetilde{\operatorname{bit}}_j(\omega^{a'})=\operatorname{bit}_j(a')$. Substituting $\widetilde{\operatorname{bit}}_j$ recovers each label bit on $H^{m_i}$. For $h=2$, zero and one correspond to $1,-1$, and $\widetilde{\operatorname{bit}}_0(z)=(1-z)/2$.

Padding introduces additional labels. Give each new input label value zero. In each layer, fix a gate whose value is zero for every input in the specified domain, and let each new gate multiply that fixed gate in the next layer by itself. For an old parent, require the padding bits of both children to be zero. Each layer grows by a factor less than $h$, so its size remains $2^{O(s)}$ because $t=O(s)$. These tests involve $O(s)$ bits and add polynomially many leaves to the Boolean formulas.

Replace NOT by $1-z$, AND by multiplication, and OR by $1-\prod_j(1-z_j)$, then substitute the bit polynomials. The resulting polynomials agree with the wiring predicates on the entire $H$-grid, including padded gates. Duplicate any reused subformula and count each occurrence of a bit or constant as a separate leaf. Constant depth gives a leaf bound $F_{\mathrm{leaf}}=\poly(s)$. Each bit polynomial has degree at most $h-1$, so the resulting polynomial has total and individual degree at most $(h-1)F_{\mathrm{leaf}}$.

To bound coefficient growth, use $\|\Lag_{\omega^a}\|_1=1$ and $\|\widetilde{\operatorname{bit}}_j\|_1\le h/2$. At every arithmetic operation, subadditivity, submultiplicativity, and $\|1-f\|_1\le1+\|f\|_1$ give intermediate coefficient norm at most $(2h)^{O(F_{\mathrm{leaf}})}$. Evaluating the bit polynomials and the arithmetic formulas requires $\poly(s,h)$ arithmetic operations. \cref{sec:agkr} allows degrees above $h-1$, since \cref{eq:layer-sum} requires agreement with the predicates only on the grid.

\end{proof}

\noindent
We now apply approximate GKR (\cref{thm:agkr}) to the circuit of \cref{lem:space-circuit}, with the verifier evaluating the polynomial extensions of the wiring predicates through the formulas above. On Boolean inputs, the circuit outputs exactly $1$ if the machine accepts and $0$ otherwise, so thresholds ${\Tol}<\Delta=1/4$ separate the two cases, and all numerical parameters are polynomial in $s$. This gives the following bit-complexity bounds.

\begin{theorem}\label{thm:space-proof}
A language decided by a fixed nondeterministic Turing machine in space
$s\ge\ceil{\log(n+2)}$ has a public-coin interactive proof over
$\C$ with perfect completeness and soundness error~$\eps$.
For $0<\eps<1$ and $r_\eps=\max\{1,\ceil{\log_{10}(1/\eps)}\}$, its bit complexities are
\[
 \ProverTime=2^{O(s)}r_\eps,\qquad
 \VerifierTime=n\poly(s)r_\eps,\qquad
 \Communication=\poly(s)r_\eps.
\]
The verifier uses $\poly(s)+O(\log r_\eps)$ space.
\end{theorem}

\begin{proof}[Proof of~\cref{thm:space-proof}]

Apply Theorem~\ref{thm:agkr} to the circuit and extensions of Lemma~\ref{lem:space-circuit}. Take $h=2$ and $H=\{1,-1\}$. The output claim $1$ is correct on accepting inputs and has error one otherwise. Set $\Delta=1/4$ and first take soundness error $1/10$.

We have $m=O(s)$, at most $\poly(s)$ rounds and degree, $\BV=1$, and $\log\BW=\poly(s)$. The bound in~\cref{eq:bounds} gives $\log(1+K),\log\CoefficientBound=\poly(s)$. With $\Delta=1/4$, the choice of ${\Tol}$ in~\cref{eq:precision-choice} therefore gives $\log(1/{\Tol})=\poly(s)$.

This requires only $\poly(s)$-bit arithmetic. Gate values, evaluations of $\AddPoly_i$ and $\MulPoly_i$, and honest message coefficients can be computed at polynomial arithmetic depth with intermediate magnitudes $2^{\poly(s)}$. Balanced addition trees keep sums over exponentially many $H$-coordinates at polynomial depth. The same bounds hold for the verifier. In particular, Horner evaluation on the unit circle has intermediate magnitudes at most $(D+1)\CoefficientBound$ for every message passing the encoding, length, and magnitude checks. The rounding bound in \cref{sec:completeness} thus applies to both honest and adversarial messages.

For the final input check, only the $n$ input bits and three constants contribute to the extension over $H$. Across both evaluation points there are at most $2(n+3)$ summands. Error at most ${\Tol}/[8(n+3)]$ per weighted summand and addition gives accumulated error below ${\Tol}/2$, leaving the rest for the final subtraction. This requires $O(\log n)$ extra precision. Theorem~\ref{thm:agkr} and \cref{sec:completeness} now give completeness for every challenge sequence.

The prover uses $2^{O(s)}$ bit operations and sends $\poly(s)$ bits. The verifier evaluates the arithmetic formulas and scans the input, using $n\poly(s)$ time and $\poly(s)$ space, including the current message.

Finally, repeat the protocol independently $r_\eps$ times. Accept only if every execution accepts. Conditional on earlier transcripts, each execution has soundness error at most $1/10$. The resulting error is at most $10^{-r_\eps}$. Reusing workspace between executions gives the stated bounds.

\end{proof}

For constant soundness error, taking $s=O(\log n)$ gives polynomial prover time, $n\polylog(n)$ verifier time, and polylogarithmic communication and space, including precision. This space bound is weaker than the $O(\log n)$ bound in~\cite[Corollary~4.5]{GKR}.

\subsection{Uniform Circuits}\label{sec:uniform}

For a general logspace-uniform circuit, the wiring predicates need not have small arithmetic formulas, so the verifier cannot evaluate $\widetilde{\mathrm{add}}_i$ and $\widetilde{\mathrm{mult}}_i$ directly. Instead, the prover sends the binary encoding of each required evaluation, computed by a fixed approximation algorithm, and proves that every bit of this encoding is correct. Each bit can be computed in $O(\log S + \log b)$ space, where $b$ is the numerical precision, so its correctness can be verified with the interactive proof for space-bounded computations of \cref{thm:space-proof}.

\subsubsection{Verifying polynomial evaluations}

Given a gate $g$, its type and its left and right children
$\mathrm{left}(g), \mathrm{right}(g)$ can be computed in $O(\log S)$ space and $\poly(S)$ time. Write $\widetilde W_i=\AddPoly_i$ for addition gates and $\widetilde W_i=\MulPoly_i$ for multiplication gates. Interpolation over the $H$-grid gives
\begin{equation}\label{eq:uniform-extensions}
 \widetilde W_i(z,u,v)=
 \sum_{g\text{ of that type}}
       \chi_g(z)\chi_{\operatorname{left}(g)}(u)
                  \chi_{\operatorname{right}(g)}(v).
\end{equation}
On $H^{m_i}\times H^{m_{i+1}}\times H^{m_{i+1}}$, a summand is one at the triple of labels for its gate and ordered children, and zero at every other triple. Thus the polynomial agrees with the wiring predicate at every triple of labels. Each $\chi$ is a product of the Lagrange basis polynomials $\Lag_{\omega^a}$ from \cref{eq:grid-selectors}. Each summand has $k_i=m_i+2m_{i+1}$ factors. Each padded layer has at most $S$ labels, giving $h^{k_i}\le S^3$ and $k_i\log h\le3\log S$. The sum has at most $S$ summands. Individual degree at most $h-1$ gives sum-check degree $D=2(h-1)$ in \cref{sec:agkr}. Since each summand has coefficient norm one, $\BW=S$ suffices.

Fixing the root approximations, arithmetic, and rounding rule determines the string the prover must send. The next lemma bounds its error and the space needed to compute each bit.

\begin{lemma}\label{lem:evaluation-bits}
Let $p,q\ge1$ be integers and let a query $\mathsf q$ specify the layer $i$,
gate type, and $k_i\ge1$ root indices, each of length $\ell=\log N$.
Write $(z,u,v)$ for the evaluation point specified by these indices.
Approximate each query root $\zeta$${}=\exp(2\pi\ii j/N)$ and each
label root $a=\omega^{j'}\in H$, where $\omega=\exp(2\pi\ii/h)$, by $\widehat\zeta=\Root_p(j,N)$ and
$\widehat a=\Root_p(j',h)$, using the algorithm of Lemma~\ref{lem:root-bits}.
Compute
\[
 \widehat\Lag_a(\zeta)=\frac1h\sum_{d=0}^{h-1}
                   (\widehat\zeta\,\overline{\widehat a})^d
\]
exactly on these binary approximations. Evaluate~\cref{eq:uniform-extensions}
with these factors and apply $\RoundDown_q$ to both final components, rounding them down to $2^{-q}\Z$. Write $\EvaluationString_{p,q}(\mathsf q)$ for the binary
encoding of the result. If $A,B\in\Z$ are its encoded integer numerators,
the represented value is $w_{p,q}=2^{-q}(A+\ii B)$.
If $\eta=2^{1-p}\le1/(12k_ih)$, then
\[
 |w_{p,q}-\widetilde W_i(z,u,v)|
 \le6Sk_ih\eta+2^{1-q}.
\]
For $0<\xi\le1$, error at most $\xi$ is attained with
$p,q=O(\log S+\log(1/\xi))$.
Every answer bit is computable in time $\poly(S,p,q,\ell)$ and space
\[
 \sigma=O\bigl(\log S+\log(p+q+\ell+2)\bigr).
\]
The answer length is $b_{\mathrm w}=O(q+\log S)$. The public input for proving
one specified answer bit has length
$t_{\mathrm w}=O(k_i\ell+\log S+\log(p+q+2))$.
\end{lemma}

The proof is in \cref{sec:evaluation-arithmetic}. Applying Theorem~\ref{thm:space-proof} to each bit gives the following protocol.

\begin{theorem}\label{thm:uniform}
For a logspace-uniform circuit, use the extensions $\AddPoly_i,\MulPoly_i$
defined in~\cref{eq:uniform-extensions} in Theorem~\ref{thm:agkr}. The resulting
protocol does not assume oracle access to $\AddPoly_i$ and $\MulPoly_i$.
It retains perfect completeness and, for the same parameters ${\Tol}$ and $\Delta$, has soundness error at most
$\eps_{\mathrm c}+\eps_{\mathrm w}$, where
$\eps_{\mathrm c}$ is the error of the protocol in \cref{sec:agkr}
and $0<\eps_{\mathrm w}<1$.
Let $Q=2L$, let $b_{\mathrm w},p,q,\sigma,t_{\mathrm w}$ be the maximum bounds in
Lemma~\ref{lem:evaluation-bits}, and put
$r_{\mathrm w}=\ceil{\log_{10}(Qb_{\mathrm w}/\eps_{\mathrm w})}$. The additional bit costs are
\[
 \ProverTime^{\mathrm{aux}}=Qb_{\mathrm w}r_{\mathrm w}\,\poly(S,p,q,\ell),\qquad
 \VerifierTime^{\mathrm{aux}}=Qb_{\mathrm w}r_{\mathrm w}\,t_{\mathrm w}\poly(\sigma),\qquad
 \Communication^{\mathrm{aux}}=Qb_{\mathrm w}r_{\mathrm w}\,\poly(\sigma).
\]
The additional verifier space is $\poly(\sigma)+O(\log r_{\mathrm w})$,
apart from the retained answer and query.
\end{theorem}

\begin{proof}[Proof of~\cref{thm:uniform}]

At each layer, the prover sends fixed-length answer strings for $\AddPoly_i$ and $\MulPoly_i$. The verifier rejects answers whose represented value has modulus exceeding $\BW+1$ and stores the rest before the interactive proofs below. For each query $\mathsf q$ and received string $\mathsf e$, the parties run an interactive proof of
\begin{equation}\label{eq:wiring-answer-bit}
 \mathsf e[\iota]=\EvaluationString_{p,q}(\mathsf q)[\iota]
 \qquad (1\le\iota\le |\EvaluationString_{p,q}(\mathsf q)|).
\end{equation}

For each statement, impose the work-tape bound $c[\log S+\log(p+q+\ell+2)]$ on the bit algorithm, with rejection if the bound is exceeded. Lemma~\ref{lem:evaluation-bits} provides a suitable constant $c$. The input-head position also fits because $\log t_{\mathrm w}=O(\sigma)$. Use the protocol of Theorem~\ref{thm:space-proof} with $h=2$ to verify the statement. The circuit deciding~\cref{eq:wiring-answer-bit} has size $2^{O(\sigma)}=\poly(S,p,q,\ell)$. Its wiring predicates have polynomial extensions given by small arithmetic formulas, and its input is the public statement. The verifier evaluates these extensions directly. It uses the stored answers in the main check of $Z_i-w_i^+(a_i+b_i)-w_i^\times a_ib_i$ only after all these interactive proofs accept.

For completeness, allocate error $\xi_i\le1$ to each evaluation at layer $i$. Correct answers pass the magnitude check. Errors of at most $\xi_i$ in the two evaluations change the gate expression by at most
\[
 |\ParentWeight_i(g_i)|(2U+U^2)\xi_i
 \le2(2U+U^2)\xi_i,
\]where $U$ is the bound in~\cref{eq:bounds}.
Allocate the remaining error to $\ParentWeight_i$ and the other arithmetic. Honest answers have the required bits and accuracy. These interactive proofs have perfect completeness.

For soundness, set the soundness error of each proof of~\cref{eq:wiring-answer-bit} to at most $\eps_{\mathrm w}/(Qb_{\mathrm w})$. Given any preceding transcript, an incorrect bit is accepted with probability at most this amount. A union bound then gives probability at most $\eps_{\mathrm w}$ that any incorrect answer passes these proofs.

Modify the execution to reject whenever an incorrect answer would pass. Every main check of $Z_i-w_i^+(a_i+b_i)-w_i^\times a_ib_i$ then uses the values represented by $\EvaluationString_{p,q}(\mathsf q)$, and subsequent challenges remain independent. An unbounded prover can simulate these interactive proofs, and rejection can only reduce acceptance. Thus Theorem~\ref{thm:agkr} bounds false acceptance in this modified execution by $\eps_{\mathrm c}$.

Run the actual and modified executions with identical random choices. They differ only when an incorrect answer passes, with probability at most $\eps_{\mathrm w}$. Hence their acceptance probabilities differ by at most this amount, giving $\eps_{\mathrm c}+\eps_{\mathrm w}$. Conditioning on all answers being correct would not justify this comparison because it could bias later challenges.

There are $Qb_{\mathrm w}r_{\mathrm w}$ constant-error executions of the proof of~\cref{eq:wiring-answer-bit}. Their costs follow from Theorem~\ref{thm:space-proof}, and their workspace can be reused. For total error $\eps$, take $\eps_{\mathrm c}=\eps_{\mathrm w}=\eps/2$.

\end{proof}

\subsubsection{Arithmetic in logarithmic space}\label{sec:evaluation-arithmetic}

To prove Lemma~\ref{lem:evaluation-bits}, we need small-space algorithms for approximating roots and multiplying Gaussian integers. We then evaluate the polynomial exactly on the rounded roots and round the result.

Hesse, Allender, and Barrington~\cite{HAB} give DLOGTIME-uniform constant-depth threshold circuits for division and iterated multiplication. Recursive evaluation counts the true inputs at each threshold gate, using constant stack depth and logarithmic-length gate indices and counters. Uniformity determines the connections in logarithmic time, so each output bit is computable in logarithmic space and polynomial time. Addition and a fixed number of compositions have the same bounds, recomputing intermediate bits as needed.

\begin{lemma}\label{lem:root-bits}
For integers $p\ge1$, $N\ge2$, and $0\le j<N$, there is a fixed
algorithm returning a complex number $\Root_p(j,N)$ with components
in $2^{-p}\Z$ such that
\[
 |\Root_p(j,N)-\exp(2\pi\ii j/N)|<2^{1-p}.
\]
Each output bit is computable in space
$O(\log(p+\log N+2))$ and time $\poly(p+\log N)$.
\end{lemma}

\begin{proof}[Proof of~\cref{lem:root-bits}]

To approximate the angle, use $\pi=16\arctan(1/5)-4\arctan(1/239)$. The double-angle formula gives $\tan(4\arctan(1/5))=120/119$. Subtracting $\arctan(1/239)$ gives an angle in $(0,\pi/2)$ with tangent one, proving the identity. Truncate each alternating arctangent series after $2(p+20)$ terms. The first omitted term bounds the error, so the resulting rational $\pi_p$ satisfies $|\pi_p-\pi|<2^{-p-20}$.

Set $\psi=2\pi_pj/N$ and evaluate the Taylor polynomials of order $M=128(p+20)$ for sine and cosine. Use exact rational arithmetic. Since $|\psi|<7$, with $M'=M+1$ each Taylor remainder is at most
\[
 \frac{7^{M'}}{M'!}\le(98/M')^{M'/2}\le2^{-M'/2}<2^{-p-20},
\]
where $M'!\ge(M'/2)^{M'/2}$ and $M'\ge256$. The angle error is less than $2^{-p-19}$. Since sine and cosine are 1-Lipschitz on the real line, this also bounds the error from changing the angle. Apply $\RoundDown_p$ to both rational values, rounding them down to $2^{-p}\Z$. The total complex error is less than $2^{1-p}$, as required. These choices determine every output bit.

There are $O(p)$ rational terms of bit length polynomial in $p+\log N$. For each sum, use the product of denominators as a common denominator. Powers and factorials are products whose factors are generated from their indices. Thus the algorithm has a fixed number of sum, product, and division steps on polynomial-length integers. The final rounding uses signed floor division. The integer algorithms above compute each output bit in $O(\log(p+\log N+2))$ space and $\poly(p+\log N)$ time.

\end{proof}

Clearing the denominators of rounded roots reduces multiplication to the following problem over Gaussian integers.

\begin{lemma}\label{lem:gaussian-product-bits}
Given $n_{\mathrm f}\ge1$ Gaussian integers with at most $n_{\mathrm{bit}}\ge1$ bits per component,
each bit of their product is computable in space $O(\log(n_{\mathrm f}+n_{\mathrm{bit}}+2))$
and time $\poly(n_{\mathrm f}+n_{\mathrm{bit}})$. The algorithm accesses individual bits of its inputs.
\end{lemma}

\begin{proof}[Proof of~\cref{lem:gaussian-product-bits}]

Write the factors as $A_j+\ii B_j$. Their product is $f(\ii)$ for the integer polynomial
\[
 f(X)=\prod_{j=1}^{n_{\mathrm f}}(A_j+B_jX)=\sum_{k=0}^{n_{\mathrm f}} f_kX^k,
 \qquad \Gamma=2^{n_{\mathrm f}(n_{\mathrm{bit}}+1)},\quad \Upsilon=2^{n_{\mathrm f}(n_{\mathrm{bit}}+1)+3}.
\]
The integer $f(\Upsilon)$ determines all coefficients of $f$, since $|f_k|\le\Gamma$ and the base $\Upsilon$ separates them. Iterated signed integer multiplication computes $f(\Upsilon)$, whose bit length is $O(n_{\mathrm f}^2(n_{\mathrm{bit}}+1))$.

To recover $f_k$, divide by $\Upsilon^k$. The lower-degree terms contribute less than $\Gamma/(\Upsilon-1)<1/2$ in absolute value. Rounding $f(\Upsilon)/\Upsilon^k$ to the nearest integer therefore gives exactly $\sum_{a\ge k}f_aR^{a-k}$. Reduce this integer modulo $\Upsilon$ and take its representative in $[-\Upsilon/2,\Upsilon/2)$. This removes the higher-degree terms and leaves $f_k$. The real and imaginary parts are then
\[
 \operatorname{Re} f(\ii)=\sum_{k\text{ even}}(-1)^{k/2}f_k,
 \qquad
 \operatorname{Im} f(\ii)=\sum_{k\text{ odd}}(-1)^{(k-1)/2}f_k.
\]
Each $f_k$ is obtained from $f(\Upsilon)$ by one division, one rounding, and one reduction modulo $\Upsilon$, and $\operatorname{Re} f(i)$ and $\operatorname{Im} f(i)$ are signed sums of the $f_k$. These are a constant number of integer operations on $\poly(n_{\mathrm f} + n_{\mathrm{bit}})$-bit integers, each computable bitwise in logarithmic space. Composing them and recomputing intermediate bits as needed, each bit of the product is computable in $O(\log(n_{\mathrm f} + n_{\mathrm{bit}} + 2))$ space and $\poly(n_{\mathrm f} + n_{\mathrm{bit}})$ time, given access to individual bits of the factors.

\end{proof}

\begin{proof}[Proof of~\cref{lem:evaluation-bits}]

Only root approximation and final rounding introduce error. The intervening arithmetic is exact on binary fractions. On the unit circle, $|\Lag_a(\zeta)|\le1$. If $|\zeta|=|a|=1$, then
\[
 |\widehat\zeta\overline{\widehat a}-\zeta\overline a|\le3\eta,
 \qquad |\widehat\zeta\overline{\widehat a}|\le(1+\eta)^2.
\]
Expanding the difference of two powers bounds the error in one Lagrange basis polynomial by
\[
 |\widehat\Lag_a(\zeta)-\Lag_a(\zeta)|
 \le\frac{3\eta}{h}\sum_{d=1}^{h-1}d(1+\eta)^{2(d-1)}
 \le\frac32(h-1)\eta e^{2h\eta}
 \le3h\eta.
\]
Replacing the $k_i$ factors in a gate summand one at a time gives total error at most $3k_ih\eta(1+3h\eta)^{k_i-1}\le6k_ih\eta$. Summing over at most $S$ gates and adding the rounding error gives the stated bound. It suffices to choose $p\ge\ceil{\log(24Sk_ih/\xi)}$ and $q\ge\ceil{\log(4/\xi)}$. The unrounded approximate sum has modulus at most $S(1+3h\eta)^{k_i}<2S$. We therefore fix a signed encoding with $O(\log S)$ integer bits per component.

For the space bound, compute each bit without storing intermediate values. Since $h$ is a power of two, each approximate basis value has denominator dividing $2^E$, where $E=\log h+2p(h-1)$. Their product in each gate summand has denominator dividing $2^{k_iE}$. Expanding its $k_i$ sums gives $h^{k_i}\le S^3$ terms. Each term multiplies at most $2k_i(h-1)$ numerators of rounded roots and is then expressed over the common denominator. Across all gates there are at most $S^4$ terms, and the sum with this denominator has $O(pk_ih+k_i\log h+\log S)$ numerator bits.

Enumerating one exponent per Lagrange basis polynomial requires $k_i\log h\le3\log S$ counter bits. \Cref{lem:root-bits} computes each bit of a rounded root, and
\Cref{lem:gaussian-product-bits} computes each bit of a product of rounded roots. Empty products have value one. Summing the terms adds a counter whose length is logarithmic in their number. The expansion in~\cref{eq:uniform-extensions} uses a fixed number of product, sum, and division steps. Since $h\le S$, these operations take polynomial time and logarithmic space in the stated parameters.

Combining this arithmetic with the algorithm for a gate's type and children gives space $\sigma$ and polynomial time for each answer bit. The public statement records the query, the layer, the gate type, $n$, $h,p,q$, the circuit dimensions, the bit index, and the proposed bit, with all integer parameters written in binary. If the generator expects $1^n$, we retain only its input-head position. Since $n\le S$, the statement has length $t_{\mathrm w}$.

\end{proof}

\subsubsection{Boolean circuits}

We apply Theorem~\ref{thm:uniform} to a logspace-uniform fan-in-two Boolean circuit of size $S_{\mathrm B}$ and depth $L_{\mathrm B}$. Take $h=2$ and $H=\{1,-1\}$. For the verifier bounds, we need a layered arithmetic circuit whose inputs occupy fixed positions.

Fix positions for the $n$ inputs and two
shared constants, redirecting constant
references accordingly.
For stages $t=1,\ldots,L_{\mathrm B}$,
introduce a copy of every gate.
At stage $t$, each noninput gate applies
its original operation to its children's
values at stage $t-1$, while input and
constant positions retain their initial
values.
At stage $0$, initialize all other
positions to zero.
Thus, after stage $t$, every gate of
original depth at most $t$ has its
correct value.
The resulting circuit has size
$O(S_{\mathrm B}L_{\mathrm B})$ and depth
$O(L_{\mathrm B})$.
Enumerating gates and counting preceding
gates suffices to assign labels,
preserving logspace uniformity.

Convert AND, OR, and NOT as in the proof of Lemma~\ref{lem:space-circuit}. Add a shared $-1$ input and propagate constants through the new layers. The final input check reads only the $n$ input bits and three constants. Every other initial position is zero and can be skipped.

Every arithmetic gate value lies in $\{-1,0,1\}$. The padded arithmetic circuit has $L=O(L_{\mathrm B})$ and $\log S=O(\log S_{\mathrm B})$. The logarithms of the polynomial and message bounds are $O(\log S)$, and \cref{sec:agkr} gives $\log(1/{\Tol})=\poly(L,\log S)$. Hence $p,q,b_{\mathrm w}=\poly(L,\log S)$ and $\sigma=O(\log S)$. At constant soundness error, Theorem~\ref{thm:uniform} gives polynomial prover time, verifier time $n\poly(L,\log S)$, and communication and space $\poly(L,\log S)$.

To prove language membership, let the prover claim that the output is one. Every language in uniform $\NC$ then has verifier time $n\polylog(n)$ and polylogarithmic communication and space. For $\Pclass$, the bounds retain their dependence on depth. For space-bounded computations, we apply the configuration-circuit construction directly. Theorem~\ref{thm:space-proof} has honest-prover time $2^{O(s)}$ for space-$s$ computations. Taking $s=\poly(n)$ gives polynomial-time verifiers and polynomial communication for every language in $\PSPACE$.

\subsection{Preprocessing}\label{subsec:non-uniform-circ}

If the circuit is fixed before the input arrives, the verifier can perform these evaluations in advance. The query points for $\AddPoly_i$ and $\MulPoly_i$ depend only on its random challenges. This holds for nonuniform circuits as well. To preserve soundness, each challenge remains private until the corresponding prover message has been received, and the stored challenges and evaluations are used for a single execution.

\begin{samepage}
\begin{theorem}\label{thm:preprocessing}
Fix a circuit, a public input domain, and protocol parameters
whose numerical bounds hold throughout that domain, using the
extensions $\AddPoly_i,\MulPoly_i$ of~\cref{eq:uniform-extensions}. There is
an offline/online implementation of Theorem~\ref{thm:agkr} with
the same completeness, parameters ${\Tol},\Delta$, and soundness error. Its
stored challenges, wiring evaluations, and constant-input contributions are used for only one execution. If $k$ is the number
of sum-check rounds, $\ell=\log N$, and $b$ bounds the bit-length
of each stored value, these stored values occupy $O\bigl((k+L)\ell+Lb\bigr)$ bits, in addition to public parameters. Preprocessing takes
polynomial time in the padded size $S$, the circuit-description length including
constant encodings, the numerical precision, and the challenge length.
Online communication is unchanged. Online verifier work omits
the evaluations of $\AddPoly_i$ and $\MulPoly_i$ and the constant-input contributions.
\end{theorem}
\end{samepage}

\begin{proof}[Proof of~\cref{thm:preprocessing}]

Before receiving the input, sample the random tape and use it to compute the $2L$ evaluations of $\AddPoly_i$ and $\MulPoly_i$ to the required accuracies. Store the tape and values privately.

Using the fixed public input layout, also store the constants' contributions to the two final input evaluations. Include their errors in the final comparison's error bound. Without this layout, the input-position map requires separate storage.

When the input arrives, compute the variable contribution and perform the checks using the stored values. Compute $\ParentWeight_i$ from the revealed challenges. Reveal each stored challenge after its corresponding prover message. Discard these stored values when execution ends, including acceptance, rejection, or abort.

Compare with online evaluation using the same challenges, rounding rules, and summation orders for the constant and variable input contributions. At each transcript prefix, no check or disclosure depends on an unrevealed challenge. The next challenge therefore remains independent and uniform. Both verifiers perform identical checks and induce the same transcript distribution. This preserves completeness, the parameters ${\Tol},\Delta$, and soundness. Publishing or reusing these stored values could reveal later challenges before earlier prover messages, invalidating this comparison.

The interpolation formula has one term per gate, each computable in polynomial time. Scanning the constants and applying the arithmetic of \cref{sec:completeness}, including the Lagrange basis polynomials for $H$, yields polynomial preprocessing time. The verifier stores $k+L$ indices and $O(L)$ values. Online work still includes input encodings and arithmetic, but omits the stored evaluations and constant contributions.

\end{proof}

\section{Zero-Knowledge Approximate GKR}\label{sec:zk}

We extend the approximate GKR protocol of~\cref{sec:agkr} to computations whose input layer contains inputs determined by a witness. The difficulty is that the prover's messages may reveal information about these entries. Following~\cite[Section~5]{GKR} and~\cite{BGG}, the prover commits to the witness before the first challenge and replaces each witness-dependent GKR message by a commitment. It then proves in zero-knowledge that the committed values define a transcript accepted by the finite-precision GKR verifier. In particular, the final proof checks the verifier's acceptance predicate, rather than evaluating the delegated circuit. For Boolean NP relations computable by polynomial-size, polylogarithmic-depth circuits, this yields communication nearly linear in the witness length, up to polynomial factors in the security parameter and polylogarithmic factors in the circuit size, under the encoding and precision assumptions stated below.

We consider a layered arithmetic circuit $C$ with $n_\mathrm{p}$ input entries specified by $x$ (including constants) and $n_\mathrm{w}$ input entries determined by a witness $w\in\mathcal W\subseteq\bits^{\WitLen}$, at fixed, disjoint, publicly known labels; all remaining input entries are zero. The public instance is $(C,x,y)$, where $y\in\C$ is the claimed output; $w$ is given to the prover. For each $j$, a fixed Boolean circuit computes binary encodings of integers $A_j(w),B_j(w)$, and the corresponding circuit input is $\WitEntry{j}=(A_j(w)+\ii B_j(w))2^{-\WitPrec{j}}$. The nonnegative integers $\WitPrec{j}$, integer bit lengths, and signed-integer encodings are fixed publicly. We write $C(x,w)$ for the exact output on these inputs. A Boolean circuit of size $F_{\mathcal W}$ tests $w\in\mathcal W$ and computes the integer encodings, with the output bits counted in its size. The gate-value bound $\BV$ of~\cref{sec:agkr} is required to hold uniformly over all $w\in\mathcal W$.

We use the unique polynomial extensions of the wiring predicates of degree below $h$ in each variable of~\cref{eq:uniform-extensions}. Fix $\eps_\mathrm{c}\le1/8$ and a parameter $\Delta>0$, and choose the parameter ${\Tol}$ and numerical precision as in~\cref{sec:agkr}. These parameters depend only on the public circuit description and the uniform bounds, not on the witness. We consider the promise problem $(\mathcal Y,\mathcal N)$ defined by
\begin{align}
    \mathcal Y&=\{(C,x,y):\exists w\in\mathcal W,\ |y-C(x,w)|\le {\Tol}\},\label{eq:zk-yes}\\
    \mathcal N&=\{(C,x,y):\forall w\in\mathcal W,\ |y-C(x,w)|>\Delta\}.\label{eq:zk-no}
\end{align}
No guarantee is required outside $\mathcal Y\cup\mathcal N$. For a Boolean witness relation, use the arithmetization of~\cref{sec:uniform} with $y=1$ and $\Delta=1/4$. Since the circuit output is in $\bits$ on Boolean inputs and ${\Tol}<\Delta$, the promise then coincides with ordinary language membership.

Fix a security parameter $\kappa\ge6$, and assume one-way functions secure against nonuniform polynomial-size circuits. We consider families for which the public descriptions, witness lengths, circuit sizes, and numerical precision are polynomially bounded in $\kappa$. Zero-knowledge (\cref{def:zk}) is computational against classical probabilistic polynomial-time verifiers with arbitrary auxiliary input; soundness holds against computationally unbounded provers.

We present the protocol and state its security and complexity guarantees in~\cref{sec:zk-protocol}, and prove these guarantees in~\cref{sec:zk-security}. In~\cref{sec:zk-uniform}, we give a variant for logspace-uniform circuits that replaces direct evaluation of the public polynomial extensions of the wiring predicates with interactive proofs of~\cref{eq:wiring-answer-bit}, and establish the resulting verifier-efficiency bounds.

\subsection{Protocol}\label{sec:zk-protocol}
We first specify the commitment scheme and the zero-knowledge proof for NP used in the construction. We then separate the public computations from the checks on committed values, describe the interaction, and state its guarantees.

\medskip\noindent{\bf Commitments.}
To hide $w$ and the GKR messages that depend on it, the prover commits to each value before the challenge that tests it. We instantiate these commitments using Naor's statistically binding and computationally hiding scheme~\cite{Naor} (\cref{def:commitment}). Let ${\mathsf G}:\bits^\kappa\to\bits^{3\kappa}$ be a pseudorandom generator (\cref{def:prg}) obtained from one-way functions~\cite{HILL}. At the outset, the verifier samples $R\leftarrow\bits^{3\kappa}$ and sends it to the prover. To commit to a bit $\beta\in\bits$, the prover samples $s\leftarrow\bits^\kappa$ and sends
\[
 \Com_R(\beta;s)={\mathsf G}(s)\oplus \beta R.
\]
An opening consists of $(\beta,s)$, which the verifier checks against the commitment. Let $\mathsf{msg}_1,\ldots,\mathsf{msg}_t$ be the binary encodings of the prover's GKR messages and $\mathsf m=w\mathbin\Vert \mathsf{msg}_1\mathbin\Vert\cdots\mathbin\Vert \mathsf{msg}_t$. Write $\mathbf d=(d_j)_{j=1}^{|\mathsf m|}$, where $d_j=\Com_R(\mathsf m_j;s_j)$, with independent seeds $s_j$. These commitments are sent at their respective protocol steps, using the same $R$.

The commitment is statistically binding over the verifier's choice of $R$, even against computationally unbounded provers. Indeed, openings to both bit values would imply $R={\mathsf G}(s_0)\oplus {\mathsf G}(s_1)$ for some $s_0,s_1\in\bits^\kappa$. Since at most $2^{2\kappa}$ strings admit such a representation, a uniform $R$ excludes every double opening with probability at least $1-2^{-\kappa}$. In particular, this single event ensures binding simultaneously for all commitments, irrespective of how they are chosen. Conversely, hiding holds for every fixed $R$, including one chosen by a malicious verifier: the pseudorandomness of ${\mathsf G}(s)$ makes $\Com_R(\beta;s)$ computationally indistinguishable from uniform for either value of $\beta$. A hybrid argument over independently sampled seeds extends this guarantee to polynomially many adaptively chosen commitments.

\medskip\noindent{\bf Zero-knowledge proof for NP.}
Let ${\mathsf B}({\mathsf x},{\mathsf w})\in\bits$ denote the final Boolean verification circuit of size $|{\mathsf B}|$, with public input ${\mathsf x}$ and witness ${\mathsf w}$. To prove $\exists{\mathsf w}\,{\mathsf B}({\mathsf x},{\mathsf w})=1$, we instantiate the MPC-in-the-head transformation of~\cite[Theorem~3.2]{IKOS} with a perfectly correct and perfectly $2$-private five-party MPC protocol in the semi-honest model (\cref{def:mpc}) (e.g., five-party BGW~\cite{STOC:BGW88}), applied to the functionality $f_{\mathsf x}({\mathsf w}_1,\ldots,{\mathsf w}_5)={\mathsf B}({\mathsf x},{\mathsf w}_1\oplus\cdots\oplus{\mathsf w}_5)$. Five parties suffice for this privacy against coalitions of at most two parties, since $5>2\cdot2$. The resulting proof has perfect completeness and communication $O(|{\mathsf B}|\poly(\kappa))$. In each repetition, given an accepting witness ${\mathsf w}$, the prover samples a fresh, uniformly random tuple of five bit-string shares whose XOR is ${\mathsf w}$ and locally executes the MPC protocol with fresh random tapes. The sharing and all MPC randomness are independent across repetitions; the witness itself remains fixed. It records each party's view, consisting of its input share, random tape, sent and received messages, and output. We fix public lengths and encodings for all MPC views and implement the Boolean functionality $f_{\mathsf x}$ on binary input shares.\footnote{For example, five-party BGW~\cite{STOC:BGW88} can be instantiated over $\mathbb F_8$, which contains enough distinct nonzero evaluation points. Boolean gates are implemented exactly over its subfield $\mathbb F_2$. This field is used only to evaluate the Boolean verification circuit on binary encodings; the delegated circuit and approximate GKR identities remain over $\C$.} The local verification checks enforce that each input share is a bit string and that the random tapes and messages have the specified lengths and encodings. By \cref{def:mpc}, perfect $2$-privacy guarantees that the joint views of any two parties can be simulated exactly from their shares, the public statement, and the output $1$. As any two of the five XOR shares are independent uniform strings, their joint distribution is independent of the witness.

The verifier samples $R'\leftarrow\bits^{3\kappa}$ for the commitments in this NP proof, independently of $R$. The string $R'$ is reused across repetitions, with fresh commitment seeds for every bit. In each repetition, the prover commits to the five newly generated views before receiving a uniformly random four-bit challenge $\mathsf{ch}\in\bits^4$. A fixed surjection $\mathsf{sel}:\bits^4\to\binom{[5]}{2}$ selects the pair of views to be opened. The verifier checks their commitment openings, local computations, consistency of exchanged messages, and outputs equal to $1$.

For soundness, condition on the event that $R'$ binds all commitments to MPC views. On a false statement, at least one pair of committed views must fail these checks. Indeed, if every pair were accepted, the five views would be consistent with the parties' local computations and pairwise consistent, and hence would define an MPC execution with output $1$, contradicting perfect correctness. Since every pair has at least one preimage under $\mathsf{sel}$, the verifier selects a rejecting pair with probability at least $1/16$. This bound holds conditioned on any preceding transcript, because each challenge is sampled independently after the corresponding commitments are fixed. Thus, after $r_\mathrm{NP}$ sequential repetitions, the soundness error is at most
\[
 \eps_\mathrm{NP}(r_\mathrm{NP})\le (15/16)^{r_\mathrm{NP}}+2^{-\kappa},
\]
where the last term bounds the probability that $R'$ fails to bind. We fix $r_\mathrm{NP}=35$: for every $\kappa\ge6$,
\[
 (15/16)^{35}+2^{-\kappa}
 \le (15/16)^{35}+1/64<1/8.
\]
All verifier messages are public coins, and both parties run in time polynomial in $|{\mathsf B}|$ and $\kappa$.

Fix a polynomial $t_{\mathsf{Sim}}$. If the simulator for two parties' MPC views exceeds $t_{\mathsf{Sim}}(|\mathsf z|)$ steps on an input $\mathsf z$, or outputs views with incorrect lengths or encodings, halt and output $\bot$ to indicate simulation failure; no test of whether the statement is true is performed. To simulate a repetition against a polynomial-time verifier, guess its four-bit challenge uniformly, sample two independent uniform input shares, and invoke the MPC simulator with public output $1$ to generate the corresponding views. Commit to these views and to dummy strings with the same lengths as the respective MPC views for the other three parties, using fresh seeds. If the challenge matches the guess, open the selected views and retain the resulting interaction. Otherwise, restore the verifier's state, including its random tape, to the start of the repetition and retry. If the verifier aborts or sends a malformed challenge before the openings, retain that outcome with independent probability $1/16$ and retry otherwise. A verifier message $R'$ of incorrect length is handled by the same immediate abort as in the honest protocol.

For a true statement, compare this procedure with a hybrid that generates fresh genuine MPC views and genuine commitments in every attempt, while using the same guessing and retention rules. Conditional on any resulting verifier transcript, the independent guess matches a well-formed challenge with probability exactly $1/16$; the retention probability for an abort or malformed challenge is also $1/16$. Thus retention does not bias the verifier's view. Capping each repetition at $\kappa^2$ attempts gives truncation probability at most $35(15/16)^{\kappa^2}$ in this hybrid. Privacy against coalitions of two parties and commitment hiding, applied over the polynomially many attempts, justify replacing genuine views by simulated views and unopened views by dummy strings. Consequently, on true statements the actual simulator is computationally indistinguishable from the real interaction, and its truncation probability is at most
\[
 35(15/16)^{\kappa^2}+\negl(\kappa).
\]
Every subroutine is terminated at its fixed polynomial step bound. The simulator therefore runs in strict polynomial time on every input and random tape, including for false statements; no simulation guarantee on false statements is asserted.

\medskip\noindent{\bf Public computation.}
The proof of $\exists{\mathsf w}\,{\mathsf B}({\mathsf x},{\mathsf w})=1$ checks acceptance by the finite-precision GKR verifier, rather than the delegated circuit. We separate this check into quantities determined entirely by the public instance and the verifier's challenges, and a Boolean predicate that checks the committed messages using those quantities. All message lengths, numerical algorithms, rounding rules, and intermediate word bounds are fixed in advance. The numerical error bounds stated in~\cref{sec:protocol} and below hold uniformly for every $w\in\mathcal W$ and every transcript satisfying the encoding, length, and magnitude checks, not only for honest transcripts. A malformed encoding, a failed magnitude check, or an arithmetic overflow sets a rejection flag; subsequent arithmetic is performed on fixed dummy operands where necessary. This makes the predicate well defined on every bit string. During the interaction, each challenge is sampled after the corresponding prover commitments, while the numerical checks are deferred to the final predicate.

Among the verifier's numerical checks, only the input-layer evaluation requires the inputs $\WitEntry{j}$; the remaining checks use the committed GKR messages and public quantities. Let $g^{\mathrm p}_1,\ldots,g^{\mathrm p}_{n_\mathrm{p}}$ and $g^{\mathrm w}_1,\ldots,g^{\mathrm w}_{n_\mathrm{w}}$ denote the publicly known labels of the inputs specified by $x$ and $w$, respectively, and let $\WitEntry{j}=(A_j(w)+\ii B_j(w))2^{-\WitPrec{j}}$. Writing $\rho_a=\rho_{L,a}$ and $c_a=c_{L,a}$ for $a\in\bits$, the exact input claim of~\cref{eq:claim} is
\[
 T_L=\sum_{a=0}^1 c_a\left(t_a^*+\sum_{j=1}^{n_\mathrm{w}}\chi_{g^{\mathrm w}_j}(\rho_a)\WitEntry{j}\right),\qquad
 t_a^*=\sum_{j=1}^{n_\mathrm{p}}\chi_{g^{\mathrm p}_j}(\rho_a)x_j.
\]
Both $t_a^*$ and $\chi_{g^{\mathrm w}_j}(\rho_a)$ depend only on public data and challenge indices. Using deterministic evaluation algorithms fixed before the interaction, compute binary approximations $t_a$ and $\widehat\chi_{a,j}$ satisfying
\begin{equation}\label{eq:zk-input-precision}
 |t_a-t_a^*|\le {\Tol}/8,\qquad
 |\widehat\chi_{a,j}-\chi_{g^{\mathrm w}_j}(\rho_a)|
 \le\frac{{\Tol}}{8\max\{1,n_\mathrm{w}\}\BV}.
\end{equation}
We define the input-value approximation
\[
 T_L^{\mathrm{app}}
 :=\sum_{a=0}^1c_a\left(
     t_a+\sum_{j=1}^{n_\mathrm{w}}
     \widehat\chi_{a,j}\WitEntry{j}\right).
\]
The bounds $|c_a|\le1$ and $|\WitEntry{j}|\le\BV$ imply
\[
 \begin{aligned}
 |T_L^{\mathrm{app}}-T_L|
 &\le\sum_{a=0}^1|c_a|\left(
      |t_a-t_a^*|
      +\sum_{j=1}^{n_\mathrm{w}}
       |\widehat\chi_{a,j}-\chi_{g^{\mathrm w}_j}(\rho_a)|
       \,|\WitEntry{j}|\right)\\
 &\le {\Tol}/2.
 \end{aligned}
\]
Let $\widehat T_L$ be the carried claim. Choose the working
precision so that the computed input residual $e_L$ satisfies
\[
 \left|e_L-(\widehat T_L-T_L^{\mathrm{app}})\right|
 \le {\Tol}/2,
\]
including the error from approximating the coefficients $c_a$.
The parameter $b$ below includes the working bits required by this
decomposition. Consequently,
\[
 \begin{aligned}
 |e_L-(\widehat T_L-T_L)|
 &\le |e_L-(\widehat T_L-T_L^{\mathrm{app}})|
      +|T_L^{\mathrm{app}}-T_L|\\
 &\le {\Tol}.
 \end{aligned}
\]
This is the error guarantee required by the input test
of~\cref{sec:protocol}; its acceptance threshold remains $4{\Tol}$.

Let $v$ consist of $t_a$, $\widehat\chi_{a,j}$, the rounded evaluations of $\AddPoly_i$ and $\MulPoly_i$ at the queried points, the rounded challenge values, and the rounded values of $\ParentWeight_i(g_i)$ used to compute $Z_i-w_i^+(a_i+b_i)-w_i^\times a_ib_i$. In the direct-evaluation setting, both parties compute $v$ deterministically from the public instance and challenge indices. The final Boolean predicate takes $v$ as public input and verifies the committed transcript, without recomputing the public input sums or evaluations of the polynomial extensions of the wiring predicates. The prover checks that the statement supplied to the proof for ${\mathsf B}$ contains this vector and the recorded challenges, aborts otherwise. For the logspace-uniform variant of~\cref{sec:zk-uniform}, the prover supplies the wiring evaluations and proves~\cref{eq:wiring-answer-bit} using the protocol of~\cref{sec:uniform}; the same stored strings are used in the final predicate after all checks accept. Computing $v$ is charged to the public preprocessing time, rather than to the size of ${\mathsf B}$.

\medskip\noindent{\bf Interaction.}
We retain the interaction structure of~\cref{sec:protocol}, but replace every witness-dependent prover message by bitwise commitments. Write $\mathbf c$ for the sequence of verifier challenge indices; $\mathbf d$ is defined above. The protocol proceeds as follows.
\begin{enumerate}[leftmargin=1.5em]
 \item \emph{Commit to the witness.} The verifier samples $R\leftarrow\bits^{3\kappa}$ and sends it to the prover. Before the first GKR challenge, the prover commits, using independent seeds, to every bit of a witness $w\in\mathcal W$.
 \item \emph{Commit to the GKR transcript.} At each sum-check round, the prover commits to the complete fixed-length coefficient vector of the round polynomial before the verifier samples the corresponding challenge. At layer $i$, the prover also commits to both claimed evaluations $a_i,b_i$ before the verifier samples the combining challenge $\lambda_i$. The verifier samples all challenge indices independently and uniformly, in precisely the order specified in~\cref{sec:protocol}. The numerical and magnitude checks are deferred until the final verification; the transcript contains commitments in place of the prover's numerical messages.
 \item \emph{Compute $v$.} In the direct-evaluation setting, both parties compute $v$ from the public instance and $\mathbf c$. In the logspace-uniform variant, its wiring-evaluation entries are the stored strings for which all checks in~\cref{sec:zk-uniform} accepted, while the remaining entries are computed directly. The verifier retains $(R,\mathbf d,\mathbf c,v)$; neither variant requires evaluating the delegated circuit.
 \item \emph{Prove acceptance.} The parties invoke the NP zero-knowledge protocol described above on the public statement
 \[
    {\mathsf x}=(C,x,y,R,\mathbf d,\mathbf c,v).
 \]
 An NP witness ${\mathsf w}$ specifies the bit strings and seeds opening every commitment in $\mathbf d$. The Boolean predicate ${\mathsf B}({\mathsf x},{\mathsf w})$ verifies all openings, checks that the initially committed witness belongs to $\mathcal W$, and runs the finite-precision GKR verifier defined in~\cref{sec:protocol} on the opened messages using $\mathbf c$ and $v$. In particular, the final input-layer check uses the inputs $\WitEntry{j}$ determined by that same $w$, without an additional claimed input-extension value. The verifier accepts exactly when the protocol reaches the NP proof and that proof accepts.
\end{enumerate}
The numerical parameters and the circuit computing the integer encodings from $w$ are part of the fixed description of ${\mathsf B}$. The prover checks that the public statement used in the final proof contains the actual commitments, challenges, and $v$; any inconsistency or malformed public verifier message causes an abort. Syntactic or numerical defects in committed values are handled by ${\mathsf B}$ according to the fixed rejection conventions above.

For the complexity bounds, let $\mu$ be the total bit length of the GKR prover messages and let $r_\mathrm{c}$ be the number of challenge bits. From~\cref{thm:agkr}, with $k$, $D$, $m$, $b$, and $\ell=\log N$ as in~\cref{sec:agkr},
\[
 \mu=O(kDb),\qquad r_\mathrm{c}=(k+L)\ell.
\]
Let $T_{\mathsf G}(\kappa)\ge\kappa$ be a polynomial bound on the Boolean circuit size of the pseudorandom generator, including its input and output lengths, and let $T_{\mathrm{op}}(b+\ell)=\poly(b+\ell)$ bound the size of each arithmetic operation or comparison used by the verifier on the specified binary encodings, including intermediate widths and rounding. The final predicate has size
\begin{equation}\label{eq:zk-circuit-size}
 |{\mathsf B}|=O\left((\WitLen+\mu)T_{\mathsf G}(\kappa)
       +(kD+n_\mathrm{w})T_{\mathrm{op}}(b+\ell)+F_{\mathcal W}+r_\mathrm{c}\right).
\end{equation}
Indeed, the first term checks the $\WitLen+\mu$ bit commitments, the second implements the sum-check, gate, and input-layer tests using $v$, and the remaining terms account for computing the integer encodings and testing $w\in\mathcal W$, and the challenge indices. In particular, computing public sums and evaluations of the polynomial extensions of the wiring predicates is not charged to $|{\mathsf B}|$: those values are supplied as public inputs to ${\mathsf B}$. The predicate uses explicit Boolean circuits for arithmetic on the specified binary encodings, without a computation-tableau reduction.

\begin{theorem}[Zero-Knowledge Approximate GKR]\label{thm:zk-direct}
Assume one-way functions secure against nonuniform polynomial-size circuits. Fix $\kappa\ge6$, a publicly specified circuit, set $\mathcal W$, and map $w\mapsto(A_j(w),B_j(w))_j$, and numerical parameters satisfying the conditions above. Suppose the public quantities required by the finite-precision GKR verifier can be computed deterministically with errors satisfying the bounds above. Then the promise problem $(\mathcal Y,\mathcal N)$ defined in~\cref{eq:zk-yes,eq:zk-no} admits a public-coin interactive proof with the following properties:
\begin{itemize}[leftmargin=*]
 \item \textbf{Completeness.} For every $(C,x,y)\in\mathcal Y$ and every $w\in\mathcal W$ satisfying $|y-C(x,w)|\le {\Tol}$, the honest prover on witness $w$ is accepted with probability $1$.

 \item \textbf{Soundness.} For every instance in $\mathcal N$ and every, possibly computationally unbounded, prover, the verifier accepts with probability at most $\eps_\mathrm{c}+(15/16)^{35}+2^{1-\kappa} <\eps_\mathrm{c}+1/8+2^{-\kappa}<1/2$.

 \item \textbf{Zero-knowledge.} The protocol is computationally zero-knowledge against classical probabilistic verifiers running in polynomial time, with arbitrary auxiliary input.

 \item \textbf{Communication.} With $\mu$, $r_\mathrm{c}$, and $|{\mathsf B}|$ as defined above, the communication is
 \begin{equation}\label{eq:zk-communication}
    O\bigl(\kappa(\WitLen+\mu)+r_\mathrm{c}+|{\mathsf B}|\poly(\kappa)\bigr).
 \end{equation}
 If the circuit computing the integer encodings and testing $w\in\mathcal W$ has size $F_{\mathcal W}\le(\WitLen+1)\poly(b)$ and $n_\mathrm{w}=O(\WitLen+1)$, then the parameter bounds of~\cref{sec:agkr} give communication $(\WitLen+1)\poly(\kappa,L,h,\log S,b)$.

 \item \textbf{Efficiency.} The honest prover runs in time polynomial in the circuit size and the stated parameters. The verifier runs in time $T_\mathrm{pub}+\poly(|{\mathsf B}|,\kappa)$ and uses $S_\mathrm{pub}+\poly(|{\mathsf B}|,\kappa,b+\ell+\log S)$ workspace, excluding read-only public data. Here $T_\mathrm{pub}$ and $S_\mathrm{pub}$ are the time and working space of the deterministic public evaluations, with storage of the output vector $v$ counted separately in the displayed workspace bound. For explicitly given unique polynomial extensions of the wiring predicates of degree below $h$ in each variable, the public evaluations can be implemented with $T_\mathrm{pub}=\poly(S,D,b+\ell),\; S_\mathrm{pub}=\poly(b+\ell+\log S)$. Thus, in this setting, the total verifier workspace is $\poly(|{\mathsf B}|,\kappa,b+\ell+\log S)$.
\end{itemize}
\end{theorem}

\subsection{Security Analysis}\label{sec:zk-security}

\begin{proof}[Proof of~\cref{thm:zk-direct}]
Throughout, we use the fixed encodings, numerical algorithms, and rejection conventions specified in~\cref{sec:zk-protocol}.

\medskip\noindent{\bf Completeness.}
Fix $(C,x,y)\in\mathcal Y$ and $w\in\mathcal W$ satisfying $|y-C(x,w)|\le {\Tol}$. The uniform gate-value bounds allow us to apply the completeness analysis of~\cref{thm:agkr} to the inputs $x$ and $\WitEntry{j}=(A_j(w)+\ii B_j(w))2^{-\WitPrec{j}}$. Its honest messages pass the encoding, length, and magnitude checks, the sum-check consistency checks, and the checks of $Z_i-w_i^+(a_i+b_i)-w_i^\times a_ib_i$ for every sequence of challenge indices in the specified sets. For the final input check, let $e_L$ denote the computed residual. The decomposition into the terms depending on $x$ and $w$ and~\cref{eq:zk-input-precision} give
\[
 |e_L-(\widehat T_L-T_L)|\le {\Tol}.
\]
The honest combined claim satisfies $|\widehat T_L-T_L|\le3{\Tol}$, so $|e_L|\le4{\Tol}$, as required. This argument is pointwise in the challenges; it does not require that they be sampled uniformly.

Every honest opening satisfies $d_j=\Com_R(\mathsf m_j;s_j)$, and the committed witness belongs to $\mathcal W$. Thus the opened witness and messages, together with their commitment seeds, satisfy ${\mathsf B}({\mathsf x},{\mathsf w})=1$. This remains true for every $R\in\bits^{3\kappa}$, whether or not it is binding. Perfect correctness of the MPC protocol and perfect completeness of the proof for ${\mathsf B}$ therefore imply acceptance with probability one by the honest verifier.

\medskip\noindent{\bf Soundness.}
Fix $(C,x,y)\in\mathcal N$ and an arbitrary prover $P^*$. Define
\[
 \mathsf{Bad}_R
 =\bigl\{R\in\{{\mathsf G}(s_0)\oplus {\mathsf G}(s_1):s_0,s_1\in\bits^\kappa\}\bigr\}.
\]
The honest verifier chooses $R$ uniformly, so $\Pr[\mathsf{Bad}_R]\le2^{-\kappa}$. For any fixed $R$ outside this event, every commitment has at most one openable bit. The opening seed need not be unique. Define a computationally unbounded prover $P'$ that runs $P^*$ and, on receiving $d_j$, enumerates $(\beta,s)\in\bits\times\bits^\kappa$ and tests $d_j={\mathsf G}(s)\oplus \beta R$. A commitment with no opening makes $\exists{\mathsf w}\,{\mathsf B}({\mathsf x},{\mathsf w})=1$ false.

In particular, the witness is determined before the first GKR challenge. Condition on the entire prefix up to this point, including the prover's state. If the recovered witness is not in $\mathcal W$, the final predicate rejects. Otherwise, fix this $w\in\mathcal W$. Since the instance is in $\mathcal N$, we have $|y-C(x,w)|>\Delta$. $P'$ assembles the recovered bits into each GKR message and sends it before receiving the corresponding challenge, yielding an adaptive prover for approximate GKR on input $(x,w)$. Its future challenges remain independent and uniform. The public numerical bounds hold for this witness by the uniform bounds.

Use the finite-precision implementation defined in~\cref{sec:zk-protocol} with the decomposition into terms depending on $x$ and $w$. Its error bound holds on every transcript passing the encoding, length, and magnitude checks, so acceptance of the final input check implies
\[
 |\widehat T_L-T_L|\le |e_L|+{\Tol}\le5{\Tol},
\]
exactly the terminal condition used in the soundness proof of~\cref{thm:agkr}. The remaining checks and error budgets are unchanged. Deferring a numerical rejection does not create an accepting transcript: once the rejection flag is set, ${\mathsf B}$ rejects regardless of subsequent messages. Hence the recovered messages and challenges coincide with the interaction between $P'$ and the approximate-GKR verifier up to its first rejection.

Let $\mathsf{Reach}$ be the event that the GKR interaction reaches the NP proof with a true statement, namely $\exists{\mathsf w}\,{\mathsf B}({\mathsf x},{\mathsf w})=1$. If $\mathsf{Bad}_R$ does not occur, $\mathsf{Reach}$ implies acceptance of $P'$ by the approximate-GKR verifier for the uniquely determined $w$. Applying~\cref{thm:agkr} for each fixed initial prefix and averaging over these prefixes and choices of $R$ gives
\[
 \Pr[\mathsf{Reach}\wedge\neg\mathsf{Bad}_R]\le\eps_\mathrm{c}.
\]
This argument conditions only on prefixes preceding the challenges in question; it does not condition on $\mathsf{Reach}$ or on the correctness of a later transcript.

For any fixed GKR transcript with a false final statement, the proof for ${\mathsf B}$ starts with an independent uniform $R'$ and fresh challenges. Its soundness bound therefore applies conditional on that transcript, regardless of the prover's retained state. Writing $\mathsf{Acc}$ for acceptance of the complete protocol, we obtain
\[
 \begin{aligned}
 \Pr[\mathsf{Acc}]
 &\le \Pr[\mathsf{Bad}_R]
       +\Pr[\mathsf{Reach}\wedge\neg\mathsf{Bad}_R]
       +\Pr[\mathsf{Acc}\wedge\neg \mathsf{Reach}]\\
 &\le 2^{-\kappa}+\eps_\mathrm{c}
       +(15/16)^{35}+2^{-\kappa}.
 \end{aligned}
\]
The two binding terms correspond to $R$ and $R'$, not to individual commitments or repetitions. No efficient extraction or computational assumption is used in this soundness argument.

\medskip\noindent{\bf Zero-Knowledge.}
Fix a classical polynomial-time verifier $V^*$ with polynomial-length auxiliary input, and consider a YES instance with a witness $w\in\mathcal W$ satisfying $|y-C(x,w)|\le {\Tol}$. Let $\mathsf{Sim}_\mathrm{NP}$ be the strict polynomial-time simulator specified in~\cref{sec:zk-protocol}. It runs in polynomial time on every statement and verifier state; its simulation guarantee applies to true statements. On those statements, its truncation probability is at most $35(15/16)^{\kappa^2}+\negl(\kappa)$.

Our simulator samples a random tape for $V^*$, runs the GKR interaction, and commits to zero in place of every witness bit and every GKR-message bit, using independent seeds. It records the challenge indices and computes $v$. On receiving $R$ or $R'$, check that it belongs to $\bits^{3\kappa}$. Check that each approximate-GKR challenge belongs to $\bits^{\ell}$. Before the final proof, check that its input equals $(C,x,y,R,\mathbf d,\mathbf c,v)$, using the commitments and challenges exchanged and the computed $v$. Abort if any check fails. In particular, it does not attempt to test whether a correctly sized commitment receiver message is binding. If the final proof is reached, it invokes $\mathsf{Sim}_\mathrm{NP}$ on the resulting ${\mathsf x}$, with the current verifier state and random tape. All rewinding is confined to this final proof. The simulator outputs the GKR transcript prefix together with the retained interaction in the proof for ${\mathsf B}$ as the view of the same execution of $V^*$; a simulator timeout produces $\bot$. Its running time is polynomial even if the resulting NP statement is false.

First replace only the interaction in the proof for ${\mathsf B}$ in the real execution by $\mathsf{Sim}_\mathrm{NP}$, leaving $\mathbf d$ genuine. Whenever this interaction is reached, the resulting statement is true by the completeness argument above, which holds for every sequence of challenge indices in the specified sets, including those chosen adaptively by $V^*$. Auxiliary-input zero-knowledge therefore justifies this replacement, with the preceding verifier state and GKR transcript included as auxiliary information. The abort decisions before this proof are unchanged.

Next let $M=\WitLen+\mu$ and order the bit commitments in $\mathbf d$ chronologically. For $j=0,\ldots,M$, let $\mathsf H_j$ commit to zero in the first $j$ positions and to the genuine bits thereafter, always generating the final interaction with $\mathsf{Sim}_\mathrm{NP}$. In these hybrids, the underlying honest GKR messages are computed from the fixed witness $w$ and the challenges actually received, irrespective of which commitments have been replaced. All message lengths are public and fixed.

Adjacent hybrids differ in one commitment. To reduce a distinguisher for them to commitment hiding, supply $w$ as polynomial-length nonuniform advice, compute the genuine bit at the selected position, and continue the interaction after receiving a commitment to either this bit or zero. The reduction completes the remaining interaction in polynomial time, without knowing the bit or seed used to compute the selected commitment: the commitments in $\mathbf d$ are never opened, and the final interaction is simulated. Later verifier challenges may depend on earlier commitments; the reduction simply runs $V^*$ to generate them. Furthermore, rewinding in the proof for ${\mathsf B}$ never crosses the selected commitment in $\mathbf d$. A polynomial-time algorithm distinguishing $\mathsf H_{j-1}$ from $\mathsf H_j$ with nonnegligible advantage would, through the reduction above, distinguish a commitment to $\mathsf m_j$ from a commitment to $0$ with the same advantage. This contradicts computational hiding.

The intermediate NP statements may be false. This causes no difficulty: this hybrid step uses only the unconditional polynomial running time of $\mathsf{Sim}_\mathrm{NP}$, not a simulation guarantee on false statements. There are polynomially many commitments, so $\mathsf H_0$ and $\mathsf H_M$ are computationally indistinguishable. In $\mathsf H_M$, the internal computation of genuine GKR messages can be omitted without changing any transmitted message or abort decision. The resulting experiment is exactly the witness-free simulator above. This proves computational zero-knowledge.

\medskip\noindent{\bf Complexity.}
The $M=\WitLen+\mu$ bit commitments in $\mathbf d$ contribute $3\kappa M$ bits, $R$ and $R'$ contribute $6\kappa$ bits, and the GKR challenges contribute $r_\mathrm{c}$ bits. The witness ${\mathsf w}$ for ${\mathsf B}$ consists of $M$ committed bits and $M$ seeds, hence has length $(\kappa+1)M$. Because $T_{\mathsf G}(\kappa)\ge\kappa$, these inputs and their opening checks are accounted for in~\cref{eq:zk-circuit-size}. With five MPC parties and $35$ repetitions, the final proof contributes $O(|{\mathsf B}|\poly(\kappa))$ bits. The vector $v$ and the final statement are determined locally from public data and the recorded interaction; they need not be retransmitted. This proves~\cref{eq:zk-communication}.

The honest prover generates the finite-precision GKR messages, the public data, and fresh MPC executions in polynomial time. The verifier spends $T_\mathrm{pub}$ time generating the public data and $\poly(|{\mathsf B}|,\kappa)$ time on the remaining checks. Its workspace is the public evaluator's working space $S_\mathrm{pub}$ plus storage for the commitments, challenges, vector $v$, and verification of the proof for ${\mathsf B}$, which is bounded by $\poly(|{\mathsf B}|,\kappa,b+\ell+\log S)$.

For explicitly given unique polynomial extensions of the wiring predicates of degree below $h$ in each variable, the bitwise evaluation algorithm underlying~\cref{lem:evaluation-bits} can obtain each gate's operation and ordered input labels from the read-only circuit description instead of running a uniform circuit generator. It generates each bit of $\EvaluationString_{p,q}(\mathsf q)$ in polynomial time and $\poly(b+\ell+\log S)$ working space. The public input sums, interpolation coefficients, roots, and values of $\ParentWeight_i$ can likewise be generated using arithmetic on the binary encodings fixed above and counters, with their output strings stored in $v$. With the stipulated working precision, this gives the displayed bounds on $T_\mathrm{pub}$ and $S_\mathrm{pub}$.

Finally, for these extensions, $D=2(h-1)$, $m=O(\log S)$, and $k=O(L\log S)$. Under the additional encoding assumptions in the theorem, $n_\mathrm{w}=O(\WitLen+1)$ and $F_{\mathcal W}\le(\WitLen+1)\poly(b)$. Substituting into~\cref{eq:zk-circuit-size,eq:zk-communication}, with the parameters chosen in~\cref{sec:agkr}, gives communication $(\WitLen+1)\poly(\kappa,L,h,\log S,b)$. The precision cost remains explicit through $b$.
\end{proof}

\subsection{Extension to Uniform Circuits}\label{sec:zk-uniform}

For circuits from a fixed logspace-uniform family, the verifier need not evaluate the polynomial extensions of the wiring predicates by summing over the gates. Instead, we use the interactive proofs of~\cref{eq:wiring-answer-bit} from~\cref{sec:uniform}. This changes only the computation of the public vector $v$: the witness commitments, the committed GKR interaction, and the final Boolean predicate remain as in~\cref{sec:zk-protocol}. The additional proofs concern deterministic functions of the public circuit parameters and the recorded challenges, not the witness or commitment openings.

\medskip\noindent{\bf Verifying wiring evaluations.}
After all GKR challenges have been sent, the recorded challenge indices determine $Q=2L$ wiring queries, one for each of $\AddPoly_i$ and $\MulPoly_i$ at each layer. Write $\mathsf q_j$ for the public description of query $j$, including its layer, gate type, and root indices. Fix public precisions $p,q$ and the deterministic evaluation algorithm of~\cref{lem:evaluation-bits}, and let
\[
 \mathsf e_j^*=\EvaluationString_{p,q}(\mathsf q_j)
 \qquad (1\le j\le Q)
\]
be the algorithm's output string. Let $b_{\mathrm w,j}=|\mathsf e_j^*|\le b_{\mathrm w}$ be the encoding length fixed by the public parameters. Each answer is transmitted with padding to the common public length $b_{\mathrm w}$; write $\mathsf e_j\in\bits^{b_{\mathrm w}}$ for the received string. The verifier checks the padding directly. Let $\sigma$ and $t_{\mathrm w}$ be the maximum space bound and public-statement length from~\cref{lem:evaluation-bits}.

Choose an evaluation-accuracy target, with $U$ the bound on claimed evaluations in~\cref{eq:bounds},
\begin{equation}\label{eq:zk-wiring-precision}
 0<\xi\le
 \min\left\{1,\frac{{\Tol}}{4(2U+U^2)}\right\}.
\end{equation}
Choose $p,q$ so that the hypotheses of~\cref{lem:evaluation-bits} hold and the value represented by each $\mathsf e_j^*$ has absolute error at most $\xi$. These precisions are fixed from the public bounds before the interaction. For any claimed evaluations satisfying $|a_i|,|b_i|\le U$, replacing both exact wiring values by these approximations changes the gate expression by at most
\[
 |\ParentWeight_i(g_i)|(2U+U^2)\xi
 \le 2(2U+U^2)\xi\le {\Tol}/2.
\]
Choose the precision for $\ParentWeight_i(g_i)$ and the remaining arithmetic so that their total contribution to the residual error is at most ${\Tol}/2$. Thus the computed residual for $Z_i-w_i^+(a_i+b_i)-w_i^\times a_ib_i$ retains the error bound ${\Tol}$ used in~\cref{sec:agkr}, uniformly over all committed messages passing the encoding, length, and magnitude checks. As in~\cref{sec:zk-protocol}, $b$ includes the working precision and the lengths of the supplied numerical encodings; the thresholds ${\Tol}$ and $\Delta$ are unchanged.

The prover sends all $Q$ answer strings before any challenge in the interactive proofs below. The verifier stores them, checks their lengths, encodings, padding, and decoded magnitudes against $\BW+1$, and rejects on failure. For $j=1,\ldots,Q$ and $\iota=1,\ldots,b_{\mathrm w,j}$, the parties apply the space-bounded proof of~\cref{thm:space-proof} to the public statement
\begin{equation}\label{eq:zk-wiring-answer-bit}
 \mathsf e_j[\iota]=\EvaluationString_{p,q}(\mathsf q_j)[\iota].
\end{equation}
Each statement contains only the query description, the bit position, and the claimed bit, not the full vector of answers. The bit algorithm and its space bound are the ones used in~\cref{thm:uniform}; in particular, the polynomial extensions of the wiring predicates of the circuit deciding~\cref{eq:zk-wiring-answer-bit} are evaluated directly from their small formulas. Fix $0<\eps_\mathrm{w}\le1/8$ and set
\begin{equation}\label{eq:zk-wiring-repetitions}
 r_\mathrm{w}=\ceil{\log_{10}(Qb_{\mathrm w}/\eps_\mathrm{w})}.
\end{equation}
For each statement in~\cref{eq:zk-wiring-answer-bit}, run $r_\mathrm{w}$ sequential copies of the constant-error proof with soundness error $1/10$, using fresh verifier challenges, and reject unless every copy accepts. The parameter $r_\mathrm{w}$ is unrelated to the $35$ repetitions in the proof for ${\mathsf B}$.

Only after all these interactive proofs accept do the parties invoke the proof for ${\mathsf B}$. The wiring entries of its public vector $v$ are the stored answer strings themselves; they cannot be replaced by separately supplied values. The other entries of $v$ are computed directly as in~\cref{sec:zk-protocol}. Both parties derive each statement in~\cref{eq:zk-wiring-answer-bit} and the statement $\exists{\mathsf w}\,{\mathsf B}({\mathsf x},{\mathsf w})=1$ from the recorded interaction, and the honest prover aborts on any inconsistent public statement. The verifier accepts only if all these checks and the proof for ${\mathsf B}$ accept. The transcripts of these interactive proofs need not be included in the Boolean predicate ${\mathsf B}$, whose size remains bounded by~\cref{eq:zk-circuit-size}.

\begin{theorem}[Zero-Knowledge Approximate GKR for Uniform Circuits]
\label{thm:zk-uniform}
Assume one-way functions secure against nonuniform polynomial-size circuits, and retain the witness encodings and numerical assumptions of~\cref{thm:zk-direct}. Let $C$ belong to a fixed logspace-uniform circuit family, and use the unique extensions of degree below $h$ in each variable given in~\cref{eq:uniform-extensions}. Fix $\kappa\ge6$ and $0<\eps_\mathrm{c},\eps_\mathrm{w}\le1/8$, with $\log(1/\eps_\mathrm{w})$ polynomially bounded in $\kappa$. Then the promise problem $(\mathcal Y,\mathcal N)$ defined in~\cref{eq:zk-yes,eq:zk-no} admits a public-coin interactive proof with the following properties:
\begin{itemize}[leftmargin=*]
 \item \textbf{Completeness.}
 For every $(C,x,y)\in\mathcal Y$ and every $w\in\mathcal W$ satisfying $|y-C(x,w)|\le {\Tol}$, the honest prover on witness $w$ is accepted with probability $1$.

 \item \textbf{Soundness.}
 For every instance in $\mathcal N$ and every, possibly computationally unbounded, prover, the verifier accepts with probability at most
 \begin{equation}\label{eq:zk-uniform-soundness}
  \eps_\mathrm{c}+\eps_\mathrm{w}+(15/16)^{35}+2^{1-\kappa}<1/2.
 \end{equation}

 \item \textbf{Zero-knowledge.}
 The protocol is computationally zero-knowledge against classical polynomial-time verifiers with arbitrary auxiliary input.

 \item \textbf{Communication.}
 With $\mu,r_\mathrm{c},|{\mathsf B}|$ as in~\cref{sec:zk-protocol} and $Q,b_{\mathrm w},\sigma,r_\mathrm{w}$ as above, the communication is
 \begin{equation}\label{eq:zk-uniform-communication}
  O\bigl(\kappa(\WitLen+\mu)+r_\mathrm{c}+|{\mathsf B}|\poly(\kappa)
       +Qb_{\mathrm w}r_\mathrm{w}\poly(\sigma)\bigr).
 \end{equation}

 \item \textbf{Efficiency.}
 The honest prover runs in time polynomial in the circuit size and the stated parameters. The verifier does not evaluate the polynomial extensions of the wiring predicates by enumerating the gates. Its running time is
 \begin{equation}\label{eq:zk-uniform-time}
     \begin{aligned}
        O\bigl((n_\mathrm{p}+n_\mathrm{w})mh+k+Lmh\bigr)\poly(b+\ell)+Qb_{\mathrm w}r_\mathrm{w}t_{\mathrm w}\poly(\sigma)+\poly(|{\mathsf B}|,\kappa),
     \end{aligned}
 \end{equation}
 and its workspace, excluding read-only public data, is
 \begin{equation}\label{eq:zk-uniform-space}
  \poly(|{\mathsf B}|,\kappa,b+\ell+\log S)
  +O(Qb_{\mathrm w})+\poly(\sigma)+O(\log r_\mathrm{w}).
 \end{equation}
 The $O(Qb_{\mathrm w})$ term explicitly accounts for the retained answer strings; completed transcripts of the proofs of~\cref{eq:zk-wiring-answer-bit} may be discarded.
\end{itemize}
For constant $\eps_\mathrm{c},\eps_\mathrm{w}$, if
$F_{\mathcal W}\le(\WitLen+1)\poly(b)$ and $n_\mathrm{w}=O(\WitLen+1)$,
then communication and verifier time are, respectively,
\[
 (\WitLen+1)\poly(\kappa,L,h,\log S,b)
 \quad\text{and}\quad
 (n_\mathrm{p}+1)\poly(\WitLen,\kappa,L,h,\log S,b).
\]
\end{theorem}

\begin{proof}[Proof of~\cref{thm:zk-uniform}]
Throughout the proof, $\mathsf e_j^*=\EvaluationString_{p,q}(\mathsf q_j)$ for $j=1,\ldots,Q$.

\medskip\noindent{\bf Completeness.}
Fix a witness $w\in\mathcal W$ with $|y-C(x,w)|\le {\Tol}$. Every honest answer consists of $\mathsf e_j^*$ with padding and passes the public encoding, length, and magnitude checks: its decoded modulus is at most $\BW+\xi\le\BW+1$. Every equality in~\cref{eq:zk-wiring-answer-bit} holds, so its proofs have perfect completeness. The numerical bound above preserves the checks of $Z_i-w_i^+(a_i+b_i)-w_i^\times a_ib_i$, and the input check is unchanged. The completeness argument of~\cref{sec:zk-security} therefore gives ${\mathsf B}({\mathsf x},{\mathsf w})=1$ for the honest openings and every sequence of challenge indices in the specified sets. Perfect completeness of the final proof completes the argument. Each constituent proof is public coin, as are the two commitment setups.

\medskip\noindent{\bf Soundness.}
Fix a NO instance and an arbitrary prover. Each statement in~\cref{eq:zk-wiring-answer-bit} is fixed before the fresh challenges of its proof. Conditional on any preceding transcript for which that statement is false, all $r_\mathrm{w}$ repetitions accept with probability at most $10^{-r_\mathrm{w}}$, even if the prover adapts its strategy across repetitions. Let $\mathsf{Bad}_\mathrm{w}$ be the event that a false statement in~\cref{eq:zk-wiring-answer-bit} passes all its repetitions. There are at most $Qb_{\mathrm w}$ statements, so
\begin{equation}\label{eq:zk-uniform-bit-error}
 \Pr[\mathsf{Bad}_\mathrm{w}]
 \le Qb_{\mathrm w}10^{-r_\mathrm{w}}\le\eps_\mathrm{w}.
\end{equation}
This union bound does not require independence between these interactive proofs.

Let $\mathsf{Bad}_R$ be the failure of the global binding event for $R$, with probability at most $2^{-\kappa}$. On its complement, recover the committed witness and messages online exactly as in~\cref{sec:zk-security}. In particular, the witness is fixed before the first GKR challenge and each message before its corresponding challenge. Let $\mathsf{Good}$ be the event that these strings are well formed, the witness belongs to $\mathcal W$, and the numerical verifier accepts them when supplied with $v$ computed by the specified algorithms, regardless of whether the later proofs of~\cref{eq:zk-wiring-answer-bit} accept. This event is determined by the GKR transcript prefix alone. Applying approximate-GKR soundness to $P'$ from~\cref{sec:zk-security}, with the error bound $\xi$ established above, gives
\[
 \Pr[\mathsf{Good}\wedge\neg\mathsf{Bad}_R]\le\eps_\mathrm{c}.
\]
These interactive proofs take place after that prefix, so this application of approximate GKR conditions on neither their acceptance nor correctness of the later answer strings.

Let $\mathsf{Reach}$ be the event that the proof for ${\mathsf B}$ is reached with a true statement. If neither $\mathsf{Bad}_R$ nor $\mathsf{Bad}_\mathrm{w}$ occurs, reaching that proof means that each stored answer consists of $\mathsf e_j^*$ with the required padding. The same strings enter ${\mathsf B}$, and binding fixes the opened witness and GKR messages. Hence
\[
 \mathsf{Reach}\wedge\neg\mathsf{Bad}_R\wedge\neg\mathsf{Bad}_\mathrm{w}
 \ \subseteq\ \mathsf{Good}\wedge\neg\mathsf{Bad}_R.
\]
For each fixed transcript reaching a false final statement, the NP proof starts with an independent $R'\leftarrow\bits^{3\kappa}$ and fresh challenges. Its conditional soundness error is at most $(15/16)^{35}+2^{-\kappa}$. Writing $\mathsf{Acc}$ for acceptance of the complete protocol, we conclude that
\[
 \begin{aligned}
 \Pr[\mathsf{Acc}]
 &\le \Pr[\mathsf{Bad}_R]+\Pr[\mathsf{Bad}_\mathrm{w}]
      +\Pr[\mathsf{Good}\wedge\neg\mathsf{Bad}_R]
      +\Pr[\mathsf{Acc}\wedge\neg \mathsf{Reach}]\\
 &\le \eps_\mathrm{c}+\eps_\mathrm{w}+(15/16)^{35}+2^{1-\kappa}.
 \end{aligned}
\]
Malformed answers or rejected proofs of~\cref{eq:zk-wiring-answer-bit} terminate the protocol and cannot increase acceptance.

\medskip\noindent{\bf Zero-Knowledge.}
Fix a classical polynomial-time verifier $V^*$ with auxiliary input and a witness satisfying the completeness condition. Extend the simulator of~\cref{sec:zk-security} as follows. After generating the commitments to zero at the positions of $\mathbf d$ and recording the verifier's challenge indices in the specified sets, compute $\mathsf e_j^*=\EvaluationString_{p,q}(\mathsf q_j)$, send each with the required padding, and execute the honest prover for~\cref{eq:zk-wiring-answer-bit} against $V^*$. Both the answers and this prover's next-message computation use only public data and the interaction so far. The bit algorithm is deterministic, and its accepting computation, including any information needed by the space-bounded prover, can be generated in polynomial time. Thus these interactive proofs can be run without the witness, including against a verifier that chooses challenge indices adaptively within the specified sets. The simulator applies the same message checks and abort rules as the honest prover. If the proof for ${\mathsf B}$ is reached, invoke the strict polynomial-time simulator $\mathsf{Sim}_\mathrm{NP}$ from~\cref{sec:zk-protocol}; its rewinding remains confined to that proof.

To justify the simulation, first replace only the interaction in the proof for ${\mathsf B}$ in the real execution by $\mathsf{Sim}_\mathrm{NP}$. Commitments in $\mathbf d$ to the honest bits and the honest prover for~\cref{eq:zk-wiring-answer-bit} yield a true final statement whenever it is reached, even for GKR challenge indices chosen by $V^*$ within the specified sets. The accumulated verifier state, including the transcripts of these interactive proofs, is supplied as auxiliary input to the NP simulator. Next replace $\mathbf d$ by commitments to zero one bit at a time. In every hybrid, generate the honest interaction for~\cref{eq:zk-wiring-answer-bit} for the challenges actually received. These proof messages can be generated in polynomial time from the public circuit parameters and the messages received so far, without $w$ or the pairs $(\mathsf m_j,s_j)$ satisfying $d_j={\mathsf G}(s_j)\oplus\mathsf m_jR$. Commitment hiding therefore justifies the same hybrids as in~\cref{sec:zk-security}. The statements for ${\mathsf B}$ in intermediate hybrids may be false; only the simulator's polynomial running time, not simulation correctness on false statements, is used there. The last hybrid is the witness-free simulator just described. No zero-knowledge property of the proof of~\cref{eq:zk-wiring-answer-bit} is required.

\medskip\noindent{\bf Complexity.}
There are at most $Qb_{\mathrm w}r_\mathrm{w}$ constant-error executions of the proof of~\cref{eq:zk-wiring-answer-bit}. By~\cref{thm:space-proof,lem:evaluation-bits}, their total prover time, verifier time, and communication are, respectively,
\[
 Qb_{\mathrm w}r_\mathrm{w}\poly(S,p,q,\ell),\qquad
 Qb_{\mathrm w}r_\mathrm{w}t_{\mathrm w}\poly(\sigma),\qquad
 Qb_{\mathrm w}r_\mathrm{w}\poly(\sigma).
\]
Sending the answer strings costs another $Qb_{\mathrm w}$ bits, absorbed by the last bound because $r_\mathrm{w}\ge1$. The verifier reuses its workspace between these executions and retains only the answer strings, the GKR challenges determining their queries, and the acceptance status. In particular, these proofs do not enlarge ${\mathsf B}$ by the size of their transcripts. Adding the communication for $\mathbf d$ and the proof for ${\mathsf B}$ gives~\cref{eq:zk-uniform-communication}.

The directly computed public input sums and interpolation coefficients require $O((n_\mathrm{p}+n_\mathrm{w})mh)$ arithmetic operations; challenge roots and values of $\ParentWeight_i$ require $O(k+Lmh)$. Implementations using these binary encodings, the proofs of~\cref{eq:zk-wiring-answer-bit}, and the verification of the proof for ${\mathsf B}$ give~\cref{eq:zk-uniform-time,eq:zk-uniform-space}. The honest prover for~\cref{eq:zk-wiring-answer-bit} and the simulation described above have polynomial running time under the standing parameter bounds. Finally, substitute $D=2(h-1)$, $k=O(L\log S)$, the bounds of~\cref{lem:evaluation-bits}, and the stated witness-encoding assumptions into~\cref{eq:zk-circuit-size,eq:zk-uniform-communication,eq:zk-uniform-time} to obtain the specialized bounds.
\end{proof}

\medskip\noindent{\bf Sequential Amplification.}
Both~\cref{thm:zk-direct,thm:zk-uniform} admit amplification without changing the thresholds ${\Tol}$ and $\Delta$. For $0<\eps<1$ with $\log(1/\eps)$ polynomially bounded in $\kappa$, run
\[
 r_\mathrm{amp}=\ceil{\log(1/\eps)}
\]
complete executions sequentially and accept only if all accept. Each execution uses fresh commitment receiver messages $R,R'$, commitment seeds, MPC sharings and random tapes, and verifier challenges, including those in the proofs of~\cref{eq:zk-wiring-answer-bit} when present. On a NO instance, the conditional acceptance probability of each execution is at most $1/2$, whatever the preceding transcript; thus the total soundness error is at most $2^{-r_\mathrm{amp}}\le\eps$. Perfect completeness is preserved. Sequential simulation supplies the accumulated verifier state as auxiliary input at each execution, and a polynomial number of hybrids preserves computational zero-knowledge. Time and communication increase by a factor $r_\mathrm{amp}$; the honest verifier reuses its workspace with an additional $O(\log r_\mathrm{amp})$-bit counter. The base parameters, and hence ${\Tol}$, $\Delta$, and the promise problem, remain fixed.
\section{One-Round Arguments for Approximate Computation}\label{sec:one-round}

We transform the public-coin protocols of~\cref{sec:agkr,sec:implementation} into one-round (two-message) arguments for approximate circuit evaluation. For a public instance $(C,x,y)$ fixed independently of the verifier's randomness, completeness holds when $|y-C(x)|\le {\Tol}$, and computational soundness holds when $|y-C(x)|>\Delta$. We apply the private-information-retrieval (PIR) transformation of Kalai and Raz~\cite{KR}, following~\cite[Section~6]{GKR}. The main requirement for an efficient honest prover is a short description of the challenges determining each response. We establish this property for both approximate GKR and its wiring-evaluation subprotocols. The verifier's first message depends only on the fixed public protocol parameters and fresh randomness; checking the response additionally requires the verifier's stored random tape, PIR queries, and recovery information.

In~\cref{sec:one-round-protocol}, we specify the transformation and its security and complexity guarantees. In~\cref{sec:one-round-agkr}, we establish the bound on the number of verifier bits determining each response for approximate GKR on logspace-uniform circuits, including the wiring-evaluation proofs, and derive the resulting one-round argument.

\subsection{Protocol}\label{sec:one-round-protocol}

We present the transformation of Kalai and Raz~\cite{KR} for public-coin interactive proofs whose honest responses depend on a bounded number of preceding verifier bits. We first transform the interaction so that each honest response is determined by the instance, the round index, and the current verifier message. We then use PIR to retrieve the corresponding database entries in a single exchange. Under the quantitative PIR assumption stated below, this yields a private-coin argument (\cref{def:ip}) with negligible losses in completeness and soundness, where soundness holds against computationally bounded provers. We establish the communication and running-time bounds, making explicit the cost of constructing the databases. The verifier's first message depends only on the fixed public protocol parameters and fresh randomness; verification additionally requires its stored random tape, PIR queries, and recovery information.

\medskip\noindent{\bf Protocol parameters and response functions.}
Fix public protocol parameters specifying the common-input length $n_\mathrm{inst}$, the message lengths and order, and all numerical bounds and precisions. Consider a public-coin interactive proof $(P,V)$ on an instance $I\in\bits^{n_\mathrm{inst}}$ with these parameters. The honest prover is deterministic, and all verifier randomness is sent as fresh challenge bits at positions fixed by the protocol parameters. Numerical algorithms, rounding rules, and rejection conventions are fixed in advance. After a failed check, the verifier sets a permanent rejection flag but completes the remaining rounds with the same message lengths, message order, and challenge generation, using fixed dummy operands where necessary. Thus rejection changes neither subsequent message lengths nor challenge generation. Let $\mu_\mathrm{tot}$ and $r_\mathrm{v}$ be the total numbers of prover and verifier bits in this underlying proof, including any subprotocols, with $\mu_\mathrm{tot}+r_\mathrm{v}\ge1$. Since all verifier bits are fresh challenges, $r_\mathrm{v}$ is also its random-tape length.

Fix $\Lambda\ge1$. For each honest response bit $t$, fix a set $\mathcal S_t$ of preceding verifier-bit positions with $|\mathcal S_t|\le\Lambda$. If $r$ is the verifier's random tape, the response is a function of $(I,r|_{\mathcal S_t})$, where $r|_{\mathcal S_t}$ lists the bits of $r$ at these positions. The sets $\mathcal S_t$ are fixed by the protocol parameters and need not consist of consecutive positions. We require that the position lists, without repetitions, and the message order can be generated from the protocol parameters in time $\poly(\mu_\mathrm{tot}+r_\mathrm{v},\Lambda,\log(n_\mathrm{inst}+2))$. For every assignment to the bits at positions $\mathcal S_t$, the response function must be defined and computable in $\poly(\ProverTime)$ time, where $\ProverTime$ bounds the honest prover's running time. These restrictions concern only the honest prover; a dishonest prover may use its entire preceding view.

\medskip\noindent{\bf Transforming the interaction.}
Split every prover message into bits. Immediately before each response bit, the verifier resends the challenge bits at positions $\mathcal S_t$, padded with zeros to length $\Lambda$. Retain the original verifier bits in their original order, sending each in a separate $\Lambda$-bit padded message followed by an ignored dummy response. Every round of the transformed protocol therefore consists of a $\Lambda$-bit verifier message followed by one prover bit. If round $j$ carries response bit $t$, its verifier message $z_j$ is $r|_{\mathcal S_t}$ padded in this way. We call the transformed protocol \emph{normalized}. There are $n_{\mathrm r}=\mu_\mathrm{tot}+r_\mathrm{v}$ rounds of the transformed protocol, with total communication
\begin{equation}\label{eq:history-communication}
 \Communication_\mathrm{norm}=n_{\mathrm r}(\Lambda+1)
   =(\mu_\mathrm{tot}+r_\mathrm{v})(\Lambda+1).
\end{equation}
This transformation adds no randomness, so the random tape still has length $r_\mathrm{v}\le\Communication_\mathrm{norm}$.

No challenge bit is disclosed earlier than in the original proof. An unbounded prover in that proof can therefore simulate the inserted messages and dummy rounds, preserving soundness. The honest transcript is unchanged after deleting the inserted messages, preserving completeness. Write $f_j(I,z)\in\bits$ for the honest response in round $j$ of the transformed protocol to $z\in\bits^\Lambda$, with dummy responses and responses to invalid padding set to zero. The transformed proof is \emph{history-ignorant} in the sense of~\cite[Definition~4]{KR}: each honest response depends only on the instance, the round index, and the current verifier message. The verifier messages are functions only of its random tape and the protocol parameters, although repeated challenge bits make these messages correlated.

\medskip\noindent{\bf PIR assumption.}
We assume a uniform PIR scheme with the quantitative guarantees of~\cite[Definition~3]{KR}. That is, the scheme satisfies \cref{def:pir} with $\mathrm{err}(\tau)=\mathrm{adv}(\tau)=2^{-\tau^3}$ and $\mathrm{size}(\tau)=2^{\tau^3}$; for an $M$-bit database, query generation and recovery take $\poly(\tau,\log M)$ time, the server takes $\poly(\tau,M)$ time, and communication is $\poly(\tau,\log M)$. These guarantees are assumed for all $M$ and $\tau$; polynomial-time privacy alone is not the hypothesis of the transformation.

Set $M=2^\Lambda$ and index the database $\mathsf{db}\in\bits^M$ by $\bits^\Lambda$. We write
\[
 \begin{aligned}
  (\mathsf{qry},\mathsf{st})&\leftarrow\PIRQuery(\tau,M,z),\\
  \mathsf{ans}&\leftarrow\PIRAnswer(\tau,\mathsf{db},\mathsf{qry}),\\
  a&=\PIRRecover(\tau,M,z,\mathsf{qry},\mathsf{st},\mathsf{ans})
 \end{aligned}
\]
for query generation, answering, and recovery. Here $\mathsf{st}$ is private recovery information, and recovery is deterministic. All messages have fixed padded encodings and length bounds fixed by the PIR scheme. The parameter $\tau$ controls PIR security; in the application below, it may exceed the target security parameter $\kappa$.

\medskip\noindent{\bf Two-message protocol.}
Both parties hold the instance $I$, fixed independently of the verifier's randomness. The honest database for round $j$ of the transformed protocol is
\[
 \mathsf{db}_j[z]=f_j(I,z),\qquad z\in\bits^\Lambda.
\]
The interaction consists of the following two messages.
\begin{enumerate}[leftmargin=*]
 \item \emph{Verifier message.} The verifier samples the original random tape and computes all messages of the transformed protocol $z_1,\ldots,z_{n_{\mathrm r}}$. Independently for each $j$, it generates
 \[
  (\mathsf{qry}_j,\mathsf{st}_j)\leftarrow\PIRQuery(\tau,M,z_j).
 \]
 It sends $(\mathsf{qry}_1,\ldots,\mathsf{qry}_{n_{\mathrm r}})$ and retains the tape, queries, and recovery information. The indices $z_j$ can be recomputed from the tape.

 \item \emph{Prover message.} The prover computes each database $\mathsf{db}_j$ and forms
 \[
  \mathsf{ans}_j\leftarrow\PIRAnswer(\tau,\mathsf{db}_j,\mathsf{qry}_j)
 \]
 with fresh randomness for each invocation. It sends $(\mathsf{ans}_1,\ldots,\mathsf{ans}_{n_{\mathrm r}})$.
\end{enumerate}
The verifier rejects malformed responses or recovery outputs outside $\bits$. Otherwise, it recovers the response bits, discards the dummy responses, and applies the original acceptance test using its saved tape. The databases specify the honest algorithm only: a dishonest prover may produce its answers jointly without constructing any databases. The verifier does not disclose its tape or private recovery information.

\begin{theorem}[Kalai--Raz Transformation]\label{thm:kr}
Under the preceding protocol and PIR assumptions, let $(P,V)$ have completeness $c$, soundness $s$ against arbitrary provers, honest-prover time $\ProverTime$, and verifier time $\VerifierTime=\poly(n_\mathrm{inst})$. Let $\Communication_\mathrm{norm}$ be as in~\cref{eq:history-communication}. For every integer
\[
 \tau\ge\max\{\tau_0,\Communication_\mathrm{norm},
              \ceil{\log(n_\mathrm{inst}+2)}\},
\]
where $\tau_0\ge2$ is a sufficiently large constant depending only on the fixed algorithms, the compiled protocol is a two-message private-coin argument with the following properties:
\begin{itemize}[leftmargin=*]
 \item \textbf{Completeness.} Its completeness is at least $c-2^{-\tau^2}$.

 \item \textbf{Soundness.} Its soundness is at most $s+2^{-\tau^2}$ against nonuniform provers of size at most $2^\tau$.

 \item \textbf{Complexity.} Communication is $\poly(\tau)$, honest-prover time is $\poly(\ProverTime,\tau,2^\Lambda)$, and verifier time is
 \[
  (\VerifierTime+\Communication_\mathrm{norm})\poly(\tau).
 \]

 \item \textbf{First message.} It can be generated in $\poly(\tau)$ time from the protocol parameters and fresh randomness, independently of the instance values. Verification uses the stored random tape, PIR queries, and recovery information.
\end{itemize}
The guarantees hold also for promise problems, with no requirement on instances outside the promise.
\end{theorem}

\begin{proof}[Proof of~\cref{thm:kr}]
The transformed protocol is history-ignorant, its verifier messages depend only on its random tape and the protocol parameters, and its random-tape length is at most $\Communication_\mathrm{norm}$. Since $\Lambda\ge1$, this also gives $r_\mathrm{v}=O(\Communication_\mathrm{norm}\Lambda)$. Hence \cite[Lemma~4.2]{KR} applies with communication bound $\Communication_\mathrm{norm}$ and maximum verifier-message length $\Lambda$. It gives soundness $s+2^{-\tau^2}$ even when the prover generates correlated PIR answers. The reduction fixes the instance and invokes the underlying soundness only on that instance; it therefore applies equally to NO instances of a promise problem.

For completeness, fix the original verifier tape. Each queried database entry is the corresponding bit of the original honest transcript. A union bound over the $n_{\mathrm r}\le\Communication_\mathrm{norm}\le\tau$ recoveries gives
\[
 \Pr[\text{some recovery fails}]
 \le n_{\mathrm r}2^{-\tau^3}\le2^{-\tau^2}.
\]
Without a recovery error, the acceptance decision is unchanged. Averaging over the tape gives completeness at least $c-2^{-\tau^2}$.

Each of the $n_{\mathrm r}$ databases has $2^\Lambda$ entries, computable individually in $\poly(\ProverTime)$ time. Together with the PIR server cost, this gives the prover bound. Since $n_{\mathrm r},\Lambda\le\Communication_\mathrm{norm}\le\tau$ and $\log(n_\mathrm{inst}+2)\le\tau$, generating the messages of the transformed protocol, forming the queries, and recovering the answers take $\poly(\tau)$ time and communication. The verifier then reconstructs the original transcript and runs its acceptance test. This gives the stated verifier bound and the instance-independent first-message algorithm.
\end{proof}

\subsection{One-Round Approximate GKR}\label{sec:one-round-agkr}

We apply the transformation of~\cref{sec:one-round-protocol} to approximate GKR for logspace-uniform circuits, including the interactive proofs of~\cref{eq:wiring-answer-bit}. The main step is to choose interpolation sets for which every encoded honest response depends on only logarithmically many challenge bits, measured in the circuit size and numerical precision. We establish this property for both approximate GKR and these interactive proofs, obtaining polynomial-size databases and hence a polynomial-time honest prover. We then derive the security and bit-complexity bounds: communication is polynomial in the circuit depth, logarithmic circuit size, numerical precision, and security parameter, while verifier time has an additional linear dependence on the number of input gates. The transformation preserves the numerical thresholds ${\Tol}$ and $\Delta$ and incurs only negligible additional completeness and soundness losses, with soundness becoming computational.

\medskip\noindent{\bf Parameter choices.}
Fix a logspace-uniform circuit family and its generator, as in~\cref{sec:uniform}. A circuit $C$ is specified by its public family parameter; the verifier need not read an explicit gate list. Let $S_0\ge2$ be its size before padding, $1\le L\le S_0$ its depth, and $n$ the number of input gates, including constants. We retain the notation of~\cref{sec:protocol,thm:agkr}: $S$ is the total size after padding, and $\ell=\log N$ is the length of a challenge index. In particular, $n$ counts complex input entries, whereas $n_\mathrm{inst}$ denotes the bit length of the encoded common input used by the transformation. Use the unique polynomial extensions of the wiring predicates of degree below $h$ in each variable of~\cref{eq:uniform-extensions}, with
\begin{equation}\label{eq:one-round-h}
 h=2^{\ceil{\log\max\{2,L,\log(S_0+2)\}}}.
\end{equation}
Set $\eps_\mathrm{c}=\eps_\mathrm{w}=1/8$ for the soundness errors of approximate GKR and all wiring-evaluation proofs, respectively. For $\Delta>0$, choose ${\Tol}$ as in~\cref{sec:soundness} and the evaluation accuracies as in~\cref{thm:uniform}. Let $b$ bound the input, output-claim, message, and working bit lengths, including the numerical parameters and the precisions of the public wiring-evaluation algorithms. These parameters and bounds are fixed uniformly over the specified input domain. The promise is $|y-C(x)|\le {\Tol}$ versus $|y-C(x)|>\Delta$; no guarantee is required in the intervening gap.

\begin{theorem}[One-Round Approximate GKR]\label{thm:one-round-agkr}
Assume the PIR scheme of~\cref{sec:one-round-protocol} and the circuit and numerical conditions above. For every integer $\kappa\ge2$, there is a two-message private-coin argument with a PIR security parameter $\tau\ge\kappa$ satisfying~\cref{thm:kr} and bounded by $\poly(\kappa,L,\log S_0,b)$. For every instance $(C,x,y)$ with the fixed parameters, chosen independently of the verifier's randomness, it has the following properties:
\begin{itemize}[leftmargin=*]
 \item \textbf{Completeness.} If $|y-C(x)|\le {\Tol}$, the honest prover is accepted with probability at least $1-2^{-\tau^2}$.

 \item \textbf{Soundness.} If $|y-C(x)|>\Delta$, every nonuniform prover of size at most $2^\kappa$ is accepted with probability at most
 \[
  1/4+2^{-\tau^2}<1/2.
 \]

 \item \textbf{Communication and verifier time.} These costs are, respectively,
 \begin{equation}\label{eq:one-round-cost}
  \poly(\kappa,L,\log S_0,b),\qquad
  (n+1)\poly(\kappa,L,\log S_0,b).
 \end{equation}

 \item \textbf{Prover time.} The honest prover runs in $\poly(S_0,b,\kappa)$ time.

 \item \textbf{First message.} It is generated in $\poly(\kappa,L,\log S_0,b)$ time using only the protocol parameters and fresh randomness. Verification additionally requires the stored random tape, PIR queries, and recovery information.
\end{itemize}
\end{theorem}

\begin{proof}[Proof of~\cref{thm:one-round-agkr}]
We bound the challenge dependence of the honest GKR messages and the messages in the proofs of~\cref{eq:wiring-answer-bit}, account for their costs, and apply~\cref{thm:kr}. All honest numerical algorithms use the fixed evaluation orders, precisions, and rounding rules of~\cref{sec:completeness}. They are terminated at their fixed polynomial step bounds and return fixed dummy strings on malformed inputs, making every database entry well defined.

\medskip\noindent{\bf Verifier bits determining each GKR response.}
Padding increases each layer's size by at most a factor $h$. Since $h=O(S_0)$ and $S_0\ge2$, the padded size and maximum number of coordinates per gate label satisfy
\[
 S=O(hS_0),\qquad
 m=O\!\left(1+\frac{\log S_0}{\log h}\right)
   =O\!\left(\frac{\log S_0}{\log h}\right).
\]
The polynomial extensions of the wiring predicates have individual degree $\delta=h-1$, so $D=\delta+h-1=2(h-1)$ and $k=O(Lm)$. Since $L\le h$ and $m=O(h)$, the constant-error choice in~\cref{eq:parameters} gives
\[
 N=O(kD+L)=\poly(h),\qquad
 \ell=O(\log h),\qquad m\ell=O(\log S_0).
\]

At layer $i>0$, the polynomial $q_j$ is determined by the two evaluation points $\rho_{i,0},\rho_{i,1}$, the preceding combining challenge, and the current sum-check prefix. These require at most $(2m_i+1+k_i)\ell$ bits. At layer zero, the initial points and coefficients are fixed. An evaluation of $\GateExtension_{i+1}$ depends only on its current point and the instance. The canonical prover recomputes these quantities from the selected indices, without using earlier rounded claims or testing earlier verifier residuals. Thus equal selected indices give equal \emph{encoded} responses, not only equal exact arithmetic values. Consequently,
\begin{equation}\label{eq:outer-history}
 \Lambda_\mathrm{out}\le(5m+1)\ell=O(\log S_0).
\end{equation}
Layer indices, round numbers, coefficient positions, and precisions are fixed by the protocol parameters and do not enlarge the database index.

\medskip\noindent{\bf Verifier bits determining each wiring-evaluation response.}
Use the notation of~\cref{lem:evaluation-bits}: $p,q$ are the root and output precisions, $b_{\mathrm w}$ bounds the answer length, $t_{\mathrm w}$ bounds the length of the statement in~\cref{eq:wiring-answer-bit}, and $\sigma$ bounds the evaluation space. Take common bounds over all queries. We take $\sigma\ge2$ large enough to include the input-head position, so that $\sigma\ge\ceil{\log(t_{\mathrm w}+2)}$. The required accuracies and the definition of $b$ permit
\[
 p,q=O(b+\log S),\qquad
 \sigma=O\bigl(\log S+\log(b+\ell+2)\bigr).
\]
Each wiring query is specified by $O(m\ell)$ GKR challenge bits. For a query $\mathsf q$, the fixed evaluation algorithm computes $\EvaluationString_{p,q}(\mathsf q)$ from these bits and the public circuit parameters. Hence each honest answer bit has this same challenge dependence.

For the proofs of~\cref{eq:wiring-answer-bit}, use the space-bounded construction with
\[
 h_\mathrm{aux}=2^{\ceil{\log(\sigma+2)}}.
\]
Although the proof of~\cref{thm:space-proof} uses $h_\mathrm{aux}=2$, \cref{lem:space-circuit} also allows this larger choice. The circuit used to verify each answer bit has size $2^{O(\sigma)}$, depth $\poly(\sigma)$, and number of coordinates per gate label $O(\sigma/\log(\sigma+2))$. Its gate values have modulus at most one, and its polynomial extensions of the wiring predicates have degree $\poly(\sigma)$ and coefficient norms $2^{\poly(\sigma)}$. At soundness error $1/10$, \cref{eq:parameters,eq:precision-choice} therefore give challenge-index length $O(\log(\sigma+2))$ and working precision $\poly(\sigma)$. The analysis of~\cref{thm:space-proof} gives prover time $2^{O(\sigma)}$, verifier time $t_{\mathrm w}\poly(\sigma)$, and communication $\poly(\sigma)$.

The preceding argument for $q_j$ now bounds the number of verifier bits determining a response in the proof of~\cref{eq:wiring-answer-bit} by $O(\sigma)$ challenge bits in that proof, together with the $O(m\ell)$ GKR challenge bits specifying its query. For each database entry, the honest prover recomputes $\EvaluationString_{p,q}(\mathsf q)$ and its asserted bit from the query; neither a previously sent answer nor a rounded claim is an additional index. Bit positions, repetition numbers, circuits deciding~\cref{eq:wiring-answer-bit}, and precision bounds are fixed by the protocol parameters. The circuit deciding~\cref{eq:wiring-answer-bit} and numerical bounds cover every query and either asserted bit value, using the input lengths fixed by the protocol parameters. Only the honest response function uses $\EvaluationString_{p,q}(\mathsf q)[\iota]$ for the asserted bit: the verifier always checks the bit actually supplied by the prover. Thus, for the complete proof, we take
\begin{equation}\label{eq:all-history}
 \begin{aligned}
  \Lambda&=O(m\ell+\sigma)
          =O\bigl(\log S_0+\log(b+2)\bigr),\\
  2^\Lambda&=\poly(S_0,b).
 \end{aligned}
\end{equation}
Every position in $\mathcal S_t$ precedes response $t$. Their locations are obtained from the layer, round, bit, and repetition indices, so the position lists have the generation bound required in~\cref{sec:one-round-protocol}. Intervening proofs of~\cref{eq:wiring-answer-bit} need not be included in a GKR message's database index.

\medskip\noindent{\bf Underlying security and complexity.}
The numerical error allocation of~\cref{thm:uniform} remains valid for these interpolation sets. Let $\xi$ bound the absolute numerical error of the values represented by $\EvaluationString_{p,q}(\mathsf q)$; this is distinct from the soundness error of their interactive proofs, $\eps_\mathrm{w}$. For claimed evaluations satisfying the magnitude bound $U$ of~\cref{eq:bounds}, the resulting perturbation of a gate expression is at most $2(2U+U^2)\xi$. Choose $0<\xi\le\min\{1,{\Tol}/[4(2U+U^2)]\}$ and allocate the remaining ${\Tol}/2$ to the other numerical operations. Each computed residual for $Z_i-w_i^+(a_i+b_i)-w_i^\times a_ib_i$ then has error at most ${\Tol}$, for all messages satisfying the encoding, length, and magnitude bounds, including $|a_i|,|b_i|\le U$.

There are $Q=2L$ evaluations, with at most $b_{\mathrm w}$ bits per answer. Each answer string is fixed before the proofs of~\cref{eq:wiring-answer-bit}, and the verifier retains that same string for the GKR check of $Z_i-w_i^+(a_i+b_i)-w_i^\times a_ib_i$. Repeat each constant-error interactive proof independently
\[
 r_\mathrm{w}=\ceil{\log_{10}(Qb_{\mathrm w}/\eps_\mathrm{w})}
\]
times. Here $r_\mathrm{w}$ is the repetition count denoted by $r$ in~\cref{thm:uniform}, as in~\cref{sec:zk-uniform}. Conditional on any preceding transcript, a false statement in~\cref{eq:wiring-answer-bit} passes all repetitions with probability at most $10^{-r_\mathrm{w}}\le\eps_\mathrm{w}/(Qb_{\mathrm w})$. A union bound and the comparison with the values represented by $\EvaluationString_{p,q}(\mathsf q)$ in~\cref{thm:uniform} give soundness at most $\eps_\mathrm{c}+\eps_\mathrm{w}=1/4$, without conditioning the GKR challenges on acceptance of the later interactive proofs. The honest messages pass every check for every sequence of challenge indices in the specified sets. Continuing the interaction after a permanent rejection flag changes neither guarantee.

As in~\cref{sec:zk-protocol}, let $\mu$ and $r_\mathrm{c}$ count only the GKR prover bits and challenge bits, respectively:
\[
 \mu=O(kDb),\qquad r_\mathrm{c}=(k+L)\ell.
\]
The complete underlying proof also includes the wiring-answer strings and the proofs of~\cref{eq:wiring-answer-bit}. Thus its total prover communication $\mu_\mathrm{tot}$, verifier randomness $r_\mathrm{v}$, and verifier time satisfy
\begin{align*}
 \mu_\mathrm{tot}&\le\mu+O(Qb_{\mathrm w})+Qb_{\mathrm w}r_\mathrm{w}\poly(\sigma),\\
 r_\mathrm{v}&\le r_\mathrm{c}+Qb_{\mathrm w}r_\mathrm{w}\poly(\sigma),\\
 \VerifierTime
 &=O\bigl((n+1)mh+kD+Lmh\bigr)\poly(b+\ell)\\
 &\quad+Qb_{\mathrm w}r_\mathrm{w}t_{\mathrm w}\poly(\sigma).
\end{align*}
Since $r_\mathrm{w}\ge1$, the transmitted answer strings are absorbed by the communication term for these interactive proofs. Substitution yields
\[
 \begin{aligned}
  \mu_\mathrm{tot},r_\mathrm{v},\Communication_\mathrm{norm}&=\poly(L,\log S_0,b),\\
  \VerifierTime&=(n+1)\poly(L,\log S_0,b),\\
  \ProverTime&=\poly(S_0,b).
 \end{aligned}
\]

\medskip\noindent{\bf One-round compilation.}
To meet the input-length convention in~\cref{thm:kr}, consider the same promise problem with the common input extended by an ignored suffix to a fixed length $n_\mathrm{inst}=\poly(S_0,b)$ large enough for the original encoding and at least $S_0+b$; below, $n_\mathrm{inst}$ denotes this padded length. Both sides of the promise ignore this suffix. It is used only in the analysis: the implemented protocol neither materializes, communicates, nor reads it. Thus $\VerifierTime=\poly(n_\mathrm{inst})$ without an additional scan of length $S_0$. Set
\[
 \tau=\max\{\tau_0,\kappa,\Communication_\mathrm{norm},
             \ceil{\log(n_\mathrm{inst}+2)}\}.
\]
Then $\tau=\poly(\kappa,L,\log S_0,b)$. The protocol parameters specify all challenge blocks and sets $\mathcal S_t$, so messages of the transformed protocol are generated in $\poly(\tau)$ time without evaluating the circuit or its polynomial extensions of the wiring predicates. Applying~\cref{thm:kr} gives completeness at least $1-2^{-\tau^2}$ and soundness at most $1/4+2^{-\tau^2}$ against size-$2^\tau$ provers, and hence against size-$2^\kappa$ provers. \Cref{eq:all-history} gives polynomial honest-prover time, and the remaining costs give~\cref{eq:one-round-cost}. The compiled verifier applies the same finite acceptance test, so compilation introduces no further numerical error and leaves ${\Tol}$ and $\Delta$ unchanged.
\end{proof}

\medskip\noindent{\bf Efficiency and scope.}
When $S_0=\poly(n)$ and $\kappa,L,b=\polylog(n+2)$, communication is polylogarithmic and verifier time is $(n+1)\polylog(n+2)$. The bound on $b$ remains a separate hypothesis for computation over $\C$; low depth alone does not ensure it. Perfectly correct PIR gives completeness one. The first message can be prepared before the instance is known, but the instance must be chosen independently of that message and the stored random tape, PIR queries, and recovery information. Every execution uses fresh verifier randomness and PIR queries. No soundness guarantee is asserted for challenge reuse or adaptive instance selection after the challenge is revealed. The argument is not publicly verifiable, and no zero-knowledge guarantee is claimed.
\section{Interactive PCPs}\label{sec:ipcp}

The verifier in \cref{sec:agkr} checks a computation by reducing its output claim to a few evaluations of the input polynomial. Here the input includes a private bit witness. The prover therefore supplies a fixed proof from which the verifier can obtain these evaluations. Following GKR~\cite[Section~7]{GKR}, we encode the witness by the values of a low-degree polynomial on a larger grid. The prover fixes this proof before the verifier samples any coins.

The verifier has three claims to check. First, the proof must be close to the evaluations of one low-degree polynomial. The values of this polynomial at the witness positions must be close to bits. Finally, those bits must give the claimed circuit output, up to the specified error. We prove these claims in that order. The first is checked by restricting the proof to two coordinates. The other two are checked by approximate GKR.

The second step requires an additional numerical bound. A polynomial close to the proof need not have the proof's bounded coefficient encoding. Lemma~\ref{pcp:decoded} bounds its values so that approximate GKR applies. All proof lengths and query counts below are measured in bits.

\subsection{Encoding}

The smaller grid $H^m$ holds the witness. The larger grid $G^m$ holds the proof and gives the verifier many possible query points. Let ${\WitLen}\ge1$ be the witness length. Choose powers of two $h,N,m$ with $h,m\ge2$, $h\mid N$, $N\ge16mh$, and $h^m\ge {\WitLen}$. Write
\[
 H=\Omega_h,\qquad G=\Omega_N,\qquad
 \PCPField=\mathbb Q(\omega_N),\qquad
 \omega_N=\exp(2\pi\ii/N),
\]
and recall the unit circle $\UnitCircle$ from \cref{sec:overview-notation}. Place $w\in\bits^{\WitLen}$ at fixed positions of $H^m$ and set all other entries to zero. The honest proof $\pi_w$ lists
\[
 f_w(r)=\sum_{a\in H^m}w_a\chi_a(r),
 \qquad r\in G^m,
\]
in a fixed order, using the interpolation basis from~\cref{eq:lagrange}. Thus $f_w$ has individual degree below $h$ and recovers the witness on $H^m$. We write $\pi$ for the proof actually supplied, which may be dishonest, and $\pi[r]$ for its entry at $r\in G^m$.

The proof must be a finite bit string even though its entries are complex numbers. Roots of unity give an exact finite representation. The next lemma specifies that representation and bounds every string the prover could submit, including a dishonest one.

\begin{samepage}
\begin{lemma}[Encoding]\label{pcp:encoding}
Each proof symbol consists of $N/2$ integers in fixed-width
two's-complement notation, with coefficient width and total bit length
\begin{align*}
 \PCPCoeffBits=1+\ceil{\log({\WitLen}h^m+1)},\quad
 \OracleBits=\frac N2\PCPCoeffBits
 =\Theta\bigl(N(\log({\WitLen}+1)+m\log h)\bigr).
\end{align*}
Every bit pattern represents an element of $\PCPField$ of modulus at most ${\EntryBound}=2N({\WitLen}+1)$. Coefficients of honest bivariate restrictions use the same encoding. Their values on $G^2$ can be checked exactly in $\poly(N,h,\OracleBits)$ time. An encoded value can be approximated within $\xi>0$ in $\poly(N,\OracleBits,\log^+(1/\xi))$ time.
\end{lemma}
\end{samepage}
\begin{proof}[Proof of~\cref{pcp:encoding}]
Let $Q=h^m$. At a point of $G$, each one-variable interpolation polynomial evaluates to a sum of $h$ powers of $\omega_N$ divided by $h$. A product over $m$ coordinates therefore has $Q$ terms divided by $Q$. Since at most ${\WitLen}$ witness positions are nonzero, expanding $f_w(r)$ gives at most ${\WitLen}Q$ root powers divided by $Q$.

We store these powers in a fixed basis. For power-of-two $N$, the minimal polynomial of $\omega_N$ is $X^{N/2}+1$. Indeed, after substituting $X+1$, the leading coefficient is $1$, every other nonconstant coefficient is even, and the constant coefficient is $2$. Eisenstein's criterion at
$2$~\cite[Section~9.4]{DummitFoote2004}
therefore gives irreducibility. Hence $1,\omega_N,\ldots,\omega_N^{N/2-1}$ is a basis of $\PCPField$ over $\mathbb Q$, and a proof symbol has the form
\[
 Q^{-1}\sum_{j=0}^{N/2-1}u_j\omega_N^j.
\]
Reduction modulo $X^{N/2}+1$ changes only the signs of root powers. Consequently $|u_j|\le {\WitLen}Q$ for an honest symbol. The least two's-complement width containing $[-{\WitLen}Q,{\WitLen}Q]$ is $\PCPCoeffBits=1+\ceil{\log({\WitLen}Q+1)}$. The $N/2$ coefficients occupy exactly $\OracleBits$ bits.

The same width also bounds dishonest symbols. Every bit pattern satisfies
\[
 |u_j|\le2^{\PCPCoeffBits-1}<2({\WitLen}Q+1)\le4({\WitLen}+1)Q.
\]
Since every root has modulus one, the represented value satisfies
\[
 \left|Q^{-1}\sum_{j=0}^{N/2-1}u_j\omega_N^j\right|
 \le\frac1Q\sum_{j=0}^{N/2-1}|u_j|
 \le\frac{N}{2Q}\,4({\WitLen}+1)Q=2N({\WitLen}+1).
\]

The low-degree test will ask the prover for a polynomial in two variables. Fixing all other coordinates of $f_w$ leaves coefficients that are sums of at most ${\WitLen}h^{m-2}$ root powers divided by $Q$, so the same representation suffices. To evaluate this polynomial on $G^2$, multiply its coefficients by root powers and add. Multiplication by a root power permutes the basis coordinates and changes signs. Thus exact equality can be checked with integer arithmetic in the stated time.

Approximate GKR also needs numerical approximations to these exact values. Choose binary fractions $\widetilde\omega_j$ with
$|\widetilde\omega_j-\omega_N^j|\le\xi/[8N({\WitLen}+1)]$ and compute
$\widetilde v=Q^{-1}\sum_j u_j\widetilde\omega_j$ exactly in binary arithmetic.
This is possible because $Q=h^m$ is a power of two. If $v$ is the encoded value, then
\[
 |v-\widetilde v|
 \le\sum_{j=0}^{N/2-1}\frac{|u_j|}{Q}\frac{\xi}{8N({\WitLen}+1)}
 \le\frac N2\,4({\WitLen}+1)\frac{\xi}{8N({\WitLen}+1)}=\frac\xi4<\xi.
\]
Multiplying by an encoded integer and summing $N/2$ terms increases the numerator length by at most the coefficient width plus $O(\log N)$. The root algorithms of \cref{sec:uniform} therefore give the stated polynomial bit cost.
\end{proof}

Let $\PCPCode\subseteq\PCPField^N$ be the evaluation code consisting of the values on $G$ of polynomials of degree below $h$. Two distinct polynomials of this kind agree on at most $h-1$ points, hence the code has relative distance at least $1-(h-1)/N$. The tensor code $\PCPCode^{\otimes m}$ consists of evaluations on $G^m$ of polynomials of individual degree below $h$. A word in this code is an array whose every coordinate line belongs to $\PCPCode$.

We write $\PCPDistance$ for relative Hamming distance, the fraction of entries at which two arrays differ. Distance to the code is minimized over all codewords, without restricting their coefficient widths. This distinction will matter when we bound the magnitude of a nearby polynomial.

\subsection{Low-degree testing}

The verifier tests a small restriction of the proof rather than reading the whole array. Fixing one coordinate gives a coordinate hyperplane. Repeating this operation eventually leaves only two coordinates, where a low-degree restriction has $h^2$ coefficients. We need to know that these restrictions retain enough of the original array's distance from the code. The following tensor-code lemma, following~\cite{BSS04}, provides this guarantee.

\begin{samepage}
\begin{lemma}[Tensor codes]\label{pcp:tensor}
Let $E\subseteq\mathbb F^{n_E}$ be a linear code over any field, with minimum distance $d_0$ and relative distance $\delta_E=d_0/n_E$. For ${\TenDim}\ge3$ suppose $\delta_E^{\TenDim}\ge7/8$. If $v\in\mathbb F^{[n_E]^{\TenDim}}$ and ${\mathcal P}$ is a uniformly random coordinate hyperplane, then
\[
 \Expect_{\mathcal P}\PCPDistance(v|_{\mathcal P},E^{\otimes({\TenDim}-1)})
 \ge2^{-14}\PCPDistance(v,E^{\otimes {\TenDim}}).
\]
\end{lemma}
\end{samepage}
\begin{proof}[Proof of~\cref{pcp:tensor}]
Write ${\mathcal P}_{j,b}=\{\mathbf x\in[n_E]^{\TenDim}:\mathbf x_j=b\}$ for the hyperplane with coordinate $j$ fixed to $b$. There are ${\TenDim}n_E$ such hyperplanes. Choose a nearest codeword $c_{\mathcal P}\in E^{\otimes({\TenDim}-1)}$ to each restriction $v|_{\mathcal P}$, and put
\[
 {\LocErr}=\Expect_{\mathcal P}\PCPDistance(v|_{\mathcal P},c_{\mathcal P}).
\]
A nearest codeword exists because the possible numbers of agreements form a finite set. The words $c_{\mathcal P}$ are local descriptions of the array. They need not agree where two hyperplanes intersect. We will show that small ${\LocErr}$ forces enough of them to agree to determine a nearby global codeword.

We use two elementary properties of tensor codes. An array belongs to $E^{\otimes {\TenDim}}$ exactly when each coordinate line belongs to $E$. Indeed, choose an information set of $E$ and extend from it in each coordinate direction. These linear extension maps commute. The same description gives minimum distance at least $d_0^{\TenDim}$. A nonzero coordinate line meets at least $d_0$ nonzero slices, and induction gives at least $d_0^{{\TenDim}-1}$ nonzero entries in each slice.

\medskip\noindent{\bf Distance to a nearby codeword.}
First suppose that a global codeword $c$ is already within distance $1/4$ of $v$. We show that its distance is at most $8{\LocErr}$. Consider the bipartite incidence graph between points and hyperplanes, with an edge when a point lies on a hyperplane. Its uniform edge is obtained by choosing a uniform point $\mathbf X$ and a uniform coordinate $\mathbf J$, then taking the hyperplane ${\mathcal P}_{\mathbf J,\mathbf X_{\mathbf J}}$. For a set of points of density $\alpha\le1/4$ and a set of hyperplanes of density $\alpha'$, at least an $(\alpha+\alpha')/8$ fraction of edges have exactly one endpoint in these sets.

Let $\mathcal A$ be the marked point set and $\mathcal B$ the marked hyperplane set. To prove the incidence bound, subtract their densities from their indicators,
\[
 \zeta(\mathbf X)=\Indicator_{\mathcal A}(\mathbf X)-\alpha,
 \qquad \psi(j,b)=\Indicator_{\mathcal B}({\mathcal P}_{j,b})-\alpha'.
\]
Both variables have mean zero, and their second moments are $\alpha(1-\alpha)$ and $\alpha'(1-\alpha')$. Put $\zeta_j(\mathbf X_j)=\Expect[\zeta(\mathbf X)\mid \mathbf X_j]$. For distinct $j,j'$, the variables $\mathbf X_j,\mathbf X_{j'}$ are independent and $\zeta_j,\zeta_{j'}$ have mean zero, so $\Expect[\zeta_j\zeta_{j'}]=0$. Also,
\[
 \Expect[\zeta\zeta_j]
 =\Expect\bigl[\zeta_j\Expect[\zeta\mid \mathbf X_j]\bigr]
 =\Expect[\zeta_j^2].
\]
Expanding a nonnegative square now gives
\[
 0\le\Expect\left[\left(\zeta-\sum_j \zeta_j\right)^2\right]
 =\Expect[\zeta^2]-\sum_j\Expect[\zeta_j^2].
\]
For $\psi_j(b)=\psi(j,b)$, conditioning on $\mathbf J$ and then $\mathbf X_j$ yields
\begin{align*}
 |\Expect[\zeta(\mathbf X)\psi(\mathbf J,\mathbf X_{\mathbf J})]|
 &=\left|\frac1{\TenDim}\sum_j\Expect[\zeta_j(\mathbf X_j)\psi_j(\mathbf X_j)]\right|\\
 &\le\frac1{\TenDim}\left(\sum_j\Expect[\zeta_j^2]\right)^{1/2}
                \left(\sum_j\Expect[\psi_j^2]\right)^{1/2}\\
 &\le {\TenDim}^{-1/2}\sqrt{\alpha(1-\alpha)\alpha'(1-\alpha')}.
\end{align*}
Here the middle line is Cauchy--Schwarz, and
$\sum_j\Expect[\psi_j^2]={\TenDim}\Expect[\psi(\mathbf J,\mathbf X_{\mathbf J})^2]={\TenDim}\alpha'(1-\alpha')$.

Let $\beta$ be the fraction of edges with exactly one marked endpoint. Expanding the two indicators gives
\[
 \beta=\alpha+\alpha'-2\alpha\alpha'-2\Expect[\zeta(\mathbf X)\psi(\mathbf J,\mathbf X_{\mathbf J})].
\]
Set $\alpha_*=(\alpha+\alpha')/2$. We have $\alpha\alpha'\le\alpha_*^2$ and
\[
 \sqrt{\alpha(1-\alpha)\alpha'(1-\alpha')}
 \le\frac{\alpha(1-\alpha)+\alpha'(1-\alpha')}2\le\alpha_*(1-\alpha_*).
\]
Substituting these inequalities gives
\begin{align*}
 \beta&\ge2(1-{\TenDim}^{-1/2})\alpha_*(1-\alpha_*)\\
 &=(1-{\TenDim}^{-1/2})(\alpha+\alpha')(1-(\alpha+\alpha')/2)\ge(\alpha+\alpha')/8.
\end{align*}
For the last step, ${\TenDim}\ge3$ gives $1-{\TenDim}^{-1/2}\ge1/3$, while
$\alpha+\alpha'\le5/4$ gives $1-(\alpha+\alpha')/2\ge3/8$.

Apply this bound by marking each point where $v\ne c$ and each hyperplane where $c_{\mathcal P}\ne c|_{\mathcal P}$. Their densities are $\alpha=\PCPDistance(v,c)$ and $\alpha'$, respectively. A boundary edge from an erroneous point to an unmarked hyperplane witnesses a disagreement between $v$ and $c_{\mathcal P}$. On a marked hyperplane, the two codewords $c_{\mathcal P}$ and $c|_{\mathcal P}$ disagree on at least a $\delta_E^{{\TenDim}-1}$ fraction of points. Thus at most $\alpha'(1-\delta_E^{{\TenDim}-1})\le \alpha'/8$ of all edges are boundary edges that fail to witness a disagreement. It follows that
\[
 {\LocErr}\ge(\alpha+\alpha')/8-\alpha'/8=\alpha/8.
\]

\medskip\noindent{\bf Consistency of the local codewords.}
The preceding argument bounds the distance to any codeword within $1/4$ of $v$. We now construct such a codeword when $0<{\LocErr}\le2^{-14}$. Put ${\LocErr}_{\mathrm{bad}}=\sqrt{\LocErr}$. For each hyperplane ${\mathcal P}$, write
${\LocErr}_{\mathcal P}=\PCPDistance(v|_{\mathcal P},c_{\mathcal P})$. Call ${\mathcal P}$ bad when
${\LocErr}_{\mathcal P}>{\LocErr}_{\mathrm{bad}}$, and let $\beta_\mathrm{bad}$ be the fraction of bad
hyperplanes. Their contribution to the average error gives
\[
 {\LocErr}=\frac1{{\TenDim}n_E}\sum_{\mathcal P}{\LocErr}_{\mathcal P}
 \ge\frac1{{\TenDim}n_E}\sum_{{\mathcal P}\text{ bad}}{\LocErr}_{\mathcal P}
 \ge{\LocErr}_{\mathrm{bad}}\beta_\mathrm{bad}.
\]
Since ${\LocErr}={\LocErr}_{\mathrm{bad}}^2$, at most a ${\LocErr}_{\mathrm{bad}}$ fraction of hyperplanes are bad.

Form an inconsistency graph whose vertices are the ${\TenDim}n_E$ hyperplanes. Join each bad vertex to every other vertex. Join two good hyperplanes in different directions when their local codewords disagree on their intersection. An independent set with at least two vertices therefore consists of good hyperplanes with mutually consistent local words.

Let $\beta_\mathrm{edge}$ be the probability that two independent uniform vertices
form an edge. The probability that at least one is bad is at most
$2\beta_\mathrm{bad}\le2{\LocErr}_{\mathrm{bad}}$. To count the remaining edges, let $\operatorname{disagr}({\mathcal P},{\mathcal Q})$ be the
relative disagreement of $c_{\mathcal P}$ and $c_{\mathcal Q}$ on ${\mathcal P}\cap {\mathcal Q}$ when the
hyperplanes have different directions, and set $\operatorname{disagr}({\mathcal P},{\mathcal Q})=0$ otherwise.
On an edge with two good endpoints, the two restrictions are distinct
words of $E^{\otimes({\TenDim}-2)}$, so $\operatorname{disagr}({\mathcal P},{\mathcal Q})\ge\delta_E^{{\TenDim}-2}$.

We bound the average of $\operatorname{disagr}$ by sampling a uniform point
$\mathbf X\in[n_E]^{\TenDim}$ and independent uniform directions $\mathbf J,\mathbf J'\in[{\TenDim}]$.
Conditional on $\mathbf J\ne \mathbf J'$, the hyperplanes
${\mathcal P}_{\mathbf J,\mathbf X_{\mathbf J}}$ and ${\mathcal P}_{\mathbf J',\mathbf X_{\mathbf J'}}$ have independent uniform fixed coordinates,
and $\mathbf X$ is uniform on their intersection. Thus
\begin{align*}
 \Expect_{{\mathcal P},{\mathcal Q}}\operatorname{disagr}({\mathcal P},{\mathcal Q})
 &=\Pr_{\mathbf X,\mathbf J,\mathbf J'}[\mathbf J\ne \mathbf J',\ c_{{\mathcal P}_{\mathbf J,\mathbf X_{\mathbf J}}}(\mathbf X)\ne c_{{\mathcal P}_{\mathbf J',\mathbf X_{\mathbf J'}}}(\mathbf X)]\\
 &\le\Pr_{\mathbf X,\mathbf J}[c_{{\mathcal P}_{\mathbf J,\mathbf X_{\mathbf J}}}(\mathbf X)\ne v(\mathbf X)]
    +\Pr_{\mathbf X,\mathbf J'}[c_{{\mathcal P}_{\mathbf J',\mathbf X_{\mathbf J'}}}(\mathbf X)\ne v(\mathbf X)]\\
 &=2{\LocErr}.
\end{align*}
The inequality holds because two values that differ cannot both equal
$v(\mathbf X)$. Each probability in the last line is ${\LocErr}$, since sampling
$\mathbf X$ and a direction selects a uniform hyperplane and then a uniform
point on it. Consequently,
\[
 \beta_\mathrm{edge}\le2{\LocErr}_{\mathrm{bad}}+\delta_E^{-({\TenDim}-2)}\Expect_{{\mathcal P},{\mathcal Q}}\operatorname{disagr}({\mathcal P},{\mathcal Q})
 \le2{\LocErr}_{\mathrm{bad}}+2{\LocErr}\delta_E^{-({\TenDim}-2)}
 \le4\delta_E^{-({\TenDim}-2)}{\LocErr}_{\mathrm{bad}}.
\]
The final inequality uses ${\LocErr}={\LocErr}_{\mathrm{bad}}^2\le{\LocErr}_{\mathrm{bad}}$ and $\delta_E\le1$.

Figure~\ref{fig:pcp-hyperplanes} illustrates why a single inconsistency forces many graph edges. Every edge has an endpoint of degree at least $({\TenDim}-2)d_0/2$. A bad vertex has degree ${\TenDim}n_E-1\ge({\TenDim}-2)d_0/2$, since $d_0\le n_E$ and ${\TenDim}\ge3$. Otherwise, the difference of the two local words is a nonzero tensor-code word on their intersection. In each of the other ${\TenDim}-2$ directions, it is nonzero on at least $d_0$ slices. The hyperplane containing any such slice must be inconsistent with at least one of the two endpoint words. These $({\TenDim}-2)d_0$ distinct hyperplanes contribute to the sum of the endpoint degrees.

\begin{figure}[ht!]
\centering

\begin{tikzpicture}[font=\small,line join=round]
  \begin{scope}[x={(1.4cm,0cm)},y={(0.63cm,0.36cm)},z={(0cm,1.28cm)}]
    \draw[black!30] (0,0,0)--(2.4,0,0)--(2.4,2,0)--(0,2,0)--cycle;
    \draw[black!30] (0,0,2.7)--(2.4,0,2.7)--(2.4,2,2.7)--(0,2,2.7)--cycle;
    \foreach \x/\y in {0/0,2.4/0,2.4/2,0/2}
      \draw[black!30] (\x,\y,0)--(\x,\y,2.7);
    \filldraw[fill=PCPBlue!14,draw=PCPBlue!70] (1.1,0,0)--(1.1,2,0)--(1.1,2,2.7)--(1.1,0,2.7)--cycle;
    \filldraw[fill=PCPBlue!5,draw=PCPBlue!70] (0,0.9,0)--(2.4,0.9,0)--(2.4,0.9,2.7)--(0,0.9,2.7)--cycle;
    \foreach \z/\lab in {0.55/R_1,1.3/R_2,2.15/R_{d_0}}{
      \draw[black!55,densely dashed] (0,0,\z)--(2.4,0,\z)--(2.4,2,\z)--(0,2,\z)--cycle;
      \node[anchor=east,font=\footnotesize] at (0,0,\z) {$\lab$};
    }
    \draw[PCPAccent,very thick] (1.1,0.9,0)--(1.1,0.9,2.7);
    \foreach \z in {0.55,1.3,2.15}
      \fill[PCPAccent] (1.1,0.9,\z) circle (1.9pt);
    \node[anchor=south] at (1.1,2,2.7) {${\mathcal P}$};
    \node[anchor=south west] at (2.4,0.9,2.7) {${\mathcal Q}$};
    \node[anchor=north,align=center] at (1.2,1,-0.4) {Coordinate hyperplanes (${\TenDim}=3$)};
  \end{scope}
  \begin{scope}[shift={(6.7cm,0.65cm)},scale=1.25]
    \node[pcp vertex] (p) at (0.45,2.25) {${\mathcal P}$};
    \node[pcp vertex] (q) at (2.65,2.25) {${\mathcal Q}$};
    \node[pcp vertex] (r1) at (0,0.45) {$R_1$};
    \node[pcp vertex] (r2) at (1.05,0.45) {$R_2$};
    \node at (1.95,0.45) {$\cdots$};
    \node[pcp vertex] (rd) at (3,0.45) {$R_{d_0}$};
    \draw[PCPAccent,thick] (p)--node[above,black,font=\footnotesize] {disagree} (q);
    \draw[PCPBlue,thick] (p)--(r1) (p)--(r2) (q)--(rd);
    \node[align=center] at (1.5,-0.55) {Inconsistency graph};
  \end{scope}
\end{tikzpicture}
\caption{A disagreement between the local codewords on ${\mathcal P}$ and ${\mathcal Q}$ (bold line, drawn for ${\TenDim}=3$) persists on at least $d_0$ slices in each other coordinate direction (dots). Each hyperplane $R_j$ through such a slice must disagree with at least one of them, so ${\mathcal P}$ and ${\mathcal Q}$ together have at least $({\TenDim}-2)d_0$ neighbors in the inconsistency graph, and one of them has degree at least $({\TenDim}-2)d_0/2$.}
\label{fig:pcp-hyperplanes}
\end{figure}
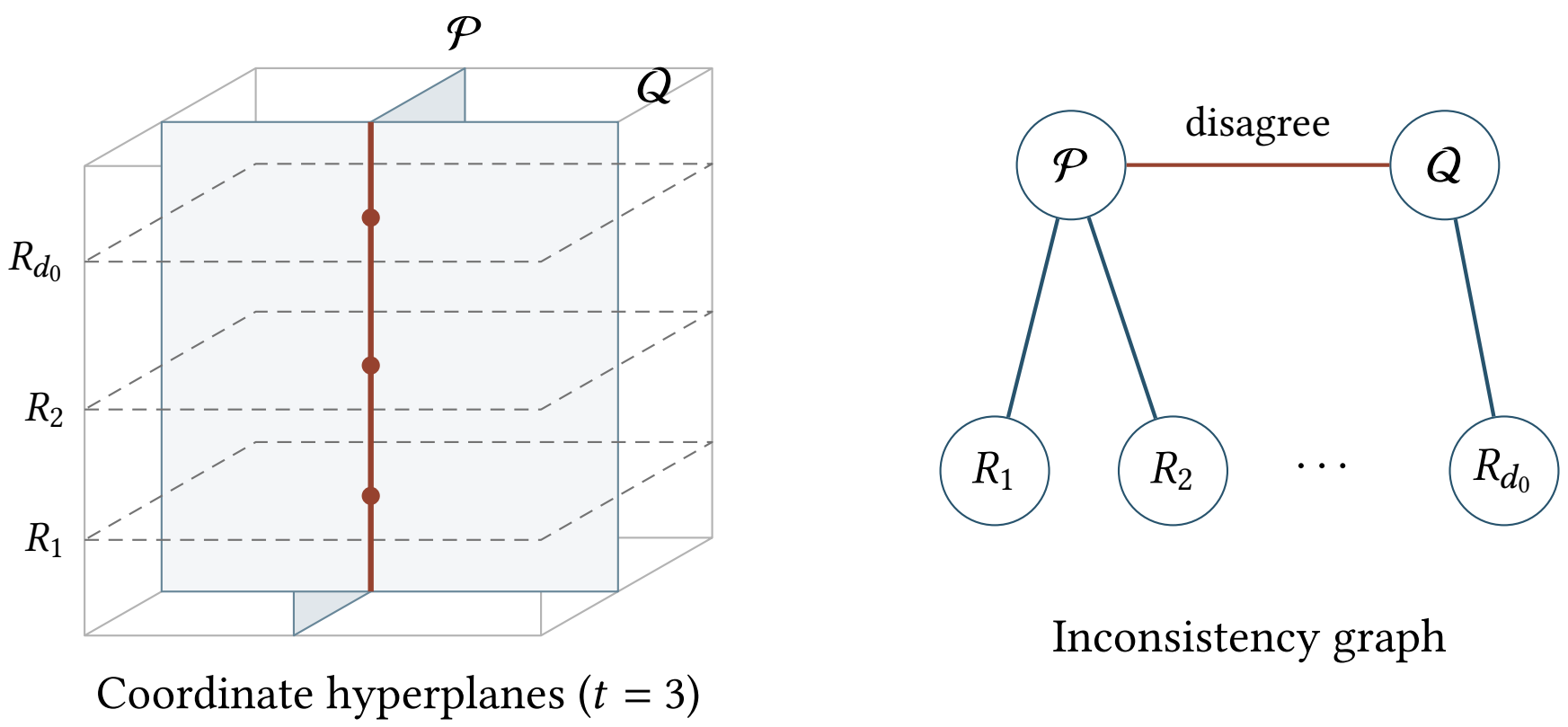

Now delete every vertex whose degree is at least $({\TenDim}-2)d_0/2$. By the preceding argument every edge loses at least one endpoint, so the vertices that remain form an independent set $\mathcal K$. Let $\beta_\mathrm{del}$ denote the fraction of vertices deleted. The graph has ${\TenDim}n_E$ vertices, and its sum of degrees
is $({\TenDim}n_E)^2\beta_\mathrm{edge}$, because each edge contributes two ordered pairs to
the probability defining $\beta_\mathrm{edge}$. The deleted vertices alone
therefore give
\[
 \beta_\mathrm{del} {\TenDim}n_E\,\frac{({\TenDim}-2)d_0}{2}
 \le\sum_{\mathcal P}\deg({\mathcal P})=({\TenDim}n_E)^2\beta_\mathrm{edge}.
\]
Dividing and substituting the bound on $\beta_\mathrm{edge}$ yields
\[
 \beta_\mathrm{del}\le\frac{2{\TenDim}n_E}{({\TenDim}-2)d_0}\beta_\mathrm{edge}
 \le\frac{8{\TenDim}}{{\TenDim}-2}\delta_E^{-({\TenDim}-1)}{\LocErr}_{\mathrm{bad}}
 \le\frac{192}{7}{\LocErr}_{\mathrm{bad}}<28{\LocErr}_{\mathrm{bad}}.
\]
Here ${\TenDim}/({\TenDim}-2)\le3$ and $\delta_E^{{\TenDim}-1}\ge\delta_E^{\TenDim}\ge7/8$.

\medskip\noindent{\bf Extending to a global codeword.}
We now extend the mutually consistent words indexed by $\mathcal K$.
The bound ${\LocErr}_{\mathrm{bad}}\le1/128$ gives $\beta_\mathrm{del}<7/32$. On the other hand,
$\delta_E\ge\delta_E^{\TenDim}\ge7/8$ and ${\TenDim}\ge3$, so
\[
 \beta_\mathrm{del}<\frac7{32}<\frac7{12}
 \le\frac{{\TenDim}-1}{{\TenDim}}\delta_E
 =\frac{({\TenDim}-1)d_0}{{\TenDim}n_E}.
\]
Hence fewer than $({\TenDim}-1)d_0$ vertices were deleted, and
$|\mathcal K|>{\TenDim}n_E-({\TenDim}-1)d_0$.

Partition $\mathcal K$ according to the fixed coordinate, and relabel coordinates so that the two largest parts have index sets $\mathcal K_1,\mathcal K_2\subseteq[n_E]$. Both sets have more than $n_E-d_0$ indices. Otherwise the second largest part has size at most $n_E-d_0$, giving
\[
 |\mathcal K|\le n_E+({\TenDim}-1)(n_E-d_0)={\TenDim}n_E-({\TenDim}-1)d_0,
\]
a contradiction. Restriction of $E$ to either $\mathcal K_1$ or $\mathcal K_2$ is injective, since a nonzero codeword cannot vanish at more than $n_E-d_0$ positions.

The local words on the selected hyperplanes agree on their intersections. They therefore define an array on
\[
 \mathcal K_1\times \mathcal K_2\times[n_E]^{{\TenDim}-2}.
\]
Its lines in the first two directions belong to the corresponding restrictions of $E$, and its other lines belong to $E$. The tensor-code characterization and the inverse restriction maps extend this array uniquely to $c\in E^{\otimes {\TenDim}}$. For each hyperplane indexed by $\mathcal K_1$, this extension agrees with its local word on the coordinates indexed by $\mathcal K_2$. Injectivity in the second direction implies agreement on the whole hyperplane. Since $|\mathcal K| \ge (1 - \beta_{\mathrm{del}})\, t n_E$ and $\mathcal K$ is
split into $t$ parts, of which $\mathcal K_1$ has the most elements,
$|\mathcal K_1| \ge (1 - \beta_{\mathrm{del}}) n_E$. The hyperplanes indexed by
$\mathcal K_1$ are pairwise disjoint, together contain at least a
$1 - \beta_{\mathrm{del}}$ fraction of the entries of the array, and are all
good. Consequently,
\[
 \PCPDistance(v,c)\le\beta_\mathrm{del}+{\LocErr}_{\mathrm{bad}}<29{\LocErr}_{\mathrm{bad}}\le29/128<1/4.
\]
The incidence argument above now gives $\PCPDistance(v,c)\le8{\LocErr}$, which is stronger than required. If ${\LocErr}>2^{-14}$, the lemma follows because relative distance is at most one. If ${\LocErr}=0$, every coordinate line belongs to $E$, so $v\in E^{\otimes {\TenDim}}$.
\end{proof}

To apply the lemma, we regard a block of coordinates as one coordinate of a larger tensor code. Each application fixes a block while preserving a constant fraction of the expected distance. Repeating this reduction gives a bivariate restriction that can be tested with one proof query.

\begin{samepage}
\begin{lemma}[Low-degree test]\label{pcp:test}
There is a public-coin test that reads one $\OracleBits$-bit proof symbol
and receives $h^2\OracleBits$ coefficient bits from the prover.
Encodings of bit witnesses always pass. If $\PCPDistance(\pi,\PCPCode^{\otimes m})>\gamma$, every prover passes $r_{\mathrm{ldt}}$ repetitions with probability at most
\[
 \exp(-r_{\mathrm{ldt}}\gamma m^{-29}).
\]
Each repetition has communication and verification cost polynomial in $N,h,m,\OracleBits$.
\end{lemma}
\end{samepage}
\begin{proof}[Proof of~\cref{pcp:test}]
The verifier first chooses the bivariate restriction. For $m=2$, no coordinates need to be fixed. Otherwise, at a recursive step with $4m'$ remaining coordinates, partition them into four equal blocks. Choose one block uniformly and fix its coordinates to a uniform point of $G^{m'}$. Of the remaining three blocks, choose a second with probabilities $1/2,1/4,1/4$ in a fixed order and fix it uniformly. These probabilities permit exact sampling with fair bits. Recurse on the remaining $2m'$ coordinates until only two remain.

We bound the expected distance after each halving. At a recursive step with $4m'$ coordinates, take $\PCPCode^{\otimes m'}$ as the base code in Lemma~\ref{pcp:tensor}. Its relative distance to the fourth power is at least
\[
 \bigl(1-(h-1)/N\bigr)^{4m'}\ge1-mh/N\ge15/16.
\]
Thus the lemma applies first in dimension four and then in dimension three. With uniform choices at both steps, these applications preserve at least a $2^{-28}$ fraction of the expected distance. Each probability in the actual second choice is at least $3/4$ of its uniform value. Since distances are nonnegative, the actual halving preserves at least a $2^{-29}$ fraction. Let $\mathsf d_j$ be the distance after $j$ halvings. Conditional on all earlier choices, the estimate just proved is
\[
 \Expect[\mathsf d_{j+1}\mid\text{earlier choices}]\ge2^{-29}\mathsf d_j.
\]
Taking expectations and iterating over $n_{\mathrm h}=\log m-1$ halvings gives
\[
 \Expect[\mathsf d_{n_{\mathrm h}}]\ge2^{-29n_{\mathrm h}}\mathsf d_0
 =2^{29}m^{-29}\mathsf d_0\ge m^{-29}\mathsf d_0,
 \qquad \mathsf d_0=\PCPDistance(\pi,\PCPCode^{\otimes m}).
\]
This also holds for $m=2$, when there is no reduction.

The verifier reveals the fixed coordinates and their values. The prover then sends the $h^2$ coefficients of a bivariate polynomial of individual degree below $h$, encoded as in Lemma~\ref{pcp:encoding}. After checking the coefficient count and encodings, the verifier samples a uniform point of $G^2$, reads the proof symbol at the corresponding point of $G^m$, and checks equality exactly. The polynomial is fixed before this final point is sampled. Its probability of disagreeing with the proof is therefore at least the restricted array's distance from the bivariate code, regardless of the prover's choice. An honest prover sends the actual restriction, which always passes.

Put $p_\mathrm{rej}=\gamma m^{-29}$. The expected-distance bound gives rejection probability at least $p_\mathrm{rej}$ in each repetition, conditional on any preceding transcript. To see why this suffices for repetition, let $\mathsf{Pass}_i$ mean that the first $i$ tests pass, and let $\mathcal F_{i-1}$ denote the preceding transcript. Then
\begin{align*}
 \Pr[\mathsf{Pass}_i]
 &=\Expect\bigl[\Indicator_{\mathsf{Pass}_{i-1}}
       \Pr[\text{test }i\text{ passes}\mid\mathcal F_{i-1}]\bigr]\\
 &\le(1-p_\mathrm{rej})\Pr[\mathsf{Pass}_{i-1}].
\end{align*}
Since $\Pr[\mathsf{Pass}_0]=1$, iteration gives
\[
 \Pr[\mathsf{Pass}_{r_{\mathrm{ldt}}}]\le(1-p_\mathrm{rej})^{r_{\mathrm{ldt}}}\le\exp(-p_\mathrm{rej}r_{\mathrm{ldt}})
 =\exp(-r_{\mathrm{ldt}}\gamma m^{-29}),
\]
where $1-p_\mathrm{rej}\le e^{-p_\mathrm{rej}}$. This argument allows the prover's messages to depend on the preceding transcript.
All random choices use finitely many fair bits, and Lemma~\ref{pcp:encoding} gives the stated verification cost.
\end{proof}

\medskip\noindent{\bf Bounding the nearby polynomial.}
The low-degree test bounds acceptance when the proof is far from the
code. In the remaining case, fix a polynomial whose evaluations are
close to the proof. Its coefficients may be far larger than those
allowed in the proof encoding. We therefore use only the fact that it
agrees with bounded proof symbols at most grid points. The next lemma
turns this agreement into a bound on the product of unit circles.

\begin{samepage}
\begin{lemma}[Magnitude bound]\label{pcp:decoded}
Let ${\EntryBound}>0$, and let $f$ have individual degree at most $d=h-1$ and agree with a ${\EntryBound}$-bounded word on a $1-\gamma$ fraction of $G^m$. If $1-\gamma-md/N>0$, then for $0<s<1/d$,
\[
 \max_{z\in\UnitCircle^m}|f(z)|
 \le {\EntryBound}\left(\frac{(1-ds)^{-m}}{1-\gamma-md/N}\right)^{1/s}.
\]
For $\gamma\le1/4$ and $N\ge4md$, this gives
\begin{equation}\label{pcp:decoded-bound}
 \max_{z\in\UnitCircle^m}|f(z)|\le\DecodedBound,
 \qquad \DecodedBound={\EntryBound}\,2^{2d(m+1)}.
\end{equation}
\end{lemma}
\end{samepage}
\begin{proof}[Proof of~\cref{pcp:decoded}]
Let $U_f$ be the maximum. There is nothing to prove if $U_f=0$, so assume
$U_f>0$ and choose a point $a$ with $|f(a)|=U_f$. We replace the coordinates
of $a$ one at a time by independent uniform points of $G$. The
inverse-moment bound from Lemma~\ref{lem:circle} controls how much the
value can decrease at each replacement. A decrease from $U_f$ to at most
${\EntryBound}$ on many of the resulting grid points will then bound $U_f/{\EntryBound}$.

Sample independent uniform $r_1,\ldots,r_m\in G$ and write
\[
 v_j=f(r_1,\ldots,r_j,a_{j+1},\ldots,a_m),
 \qquad 0\le j\le m.
\]
Thus $|v_0|=U_f$ and $v_m=f(r)$. At step $j$, the remaining dependence on
the coordinate being replaced is the univariate polynomial
\[
 f_j(X)=f(r_1,\ldots,r_{j-1},X,a_{j+1},\ldots,a_m).
\]
It has degree at most $d$ and satisfies $f_j(a_j)=v_{j-1}$.

We discard a sample if a replacement lands in the exceptional set from
Lemma~\ref{lem:circle}. Let $\mathsf{Keep}_j$ be the event that the sample has not
been discarded after $j$ replacements, with $\mathsf{Keep}_0$ the whole probability
space. On $\mathsf{Keep}_{j-1}$, we have $v_{j-1}\ne0$, so $f_j$ is nonzero.
The lemma supplies a set $\mathcal Z_j\subseteq G$ of at most $d$ points containing
all grid zeros of $f_j$. On discarded prefixes set $\mathcal Z_j=\varnothing$. Define
$\mathsf{Keep}_j=\mathsf{Keep}_{j-1}\cap\{r_j\notin \mathcal Z_j\}$.
Since $M(f_j)\ge|v_{j-1}|$, \cref{eq:inverse-moment} gives
\[
 \Expect\left[
   \Indicator_{r_j\notin \mathcal Z_j}
   \left(\frac{|v_{j-1}|}{|f_j(r_j)|}\right)^s
   \,\middle|\,r_1,\ldots,r_{j-1}\right]
 \le\frac1{1-ds}
\]
on $\mathsf{Keep}_{j-1}$. The integrand is defined to be zero on $\mathcal Z_j$.

To multiply these one-coordinate bounds, define
$W_j=(U_f/|v_j|)^s$ on $\mathsf{Keep}_j$ and $W_j=0$ otherwise. Surviving values are
nonzero, so these variables are well defined. On $\mathsf{Keep}_j$,
\[
 W_j=W_{j-1}\left(\frac{|v_{j-1}|}{|f_j(r_j)|}\right)^s.
\]
On a discarded prefix both sides of the desired conditional inequality are zero. On a surviving prefix the preceding inverse-moment estimate therefore gives
\[
 \Expect[W_j\mid r_1,\ldots,r_{j-1}]
 \le(1-ds)^{-1}W_{j-1},\qquad W_0=1.
\]
Taking expectations of both sides gives
\[
 \Expect[W_j]
 =\Expect\bigl[\Expect[W_j\mid r_1,\ldots,r_{j-1}]\bigr]
 \le(1-ds)^{-1}\Expect[W_{j-1}].
\]
Starting from $W_0=1$ and applying this inequality $m$ times yields
$\Expect[W_m]\le(1-ds)^{-m}$.

All coordinates are sampled even when a sample is discarded, so $r$ is uniform on $G^m$. Let $\mathsf{Drop}_j=\mathsf{Keep}_{j-1}\cap\{r_j\in \mathcal Z_j\}$ be the event that the first deletion occurs at step $j$. Conditional on a surviving prefix, $r_j$ is uniform and $|\mathcal Z_j|\le d$. Thus
\[
 \Pr[\mathsf{Drop}_j]
 =\Expect\left[\Indicator_{\mathsf{Keep}_{j-1}}\frac{|\mathcal Z_j|}{N}\right]
 \le\frac dN.
\]
The events $\mathsf{Drop}_j$ are disjoint and their union is $\mathsf{Keep}_m^c$, giving
$\Pr[\mathsf{Keep}_m^c]=\sum_{j=1}^m\Pr[\mathsf{Drop}_j]\le md/N$. The hypothesis gives $|f(r)|\le {\EntryBound}$ with
probability at least $1-\gamma$. Consequently,
$\Pr[\mathsf{Keep}_m\cap\{|f(r)|\le {\EntryBound}\}]\ge1-\gamma-md/N$.
On this intersection,
$W_m\ge(U_f/{\EntryBound})^s$, and hence
\[
 (1-\gamma-md/N)(U_f/{\EntryBound})^s
 \le\Expect[W_m]\le(1-ds)^{-m}.
\]
This proves the first bound. If $\gamma\le1/4$ and $N\ge4md$, the
left-hand prefactor is at least $1/2$. Substituting $s=1/(2d)$ gives
$U_f\le {\EntryBound}(2^{m+1})^{2d}$, which is~\cref{pcp:decoded-bound}.
\end{proof}

\subsection{Witness verification}

Being close to the proof does not force a low-degree polynomial to encode bits. Let $\WitnessPositions\subseteq H^m$ denote the ${\WitLen}$ witness positions, and write $z=(f(a))_{a\in\WitnessPositions}$ for the corresponding values of $f$. We first check that $z$ is
close to a bit vector. We then bound how replacing $z$ by that vector
changes the circuit output. The numerical circuit uses only these
values $f(a)$ for $a\in\WitnessPositions$, and padding and copying gates use a public zero.

For a complex number $z_a$, the expression $z_a(z_a-1)$ vanishes
exactly at $0$ and $1$. Summing these expressions would allow
cancellation. We instead sum their squared absolute values, which are
nonnegative. To compute this sum using addition and multiplication,
define the coefficientwise conjugate polynomial
\[
 \ConjugatePoly(X)=\overline{f(\overline X)}.
\]
It has the same individual degree and bound on $\UnitCircle^m$ as $f$.
Since $\ConjugatePoly(a^{-1})=\overline{f(a)}$ for $a\in H^m$, an
arithmetic circuit can compute
\begin{equation}\label{eq:pcp-booleanity}
 B(f)=\sum_{a\in\WitnessPositions}
       f(a)(f(a)-1)\ConjugatePoly(a^{-1})(\ConjugatePoly(a^{-1})-1)
     =\sum_{a\in\WitnessPositions}|z_a(z_a-1)|^2.
\end{equation}
The circuit has size $O({\WitLen})$ and depth $O(\log({\WitLen}+1))$ before layering
and padding. Its two input arrays contain $f|_{H^m}$ and
$\ConjugatePoly|_{H^m}$. The circuit pairs the input labeled $a$ in the first slice with
the input labeled $a^{-1}$ in the second, so the circuit uses only addition and
multiplication.

\begin{samepage}
\begin{lemma}\label{lem:pcp-booleanity}
For every $z\in\C^{\WitLen}$,
\[
 B(f)\ge\frac14\min_{w\in\bits^{\WitLen}}\|z-w\|_2^2.
\]
The factor $1/4$ is sharp, and $B(f)=0$ for bit values. If $|f|,|\ConjugatePoly|\le\DecodedBound$ on $\UnitCircle^m$, every gate in~\cref{eq:pcp-booleanity} has modulus at most
\[
 \BoolBound=2+({\WitLen}+1)(\DecodedBound+1)^4.
\]
\end{lemma}
\end{samepage}
\begin{proof}[Proof of~\cref{lem:pcp-booleanity}]
The points $0$ and $1$ are one unit apart. The triangle inequality
therefore gives $\max\{|z_a|,|z_a-1|\}\ge1/2$. The other factor is
the distance from $z_a$ to its nearest bit. Consequently,
\[
 |z_a(z_a-1)|^2
 \ge\frac14\min\{|z_a|^2,|z_a-1|^2\}.
\]
Summing over the coordinates proves the inequality, since a nearest
bit vector is obtained by choosing a nearest bit in each coordinate.
Equality holds when one entry is $1/2$ and all others are bits. A bit
entry contributes zero.

Each factor in a summand has modulus at most $\DecodedBound+1$.
Products have at most four such factors, and partial sums contain at
most ${\WitLen}$ products. These bounds, together with the constant and
subtraction gates, are at most $\BoolBound$.
\end{proof}

The verifier must ensure that the second input slice contains the
conjugate of the first. Both slices are therefore evaluated from the
same fixed proof. At a final query $r\in G^m$, the verifier reads the
value representing $f(r)$. It evaluates $\ConjugatePoly(r)$ by
conjugating the value at $r^{-1}$. Inversion preserves the uniform
distribution on $G^m$, so both query points are uniform. Their
independence is not needed.

For completeness, let $X_{\mathrm{sl}}\in H^{n_{\mathrm{sl}}}$ index the input slices, with private
slice labels $\mathbf h_0,\mathbf h_1$. Their combined extension is
\begin{equation}\label{eq:pcp-input-slices}
 \GateExtension_\mathrm{in}(X_{\mathrm{sl}},X)
   =\chi_{\mathbf h_0}(X_{\mathrm{sl}})f(X)+\chi_{\mathbf h_1}(X_{\mathrm{sl}})\ConjugatePoly(X)
       +K_\mathrm{pub}(X_{\mathrm{sl}},X),
\end{equation}
where $K_\mathrm{pub}$ contains the public inputs and $n_{\mathrm{sl}}$ is large enough
for all slices. Since $|\chi_{\mathbf h_j}(X_{\mathrm{sl}})|\le1$ on $\UnitCircle^{n_{\mathrm{sl}}}$, two
proof evaluations and an evaluation of $K_\mathrm{pub}$, with approximation
and arithmetic errors totaling at most ${\Tol}$ in the final residual,
suffice for each final input check. The
numerical circuit uses only the first private slice.

We apply Theorem~\ref{thm:agkr} to these fixed complex inputs. Its
soundness proof uses \cref{eq:layer-sum} over $\C$ and a final
input-claim discrepancy of at most $5{\Tol}$. Both remain valid when the
final input evaluations have certified errors within the residual
budget ${\Tol}$, under the same bounds on gate values and polynomial coefficients. For honest inputs,
Lemma~\ref{pcp:encoding} supplies the required polynomial-time
approximations. Their errors enter the completeness and bit-cost
analysis as input errors, as in \cref{sec:completeness}.

It remains to relate a nearly Boolean input $z$ to an actual bit
witness. Interpret the bits through public affine maps
\begin{equation}\label{eq:pcp-decoding}
 {\DecMap}_g(w)=d_{g0}+\sum_{j=1}^{{\WitLen}}d_{gj}w_j,
\end{equation}
whose coefficients have finite binary real and imaginary parts. This
includes two's-complement encodings. Fix $0<\eta<1/2$ and public
rational bounds
\[
 A_g\ge\Bigl(\sum_j|d_{gj}|^2\Bigr)^{1/2},
 \qquad A_g^{(1)}\ge\sum_j|d_{gj}|.
\]
Let $M_g$ bound gate $g$ of $C(x,{\DecMap}(w))$ for every $w\in\bits^{\WitLen}$.
The following recurrence bounds the change at each gate when a bit
vector $w$ is replaced by a vector $z$ within distance $\eta$.

\begin{samepage}
\begin{lemma}\label{lem:pcp-perturbation}
Suppose $\|z-w\|_2\le\eta$ for a bit vector $w$. Set $\mathcal J_g=A_g\eta$ at decoded input $g$ and $\mathcal J_g=0$ at public inputs. At subsequent gates set
\begin{align}
 \mathcal J_g&=\mathcal J_u+\mathcal J_v &&\text{if }g=u+v,\notag\\
 \mathcal J_g&=M_u\mathcal J_v+M_v\mathcal J_u+\mathcal J_u\mathcal J_v &&\text{if }g=uv.
 \label{eq:pcp-perturbation}
\end{align}
The gate value at $z$ differs from its value at $w$ by at most $\mathcal J_g$ and has modulus at most $M_g+\mathcal J_g$. Denote the output bound by $\mathcal J(\eta)$.
\end{lemma}
\end{samepage}
\begin{proof}[Proof of~\cref{lem:pcp-perturbation}]
At a decoded input, Cauchy--Schwarz gives
\[
 |{\DecMap}_g(z)-{\DecMap}_g(w)|
 =\left|\sum_jd_{gj}(z_j-w_j)\right|
 \le A_g\|z-w\|_2\le A_g\eta.
\]
Public inputs do not change. An addition gate adds the two errors.
At a multiplication gate, let $a,b$ denote the unperturbed operand
values and $e_a,e_b$ their changes. The identity
\[
 (a+e_a)(b+e_b)-ab=ae_b+be_a+e_ae_b
\]
gives the stated recurrence by the triangle inequality. Induction
over the gates proves the error bound. Adding it to the unperturbed
magnitude bound $M_g$ gives $M_g+\mathcal J_g$.
\end{proof}

This output error determines the gap needed in the soundness
statement. If a claim differs from the output for every bit witness
by more than $\Delta_{\mathrm{out}}+\mathcal J(\eta)$, where $\Delta_{\mathrm{out}}>0$ is the threshold of the output invocation in the protocol below, it still differs by more than
$\Delta_{\mathrm{out}}$ after a bit witness is replaced by any $z$ within distance
$\eta$. Thus approximate GKR can reject the resulting numerical
claim at threshold $\Delta_{\mathrm{out}}$.

The required value of $\eta$ can be chosen with polynomially many
bits. Let $A_{\max}=\max\{1,\max_g A_g\}$,
$M_{\max}=\max\{1,\max_g M_g\}$, and let $L_\mathrm{dec}$ be the depth after decoding.
If $A_{\max}\eta(2M_{\max}+1)^{L_\mathrm{dec}}\le1$, then
\[
 \mathcal J(\eta)\le A_{\max}\eta(2M_{\max}+1)^{L_\mathrm{dec}}.
\]
Indeed, while the operand errors are at most $e\le1$, addition gives
error at most $2e$ and multiplication gives at most
$2M_{\max}e+e^2\le(2M_{\max}+1)e$. The assumed bound keeps this estimate valid at
every depth. In particular, $\mathcal J(\eta)\le\mathcal J_*\le1$ can be achieved with
\[
 \log(1/\eta)=O\bigl(\log A_{\max}+L_\mathrm{dec}\log(2M_{\max}+1)+\log(1/\mathcal J_*)\bigr).
\]

\subsection{Protocol and soundness}

We now combine the tests. The prover first fixes the entire proof. The verifier then runs $r_{\mathrm{ldt}}$ repetitions of the low-degree test from Lemma~\ref{pcp:test}. Next it runs approximate GKR on the circuit for $B(f)$, with claimed output zero. Finally, it runs approximate GKR on $C(x,{\DecMap}(z))$, with claimed output $y$. Each invocation uses fresh verifier coins. Both obtain their final input evaluations from the same fixed proof through~\cref{eq:pcp-input-slices}.

The soundness argument splits into two cases, depending on the distance of the proof from $\PCPCode^{\otimes m}$. In the first case the proof is far from every low-degree codeword, so the test of Lemma~\ref{pcp:test} accepts with probability at most $\delta_{\mathrm{ldt}}=\exp(-r_{\mathrm{ldt}}\gamma m^{-29})$. A proof close to a codeword can be compared with that fixed polynomial, except when a query reaches one of the differing entries. The two approximate-GKR invocations then check whether its witness values are close to bits and whether the corresponding computation is correct.

The first invocation uses threshold $0<\Delta_{\mathrm{bool}}<\eta^2/4$, error $\eps_{\mathrm{bool}}$, and gate bound $\BoolBound$. The second uses threshold $\Delta_{\mathrm{out}}$, error $\eps_{\mathrm{out}}$, and parameter ${\TolC}$. Its bounds include $M_g+\mathcal J_g$ and $\DecodedBound$ for every input determined by $f$. At affine decoding gates, $|d_{g0}|+(1+\eta)A_g^{(1)}$ bounds every partial sum, and coefficient magnitudes bound constant gates.

Choose $\AddPoly_i$ and $\MulPoly_i$ of individual degree below $h$, giving $D_{\mathrm{bool}}=D_{\mathrm{out}}=2(h-1)$. Each extension is a sum of interpolation basis products of coefficient norm one, one per gate of the corresponding type. Its coefficient norm is at most the number of gates in that layer. Choose a common grid $N$ at least as large as required by~\cref{eq:parameters} for both invocations.

\begin{samepage}
\begin{theorem}\label{thm:pcp-native}
With the encoding and parameters above, let $0<\gamma\le1/4$ and $\delta_{\mathrm{ldt}}=\exp(-r_{\mathrm{ldt}}\gamma m^{-29})$. There is a public-coin interactive PCP (\cref{def:ipcp}) with perfect completeness for
\[
 \exists w\in\bits^{\WitLen}\quad |y-C(x,{\DecMap}(w))|\le {\TolC}
\]
and soundness at most
\begin{equation}\label{eq:pcp-soundness}
 s_0=\max\{\delta_{\mathrm{ldt}},\;6\gamma+\max\{\eps_{\mathrm{bool}},\eps_{\mathrm{out}}\}\}
\end{equation}
for the promise
\[
 \forall w\in\bits^{\WitLen}\quad |y-C(x,{\DecMap}(w))|>\Delta_{\mathrm{out}}+\mathcal J(\eta).
\]
The proof has $N^m\OracleBits$ bits, and the verifier reads at most $(r_{\mathrm{ldt}}+6)\OracleBits$ bits. The proof depends only on the bit witness and the public encoding parameters, not on the circuit's structure or coefficients.
\end{theorem}
\end{samepage}
\begin{proof}[Proof of~\cref{thm:pcp-native}]
For an honest bit witness, the low-degree test always accepts and $B(f)=0$. Both approximate-GKR invocations accept for every challenge sequence. This proves perfect completeness.

For soundness, fix an arbitrary proof and prover strategy. If the proof has distance greater than $\gamma$ from $\PCPCode^{\otimes m}$, Lemma~\ref{pcp:test} bounds acceptance by $\delta_{\mathrm{ldt}}$. It remains to consider a proof within distance $\gamma$ of the code.

\medskip\noindent{\bf Replacing the proof by a polynomial.}
Before sampling any verifier coins, fix a polynomial $f$ whose codeword is within distance $\gamma$ of the proof. Lemma~\ref{pcp:decoded} bounds $f$ by $\DecodedBound$. Use the same verifier coins and prover strategy to compare the actual execution with one that uses $f$ and $\ConjugatePoly$ for the final input evaluations. Before the first differing evaluation, the transcripts and prover messages coincide. At agreement locations retain the actual evaluation rule. At other locations use an approximation to $f$ within the same error budget. This is an execution used only for analysis, so it requires neither an efficient decoder nor a bounded-width encoding of $f$.

The first approximate-GKR invocation uses at most four proof query points and the second at most two. Each query point is marginally uniform on $G^m$, since its coordinates are verifier challenges and inversion preserves uniformity. Define all these query points even on rejecting paths by sampling the unused coins. Let $\mathcal D=\{r\in G^m:\pi[r]\ne f(r)\}$, so $|\mathcal D|\le\gamma N^m$. If $r^{(1)},\ldots,r^{(6)}$ denote these query points, including unused ones, marginal uniformity gives $\Pr[r^{(i)}\in\mathcal D]=|\mathcal D|/N^m$. The executions can first differ only at such a query point. Therefore
\[
 \Pr[\text{the executions differ}]
 \le\Pr\left[\bigcup_{i=1}^6\{r^{(i)}\in\mathcal D\}\right]
 \le\sum_{i=1}^6\frac{|\mathcal D|}{N^m}\le6\gamma.
\]
No independence between these query points is used. Since the executions have the same acceptance outcome when they agree,
\[
 \Pr[\text{actual execution accepts}]
 \le\Pr[\text{modified execution accepts}]+6\gamma.
\]

\medskip\noindent{\bf Checking the witness and the output.}
In the modified execution, put $z=(f(a))_{a\in\WitnessPositions}$. If its distance from $\bits^{\WitLen}$ exceeds $\eta$, Lemma~\ref{lem:pcp-booleanity} gives $B(f)>\eta^2/4>\Delta_{\mathrm{bool}}$. The Booleanity circuit satisfies its gate bound. Approximate-GKR soundness therefore bounds acceptance by $\eps_{\mathrm{bool}}$.

Otherwise there is a bit vector $w$ with $\|z-w\|_2\le\eta$. Lemma~\ref{lem:pcp-perturbation} gives the numerical gate bounds and
\[
 |y-C(x,{\DecMap}(z))|\ge |y-C(x,{\DecMap}(w))|-\mathcal J(\eta)>\Delta_{\mathrm{out}}.
\]
The last inequality is exactly where the promise gap $\mathcal J(\eta)$ is used. Every unused input determined by $f$ is still bounded by $\DecodedBound$. Conditional on any preceding transcript, the second invocation has fresh coins, so it accepts with probability at most $\eps_{\mathrm{out}}$.

The vector $z$ is fixed before the verifier samples its coins. Thus the two cases contribute $\max\{\eps_{\mathrm{bool}},\eps_{\mathrm{out}}\}$, rather than a sum. Adding the $6\gamma$ probability that the actual proof differs at a queried entry, and including the far-from-code case, gives~\cref{eq:pcp-soundness}.

Each low-degree test reads one symbol. The two approximate-GKR invocations read at most six more, and each symbol occupies $\OracleBits$ bits. The proof string has $N^m$ symbols, giving the stated proof length and bit-query bound. Its contents depend only on the witness and encoding parameters.
\end{proof}

\subsection{Complexity}

Choosing $\gamma=1/100$, $\eps_{\mathrm{bool}}=\eps_{\mathrm{out}}=1/20$, and $r_{\mathrm{ldt}}=\ceil{100m^{29}\ln20}$ gives $s_0\le0.11$. Repeating the interaction $\PCPReps$ times with fresh coins and the same proof gives soundness at most $0.11^{\PCPReps}$, since the bound holds conditional on every preceding transcript. Proof length is unchanged. Communication and queries increase by a factor of $\PCPReps$.

For invocation $j\in\{\mathrm{bool},\mathrm{out}\}$, let $k_j$ be its number of sum-check rounds, $D_j$ the degree bound, $L_j$ its depth, and $b_j$ its working bit length. Communication in one complete interaction, in bits, is
\[
 O\left(r_{\mathrm{ldt}}(h^2\OracleBits+m\log N+\log^2m)
   +\sum_{j\in\{\mathrm{bool},\mathrm{out}\}}\bigl(k_jD_jb_j+(k_j+L_j)\log N\bigr)\right).
\]
The verifier's bit cost is polynomial in these parameters, with evaluations of the public-input extension and of $\AddPoly_i,\MulPoly_i$ charged separately. To generate the proof, the honest prover expands at most ${\WitLen}h^m$ interpolation terms per entry. Its total work is polynomial in the two circuit sizes, $N^m$, and the required bit lengths.

For polynomial proof length, take $m$ to be the least power of two at least $\max\{2,\ceil{\log_h {\WitLen}}\}$. Then $m\le2\max\{2,\log_h {\WitLen}+1\}$, since rounding up to a power of two increases a positive number by at most a factor of two. Hence
\[
 h^m\le h^{2\max\{2,\log_h {\WitLen}+1\}}
 =\max\{h^4,h^2{\WitLen}^2\}.
\]
Choose $h\ge64$ so that both padded circuits have depth and maximum number of coordinates per gate label at most $h$, including the input slice labels. Set $N=h^5$. Since $k_j\le3h^2$ and $D_j=2(h-1)$,
\[
 \frac{4(k_jD_j+L_j)}{\eps_j}\le80(6h^3+h)\le h^5.
\]
This meets the grid bound in~\cref{eq:parameters} and gives $N\ge16mh$. The Fourier grid has size $2h$, which divides $N$. The proof length is $N^m\OracleBits=(h^m)^5\OracleBits=\poly({\WitLen},h)$.
Layering and padding increase size polynomially, so $h$ can be polynomial
in the circuit depths and logarithms of the unpadded sizes and witness
length. With $\PCPReps=O(\log(1/\eps))$, the protocol has soundness $\eps$ and reads
\[
 O(\PCPReps m^{29}\OracleBits)
 =O\!\left(\PCPReps m^{29}h^5(\log({\WitLen}+1)+m\log h)\right)
\]
proof bits. Its communication is polynomial in $h$, $\max_j b_j$, and $\PCPReps$.
The word lengths depend on the magnitude bounds and the output gap as in
\cref{sec:completeness}.

\end{document}